\documentclass[a4paper,12pt,oneside]{memoir}
\usepackage{simpler-wick}
\usepackage{graphicx}
\usepackage[utf8]{inputenc}
\usepackage{combelow}
\usepackage{indentfirst}
\usepackage{braket}
\usepackage{setspace}
\usepackage{amsmath, amsthm, amssymb, amsfonts,bm}
\usepackage[multiple]{footmisc}
\usepackage{mathtools, changepage, slashed}
\usepackage{tikz-feynman}
\usepackage{bm, mathrsfs}
\usepackage{bbold}
\usepackage{gensymb}
\usepackage[a4paper,top=3cm,left=3cm,right=2cm,bottom=2cm]{geometry}
\usepackage{epstopdf}
\usepackage{hyperref}
\usepackage{pgfplots}
\pgfplotsset{compat=1.17} 
\usepackage[sorting=none]{biblatex}
\usepackage{subcaption}
\usepackage[export]{adjustbox}
\numberwithin{equation}{section}
\usepackage[inkscapelatex=false]{svg}
\usepackage{doi}
\usepackage{pdfpages}
\hypersetup{
  colorlinks   = true, %Colours links instead of ugly boxes
  urlcolor     = black, %Colour for external hyperlinks
  linkcolor    = black, %Colour of internal links
  citecolor   = black %Colour of citations
}
\DeclareMathOperator{\Tr}{Tr}

\newcommand{\normalord}[1]{%
  {:\mathrel{\mspace{1mu}#1\mspace{1mu}}:}%
}

\OnehalfSpacing
\usepackage{lmodern}
\counterwithout*{footnote}{chapter}
\usepackage{tcolorbox}

\begin{document}

\pagestyle{empty}

\pagestyle{empty}
\begin{center}

%%%%%%%%%%%%%%%%%%%%%%%%%%%%%%%%%%%%%%%%%%%%%%%%%%%%%%%%%%%%%%%%%%%%%%%%%%%%%%%%%%%%%    
%\TÃ­tulo da Tese \ Thesis title
	{\fontsize{16}{16} \selectfont University of S\~ao Paulo \\}
	\vspace{0.1cm}
	{\fontsize{16}{16} \selectfont Physics Institute}
    \vspace{3.3cm}

	{\fontsize{22}{22}\selectfont 
     A study of string theory and the AdS/CFT correspondence\par}
    \vspace{2cm}

%%%%%%%%%%%%%%%%%%%%%%%%%%%%%%%%%%%%%%%%%%%%%%%%%%%%%%%%%%%%%%%%%%%%%%%%%%%%%%%%%%%%%    
%\Nome do Autor \ Author's name

    {\fontsize{18}{18}\selectfont  Pedro Fernandes Henriques Bairrão\par}

    \vspace{2cm}

\end{center}

%%%%%%%%%%%%%%%%%%%%%%%%%%%%%%%%%%%%%%%%%%%%%%%%%%%%%%%%%%%%%%%%%%%%%%%%%%%%%%%%%%%%%%    
%\Orientador e coorientador (se existir) \ Supervisor and co-supervisor (if there is one)
\leftskip 6cm
\begin{flushright}	
\leftskip 6cm
Supervisor: Prof. Dr. Fernando Tadeu Caldeira Brandt
\leftskip 6cm
%Se nÃ£o houver coorientador, comente a linha abaixo \ If there is no co-supervisor, comment below
%Co-supervisor: Prof. Dr.  \underline{ \hskip 5cm  } 
\end{flushright}	

    \vspace{0.8cm}    

%%%%%%%%%%%%%%%%%%%%%%%%%%%%%%%%%%%%%%%%%%%%%%%%%%%%%%%%%%%%%%%%%%%%%%%%%%%%%%%%%%%%%    
% Grau AcadÃªmico \ Degree

\par
\leftskip 6cm
\noindent {Dissertation submitted to the Physics Institute of the University of São Paulo in partial fulfillment of the requirements for the degree of Master of Science.}
\par
\leftskip 0cm
\vskip 2cm

%%%%%%%%%%%%%%%%%%%%%%%%%%%%%%%%%%%%%%%%%%%%%%%%%%%%%%%%%%%%%%%%%%%%%%%%%%%%%%%%%%%%    
% Banca Examinadora -- Primeiro nome Ã© do presidente ou do presidente da banca \ Examining committee -- The first name must be the supervisor's name or the examination committee president's name.

\noindent Examining Committee: \\
\noindent Prof. Dr. Fernando Tadeu Caldeira Brandt (IF-USP)\\
Prof. Dr.  Nathan Jacob Berkovits (IFT-UNESP)\\
Prof. Dr. Nelson Ricardo de Freitas Braga (IF-UFRJ)\\
\vspace{2.5cm}

%%%%%%%%%%%%%%%%%%%%%%%%%%%%%%%%%%%%%%%%%%%%%%%%%%%%%%%%%%%%%%%%%%%%%%%%%%%%%%%%%%%%    
%Data \ Date
{\centering
    {S\~ao Paulo \\  2025}
\clearpage
}

\newpage
\pagenumbering{gobble}

\vspace*{\fill}
\begin{minipage}[b]{\linewidth}
  \begin{flushright}
    \textit{A Alex, Rafinha e Ana Lu}
  \end{flushright}
\end{minipage}
\vspace*{3cm}
\newpage

\pagestyle{plain}

\chapter*{Agradecimentos}
\markboth{Agradecimentos}{}

A Fernando Brandt, que me orientou desde o início da graduação, pela confiança depositada em mim, e pela liberdade que me foi dada para que pudesse descobrir e seguir meus interesses.

A Nathan Berkovits, com quem tive o privilégio de ter o primeiro contato formal com a teoria de cordas. Sua precisão e rigor conceitual me são grandes inspirações.

A Manoel Robilotta, primeiramente por seu curso \textit{sui generis} de introdução ao eletromagnetismo, onde se consolidou meu amor e fascínio pela física. Em segundo lugar, pelas longas e profundas conversas que tivemos nos momentos mais incertos deste projeto.

Ao Grupo de Hádrons e Física Teórica do IFUSP (GRHAFITE), em particular à Kanchan Khenchandani, pelo acolhimento e pelo convite para apresentar ao grupo parte deste trabalho. Agradeço também ao Centro de Instrumentação e Física de Altas Energias (HEPIC), em particular ao Marcelo Munhoz e ao Lucas Ferrandi, pela abertura com que me receberam em suas reuniões durante alguns meses. Sou também grato aos grupos dos professores Eduardo Casali e Gabriel Menezes, assim com aos demais alunos do curso de teoria de cordas do IFUSP em 2025, pelas produtivas discussões sobre cordas e temas relacionados.

Agradeço a todos os professores que contribuíram para a minha formação. Não poderia deixar de mencionar alguns nomes em específico cujo papel na construção das minhas bases teóricas foi inestimável: Gustavo Burdman, João Barata, Renata Funchal, Raul Abramo, Fernando Garcia e Ana Blak.

Ao Rafael Grossi, pelas estimulantes discussões sobre teoria de cordas, pelos conselhos sobre o mestrado, e pela leitura preliminar de partes da dissertação. 

Ao Sérgio Martins Filho, pelas divertidas conversas sobre física e o meio acadêmico, e pela ajuda na hora de navegar as águas turvas da burocracia institucional. Agradeço também ao Bruno ``Hércules'' Monteiro, pela ajuda com diversas dúvidas que tive em relação a aspectos burocráticos do mestrado.

A Sylvia e Miguel, por todo o apoio em diferentes formas. Em particular, pelas longas conversas que sempre acabávamos tendo quando este trabalho aproximava-se de um ponto de inflexão. 

A Mari, pelo companheirismo nos melhores dias, e pela ternura nos piores. Obrigado por me mostrar, pela via do exemplo, que é possível estabelecer uma relação com os estudos ao mesmo tempo profunda e leve.

A Alex de Lima Barros, Rafael Andrade Pereira e  Ana Luiza Sério, que despertaram em mim o interesse pela física. A vocês dedico esta dissertação.

Ao Éber, Alessandro e Ademir, pela simpatia, prontidão e eficiência com que sempre me auxiliaram. 

Por último, sou imensamente grato a todos os amigos que tornaram a minha vida mais leve ao longo destes anos. Contribuíram com este trabalho mais do que imaginam. 

O presente trabalho foi realizado com apoio da Coordenação de Aperfeiçoamento de Pessoal de Nível Superior - Brasil (CAPES) - Código de Financiamento 001.

\newpage

%\chapter*{Disclaimer}

This version of the dissertation has undergone minor changes and corrections for the purposes of uploading it to arXiv, which did not change its content meaningfully. The original version can be found at \url{https://www.teses.usp.br/teses/disponiveis/43/43134/tde-11092025-210722/pt-br.php}. Corrections, suggestions and comments are welcome at p.bairrao@unesp.br.

\newpage

\chapter*{Abstract}
This dissertation consists of a comprehensive and pedagogical review of Maldacena's original derivation of the AdS/CFT correspondence and the main topics of string theory necessary to understand it. The large $N$ expansion of Yang-Mills theory is presented as the main motivation for seeking a string-theoretic language for gauge theories. The bosonic string is studied in conformal gauge, with an emphasis on the spectrum and the low energy effective actions for both closed and open strings. The properties of D-branes are studied via T-duality, and closed-open string duality is checked by explicitly computing the interaction amplitude between two D-branes. The type II superstring theories are studied in the RNS formalism and their massless spectrum is shown to match that of the type II ten-dimensional supergravity theories. Supersymmetric branes are discussed, including their realization as supergravity solitons, and the original derivation of the AdS/CFT duality is presented. Some of the properties of the correspondence are discussed, including the idea of the holographic dictionary.

%\bigskip

%\textbf{Keywords:} string theory; holography; quantum gravity; duality; gauge theories

%\newpage
%\chapter*{Resumo}
%Esta dissertação consiste em uma revisão abrangente e pedagógica da derivação original da correspondência AdS/CFT por Maldacena, assim como dos principais tópicos da teoria de cordas necessários para entendê-la. A expansão de grande $N$ da teoria de Yang-Mills é apresentada como principal motivação para a busca de uma linguagem de teoria de cordas para teorias de calibre. A corda bosônica é estudada no calibre conforme, com ênfase no espectro e nas ações efetivas de baixas energias para cordas fechadas e abertas. As propriedades de D-branas são estudadas via dualidade T, e a dualidade de cordas abertas e fechadas é verificada calculando explicitamente a amplitude de interação entre duas D-branas. As teorias de supercordas do tipo II são estudadas no formalismo RNS e é verificado que o seu espectro não massivo coincide com o das teorias de supergravidade do tipo II em dez dimensões. Branas supersimétricas são discutidas, incluindo a sua realização como sólitons da supergravidade, e a derivação original da dualidade AdS/CFT é apresentada. Algumas das propriedades da correspondência são discutidas, incluindo a ideia do dicionário holográfico.

%\bigskip

%\textbf{Palavras-chave:} teoria de cordas; holografia; gravidade quântica; dualidade; teorias de calibre

\newpage

\listoffigures*

\newpage

\listoftables*

\newpage

\tableofcontents*
\newpage

\pagenumbering{arabic}

\setcounter{page}{10}
%\documentclass[a4paper,12pt]{memoir}
%\usepackage{graphicx}
%\usepackage[utf8]{inputenc}
%\usepackage{indentfirst}
%\usepackage{braket}
%\usepackage{setspace}
%\usepackage{amsmath, amsthm, amssymb, amsfonts,bm}
%\usepackage[multiple]{footmisc}
%\usepackage{mathtools, changepage, slashed}
%\usepackage{tikz-feynman}
%\usepackage{bm, mathrsfs}
%\usepackage{gensymb}
%\usepackage[a4paper,top=3cm,left=3cm,right=2cm,bottom=2cm]{geometry}
%\usepackage{epstopdf}
%\usepackage{hyperref}
%\usepackage{pgfplots}
%\pgfplotsset{compat=1.18} 
%\usepackage[sorting=none]{biblatex}
%\addbibresource{refs.bib}
%\numberwithin{equation}{section}
%\usepackage[inkscapelatex=false]{svg}
%\usepackage[super]{natbib}
%\usepackage{doi}
%\hypersetup{
%  colorlinks   = true, %Colours links instead of ugly boxes
%  urlcolor     = black, %Colour for external hyperlinks
%  linkcolor    = black, %Colour of internal links
%  citecolor   = black %Colour of citations
%}
%\DeclareMathOperator{\Tr}{Tr}

%\newcommand{\normalord}[1]{%
%  {:\mathrel{\mspace{1mu}#1\mspace{1mu}}:}%
%}

\OnehalfSpacing
%\usepackage{newtx}
%\usepackage{newtxtext}
%\usepackage{lmodern}

%\title{T-duality chapter}
%\author{pedrobairrao}

%\begin{document}

%\section{Introduction}
\chapter{Introduction}\label{ch1}

I believe it is safe to say that the AdS/CFT correspondence is one of the most surprising concepts that a student of physics interested both in particle physics and gravity can hear about. The main idea is perhaps best illustrated by the opening remark of the lecture notes by Polchinski and Horowitz \cite{Horowitz:2006ct}: \textit{Hidden within every non-Abelian gauge theory, even within the weak and strong nuclear interactions, is a theory of quantum gravity}. The name AdS/CFT comes from the fact that, in the first discovered instance of such a gauge/gravity correspondence, by Maldacena in \cite{Maldacena:1997re}, the gauge theory was also a Conformal Field Theory, and the gravitational theory was defined on an Anti-de Sitter background. The correspondence, which lacks a rigorous proof despite having passed every quantitative test it has ever been subjected to, states that there is a full quantum mechanical duality between the five-dimensional gravitational theory on AdS, a solution of Einstein's equations with constant negative cosmological constant, and the four-dimensional gauge theory, taken to be defined at the asymptotic region of this spacetime. Because it relates theories on spacetimes of different dimensions, AdS/CFT is said to be a holographic duality, in analogy with a hologram that manages to encode three-dimensional information on a two-dimensional surface. The word ``duality'' here means that the two theories should be understood as different ways to describe the same underlying physical system, whose degrees of freedom organize themselves into those of a gravitational theory in some region of parameter space, and into those of a lower-dimensional gauge theory in another. The regime in which the gravitational description becomes weakly coupled, and therefore treatable with standard perturbative methods, is precisely that in which the gauge theory becomes strongly coupled.

Having undergone over 25 years of intense research since its discovery, the original AdS/CFT duality has been extended, generalized, and reinterpreted to such an extent that there are now different ways to understand it. With the benefit of hindsight, we can say that a lot of what makes AdS/CFT work is already contained on the geometric properties of AdS space, and it would not have been impossible for some form of the correspondence to have been discovered via a careful study of how to properly define a quantum gravity theory on such a space. However, this was not the case. What actually happened was that the duality was derived from within a powerful theoretical framework capable of encompassing both general relativity and gauge theories as specific limits, and of providing a conceptual bridge between the two. That framework is string theory. Not only does string theory provide the closest thing to a constructive derivation of AdS/CFT, it is also present on one side of the correspondence, since the quantum gravity theory defined over AdS is a string theory. For this reason the name gauge/string duality is also common. A more detailed statement of the original correspondence is

\begin{equation*}
    \fbox{\begin{tabular}{@{}c@{}}
        Type IIB superstring theory \\ on an AdS$_5 \times S^5$ background 
  \end{tabular}}
  =
  \fbox{\begin{tabular}{@{}c@{}}
        $\mathcal{N}=4$ super Yang-Mills theory in \\ 
        four-dimensional Minkowski space
  \end{tabular}}
\end{equation*}

\vspace{0.3cm}

Type IIB string theory is a particular supersymmetric string theory, AdS$_5 \times S^5$ is five-dimensional anti-de Sitter space times a five-sphere, and $\mathcal{N}=4$ super Yang-Mills is a supersymmetric version of Yang-Mills theory which has just the right matter content to be exactly conformally-invariant at the quantum level.

This dissertation is a review of the main topics in string theory related to AdS/CFT, and of the original derivation by Maldacena. The text is written with a reader of advanced undergraduate or graduate level in mind, who is familiar with general relativity and quantum field theory, but not string theory, conformal field theory or supersymmetry. While string theory does differ from quantum field theory in a number of ways, the large difference between the two commonly felt by the beginner is in many cases more due to language than content. String theory should, after all, reduce to field theory at low enough energies, and many field theory notions have a direct analogue in string theory. The goal of this text is to investigate the conceptual and technical features of string theory useful for gaining some intuition for AdS/CFT, while maintaining as much contact as possible with the methods of traditional field theory, in particular those of perturbative nonabelian gauge theories, at the level of the average graduate course. Following this parallel at times means taking some detours from what would be the most direct route from the basic notions of strings to holography. For example, the bosonic string is treated in the Lorentz-invariant conformal gauge, with Faddeev-Popov ghosts, instead of the manifestly unitary lightcone gauge. A special emphasis is given to the stringy phenomenon of open-closed duality, which, while not strictly necessary for discussing AdS/CFT, illustrates the fundamental properties of string theory that lead to it, and can be verified quantitatively already in the bosonic string, where the analysis is far simpler than in the supersymmetric theory. The general philosophy employed is that properties of string theory that are present for both bosonic and superstrings are worked out in detail only for the former, which is used as a kind of toy model. The discussion of the superstring, which is the actual setting of AdS/CFT, focuses on ideas that have no counterpart in the bosonic theory.

While the idea of gauge/gravity duality has been used to study a wide variety of physical systems (see for instance \cite{Năstase_2015}), the application that will serve as the main phenomenological motivation for its development will be the strong interactions, whose history is actually deeply intertwined with that of string theory. In Chapter \ref{ch2} some of these historical ties are discussed, as well as the large $N$ expansion of gauge theories, where a first hint of a connection between them and closed string theory can be seen.

Chapter \ref{ch3} is a general introduction to the bosonic string and its quantization in conformal gauge, using a combination of canonical and path integral methods. The critical dimension and normal ordering constant are extracted by demanding the absence of gauge anomalies, and the spectrum of the theory is derived via the method of ``old covariant quantization''. 

The next three chapters are dedicated to topics that play a role in AdS/CFT, but can still be examined in the context of the bosonic string.

Chapter \ref{ch4} consists of a discussion of T-duality, an important feature of string theory that emerges when one considers strings propagating not on Minkowski space, but on spacetimes that have some coordinates compactified into a circle. This duality is a consequence of strings being able to wrap around these compact directions, and it provides a way to investigate the properties of D-branes, objects that play a central role in holography.

Chapter \ref{ch5} focuses on the low-energy limit of string theory, in which the physics becomes well described by quantum field theory. The two sides of the correspondence, quantum gravity and gauge theories, emerge when such a limit is taken for systems of closed strings, or open strings and D-branes, respectively.

Chapter \ref{ch6} is a discussion of open-closed string duality, also called worldsheet duality, a property of string theory that allows for processes involving D-branes, that naturally have a description in terms of open strings, to be also described in terms of closed strings. This serves as a first indication that it is possible for closed and open strings to serve as different descriptions of the same physical system. AdS/CFT is in a sense a particular realization of this idea. The actual process studied is an interaction of two parallel D-branes, which can be understood as being either a consequence of a one-loop open string process or a tree-level closed string one. The amplitudes corresponding to both interpretations are computed and seen to match. The closed string description also involves a construction of the D-branes as closed string coherent states, setting the stage for the later identification of supersymmetric D-branes with supergravity solitons, which are closed superstring coherent states. This identification is the starting point for Maldacena's derivation.

Chapter \ref{ch7} is concerned with superstring theory, which is studied in the RNS formalism. Only the type II theories are discussed, since they are the most relevant ones for holography. Their massless spectrum is derived and seen to reproduce that of the type II ten-dimensional supergravity theories. Supersymmetric D-branes are also discussed, including their realization in terms of classical supergravity solutions. Lastly, Maldacena's original derivation of AdS/CFT is reviewed, followed by a discussion of how the parameters and observables on both sides of the correspondence are related.

Natural units $\hbar = c = 1$ are used throughout this work, and the Minkowski metric is written as $\eta = \text{diag}(-1,+1,\dots,+1)$ in any number of dimensions. The Einstein summation convention for repeated indices holds unless explicitly stated.

The figures were made using the free software Inkscape.

%\end{document}

%\documentclass[a4paper,12pt]{memoir}
%\usepackage{graphicx}
%\usepackage[utf8]{inputenc}
%\usepackage{indentfirst}
%\usepackage{braket}
%\usepackage{setspace}
%\usepackage{amsmath, amsthm, amssymb, amsfonts,bm}
%\usepackage[multiple]{footmisc}
%\usepackage{mathtools, changepage, slashed}
%\usepackage{tikz-feynman}
%\usepackage{bm, mathrsfs}
%\usepackage{gensymb}
%\usepackage[a4paper,top=3cm,left=3cm,right=2cm,bottom=2cm]{geometry}
%\usepackage{epstopdf}
%\usepackage{simpler-wick}
%\usepackage{hyperref}
%\usepackage{pgfplots}
%\pgfplotsset{compat=1.18} 
%\usepackage[sorting=none]{biblatex}
%\addbibresource{refs.bib}
%\numberwithin{equation}{section}
%\usepackage[inkscapelatex=false]{svg}
%\usepackage[super]{natbib}
%\usepackage{doi}
%\hypersetup{
%  colorlinks   = true, %Colours links instead of ugly boxes
%  urlcolor     = black, %Colour for external hyperlinks
%  linkcolor    = black, %Colour of internal links
%  citecolor   = black %Colour of citations
%}
%\DeclareMathOperator{\Tr}{Tr}

%\newcommand{\normalord}[1]{%
%  {:\mathrel{\mspace{1mu}#1\mspace{1mu}}:}%
%}

%\OnehalfSpacing
%\usepackage{newtx}
%\usepackage{newtxtext}
%\usepackage{lmodern}

%\title{History and large $N$ chapter}
%\author{pedrobairrao}

%\begin{document}

\chapter{Strings and the strong force}\label{ch2}

\section{The shared history of string theory and QCD}\label{sec21}

Before becoming a theory of quantum gravity and grand unification, string theory was a model of the strong nuclear force. The paper that is often said to mark the origin of string theory, which back then went by the name of ``dual resonance models'', is Veneziano's \textit{Construction of a Crossing-Symmetric, Regge-Behaved Amplitude for Linearly Rising Trajectories} \cite{Veneziano:1968yb}, published in 1968. The main result is the so-called Veneziano amplitude, a scattering amplitude for mesons that satisfied a number of experimentally seen properties of hadronic processes that no quantum field-theoretic model had been able to accommodate.

Veneziano's paper was considered a triumph of the S-matrix theory approach to the strong interactions, a research program that, roughly speaking, advocated for the inadequacy of quantum field theory for the description of the strong force, and elevated the S-matrix itself to the position of fundamental importance.\footnote{A more detailed account of the history of string theory and its relation to S-matrix theory can be found in \cite{rickles2014brief}, which serves as the central reference for this section. For a pedagogical introduction to S-matrix methods for the strong interactions and related topics, see \cite{Collins_1977} } One of the empirical facts that this amplitude successfully captures is Regge trajectories, linear relations seen when one plots the square of the masses of hadrons against their spins (with all other quantum numbers held fixed)\footnote{The quantities $\alpha_0$ and $\alpha^\prime$ are called the Regge intercept and the Regge slope. They continue to play a major role in modern string theory, where $\alpha^\prime$ is reinterpreted as the inverse of the string tension (which relates it to the Planck length), and $\alpha_0$ becomes a normal ordering constant. }
\begin{equation}
    \text{spin}(m^2) = \alpha(m^2) = \alpha_0 + \alpha^\prime m^2.
\end{equation}
The poles of the Veneziano amplitude, where one finds the masses of the exchanged particles in the scattering process, satisfy this relation.

\begin{figure}
\begin{center}
\includegraphics[width=0.7\columnwidth]{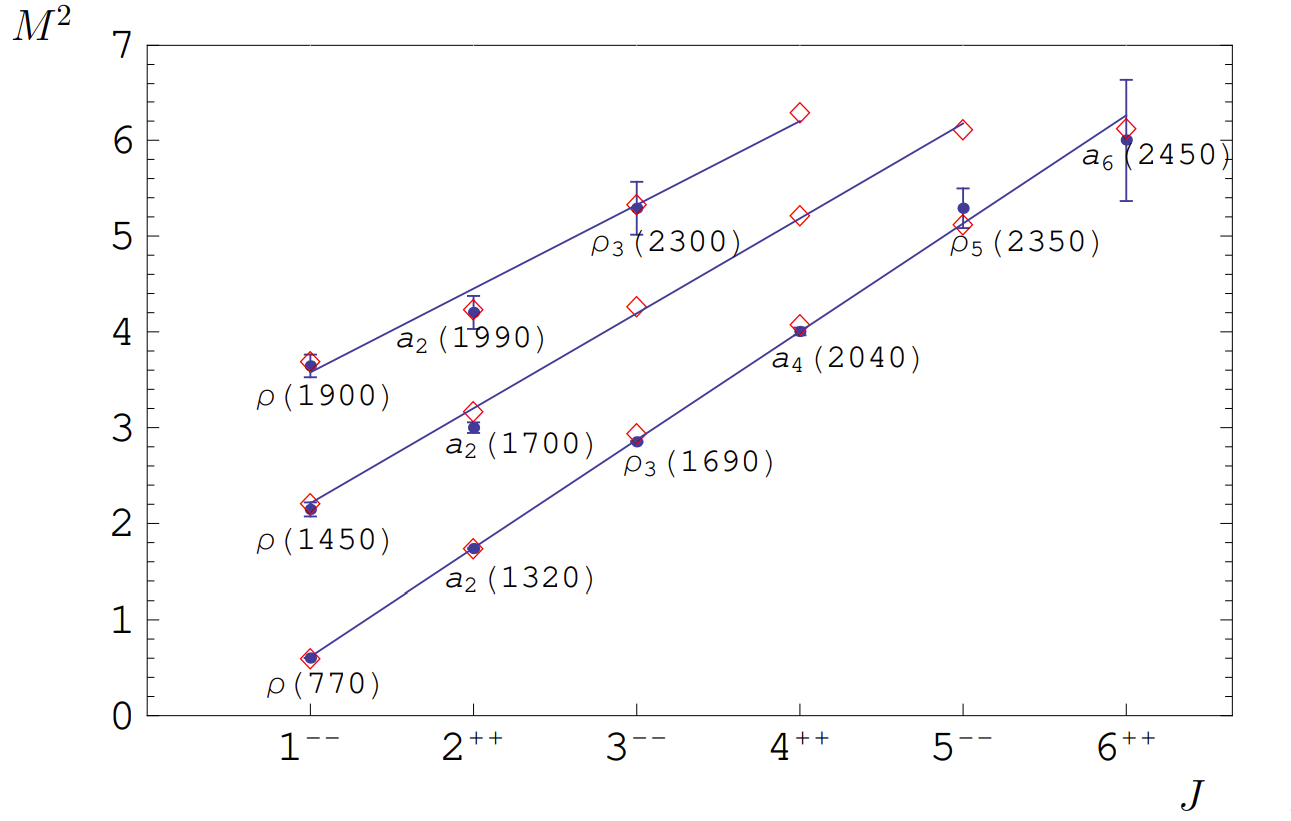}
\caption[Plot of Regge trajectories]{Plot of Regge trajectories, taken from \cite{Ebert_2009}.}
\label{Regge plot}
\end{center}
\end{figure}

If the mesons are treated as the asymptotic excitations of some quantum field theory, there is no explanation of Regge trajectories, essentially by design, as the masses and spins of each particle must be specified by hand in order to set up the theory in the first place. Such a degree of correlation between usually unrelated quantum numbers points to a structure of mesons that differs from that of elementary particles. Soon after the discovery of the Veneziano amplitude it was recognized independently by Nambu \cite{doi:10.1142/9789812795823_0024}, Susskind \cite{PhysRevLett.23.545} and Nielsen \cite{Nielsenpreprint} that it could be derived by modeling the mesons as one-dimensional extended objects, or strings, with a quark attached to one endpoint and an antiquark in the other. This provides a simple explanation of Regge trajectories, as can already be seen in the following simple classical calculation.\footnote{We follow \cite{Bali:2000gf} for this example.} Picture a rigid relativistic ``string'' of length $2l$ rotating with constant angular velocity $\omega = v(r)/r$ around its center, and with constant energy per unit length $T$, which is called the string's tension. We assume that the endpoints travel at the speed of light, so $v(l) = 1$ and therefore $v(r) = r/l$.

\begin{figure}[h]
\begin{center}
\includegraphics[width=0.35\columnwidth]{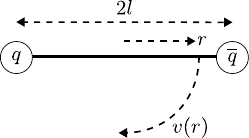}
\caption{Simplified rigid string model of a meson.}
\end{center}
\end{figure}

The total energy stored in this rotating string is 
\begin{equation}
    E = \int_{-l}^l dr \, T \gamma(v) = T \int_{-l}^l \frac{dr}{\sqrt{1-r^2/l^2}} = \pi l T.
\end{equation}
The momentum density along the string is $p = T \gamma(v) v$, so the total angular momentum carried by it is
\begin{equation}
    L = \int_{-l}^l dr \, r \, T v \, \gamma(v) = \frac{T}{l} \int_{-l}^l \frac{dr \, r^2}{\sqrt{1 - r^2 /l^2}} = \frac{\pi T l^2}{2} = \frac{E^2}{2 \pi T}.
\end{equation}
If this string is taken to be some microscopic object whose extended nature cannot be accurately inferred by experiment, it will naturally be classified as a particle, whose total energy and angular momentum will be associated with a mass $m$ and an intrinsic spin $S$. Comparing the results just calculated for $E \equiv m$ and $L \equiv S$ gives $S(m^2) = \alpha^\prime m^2$, for
\begin{equation}
    \alpha^\prime = \frac{1}{2 \pi T}.
\end{equation}
Quantization would then lead to discrete values for the angular momentum, and therefore mass squared, which explains the points in figure \ref{Regge plot}, as well as a constant zero-point energy contribution to $E^2=m^2$ which is associated to $\alpha_0$. The details will be worked out in Chapter \ref{ch2}. Note that in this picture each meson lying on the same Regge trajectory is interpreted not as an independent particle, but as a particular excited state of an open string.

Despite solving this and other puzzles of the strong interactions, these early string models suffered from a number of difficulties, such as the apparent lack of quantum consistency of the theory in $D \neq 26$ or $D \neq 10$ spacetime dimensions, depending on the particular model used. They also lacked some of the conceptual clarity by then enjoyed by more traditional quantum field theories. Eventually, the phenomenon of asymptotic freedom was understood and quantum chromodynamics (QCD) came along, quickly becoming the mainstream theory of the strong interactions \cite{GrossWilczek1973,Politzer1973}. QCD is an $SU(3)$ Yang-Mills theory minimally coupled to fermions that transform in the fundamental representation of the gauge group:
\begin{equation}
    S_\text{QCD} = \int d^4x \Big( - \frac{1}{4 g^2_{\text{YM}}} \text{Tr} ( F^{\mu \nu} F_{\mu \nu} ) -i \bar{\psi}_a (  \gamma^\mu D_\mu + m_a ) \psi_a  \Big), 
\end{equation}
where the gluon field $A_\mu$ takes values in the adjoint representation of SU(3); the generators of SU(3) are normalized as
%
%gluon field $A_\mu$ takes values in the adjoint %representation of $SU(3)$, whose generators we %normalize as 
$\text{Tr}(T^a T^b) = \delta^{a b}$, $F_{\mu \nu} = \partial_\mu A_\nu - \partial_\nu A_\mu -i[A_\mu , A_\nu] $ and $D_\mu \psi = \partial_\mu \psi - i A_\mu$ is the fundamental covariant derivative.\footnote{Another common convention for this action has no factor of $1/g^2_{\text{YM}}$ in front of the trace and explicit factors of $g$ on $F_{\mu \nu}$ and the covariant derivative. They are related by taking $A_\mu \to g A_\mu$. } The index $a$ on the fermions is a flavor index, meaning that it runs over the different species of quarks (up, down, strange, charm, bottom and top). In spite of being a field theory, QCD offers a natural explanation of the partial phenomenological success of the string models. This is because at low energies the theory's effective coupling becomes large, which is expected to produce quark confinement. In this nonperturbative regime the gluon field produced by a pair of separated quarks has been verified in lattice simulations to assume the form of a narrow flux tube between them, inducing an effective potential similar to the one produced by an open string. \cite{Baker:2018mhw}\footnote{See \cite{Greensite:2011zz} for a pedagogical introduction to confinement.} This allows the string models to be interpreted as an effective description of these QCD strings, hence their phenomenological utility. 

\begin{figure}[t]
\begin{center}
\includegraphics[width=0.5\columnwidth]{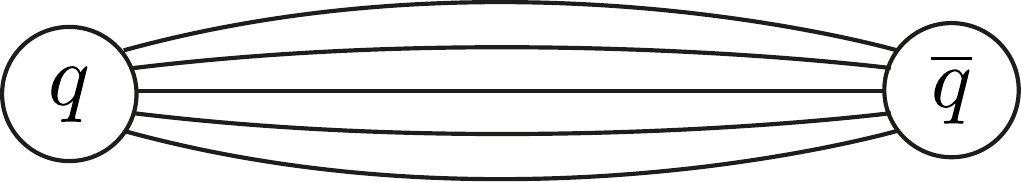}
\caption[Gluon field flux tube]{Gluon field flux tube between a quark and an antiquark, forming a shape similar to a stretched open string between them.}
\end{center}
\end{figure}

Around the same time of QCD's rise to hegemony, it was realized that the mathematical structure of the string models allowed them to be reinterpreted as quantum gravity theories \cite{Scherk:1974ca}. One of the main reasons for this is that the theory also contained massless spin two particles that behaved as gravitons, which could be associated with propagating closed strings, just like mesons were associated with open strings.\footnote{In this new interpretation of the theory it was more natural to associate the open string massless vector particles to gauge bosons, instead of mesons.} The ten-dimensional model in particular, which involves supersymmetry, also had the potential for a grand unified description of all forces.

\begin{figure}[h]
     \centering
     \begin{subfigure}[h]{0.2\textwidth}
         \centering
         \includegraphics[width=\textwidth]{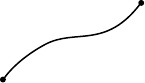}
         \caption{}
     \end{subfigure}
     \hspace{3cm}
     \begin{subfigure}[h]{0.2\textwidth}
         \centering
         \includegraphics[width=\textwidth]{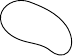}
         \caption{}
     \end{subfigure}
     \caption[Open and closed string]{An open string (a) has two endpoints, represented as dots. A closed string (b) closes in on itself. String theory includes both, but the meson interpretation only holds for the former.}
\end{figure}

This new way of looking at the theory would then lead to the modern string theory, or superstring theory, which became along the years progressively more detached from its strong interaction origins. This changed drastically in 1997, with the publishing of Maldacena's paper on AdS/CFT. The connection between string theory and the strong interactions hinted at by this work did not involve modeling any QCD degrees of freedom as open strings, like in the early string models, but instead suggested an identification with a theory of closed strings. Remarkably, that there should be some connection between gauge theories and closed string theories had been anticipated by 't Hooft via the so-called large $N$ expansion, to which we turn to now.

\section{Large $N$ Yang-Mills}\label{sec22}

The fact that the effective coupling of QCD is small in the ultraviolet (UV) and large in the infrared (IR) is one of the theory's most important properties. This in fact holds in $SU(N)$ nonabelian gauge theories for any $N>1$. The inclusion of the six fundamental quarks of QCD does not change this, so for simplicity we will ignore them and study only the pure Yang-Mills sector
\begin{equation}
    S_\text{YM} = - \frac{1}{4 g_\text{YM}^2} \int d^4x \, \text{Tr}( F^{\mu \nu} F_{\mu \nu}).
\end{equation}
A one-loop renormalization of the theory leads to the running coupling \cite{Tong_gauge}\footnote{This section is based on Tong's lecture notes on gauge theories \cite{Tong_gauge} and Guillermo Silva's Physics Latam mini-course on matrix theory \cite{Guillelectures}. Many of the figures are copies of the ones in \cite{Tong_gauge}. A more detailed, but still introductory treatment can be found in \cite{Coleman_1985}.}
\begin{equation}
    \frac{1}{g^2_{\text{YM}}(\mu)} = \frac{1}{g^2_{\text{YM}}} - \frac{11}{(4 \pi)^2} \log \left( \frac{\Lambda^2_\text{UV}}{\mu^2} \right),
\end{equation}
where $g_\text{YM} = g_\text{YM}(\Lambda_\text{UV})$, with  $\Lambda_\text{UV}$ a UV energy cutoff, and $\mu$ a renormalization scale. The energy scale at which nonperturbative effects dominate can be estimated by the value of $\mu$ for which $g_{\text{YM}}(\mu)$ diverges. This is at $\mu = \Lambda_\text{QCD}$, where\footnote{It is customary to call $\Lambda_\text{QCD}$ the ``QCD scale'' even in pure Yang-Mills theory. }
\begin{equation}
    \Lambda_\text{QCD} = \Lambda_\text{UV} \exp \left( - \frac{1}{22} \frac{(4 \pi)^2}{ g^2_{\text{YM}}} \right).
\end{equation}
Differentiating this expression with respect to $\Lambda_\text{UV}$ reveals that $\Lambda_\text{QCD}$ is independent of the cutoff, as it should be, since Yang-Mills theory is renormalizable in four dimensions. Being the only dimensional parameter in the theory, $\Lambda_\text{QCD}$ is expected to set the mass scale of the particles of the spectrum, which for pure Yang-Mills are the glueballs (massive bound states of gluons).\footnote{In full QCD one would also have the hadrons, of course.} Extracting information from the theory at energies below $\Lambda_\text{QCD}$, which for actual QCD is around the $150$ - $200$ MeV range, is a challenge, because the coupling is expected to become large and perturbation theory ceases to be reliable. One analytical approach to try to circumvent this problem is 't Hooft's large $N$ expansion. The main idea is to consider an $SU(N)$ gauge group instead of $SU(3)$ and try to use $1/N$ as an additional expansion parameter. This of course relies on the observables possessing an expansion in powers of $N$, which will be argued to be the case from a couple of examples. 

The gluon propagator in an $SU(N)$ gauge theory is
\begin{equation}
    \braket{A^i_{\mu j}(x) A^k_{\nu l}(y)} = g^2_{\text{YM}} \Delta_{\mu \nu}(x-y) \Big( \delta^i_l \delta^k_j - \frac{1}{N} \delta^i_j \delta^k_l \Big),
\end{equation}
where the indices $i,j,k,l=1, \dots , N$ go over the fundamental (upstairs) or anti-fundamental (downstairs) representation of $SU(N)$, and $\Delta_{\mu \nu}(x-y)$ is the photon propagator in some chosen gauge. The second term inside the parenthesis guarantees that $\text{Tr} A_\mu =0$, so that all $SU(N)$ group elements $g \sim e^{iA}$ have unit determinant. For simplicity we will throw this second term away, which means working with a $U(N)$ gauge theory. In an expansion in $1/N$ this term is subleading, at its contributions can be worked out as corrections after the general structure of the series is revealed. The analysis of the $N$-dependence of Feynman diagrams is facilitated by using the double line notation:
\begin{equation}
    \braket{A^i_{\mu j}(x) A^k_{\nu l}(y)} \longrightarrow \hspace{0.3cm} 
        \includegraphics[scale=1,valign=c]{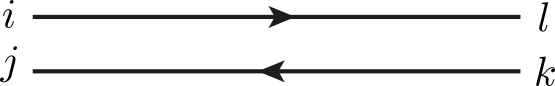}.
\end{equation}
The lines indicate group index contraction with respect to $\delta^i_l \delta^k_j$, and always point from an index in the fundamental to one in the anti-fundamental. The three gluon vertex becomes

\begin{figure}[h]
\begin{center}
\includegraphics[width= 0.3\textwidth]{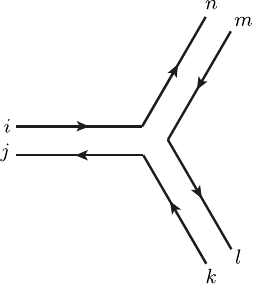}
\caption[Three gluon vertex in the double line notation.]{Three gluon vertex in the double line notation.}
\label{double line 3pt function}
\end{center}
\end{figure}
and brings to each diagram a factor of $1/g^2_{\text{YM}}$. The same goes for the four gluon vertex. Let us then examine the correlation function
\begin{equation}
    \braket{\text{Tr}A^4} \equiv \left\langle  A^i_{\mu j}(x) A^j_{\nu k}(x) A^k_{\rho l}(x) A^l_{\sigma i}(x) \right\rangle
\end{equation}
to first order in regular perturbation theory, meaning that we take the expectation value with respect to the quadratic action. Wick's theorem leads to
    \begin{align}
         \left\langle \Tr A^4 \right\rangle &= \left\langle  A^i_{\phantom{a} j} A^j_{\phantom{a} k} A^k_{\phantom{a} l} A^l_{\phantom{a} i} \right\rangle \notag\\[5pt]
         &= \wick{
                 \c1 A^i_{\phantom{a} j} \c1 A^j_{\phantom{a} k} }
                 \wick{\c1 A^k_{\phantom{a} l} \c1 A^l_{\phantom{a} i}
                 }
            +
            \wick{
                \c1 A^i_{\phantom{a} j} \c2 A^j_{\phantom{a} k} \c1 A^k_{\phantom{a} l} \c2 A^l_{\phantom{a} i}
                }
            +
            \wick{
                \c2 A^i_{\phantom{a} j} \c1 A^j_{\phantom{a} k} \c1 A^k_{\phantom{a} l} \c2 A^l_{\phantom{a} i}
                }
                \notag\\[5pt]
                &=2 \left\langle A^i_{\phantom{a} j} A^j_{\phantom{a} k} \right\rangle \left\langle A^k_{\phantom{a} l} A^l_{\phantom{a} i} \right\rangle + \left\langle A^i_{\phantom{a} j} A^k_{\phantom{a} l} \right\rangle \left\langle A^j_{\phantom{a} k} A^l_{\phantom{a} i} \right\rangle.
\end{align}
The $N$-dependence is only related to the index contractions, so we ignore both spacetime coordinates and Lorentz indices by setting $\left\langle A^i_{\phantom{a} j}A^k_{\phantom{a} l} \right\rangle \sim g^2_{\text{YM}} \delta^i_l\delta^k_j$ for the propagator. The result is
\begin{align}
        \left\langle \text{Tr} A^4 \right\rangle &=2 \left\langle A^i_{\phantom{a} j} A^j_{\phantom{a} k} \right\rangle \left\langle A^k_{\phantom{a} l} A^l_{\phantom{a} i} \right\rangle + \left\langle A^i_{\phantom{a} j} A^k_{\phantom{a} l} \right\rangle \left\langle A^j_{\phantom{a} k} A^l_{\phantom{a} i} \right\rangle \notag\\[5pt]
        &\sim 2g^4 \delta^i_k \delta^j_j \delta^k_i \delta^l_l + g^4 \delta^i_l \delta^k_j \delta^j_i \delta^l_k \notag\\[5pt]
        &= 2g^4 N^3 + g^4 N.
        \label{trace4}
    \end{align}
The first term contains three factors of $\delta^i_i=N$, which makes it dominant at large $N$ with respect to the second one, which only has one such trace. This computation is represented diagrammatically as

\begin{figure}[h]
\begin{center}
\includegraphics[width= 0.8\textwidth]{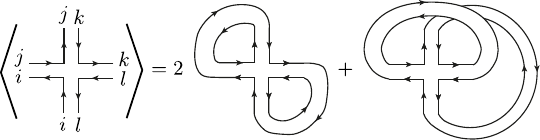}
\caption[Diagrammatic version of (2.2.8)]{Diagrammatic version of \eqref{trace4}.}
\end{center}
\end{figure}

The first diagram, which corresponds to the $O(N^3)$ contribution, is planar, meaning that it can de drawn on a plane with no self-intersections. The second one is not, since it is necessary do draw one propagator going under the other in order to achieve the correct index contractions. The double line notation makes this easy to see because every closed line, or index loop, corresponds to one factor of $\delta^i_i = N$. The first diagram above has three different index loops while the second only has one, even though they both are ``two-loop'' diagrams in the usual sense, which is that of momentum loops. 

All diagrams that contribute to some observable can be classified in powers of $N$ in this way, with the leading contribution being always planar, as this maximizes the number of index loops. It is therefore sensible to try to simplify the theory by keeping only the planar diagrams, which corresponds to the so-called large $N$ limit. For $N=3$ this would mean throwing away terms of order $1/3 \sim 0.3$ with respect to the planar ones. Finite $N$ corrections can be computed up to the desired precision by including non-planar diagrams. The large $N$ series therefore corresponds to a useful reorganization of the gauge theory's Feynman diagrams into two distinct expansions, one in terms of the coupling and one in terms of $N$. The leading term in $N$, formally given by setting $N \to \infty$ but keeping the coupling finite, is expected to be a better approximation to the nonperturbative physics than regular perturbative expansion in the coupling. 

Naively setting $N \to \infty$ in Yang-Mills does not quite work. One way to see why is via the formula for the QCD scale, which at arbitrary $N$ becomes
\begin{equation}
    \Lambda_\text{QCD} = \Lambda_\text{UV} \exp \left( - \frac{3}{22} \frac{(4 \pi)^2}{ g^2_{\text{YM}}N} \right).
\end{equation}
In the $N \to \infty$ limit we would have $\Lambda_\text{QCD} = \Lambda_\text{UV}$, spoiling the cutoff-independence of physical predictions. A better behaved limit is achieved by trading $g_\text{YM}$ for the 't Hooft coupling
\begin{equation}
    \lambda = g^2_{\text{YM}}N,
\end{equation}
and then taking $N \to \infty$ at fixed $\lambda$. This preserves the value of $\Lambda_\text{QCD}$. The action becomes
\begin{equation}
    S_\text{YM} = - \frac{N}{4 \lambda} \int d^4x \, \text{Tr}( F^{\mu \nu} F_{\mu \nu}),
\end{equation}
from which one sees that the general diagram scales as 
\begin{equation}
        \text{diagram} \sim \left( \frac{\lambda}{N} \right)^\text{number of propagators} \left( \frac{N}{\lambda} \right)^\text{number of vertices} N^\text{number of index loops}.
        \label{diagram scaling}
    \end{equation}
Consider for example the vacuum diagrams
\begin{align}
    \includegraphics[scale=0.37,valign=c]{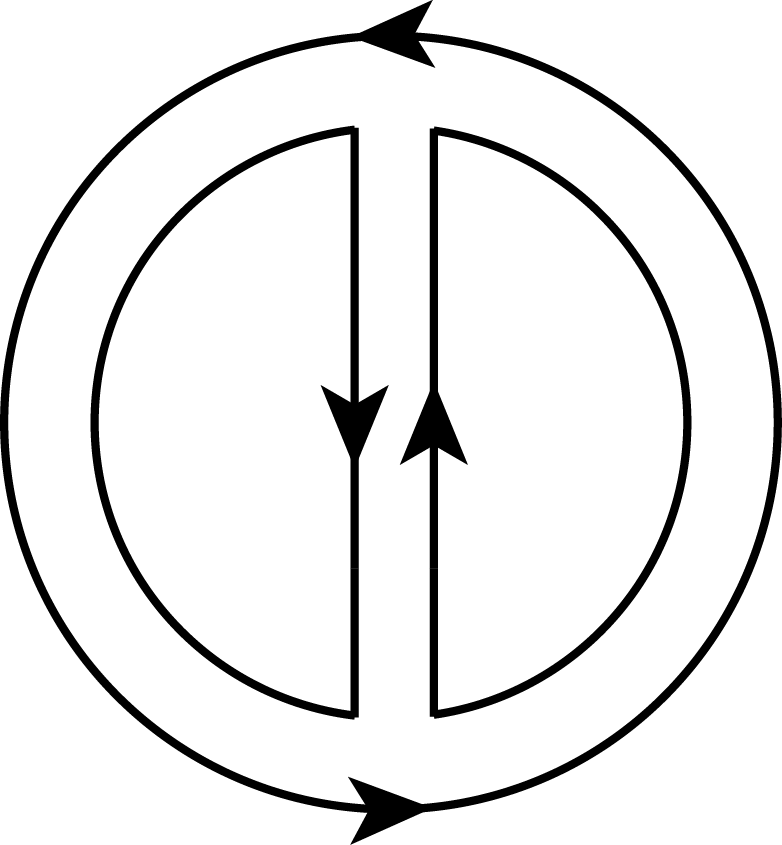}  &\sim \left( \frac{\lambda}{N} \right)^3  \left( \frac{N}{\lambda} \right)^2 N^3 = \lambda N^2 \label{bubble1} \\[0.5cm]
    \includegraphics[scale=0.37,valign=c]{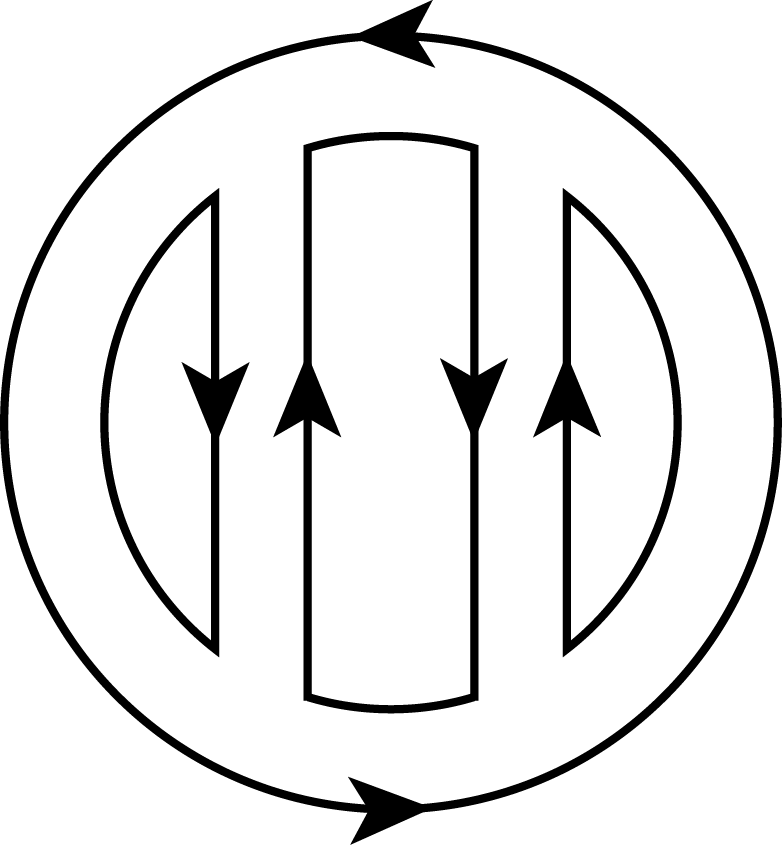}  &\sim \left( \frac{\lambda}{N} \right)^6  \left( \frac{N}{\lambda} \right)^4 N^4 = \lambda^2 N^2 \label{bubble2} \\[0.5cm]
    \includegraphics[scale=0.37,valign=c]{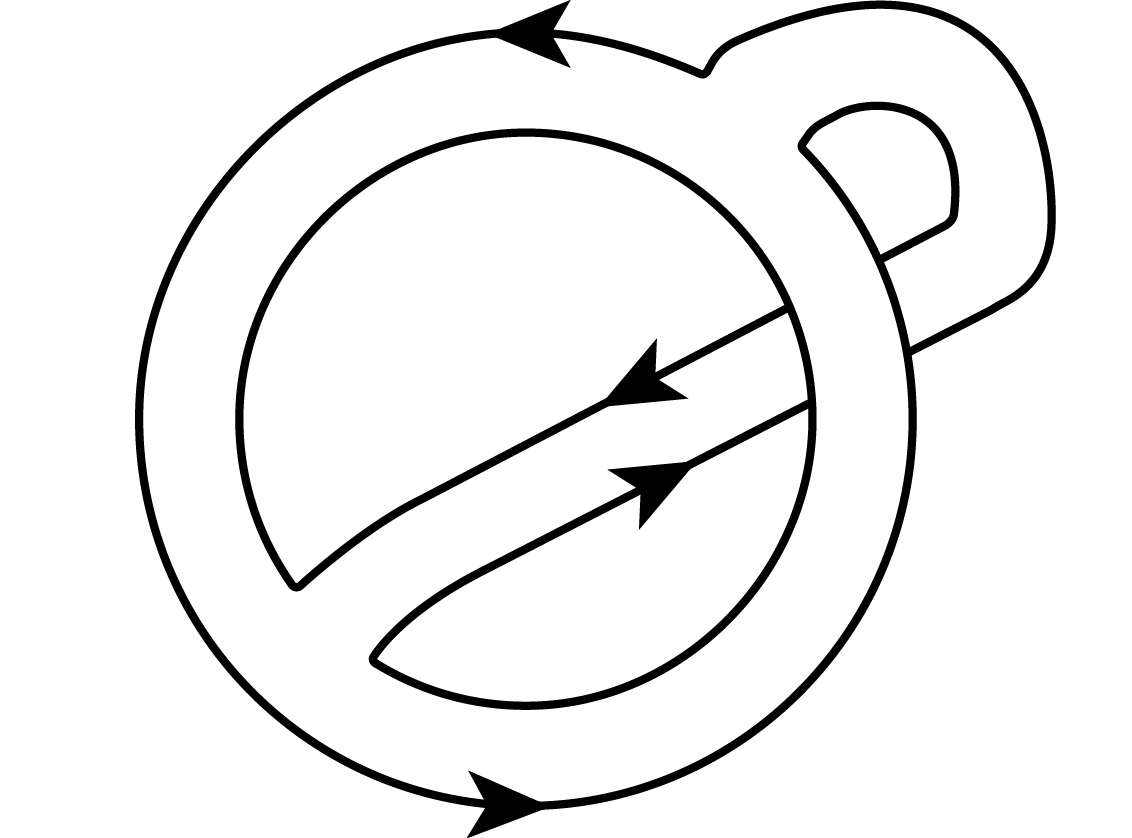}  &\sim \left( \frac{\lambda}{N} \right)^3  \left( \frac{N}{\lambda} \right)^2 N = \lambda
\end{align}
The first two are both planar and therefore come with a $N^2$, but the second one has double the amount of vertices as the first, so they differ by a factor of $\lambda \sim g^2_{\text{YM}}$. The third one has only two vertices so it has the same power of $\lambda$ as the first one, but is is non-planar and thus has a lower power of $N$. The classification of diagrams in terms of planarity which governs the large $N$ expansion has a topological interpretation. All vacuum diagrams shown can be drawn over closed two-dimensional surfaces.

\begin{figure}[h]
\begin{center}
\includegraphics[width= 0.5\textwidth]{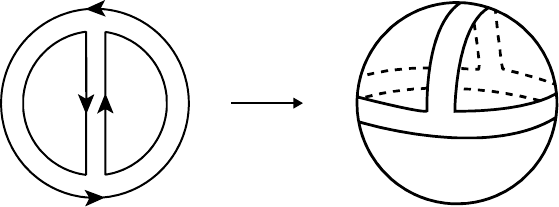}
\caption[Drawing of a planar vacuum diagram over a sphere]{Drawing of a planar vacuum diagram over a sphere. The outermost index loop becomes the sphere's lower hemisphere.}
\label{diagram_to_sphere}
\end{center}
\end{figure}

The resulting object is a closed surface with a ``skeleton'' of vertices and edges, which can be viewed as surrounding the faces of a ``curved polyhedron''. The number of vertices $V$ of this polyhedron is the number of vertices of the original diagram, the number of edges $E$ is the number of propagators, the number of faces $F$ is the number of index loops, including the outer one which closes on the lower hemisphere of the sphere in figure \ref{diagram_to_sphere}. This turns \eqref{diagram scaling} into
\begin{equation}
    \text{diagram} \sim \left( \frac{\lambda}{N} \right)^E \left( \frac{N}{\lambda} \right)^V N^F =N^\chi \lambda^{E-V},
\end{equation}
where 
\begin{equation}
    \chi = F + V - E
\end{equation}
is the polyhedron's Euler number, or Euler characteristic. The Euler number of any closed surface can also be written as
\begin{equation}
    \chi = 2 - 2h,
\end{equation}
where $h$ is the genus, or the number of holes. This is a topological invariant, a quantity whose value does not change under any continuous transformation of the polyhedron that does not alter its topology, which in this case means the genus \cite{Staessens:2010vi}. The $N^2$ dependence of the first two vacuum diagrams \eqref{bubble1} and \eqref{bubble2} reflects the fact that the sphere has no holes, so its Euler number is $\chi=2$. The third diagram cannot be drawn on a sphere, but it can be drawn over a torus

\begin{figure}[h]
\begin{center}
\includegraphics[width= 0.6\textwidth]{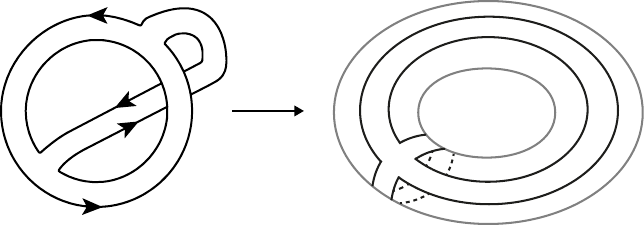}
\caption[Drawing of a nonplanar diagram over a torus]{Drawing of a nonplanar vacuum diagram with Euler number $\chi =0$ over a torus.}
\end{center}
\end{figure}

A torus has one hole, and therefore $\chi =0$. Correspondingly, the third diagram is proportional to $N^0$. Going to higher orders in the $N$ expansion means considering more negative Euler numbers, or surfaces with more holes.\footnote{This topological organization of the large $N$ expansion holds for all diagrams, not just vacuum ones. See \cite{Coleman_1985}}

If the gauge theory's partition function with no sources $Z$ is given by the sum of all vacuum diagrams, its organization into a large $N$ series is a topological expansion, in the sense that each term can be associated with a closed surface of increasing genus. 
\begin{equation}
        Z= \includegraphics[scale=0.4,valign=c]{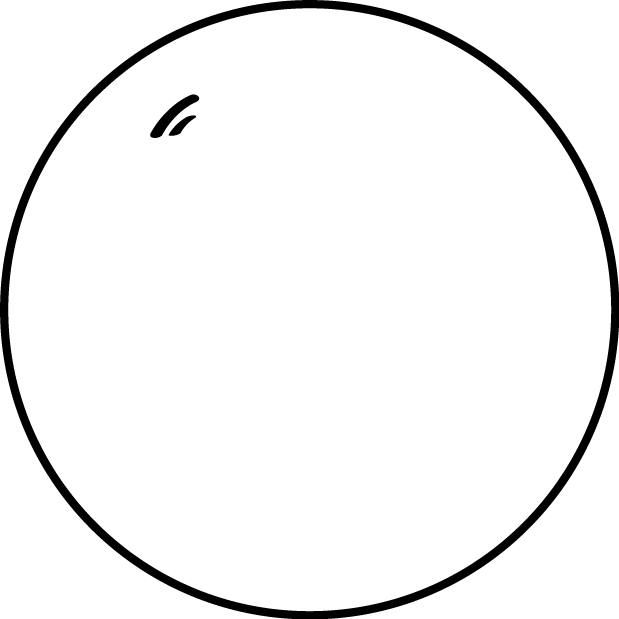} + \includegraphics[scale=0.4,valign=c]{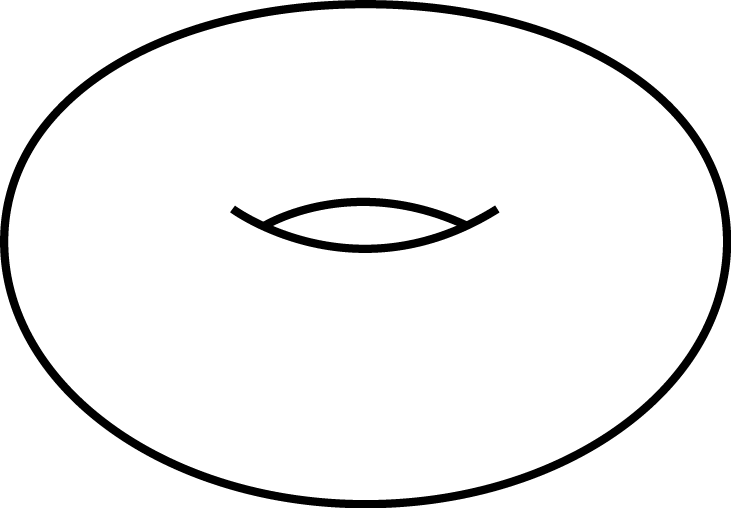} + \includegraphics[scale=0.4,valign=c]{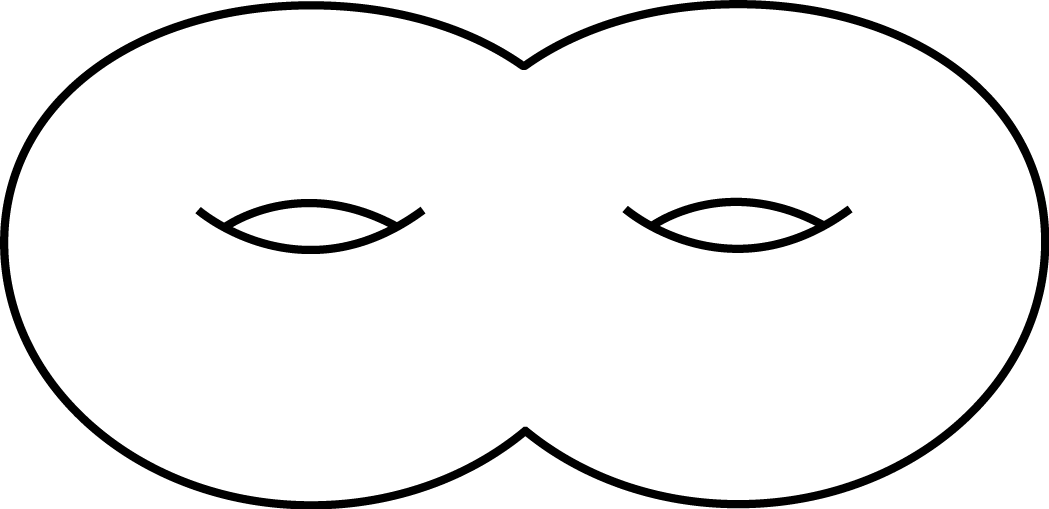} + \dots
        \label{topological series}
    \end{equation}

As will become clear in the next chapter, this is is precisely the form taken by the partition function of a theory of weakly interacting closed strings, expanded perturbatively. Since we know that the degrees of freedom in perturbative Yang-Mills (quarks and gluons) are very different from those of closed strings (gravitons, among other particles), this stringy description of Yang-Mills theory, if valid, should describe the theory in its strongly coupled regime. AdS/CFT provides a precise realization of this heuristic reasoning. 

A notable aspect of the large $N$ limit is that it is, in a sense, a classical limit. The overall factor of $N$ in front of the action makes it so that for $N \to \infty$ the path integral becomes increasingly localized around some minimum in field space, just like the $1/\hbar$ factor in front of the action localizes the path integral around the classical field configurations in the $\hbar \to 0$ limit. This too will become precise in AdS/CFT.

%\end{document}

%\documentclass[a4paper,12pt]{memoir}
%\usepackage{graphicx}
%\usepackage[utf8]{inputenc}
%\usepackage{indentfirst}
%\usepackage{braket}
%\usepackage{setspace}
%\usepackage{amsmath, amsthm, amssymb, amsfonts,bm}
%\usepackage{mathtools, changepage, slashed}
%\usepackage[multiple]{footmisc}
%\usepackage{tikz-feynman}
%\usepackage{bm, mathrsfs}
%\usepackage{gensymb}
%\usepackage[a4paper,top=3cm,left=3cm,right=2cm,bottom=2cm]{geometry}
%\usepackage{epstopdf}
%\usepackage{hyperref}
%\usepackage{pgfplots}
%\pgfplotsset{compat=1.18} 
%\usepackage[sorting=none]{biblatex}
%\addbibresource{refs.bib}
%\numberwithin{equation}{section}
%\usepackage[inkscapelatex=false]{svg}
%\svgpath{{../images/}}
%\usepackage[super]{natbib}
%\usepackage{doi}
%\hypersetup{
%  colorlinks   = true, %Colours links instead of ugly boxes
%  urlcolor     = black, %Colour for external hyperlinks
%  linkcolor    = black, %Colour of internal links
%  citecolor   = black %Colour of citations
%}
%\DeclareMathOperator{\Tr}{Tr}

%\newcommand{\normalord}[1]{%
%  {:\mathrel{\mspace{1mu}#1\mspace{1mu}}:}%
%}

%\OnehalfSpacing
%\usepackage{newtx}
%\usepackage{newtxtext}
%\usepackage{lmodern}

%\title{Bosonic string theory}
%\author{pedrobairrao}

%\begin{document}

\chapter{Bosonic strings}\label{ch3}

\section{The relativistic particle}\label{sec31}

Even though the AdS/CFT correspondence is part of superstring theory, most of the ideas required to formulate it are already present in the simpler bosonic string theory, which we therefore examine first.

A natural starting point for the formulation of this theory is to review the dynamics of a free relativistic particle, i.e. a zero-dimensional object that propagates freely through Minkowski space. The generalization to a one-dimensional object, the string, is then straightforward. We parametrize the spacetime trajectory of the particle by a function $x^\mu(\tau)$ that associates to each value of the particle's proper time $\tau$ a position along its worldline. For the later discussion of string theory it is convenient to leave the spacetime dimension $D$ arbitrary. Knowing that the equations of motion satisfied by such a free particle are
\begin{equation}
    \frac{d^2 x^\mu}{d \tau^2} = 0,
    \label{eqmo}
\end{equation}
we seek the simplest action that leads to them as the Euler-Lagrange equations. We also ask that it be invariant under spacetime Poincaré transformations 
\begin{equation}
    x^\mu(\tau) \to \Lambda^\mu_{\phantom{a} \nu} x^\nu(\tau) + a^\mu \, , \hspace{1cm} \Lambda^\mu_{\phantom{a} \nu} \in \text{SO}(1,D-1),
    \label{transfs}
\end{equation}
and worldline reparametrizations $\tau \to \tau^\prime(\tau)$. That action is 
\begin{equation}
    S_{\text{part}} = -m \int \sqrt{ - \eta_{\mu \nu} dx^\mu dx^\nu } = -m \int d \tau \sqrt{- \eta_{\mu \nu} \frac{dx^\mu}{d \tau} \frac{dx^\nu}{d \tau} }.
    \label{Spart}
\end{equation}
Since $\sqrt{ - \eta_{\mu \nu} dx^\mu dx^\nu }$ is the distance element along the worldline, $S_{\text{part}}$ measures the invariant length of the worldline times $-m$, so minimizing $S_{\text{part}}$ corresponds to finding the shortest path between the initial and final positions, a geodesic.

The global symmetry of $S_{\text{part}}$ with respect to Poincaré transformations leads to the conservation of its generators: angular momentum (in its relativistic version) and the energy-momentum tensor associated to the particle's motion. The worldline reparametrizations are local transformations and we therefore expect them to correspond to gauge symmetries. The infinitesimal version of $\tau \to \tau^\prime(\tau)$ is the worldline diffeomorphism
\begin{equation}
    \tau \to \tau + \xi(\tau),
\end{equation}
in terms of which $x^\mu(\tau)$ transforms as a scalar. The associated conserved quantity is the Hamiltonian
\begin{equation}
    H = \frac{d x^\mu}{d\tau} p_\mu -L= 0\, , \hspace{0.5cm} L= -m  \sqrt{- \eta_{\mu \nu} \frac{dx^\mu}{d \tau} \frac{dx^\nu}{d \tau} },
\end{equation}
where $p_\mu = m \dot{x}_\mu / \sqrt{-\dot{x}^2}$ are the canonical momenta conjugate to $x^\mu$.\footnote{We will often use the notation $\frac{d f}{d \tau} \equiv \dot{f}$.} From the perspective of the quantum theory, the vanishing of the Hamiltonian means that the operator that implements the proper time reparametrizations on states is the identity operator, confirming that they are indeed gauge transformations. The presence of this one-parameter gauge invariance signals that one of the $D$ degrees of freedom $x^\mu(\tau)$ is nonphysical. This is to be expected, since the momenta can be verified to satisfy
\begin{equation}
    p^2 +m^2 = 0,
\end{equation}
the momentum space equivalent of the Klein-Gordon equation, which can be used to solve for one of the $x^\mu$ in terms of the others. Note that this does not emerge as an equation of motion, but as a constraint due to a gauge symmetry, which is therefore valid off-shell.

This gauge symmetry allows us to fix $x^0(\tau)=\tau$, effectively making coordinate time $x^0 \equiv t$ the worldline parameter. One then finds in the nonrelativistic limit
\begin{equation}
    S_{\text{part}} = -m \int d t \sqrt{1- v^2} = \int  d t \left( -m + \frac{1}{2} m v^2 \right) + \mathcal{O}\left( v^4 \right),
\end{equation}
where $v^i=dx^i/d t$. The constant $m$ is therefore the particle's mass. 
An important generalization arises when the particle moves in a curved spacetime
%
%An important generalization is the case in which %the particle moves in a curved spacetime 
%
of metric $G_{\mu \nu}$ and possesses charge $q$ with respect to a $U(1)$ gauge field:
\begin{equation}
    S_{\text{part}} \to -m \int \sqrt{ - G_{\mu \nu}(x) dx^\mu dx^\nu} + q \int A_\mu(x) dx^\mu.
\end{equation}
 The equations of motion become
\begin{equation}
        \frac{d^2 x^\mu}{d \tau^2} + \Gamma^\mu_{\rho \sigma} \frac{d x^\rho}{d \tau} \frac{d x^\sigma}{d \tau} =  \frac{q}{m} F^{\mu}_{\phantom{a}\nu} \frac{d x^\nu}{d \tau}.
\end{equation}
where the $\Gamma^\mu_{\rho \sigma}$ are the Christoffel symbols of $G_{\mu \nu}$ and $F_{\mu \nu}$ is the field strength tensor of $A_\mu$:
\begin{align}
    \Gamma^\mu_{\rho \sigma} &= \frac{1}{2}G^{\mu \lambda} \left( \partial_\rho G_{\sigma \lambda} + \partial_\sigma G_{\rho \lambda} - \partial_\lambda G_{\rho \sigma} \right) \notag\\[5pt]
    F_{\mu \nu} &= \partial_\mu A_\nu - \partial_\nu A_\mu. 
\end{align}

By introducing an auxiliary field $g_{\tau \tau}(\tau)$, which plays the role of a metric for the worldline, one may construct the action
\begin{equation}
    S^\prime_\text{part} = - \frac{1}{2} \int d \tau \sqrt{-g_{\tau \tau}} \left( g^{\tau \tau}  \frac{dx^\mu}{d \tau} \frac{dx^\nu}{d \tau} \eta_{\mu \nu} +m^2 \right),
    \label{Sprime}
\end{equation}
where $g^{\tau \tau} = 1/g_{\tau \tau}$. The equation of motion from varying $g_{\tau \tau}$,
\begin{equation}
    \frac{dx^\mu}{d \tau} \frac{dx^\nu}{d \tau} \eta_{\mu \nu} - g_{\tau \tau} m^2 =0,
    \label{aovariar}
\end{equation}
shows that the field $g_{\tau \tau}$, being completely determined by $x^\mu$, is not a new degree of freedom. $S_\text{part}$ and $S^\prime_\text{part}$ are classically equivalent, as $S_\text{part}$ is obtained by inserting the solution \eqref{aovariar} for the metric into $S^\prime_\text{part}$. The coupling of this alternative action to nontrivial gravitational and $U(1)$ backgrounds is again given by adding the $qA_\mu dx^\mu$ term and substituting  $\eta_{\mu \nu} \to G_{\mu \nu}(x)$.

\section{Classical bosonic strings}\label{sec32}
    
A string of finite size is a one-dimensional object, and therefore sweeps out a two-dimensional surface as it moves through spacetime, a worldsheet. The worldsheet of a string is parameterized by a timelike coordinate $\tau$ and a spacelike one $\sigma \in [0,l]$, and the functions $X^\mu(\tau,\sigma)$, which play the role analogous to that of $x^\mu(\tau)$ for the particle, describe how the string is embedded into spacetime. The simplest action for $X^\mu$ that is invariant under spacetime Poincaré transformations and worldsheet reparametrizations is the Nambu-Goto action
\begin{equation}
    S_\text{NG} = -T \int d \tau d\sigma \sqrt{-h}, \hspace{1cm} h \equiv \det(h_{ab})
\end{equation}
where $T$ is a constant with energy dimension $[T]=2$ and
\begin{equation}
    h_{ab} = \partial_a X^\mu \partial_b X_\mu
\end{equation}
is the pullback of the ambient flat metric $\eta_{\mu \nu}$ to the worldsheet, giving the induced two-dimensional metric on it. The indices $a$,$b$ run over $(\tau,\sigma)$. The worldsheet diffeomorphism invariance of this action is manifest, since it is given by $-T$ times the total area of the Euclidean worldsheet: taking $\sigma^2=i\tau$ and $\sigma \equiv \sigma^1$, the integrand becomes 
\begin{equation}
    dA = \sqrt{g}  \, d^2\sigma,
\end{equation}
where $g_{ab}$ is the Wick-rotated metric. In this sense the Nambu-Goto action is the natural geometric generalization of the particle action. That this is a sensible action for the string can also be understood from the effective field theory perspective, since $\sqrt{-h}$ is the Lagrangian with the smallest number of derivatives that is invariant under the desired symmetries, and is therefore the most relevant interaction at low energies allowed for the theory. In this spirit, one may ask what would the next allowed terms be in the effective field theory expansion. It turns out that the next nontrivial possibility is $\sqrt{- h} R$, where $R$ is the worldsheet Ricci scalar \cite{effectivestrings}. This adds to the action an Einstein-Hilbert term, which is topological in two dimensions, and therefore does not contribute to the dynamics. This follows from the fact that in two dimensions the Riemann tensor has only one independent component. Taking into account its symmetries, this forces it to have the form
\begin{equation}
    R_{abcd} = \frac{R}{2} \Big( g_{ac}g_{bd} - g_{ad}g_{bc} \Big),
\end{equation}
which implies $R_{ab}=\frac{1}{2}g_{ab}R$ and therefore $G_{ab}=R_{ab}-\frac{1}{2}g_{ab}R=0$. The variation of the Einstein-Hilbert action in any dimension is proportional to $G_{ab}$, but since this vanishes automatically in two dimensions, the action is invariant under any local variation of the metric. Nonetheless, the fact that such a term is sensitive to the topology of the worldsheet will be relevant in string perturbation theory. Terms with an even higher number of derivatives are not included in conventional bosonic string theory. They do, however, appear in effective string theories, which model solitonic string-like objects such as the QCD flux tubes discussed in the previous chapter, or vortices in condensed matter systems \cite{effectivestrings}.

The two-parameter gauge symmetry of reparametrizations of both $\tau$ and $\sigma$ means that only $D-2$ of the $X^\mu(\tau,\sigma)$ degrees of freedom are physical. These correspond to the transverse oscillations of the string. The choice $X^0(\tau,\sigma)= R \tau \equiv t$, where $R$ is an arbitrary parameter with dimensions of length, is a partial gauge-fixing that leads to a simple geometric interpretation of the remaining functions $X^i(t,\sigma)$, $i=1,\dots,D$. For each fixed time $t$, varying $\sigma$ from 0 to $l$ in $X^i(t,\sigma)$ moves us along the curve in spacetime that corresponds to the shape of the string at that instant. It should be emphasized, however, that this is a gauge-dependent picture. The only physical information about $(\tau,\sigma)$ that holds in general is that these quantities parameterize the worldsheet associated to the motion of the string through spacetime. Still in this gauge, assuming the string to be static, so that $\partial_t X^i=0$, one finds
\begin{equation}
    S \to -T \int^{\infty}_{-\infty} dt \int_0^l d \sigma \sqrt{\left( \frac{\partial X^i}{\partial \sigma} \right)^2} = -T \int^{\infty}_{-\infty} dt L_s,
\end{equation}
where $L_s$ is the string's length. For vanishing kinetic energy the action reduces to minus the time integral of the potential energy, which is given therefore by $T L_s$. This shows that the constant $T$ has the interpretation of tension, potential energy divided by the length. The fact that the energy of the string grows linearly with its length was an important factor in favor of the early string models of the strong forces, in addition to the Regge trajectories. Potentials that grow linearly with the separation between two quarks are a common feature of phenomenological models of confinement, and are also seen in lattice simulations \cite{Greensite:2011zz}. 

As in the case of the particle, the inclusion of an auxiliary metric for the worldsheet allows the construction of a simpler action. For the string, that is the Polyakov action
\begin{equation}
    S_\text{P}= - \frac{1}{4 \pi \alpha^\prime} \int d \tau d \sigma \sqrt{-g} \, g^{a b} \partial_a X^\mu \partial_b X_\mu.
\end{equation}
The equation of motion from varying with respect to the metric is $T_{ab}=0$, where
\begin{equation}
    T_{ab} = \frac{4 \pi}{\sqrt{- g}} \frac{\delta S_\text{P}}{\delta g^{ab}} = - \frac{1}{\alpha^\prime} \bigg( \partial_a X^\mu \partial_b X_\mu - \frac{1}{2} g_{ab} \partial_c X^\mu \partial^c X_\mu \bigg)
    \label{T}
\end{equation}
is the worldsheet energy-momentum tensor. Solving this for $g_{ab}$ and inserting the result into $S_\text{P}$ leads to the Nambu-Goto action.

Aside from the usual Poincaré and diffeomorphism symmetries, the Polyakov action is also invariant under Weyl transformations
\begin{equation}
    g_{ab} \to e^{2 \omega(\tau,\sigma)} g_{ab},
\end{equation}
where $\omega(\tau,\sigma)$ is an arbitrary function. Plugging the infinitesimal version $\delta_\omega g^{ab} = - 2 \omega g^{ab}$ into \eqref{T}, one finds that
\begin{equation}
    \delta_\omega S_\text{P} = - \frac{\sqrt{-\gamma}}{2 \pi} T_{ab} \omega g^{ab} = - \frac{\sqrt{-\gamma}}{2 \pi} T^a_{\phantom{a} a} \omega = 0.
\end{equation}
Since $\omega$ is arbitrary, this shows that Weyl invariance leads to 
\begin{equation}
    T^a_{\phantom{a} a} = 0,
\end{equation}
as an off-shell constraint.

The variation of $S_\text{P}$ with respect to $X^\mu$ is given by
\begin{align}
    \delta_X S_\text{P} = \frac{1}{2 \pi \alpha^\prime} \int d\tau d \sigma \sqrt{-g} \, \delta X^\mu \nabla^2 X_\mu &- \frac{1}{2 \pi \alpha^\prime} \int_0^l d \sigma \sqrt{-g} \, \delta X^\mu \partial^\tau X_\mu \Big\rvert_{\tau = - \infty}^{\tau = \infty} \notag\\[5pt]
    &-\frac{1}{2 \pi \alpha^\prime}  \int_{- \infty}^{\infty} d \tau \sqrt{-g} \, \delta X^\mu \partial^\sigma X_\mu \Big\rvert_{\sigma = 0}^{\sigma = l}.
\end{align}
The first term gives the equation of motion
\begin{equation}
    \nabla^2 X^\mu=0.
\end{equation}
Assuming that the variation is zero at $\tau = \pm \infty$, the second term vanishes. There is more than one way to make the third term vanish. One possibility is having periodic boundary conditions
\begin{align}
X^\mu(\tau,0) = X^\mu(\tau,l)\,, && \partial^\sigma X^\mu(\tau,0) = \partial^\sigma X^\mu(\tau,l) \, ,&& g_{ab}(\tau,0) = g_{ab}(\tau,l).
\end{align}
Since $\sigma = 0,l$ corresponds to the endpoints of the string, periodicity in $\sigma$ means that the two endpoints are glued together, forming a closed string. The worldsheet of such a closed string has the topology of a cylinder.
\begin{figure}[h]
\begin{center}
\includegraphics[width= 0.5\textwidth]{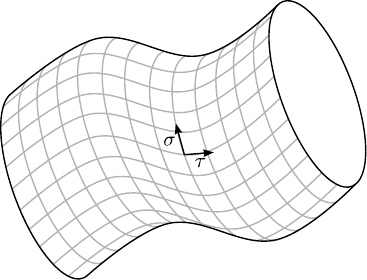}
\caption[Closed string worldsheet]{Closed string propagating from left to right. The corresponding worldsheet is a cylinder-like surface.}
\end{center}
\end{figure}

Another option is to have Neumann boundary conditions at the endpoints:
\begin{equation}
    \partial^\sigma X^\mu(\tau,0) = \partial^\sigma X^\mu(\tau,\sigma) = 0.
\end{equation}
Since there is no periodicity, this describes an open string with free endpoints, whose worldsheet has the two timelike curves $X^\mu(\tau,0)$ and $X^\mu(\tau,l)$, the trajectory of the endpoints, as boundaries. If the boundaries are not straight lines with respect to the coordinates chosen, a more covariant way to indicate Neumann boundary conditions is to say that $n^a \partial_a X^\mu=0$ at the boundaries, whose normal vector is $n^a$.
\begin{figure}[h]
\begin{center}
\includegraphics[width= 0.5\textwidth]{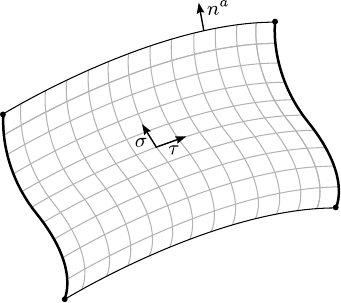}
\caption[Open string worldsheet]{Open string propagating from left to right. The corresponding worldsheet is an open, sheet-like surface.}
\end{center}
\end{figure}

The third possibility is to have $\delta X^\mu=0$ at the endpoints, so that each one is held fixed at some point in space, possibly a different point for each one. This corresponds to Dirichlet boundary conditions. It is not obvious that such seemingly artificial conditions should be allowed, since they violate spacetime Poincaré symmetry. However, it will become clear that they have a natural interpretation in terms of D-branes, higher-dimensional dynamical objects where an open string can end, whose existence and nature are not evident in the equations of motion, but which are fundamental for AdS/CFT. In general, an open string can have Neumann conditions for some directions and Dirichlet for others, except for $\mu=0$, since that would mean that the endpoints are fixed in time. One may also consider mixed boundary conditions, where one endpoint is fixed while another is free, but we will not discuss them (see for instance \cite{Blumenhagen:2013fgp}). 
\begin{figure}[h]
\begin{center}
\includegraphics[width= 0.3\textwidth]{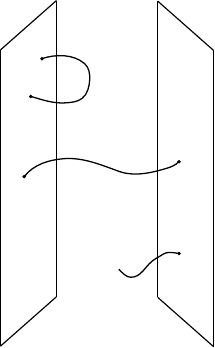}
\caption[D-branes with strings attached]{Two D-branes, represented as two-dimensional planes, with three open strings attached. One has both endpoints in the same brane, one has one endpoint in each, and one has one endpoint fixed and the other one free.}
\label{D-brane figure}
\end{center}
\end{figure}

Just as in the Nambu-Goto case, we add to the Polyakov action a gravitational term $\sqrt{-g} R$, with $R$ now being build out of the auxiliary metric. However, the Ricci scalar under a Weyl transformation is invariant only up to a total derivative of $\omega$. Describing open strings requires worldsheets with boundaries, in which case the the action will pick up a contribution from these boundary terms, breaking Weyl invariance. The correct quantity to add is
\begin{equation}
    \raisebox{2pt}{$\chi$}(W) = \frac{1}{4 \pi} \int d\tau d\sigma \sqrt{-g} R + \frac{1}{2 \pi} \int ds K,
    \label{chi}
\end{equation}
where the second integral is over the boundaries, and $K$ is the trace of the extrinsic curvature on them \cite{Johnson_2002}.
This is the so-called Gibbons-York-Hawking term. It vanishes when there is no boundary, and cancels the Weyl-dependence of the first integral when there is one. This is the standard modification of the Einstein-Hilbert action for manifolds with boundaries, in which case it can be shown that the additional term is necessary for the equations of motion to be the Einstein equations \cite{York:1986lje}. Here it appears by demanding Weyl invariance of the string action, and guarantees that $\raisebox{2pt}{$\chi$} $ is topological also in the open string case.

The Polyakov action allows for a variety of different gauges to be fixed. Note that while $g_{ab}$ couples to the $X^\mu$ fields like a two-dimensional metric, it has three independent degrees of freedom. Nothing requires it to satisfy $g^a_{\phantom{a} a}=2$. Using the two parameters from worldsheet diffeomorphisms  together with the one from Weyl transformations, we are able to fix these three metric components to whatever functions we desire, at least locally.\footnote{This is discussed in more detail in Chapter \ref{ch6}.} The choice
\begin{equation}
    g_{ab} = \eta_{ab},
\end{equation}
is called conformal gauge.\footnote{The more general choice $g_{ab} = e^{2 f(\tau, \sigma)} \eta_{ab}$ is also commonly called conformal gauge in the string theory literature. It leads to the same form of the action as the $g_{ab} = \eta_{ab}$ case (the function $f(\tau,\sigma)$ drops out due to Weyl invariance), which will suffice for our purposes.} The Polyakov action in conformal gauge becomes
\begin{equation}
    S_\text{P}= - \frac{1}{4 \pi \alpha^\prime} \int d \tau d \sigma \eta^{a b} \partial_a X^\mu \partial_b X_\mu.
    \label{PolyC}
\end{equation}
Since the metric was eliminated in the process of gauge-fixing, its equation of motion $T_{ab}=0$ must be imposed as a constraint on the dynamics \cite{dirac2013lectures}. The Polyakov action in conformal gauge therefore describes $D$ free massless scalars $X^\mu$, subjected to the so-called Virasoro constraint
\begin{equation}
    \partial_a X^\mu \partial_b X_\mu - \frac{1}{2} \eta_{ab} \eta^{cd} \partial_c X^\mu \partial_d X_\mu=0.
\end{equation}
The equations of motion are $\partial^2_\tau X^\mu = \partial^2_\sigma X^\mu$. The most general solution compatible with closed string (periodic) boundary conditions is
\begin{equation}
        X^\mu(\tau,\sigma) = x^\mu +  \frac{2 \pi \alpha^{\prime} p^\mu \tau}{l} + i \sqrt{\frac{\alpha^\prime}{2}} \sum_{n \neq 0}\frac{1}{n} \Big( \alpha^\mu_n e^{-2 \pi i n(\tau - \sigma)/l} + \tilde{\alpha}^\mu_n e^{-2 \pi i n(\tau + \sigma)/l} \Big),
        \label{closed}
    \end{equation}
where the coefficients have been normalized for later convenience. The constant 
\begin{equation}
    x^\mu = \frac{1}{l} \int_0^l d\sigma X^\mu(0,\sigma)
\end{equation}
is the spacetime position of the string's center of mass at $\tau=0$. Writing the Polyakov action as $\int d\tau d\sigma \mathcal{L}_P$, the canonical momentum density conjugate to $X^\mu$ is given by
\begin{align}
    \Pi^\mu &= \frac{\partial \mathcal{L}_P}{\partial (\partial_\tau X_\mu)} = -\frac{1}{2 \pi \alpha^\prime} \partial^\tau X^\mu \notag\\[5pt]
    &= \frac{p^\mu}{l} + \frac{1}{l\sqrt{2 \alpha^\prime}} \underset{n \neq 0}{\sum_{n=-\infty}^\infty} \Big( \alpha^\mu_n e^{-2 \pi i n(\tau - \sigma)/l} + \tilde{\alpha}^\mu_n e^{-2 \pi i n(\tau + \sigma)/l} \Big).
\end{align}
This identifies $p^\mu = \int_0^l d\sigma \Pi^\mu$ as the total spacetime momentum carried by the string.
The general solution for an open string with both endpoints free (Neumann boundary conditions) is
\begin{equation}
        X^\mu(\tau,\sigma) = x^\mu +  \frac{2 \pi \alpha^{\prime} p^\mu \tau}{l} + i \sqrt{2 \alpha^\prime} \sum_{n \neq 0} \frac{\alpha^\mu_n}{n} e^{- \pi i n \tau/l} \cos \left( \frac{\pi n  \sigma}{l} \right).
    \end{equation}
The interpretation of $x^\mu$ and $p^\mu$ are the same as before. In both cases the general structure of the solution is that of a free particle of initial position $x^\mu$ and momentum $p^\mu$, carrying an infinite amount of internal degrees of freedom described by the $\alpha^\mu_n$ modes (and $\tilde{\alpha}^\mu_n$ in the closed case). They correspond to internal motions of the string, such as rotation and vibration.

Finally, if one imposes Dirichlet condition for a given component $X^I$, the solution in this direction is
\begin{equation}
        X^I(\tau, \sigma) = y_1^I + \frac{\left( y^I_2 - y^I_1 \right) \sigma}{l} + \sqrt{2 \alpha^\prime} \sum_{n \neq 0} \frac{\alpha_n^I}{n} e^{-i \pi n \tau / l} \sin \left( \frac{\pi n \sigma}{l} \right),
        \label{Dirichlet sol}
    \end{equation}
where $y_1^I=X^I(\tau,0)$ and $y^I_2=X^I(\tau,l)$ are where each endpoint of the string is fixed. Note that for this solution the center of mass is not free to travel in the $I$th direction. Accordingly, one has $p^I=0$. Reality of $X^\mu$ demands that for all boundary conditions
\begin{equation}
    \alpha^\mu_{-n} = (\alpha^\mu_n)^\ast.
\end{equation}
The same holds for the $\tilde{\alpha}^\mu_n$ in the closed case. 

Since canonical methods will play an important role in discussing the quantum string, it is useful to collect some results on the Hamiltonian formulation of the theory in conformal gauge. The Hamiltonian associated to the Polyakov Lagrangian is
\begin{equation}
H_\text{P} = \int_0^l d \sigma \Big( \pi \alpha^\prime \Pi^\mu \Pi_\mu + \frac{1}{4 \pi \alpha^\prime} \partial_\sigma X^\mu \partial_\sigma X_\mu \Big)
\end{equation}
and the corresponding equal time Poisson brackets are
\begin{equation}
    \big\{ X^\mu(\tau,\sigma) , \Pi^\mu(\tau,\sigma^\prime) \big\}_\text{PB} = \eta^{\mu \nu} \delta(\sigma - \sigma^\prime),
\end{equation}
\begin{equation}
    \big\{ X^\mu(\tau,\sigma) , X^\nu(\tau,\sigma^\prime) \big\}_\text{PB} = \big\{ \Pi^\mu(\tau,\sigma) , \Pi^\nu(\tau,\sigma^\prime) \big\}_\text{PB}=0.
\end{equation}
In terms of the mode expansions, these translate into
\begin{equation}
    \{x^\mu,p^\nu\}_\text{PB} = \eta^{\mu \nu},
\end{equation}
\begin{equation}
    \{\alpha^\mu_n,\alpha^\nu_m\}_\text{PB} = \{\tilde{\alpha}^\mu_n,\tilde{\alpha}^\nu_m \}_\text{PB} = -i n \eta^{\mu \nu} \delta_{n+m,0}, 
    \label{Poisson}
\end{equation}
with all others vanishing, and the Hamiltonian becomes
\begin{equation}
    H_\text{P} =
    \begin{dcases}
        \frac{\pi \alpha^\prime}{l} p_\mu p^\mu + \frac{\pi}{l} \sum_{n \neq 0 }  \Big( \alpha^\mu_{-n} \alpha_{n \mu} + \tilde{\alpha}^\mu_{-n} \tilde{\alpha}_{n \mu} \Big) \hspace{0.5cm} \text{(closed)} \\
        \frac{\pi \alpha^\prime}{l} p_\mu p^\mu + \frac{\pi}{2l} \sum_{n \neq 0 } \alpha^\mu_{-n} \alpha_{n \mu} \hspace{0.5cm} \text{(NN)} \\
        \frac{\Delta y^2}{4 \pi \alpha^\prime}+ \frac{\pi}{2l} \sum_{n \neq 0 } \alpha^\mu_{-n} \alpha_{n \mu} \hspace{0.5cm} \text{(DD)}, 
        \label{Hamiltonians}
    \end{dcases}
\end{equation}
where NN means an open string with Neumann conditions imposed on both endpoints on all directions, while DD means that Dirichlet conditions are imposed on both endpoints, for some directions. In this case $\mu$ runs only over the Neumann directions and $\Delta y^2= (y_2-y_1)^I (y_2-y_1)^I$, with $I$ running over the Dirichlet directions, a notation that will be maintained for the remainder of the text. It is customary to choose $l=2\pi$ for closed strings and $l=\pi$ for open strings. From now on we adopt this convention. 

The fact that the Polyakov action in conformal gauge still describes the dynamics of the $D$ degrees of freedom $X^\mu$, of which only the $D-2$ transverse ones are physical, is a sign that there is still some gauge freedom left unfixed. This is also evident from the wrong sign in front of the kinetic term of $X^0$ in the action \eqref{PolyC}, which is responsible for the negative contributions to the Hamiltonian one finds when taking into account that $\eta_{00}=-1$, making it unbounded from below. This is the same kind of problem one runs into when trying to canonically quantize QED without gauge-fixing \cite{Peskin:1995ev}. Indeed, there is a subgroup of diff$\times$Weyl transformations that leaves the conformal gauge metric invariant, which consists of diffeomorphisms $\sigma^a \to \sigma^{\prime a}$ such that
\begin{equation}
    \eta_{ab} \to \eta^\prime_{ab} = e^{-2\Omega(\tau,\sigma)} \eta_{ab},
\end{equation}
followed by a Weyl transformation with parameter $\omega=\Omega$ that cancels the overall factor in the above, restoring the original form of the metric. These are called conformal transformations, and the invariance of the action with respect to them means that the worldsheet theory in conformal gauge is a conformal field theory (CFT). CFT is a vast subject, from which we will need only the most basic ideas.\footnote{For a general introduction to conformal field theory see \cite{DiFrancesco:1997nk}.} Conformal transformations can be seen as a particular generalization of rescalings, i.e. diffeomorphisms that rescale all coordinates as $\sigma^a \to \lambda \sigma^a$, while keeping the metric fixed. This has the effect of changing the norm of vectors, and therefore the value of areas, while preserving all angles.\footnote{The angle between two vectors $v$ and $u$ is defined as the number $\theta$ in the inner product
\begin{equation}
    v \cdot u = g_{ab}v^a u^b = |v| |u| \cos \theta,
\end{equation}
where $|v| = \sqrt{g_{ab} v^a v^b }$. Under the scaling transformation the components of each vector are multiplied by $\lambda$, so $v \cdot u \to \lambda^2 v \cdot u$ and $|u| |v| \to \lambda^2 |u| |v|$. The factor of $\lambda^2$ in both sides cancels, leaving the angle invariant.} Equivalently, one may view them as active transformations that keep the coordinates fixed, while transforming all dynamical fields via the pushforward. In a general relativistic context this includes the metric, which transforms as
\begin{equation}
    g_{ab} \to g^\prime_{ab} = \frac{\partial \sigma^{\prime c} }{\partial \sigma^a} \frac{\partial \sigma^{\prime d} }{\partial \sigma^b} g_{cd} = \lambda^2 g_{ab}.
\end{equation}
Conformal transformations are a local version of this: they are defined as diffeomorphisms under which the pushforward of the metric is $g_{ab} \to \lambda^2(\tau,\sigma) g_{ab} $, for some nonvanishing function $\lambda(\tau,\sigma) = e^{-\Omega(\tau,\sigma)}$. There is a subtle difference in how conformal transformations are defined in string theory, where the worldsheet metric is itself a dynamical field, and in other contexts where the theory is taken from the start to be defined over a manifold with fixed metric. In the latter case, the active diffeomorphisms just described are the relevant transformations, since the fact that the metric is fixed from the start means that it is not pushed forward along with the dynamical fields. In string theory, we know that the fixed metric theory (the Polyakov action with $g_{ab} \to \eta_{ab}$) emerges as a gauge-fixing of a dynamical metric to a particular form, and dynamical metrics are pushed forward along diffeomorphisms together with the other fields. An active diffeomorphism such that $\eta_{ab} \to e^{-2 \Omega} \eta_{ab}$ therefore moves us out of our gauge slice of choice. In order to have a symmetry transformation that acts only inside the fixed gauge, one adds to the definition of a conformal transformation a Weyl rescaling, whose only purpose is to undo the change of the metric brought by the diffeomorphism:
\begin{equation}
    \eta_{ab} \xrightarrow[]{\text{diff}} e^{-2\Omega} \eta_{ab} \xrightarrow[]{\text{Weyl}} e^{2\Omega} e^{-2\Omega} \eta_{ab} = \eta_{ab}.
\end{equation}
The result is the same kind of transformation that acts on theories with nondynamical metrics. They are implemented in string theory via a combined diff$\times$Weyl transformation. An important feature of conformal symmetry is that it forbids any dimensionful parameters. Whenever an operator in the Lagrangian of a field theory comes multiplied by a dimensionful constant, that means that the energy dimension of this operator does not cancel against that of the spacetime integration measure. Such a coupling therefore will not be invariant under conformal transformations. Intuitively, this is a consequence of the fact that any dimensionful parameter can be understood as natural scale, which is not allowed in a scale-invariant theory.

Complex coordinates are very useful when discussing two-dimensional conformal transformations. One can map the plane described by coordinates $(\sigma^1,\sigma^2)$ into the complex plane by defining
\begin{equation}
    w= \sigma^1 + i \sigma^2, \hspace{0.5cm}  \bar{w} = \sigma^1 - i \sigma^2,
\end{equation}
in terms of which the Euclidean metric $ds^2 = (d\sigma^1)^2+ (d \sigma^2)^2$ becomes $ds^2 = dw \, d \bar{w}$. It is then easy to see that any holomorphic mapping $w \to z(w)$ is conformal, since the metric changes according to \footnote{Note that holomorphicity requires that $z(w)$ be a function of $w$ alone, not of $(w,\bar{w})$. }
\begin{equation}
    ds^2 \to dz \,  d\bar{z} = \bigg| \frac{\partial w}{\partial z} \bigg|^{-2} dw \, d\bar{w}.
\end{equation}
The set of such holomorphic diffeomorphisms is precisely the set of conformal transformations \cite{DiFrancesco:1997nk}. Consider an infinitesimal conformal transformation $w \to z(w) = w + \epsilon(w)$, where we take $\epsilon(w)$ to be small. A function on the complex plane $f(w,\bar{w})$ varies by
\begin{equation}
    \delta f(w,\bar{w}) = f^\prime (w,\bar{w}) - f(w,\bar{w}) = -\epsilon(w) \partial_w f(w,\bar{w}) + \mathcal{O}(\epsilon^2).
\end{equation}
Expanding $\epsilon(w)$ as a Laurent series
\begin{equation}
    \epsilon(w) = \sum_{n = - \infty}^\infty c_n w^{n+1}
\end{equation}
leads to
\begin{equation}
    \delta f(w,\bar{w}) = \sum_{n = - \infty}^\infty c_n \, l_n f(w),
\end{equation}
where $l_n = - w^{n+1} \partial_w$ are the generators of infinitesimal conformal transformations, also known as Virasoro generators. By doing instead an antiholomorphic transformation $\bar{w} \to \bar{w} + \bar{\epsilon}(\bar{w})$ we find the same structure, but with the antiholomorphic $\tilde{l}_n = - \bar{w}^{n+1} \partial_{\bar{w}}$ generators instead. These are easily found to satisfy the commutation relations
\begin{align}
    [l_m,l_n] &= (m-n)l_{m+n} \notag\\[5pt]
    [\tilde{l}_m,\tilde{l}_n] &= (m-n)\tilde{l}_{m+n} \notag\\[5pt]
    [l_m,\tilde{l}_n] &= 0. 
\end{align}
This is the two-dimensional conformal algebra, called the Witt algebra. Any two-dimensional CFT is expected to furnish a representation of it with its fields.

The fact that conformal transformations appear as leftover gauge transformations allows many powerful CFT techniques to be used in string theory.\footnote{Assuming that conformal invariance is not anomalous at the quantum level, which turns out to be a nontrivial condition.} For this it is convenient to Wick rotate to an Euclidean worldsheet, which means using in place of $\tau$ the imaginary time $\sigma^2 = i \tau$. One then usually writes $\sigma \equiv \sigma^1$. The solutions discussed for the equations of motion are all analytic functions of $\tau$ and $\sigma$, so in the classical theory this Wick rotation amounts to a simple direct substitution. A particularly useful complex coordinate system for string theory is 
\begin{equation}
    z = e^{-iw} =e^{ - i ( \sigma^1 + i \sigma^2 ) } \, , \hspace{0.5cm} \bar{z}= e^{i\bar{w}} = e^{ i( \sigma^1 - i \sigma^2 ) }.
    \label{complex coords}
\end{equation}
    In terms of these, the solutions to the equations of motion are
    \begin{align}
        X^\mu(z,\bar{z}) &= x^\mu - i\frac{\alpha^{\prime} }{2}p^\mu \ln |z|^2 + i \sqrt{\frac{\alpha^\prime}{2}} \sum_{n \neq 0}\frac{1}{n} \left( \frac{\alpha^\mu_n}{z^n} + \frac{\tilde{\alpha}^\mu_n}{\bar{z}^n} \right) \hspace{1cm} \text{(closed)} \notag\\[5pt]
        X^\mu(z,\bar{z}) &= x^\mu  - i\alpha^{\prime} p^\mu \ln |z|^2 + i \sqrt{\frac{\alpha^\prime}{2}}  \sum_{n \neq 0} \frac{\alpha^\mu_n}{n}  \left( \frac{1}{z^n} + \frac{1}{\bar{z}^n} \right) \hspace{1cm} \text{(NN)} \notag\\[5pt]
        X^I(z,\bar{z}) &= y_1^I + \frac{i ( y^I_2 - y^I_1 ) }{2 \pi} \ln \left( \frac{z}{\bar{z}} \right) - i \sqrt{\frac{\alpha^\prime}{2}}  \sum_{n \neq 0} \frac{\alpha^I_n}{n}  \left( \frac{1}{z^n} + \frac{1}{\bar{z}^n} \right)  \hspace{1cm} \text{(DD)}
    \end{align}
Note that all of the above are of the form $f(z)+g(\bar{z})$, which ensures that $\partial_z X^\mu(z)$ is a function of $z$ alone (holomorphic) and $\partial_{\bar{z}}X^\mu(\bar{z})$ is a function of $\bar{z}$ alone (antiholomorphic). The equivalent statement that 
\begin{equation}
    \partial \bar{\partial} X^\mu(z,\bar{z}) = 0, \hspace{0.5cm} \text{ where } \partial \equiv \partial_z, \text{ } \bar{\partial} \equiv \partial_{\bar{z}},
\end{equation}
is in fact obtained as the equation of motion for the $X^\mu$ if one uses these complex coordinates from the start in the action. Note that $z$ in terms of the Lorentzian coordinates becomes a function only of the combination $\tau - \sigma$, whereas $\bar{z}$ becomes a function of $\tau + \sigma$. For this reason the terms right and left-moving will be used interchangeably with holomorphic and antiholomorphic. Likewise, mode operators such as $\alpha^\mu_n$, which come from the Laurent expansion of a right-moving field, will also be referred to as right-movers, and conversely for the left-movers $\tilde{\alpha}^\mu_n$.

The flat worldsheet metric $g_{\alpha \beta}$, where the greek indices go over $z$ and $\bar{z}$, is given by
\begin{equation}
    g_{\alpha \beta} = 
    \begin{pmatrix}
        0 & \frac{1}{2|z|^2} \\
        \frac{1}{2|z|^2} & 0
    \end{pmatrix},
    \hspace{0.5cm} 
    g^{\alpha \beta} = 
    \begin{pmatrix}
        0 & 2|z|^2 \\
        2|z|^2 & 0
    \end{pmatrix}.
\end{equation}
The tracelessness of the energy-momentum tensor in complex coordinates amounts to
    \begin{equation}
        g^{\alpha \beta} T_{\alpha \beta} = 2 |z|^2 (T_{z \bar{z}} + T_{\bar{z} z}) = 0,
    \end{equation} 
which means that $T_{z \bar{z}} = T_{\bar{z} z}=0$, since $T_{\alpha \beta}$ is symmetric by definition. From \eqref{T} one finds that the two remaining components are
\begin{align}
    T_{zz} &\equiv T(z) = -\frac{1}{\alpha^\prime} \partial X^\mu(z) \partial X_\mu(z) \notag\\[5pt]
    T_{\bar{z} \bar{z} } &\equiv \tilde{T}( \bar{z}) = -\frac{1}{\alpha^\prime} \bar{\partial} X^\mu(\bar{z}) \bar{\partial} X_\mu(\bar{z}).
    \label{complex T}
\end{align}
For all boundary conditions, the derivative of $X^\mu$ is given by
\begin{equation}
     \partial X^\mu(z) = \mp i \sqrt{\frac{\alpha^\prime}{2}} \sum_{n = - \infty}^\infty \frac{\alpha^\mu_n}{z^{n+1}},
     \label{derivatives X}
\end{equation}
where
\begin{equation}
    \alpha^\mu_0 \equiv 
    \begin{dcases}
        \sqrt{\dfrac{\alpha^\prime}{2}} p^\mu \hspace{0.5cm} \text{(closed)} \\[5pt]
        \sqrt{2 \alpha^\prime} p^\mu \hspace{0.5cm}  \text{(NN)} \\[5pt]
        \dfrac{y_2^I - y_1^I}{\pi \sqrt{2 \alpha^\prime}} \hspace{0.5cm} \text{(DD)}
    \end{dcases}
\end{equation}
and with the minus sign for the closed and NN conditions and the plus sign for DD. In all cases one finds for the holomorphic component of the energy-momentum tensor
\begin{equation}
    T(z) = \frac{1}{2} \sum_{n,m=-\infty}^\infty \frac{\alpha^\mu_n \alpha_{m \mu}}{z^{n+m+2}} \equiv \sum_{n=-\infty}^\infty \frac{L_n}{z^{n+2}},
    \label{HoloT}
\end{equation}
with
\begin{equation}
    L_n = \frac{1}{2} \sum_{m = -\infty}^\infty \alpha^\mu_{n-m} \alpha_{m \mu}.
    \label{Ln}
\end{equation}

A completely analogous computation for the antiholomorphic component $\tilde{T}(\bar{z})$ leads to 
\begin{equation}
    \tilde{T}(\bar{z}) = \sum_{n=-\infty}^\infty \frac{\tilde{L}_n}{\bar{z}^{n+2}}
\end{equation}
with\footnote{Assuming of course that the $\tilde{\alpha}^\mu_n$ are present, which only happens for closed strings. For open strings, one finds $\tilde{L}_n=L_n$.}
\begin{equation}
    \tilde{L}_n = \frac{1}{2} \sum_{m = -\infty}^\infty \tilde{\alpha}^\mu_{n-m} \tilde{\alpha}_{m \mu}.
\end{equation}

Using the Poisson brackets of \eqref{Poisson}, we find that
\begin{align}
    \{ L_m , L_n \}_\text{PB} &= \frac{1}{4} \sum_{kl} \{ \alpha^\mu_{m-k} \alpha_{\mu k} , \alpha^\nu_{n-l} , \alpha_{\nu l} \}_\text{PB} \notag\\[5pt]
    &= \frac{1}{4} \sum_{kl} \Big( \alpha^\mu_{m-k} \{ \alpha_{\mu k} , \alpha^\nu_{n-l}  \}_\text{PB} \alpha_{\nu l}  + \{ \alpha^\mu_{m-k} , \alpha^\nu_{n-l} \}_\text{PB} \alpha_{\mu k} \alpha_{\nu l} \notag\\[5pt]
    & \hspace{1.5cm} + \alpha^\nu_{n-l} \alpha^\mu_{m-k} \{ \alpha_{\mu k} , \alpha_{\nu l} \}_\text{PB} + \alpha^\nu_{n-l} \{ \alpha^\mu_{m-k} , \alpha_{\nu l} \}_\text{PB} \alpha_{\mu k} \Big) \notag\\[5pt]
    & = - \frac{i}{2} \sum_l \Big( (l-n) \alpha^\mu_{m+n -l} \alpha_{\mu l} -l \alpha^\mu_{n-l} \alpha_{m+l \mu}  \Big) \notag\\[5pt]
    & =-i(m-n) L_{m+n},
 \end{align}
 where $l \to l-m$ was done to go from the third to the last line.
The same computation shows that $\{ \tilde{L}_m , \tilde{L}_n \}_\text{PB}=-i(m-n) \tilde{L}_{m+n}$, and a similar one leads to $\{ L_m , \tilde{L}_n \}_\text{PB}=0$, confirming that the $L_n$ and $\tilde{L}_n$ are the Virasoro generators of the worldsheet conformal algebra. The condition that $T_{\alpha \beta}=0$ is equivalent to 
\begin{equation}
    L_n = \tilde{L}_n =0 \text{ for all $n$}.
\end{equation}
This expresses the fact that in string theory the conformal transformations are gauge symmetries, whose generators must therefore annihilate all physical states. The $n=0$ constraint is particularly relevant, since it involves the spacetime momentum $\alpha^\mu_0 \sim p^\mu$ in the case of closed or NN strings:
\begin{align}
    L_0 &= \frac{1}{2} \sum_{m = -\infty}^\infty \alpha^\mu_{-m} \alpha_{m \mu} = \frac{1}{2} \alpha^\mu_0 \alpha_{0 \mu} + \sum_{m>0} \alpha^\mu_{-m} \alpha_{m \mu} =0 \notag\\[5pt]
    \tilde{L}_0 &= \frac{1}{2} \sum_{m = -\infty}^\infty \tilde{\alpha}^\mu_{-m} \tilde{\alpha}_{m \mu} = \frac{1}{2} \tilde{\alpha}^\mu_0 \tilde{\alpha}_{0 \mu} + \sum_{m>0} \tilde{\alpha}^\mu_{-m} \tilde{\alpha}_{m \mu} =0
\end{align}
Plugging $\alpha^\mu_0=\tilde{\alpha}^\mu_0 = \sqrt{\alpha^\prime / 2} p^\mu$ into these equations leads to
\begin{equation}
    M^2 =\frac{4}{\alpha^\prime} \sum_{m >0} \alpha^\mu_{-m} \alpha_{m \mu} =\frac{4}{\alpha^\prime} \sum_{m >0} \tilde{\alpha}^\mu_{-m} \tilde{\alpha}_{m \mu} \hspace{1cm} \text{(closed)} 
    \label{level matching}
\end{equation}
where
\begin{equation}
    M^2 = -p^\mu p_\mu
\end{equation}
is the total mass of the closed string. The second equality in \eqref{level matching}, which relates the $\alpha$ with the $\tilde{\alpha}$ excitations, is called the level matching condition. Using instead $\alpha^\mu_0 = \sqrt{2 \alpha^\prime} p^\mu$ in $L_0$ gives the mass relation for open strings
\begin{equation}
    M^2 = \frac{1}{\alpha^\prime} \sum_{m >0} \alpha^\mu_{-m} \alpha_{m \mu} \hspace{1cm} \text{(NN)}.
\end{equation}
If in some directions DD conditions are imposed, this gets modified to
\begin{equation}
    M^2 = \frac{\Delta y^2}{(2 \pi \alpha^{\prime})^2} + \frac{1}{\alpha^\prime} \sum_{m >0} \Big( \alpha^\mu_{-m} \alpha_{m \mu} + \alpha^I_{-m} \alpha^I_m  \Big) \hspace{1cm} \text{(DD)}.
\end{equation}

\bigskip

\section{The quantum bosonic string}\label{sec33}

Having studied the main properties of classical bosonic strings, we now turn to the quantum theory. Regardless of the particular quantization method employed, one should be able to understand the result as a theory consisting of a Hilbert space of physical states, which undergo unitary evolution according to the Schrödinger equation. The embedding functions $X^\mu(\tau,\sigma)$ are observables of the classical theory, so they should in the quantum theory become Hermitian operators. Among the states of the Hilbert space there are the coherent states, which are the ones in direct correspondence with the string configurations of the classical theory, in the sense that for such a state $\ket{\psi_c}$ one has
\begin{equation}
    \braket{\psi_c|X^\mu(\tau,\sigma)|\psi_c} = X^\mu_c(\tau,\sigma),
\end{equation}
where $X_c^\mu(\tau,\sigma)$ is a particular solution of the equations of motion. Assuming that the set of all coherent states forms a complete basis, the fundamental question to be answered is: given a string in some particular Heisenberg picture coherent state $\ket{\psi_i,t_i}$ at time $t_i$, what is the probability to measure at a later time $t_f > t_i$ the string to be in the coherent state $\ket{\psi_f,t_f}$? The standard answer is given by the born rule 
\begin{equation}
    P\left( \psi_f ; t_f | \psi_i ; t_i \right) = |\braket{\psi_f,t_f| \psi_i,t_i}|^2
\end{equation}
together with the Feynman prescription that the amplitude $\braket{\psi_f,t_f| \psi_i,t_i}$ is computed by summing over all possible evolutions that interpolate between the initial and final state, each weighed by the complex exponential of the associated action:
\begin{equation}
    \braket{\psi_f,t_f| \psi_i,t_i} = \underset{X(\tau_i,\sigma)=X_i(\sigma)}{\overset{X(\tau_f,\sigma)=X_f(\sigma)}{\int}} \frac{\mathcal{D}g \mathcal{D}X }{V_{\text{diff}\times \text{Weyl}}}\exp \bigg(- \frac{i}{4 \pi \alpha^\prime} \int d \tau d \sigma \sqrt{-g} g^{a b} \partial_a X^\mu \partial_b X_\mu - i \lambda \raisebox{2pt}{$\chi$} \bigg).
    \label{amplitude}
\end{equation}
In this expression the curves $X^\mu_i(\sigma)$ and $X^\mu_f(\sigma)$ are the classical counterparts of the initial and final coherent states $\ket{\psi_i,t_i}$ and $\ket{\psi_f,t_f}$. The functional integration over $\mathcal{D}X \mathcal{D} g$ is a sum over all two-dimensional worldsheets bounded by $X_i(\sigma)$ and $X_f(\sigma)$ which, according to the interpretation given in the last section, should be understood as a sum over all paths the string can take in going from the initial to the final configuration, as well as all shapes it may twist and stretch itself into along the way. For each such worldsheet, one assumes the timelike coordinate $\tau$ to lie in the range $[\tau_i,\tau_f]$, so that $\tau=\tau_i$ corresponds to the initial configuration of the string $X_i(\sigma)$, and $\tau_f$ corresponds to the final one $X_f(\sigma)$. 

\begin{figure}[h]
\begin{center}
\includegraphics[width= 1\textwidth]{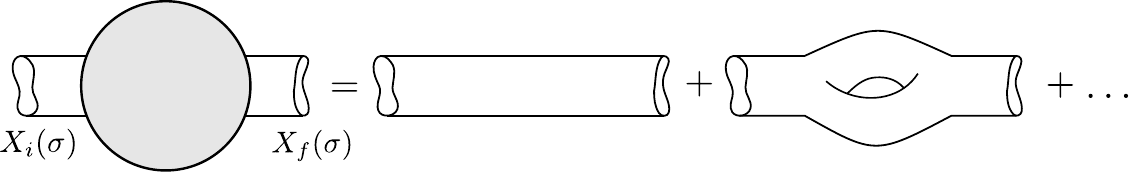}
\caption[Closed string transition amplitude]{Representation of the $X_i(\sigma) \to X_f(\sigma)$ transition amplitude for a closed string, where $X_i(\sigma)$ and $X_f(\sigma)$ are fixed initial and final curves. The first term in the right-hand side is a tree-level closed string propagator. The second one is a one-loop contribution where the string splits into two, which travel for a bit before merging together. }
\label{closed string perturbation series figure}
\end{center}
\end{figure}

The division by the volume of the gauge group $V_{\text{diff}\times \text{Weyl}}$ is meant to signify that one should fix a gauge in order to avoid overcounting due to the physically equivalent worldsheets related to one another by diffeomorphisms and Weyl transformations. This path integral formulation of string theory is therefore analogous to the worldline formalism of quantum field theory, where one computes scattering amplitudes by directly summing over all paths each individual particle may take in going from the initial to the final state \cite{Strassler_1992}. This should be contrasted with the more common second-quantized formalism of field theory, where one integrates over all field configurations to find an off-shell correlation function, which must then be fed into the LSZ formula in order to produce a scattering amplitude.\footnote{The equivalent of the second-quantized formalism in the context of string theory is called string field theory. See footnote \ref{string field theory footnote}.} Note that the topological gravitational term $\raisebox{2pt}{$\chi$} $ has been included (see \eqref{chi}), multiplied by a constant $\lambda$.

Instead of directly studying the path integral in \eqref{amplitude}, we will often work with its Euclidean version
\begin{equation}
    \int\frac{\mathcal{D}g \mathcal{D}X }{V_{\text{diff}\times \text{Weyl}}}\exp \bigg(- \frac{1}{4 \pi \alpha^\prime} \int_W d^2 \sigma \sqrt{g} g^{a b} \partial_a X^\mu \partial_b X_\mu - \lambda \raisebox{2pt}{$\chi$}(W) \bigg),
\end{equation}
which is easier to do computations with, and assume that the results can be analytically continued back to Minkowskian signature. 

As explained in the previous section, $\raisebox{2pt}{$\chi$} $ is topological and simply takes the value of the Euler number of the worldsheet over which it is computed. This allows the above expression to be organized as a sum over topologies
\begin{equation}
     \sum_\chi e^{-\lambda \raisebox{2pt}{$\chi$}} \int\frac{\mathcal{D}g \mathcal{D}X }{V_{\text{diff}\times \text{Weyl}}(\raisebox{2pt}{$\chi$})}\exp \bigg(- \frac{1}{4 \pi \alpha^\prime} \int d^2 \sigma \sqrt{g} g^{a b} \partial_a X^\mu \partial_b X_\mu \bigg),
\end{equation}
where for each value of $\raisebox{2pt}{$\chi$} $ in the sum, one integrates only over the worldsheets of the corresponding topology, and divides by the volume of the subset of the total gauge group given by the transformations compatible with it. 

\begin{figure}
     \centering
     \begin{subfigure}[h]{0.2\textwidth}
         \centering
         \includegraphics[width=\textwidth]{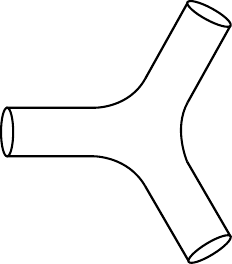}
         \caption{}
         \label{closed 3pt}
     \end{subfigure}
     \hspace{3cm}
     \begin{subfigure}[h]{0.2\textwidth}
         \centering
         \includegraphics[width=\textwidth]{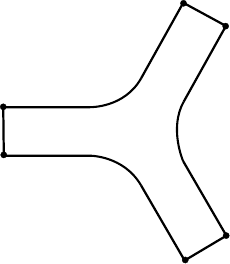}
         \caption{}
     \end{subfigure}
     \caption[Basic interaction of closed and open strings]{Basic interaction of closed and open strings.}
     \label{3 pt functions}
\end{figure}

The basic interaction of string theory is the process in which a string splits into two, as in figure \ref{3 pt functions}. Any diagram with a different number of external states can be built by combining these basic three-point functions. For instance, a closed string worldsheet like the one in \ref{closed 3pt}, but with four strings emerging from the center, is conformally equivalent to one where two three-point interactions are connected by a propagator. This is a consequence of the uniformization theorem for Riemann surfaces.\footnote{This theorem states that every compact, simply connected Riemann surface is conformally equivalent to a sphere \cite{Staessens:2010vi}. Any tree-level worldsheet for closed string interactions is one such surface, apart from the points where the external legs are inserted.} In particular, the topological expansion of the closed string zero-point function, which consists of  the sum of all vacuum diagrams of the theory, is precisely of the form \eqref{topological series} that was found in the large $N$ expansion of Yang-Mills.

Since each hole in the worldsheets that appear in \eqref{topological series} is the result of joining two three-point interactions, in Euclidean signature the amplitude to emit and then reabsorb a closed string is proportional to $e^{2 \lambda}$. The amplitude for a closed string to split into two is therefore regulated by the string coupling
\begin{equation}
    g_s = e^{\lambda}.
\end{equation}
For small $\lambda$, $g_s$ will also be small and the topological expansion takes the form of a perturbative series: the dominant closed string contribution to the zero-point function is given by the worldsheet with the topology of a sphere, and the higher genus ones are small corrections. There is a close analogy with the Feynman diagram expansion of perturbative QFT. At each genus, the same path integral over the quadratic Polyakov action must be computed, the only part that changes being the global structure of the worldsheet. 

For surfaces with boundaries, the Euler number is
\begin{equation}
    \raisebox{2pt}{$\chi$} = 2 - 2h - b,
\end{equation}
where $b$ is the number of boundaries. In an open string worldsheet, there is always one boundary given by the perimeter of the surface, and any additional one takes the form of a hole in the worldsheet, so that the open string contributions to the zero-point function are 
\begin{equation}
        Z= \includegraphics[scale=0.4,valign=c]{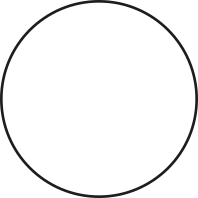} + \includegraphics[scale=0.4,valign=c]{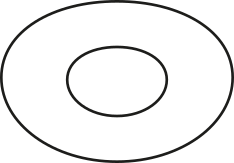} + \includegraphics[scale=0.4,valign=c]{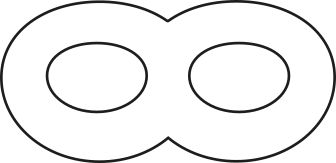} + \dots
    \end{equation}

With the path integral in hands, we now fix the gauge using the Faddeev-Popov method, following closely Section 3.3 of \cite{Polchinskivol1:1998rq}. This procedure provides a way to gauge-fix a theory without losing manifest Lorentz invariance, at the expense of introducing additional fields in the action, the Faddeev-Popov ghosts. The action for the ghosts does not depend on the genus of the worldsheet, so we will ignore the $\raisebox{2pt}{$\chi$} $ expansion and study 
\begin{equation}
    \mathcal{Z} = \int\frac{\mathcal{D}g \mathcal{D}X }{V_{\text{diff}\times \text{Weyl}}}\exp \bigg(- \frac{1}{4 \pi \alpha^\prime} \int d^2 \sigma \sqrt{g} g^{a b} \partial_a X^\mu \partial_b X_\mu \bigg),
    \label{PathIntegral}
\end{equation}
instead, with the worldsheet topology unspecified. In Section \ref{sec62} we will return to this computation and consider precisely how a dependence on the genus may arise.

As mentioned in the previous section, worldsheet diffeomorphisms plus Weyl transformations provide enough gauge freedom to completely fix the metric to any form desired:\footnote{In these and the following expressions, the $(\sigma)$-dependence of functions should always be understood as a shorthand for $(\sigma^1,\sigma^2)$.}
\begin{equation}
    g_{ab}(\sigma) \xrightarrow{\text{Diff} \times \text{Weyl}} \hat{g}_{ab}(\sigma) \, , \hspace{0.5cm} \hat{g}_{ab}(\sigma) = \text{form chosen for the metric}.
\end{equation}

The starting point for the Faddeev-Popov procedure is the trivial identity 
\begin{equation}
    1=\int \mathcal{D}h \, \delta [g-h],
    \label{FP identity}
\end{equation}
where
\begin{equation}
    \delta [g-h] = \prod_{a,b,\sigma} \delta\big(g_{ab}(\sigma)-h_{ab}(\sigma)\big)
\end{equation}
is a delta functional that enforces $h_{ab}(\sigma)=g_{ab}(\sigma)$ at every point $\sigma$. Since by assumption all metrics are gauge-equivalent, we can reexpress $h_{ab}$ as $\hat{g}_{ab}^\zeta$, the image of some fixed metric $\hat{g}_{ab}$ under a transformation $\zeta: h \to \hat{g}$, where $\zeta$ consists of a combined worldsheet reparametrization and Weyl transformation,
\begin{equation}
    \hat{g}^\zeta_{ab}(\sigma^\prime) = e^{2 \omega(\sigma)} \frac{\partial \sigma^{\prime c}}{\partial \sigma^{ a}} \frac{\partial \sigma^{\prime d}}{\partial \sigma^{b}} h_{cd}(\sigma).
    \label{gauge}
\end{equation}
With this change of variables the integral over all metrics in \eqref{FP identity} becomes an integral over all gauge transformations,
\begin{equation}
    1 = \int \mathcal{D}\zeta \det \left( \frac{\delta \hat{g}^\zeta}{\delta \zeta} \right)\Bigg|_{\hat{g}^\zeta=g} \delta \big[ g-\hat{g}^\zeta \big],
\end{equation}
where $\mathcal{D}\zeta$ is a gauge-invariant measure on the $\text{diff} \times \text{Weyl}$ gauge group. The Jacobian, in this context usually called the Faddeev-Popov determinant, is actually independent of $\zeta$. This is due to the delta functional. If $\zeta_0$ is the particular gauge transformation such that $\hat{g}^{\zeta_0}=g$, we know from the usual properties of delta functions that $\delta[g-\hat{g}^\zeta] \sim \delta[\zeta - \zeta_0]$, so the determinant should be evaluated at $\zeta=\zeta_0$, after which all $\zeta$ dependence is gone and $\det( \delta \hat{g}^\zeta / \delta \zeta )|_{\zeta = \zeta_0} \equiv \Delta_\text{FP}[g]$ can be safely pulled out of the integral:
\begin{equation}
    1 = \Delta_\text{FP}[g] \int \mathcal{D} \zeta \, \delta \big[ g - \hat{g}^\zeta \big].
    \label{FPmeasure}
\end{equation}
Since \eqref{FPmeasure} holds for every individual metric $g$ (the integral being along the gauge orbit of $g$), we may stick this factor of $1$ into the integrand in \eqref{PathIntegral} to find
\begin{align}
    \mathcal{Z} &= \int\frac{ \mathcal{D} \zeta \mathcal{D}g \mathcal{D}X }{V_{\text{diff}\times \text{Weyl}}} \Delta_\text{FP}[g] \delta\left[ g - \hat{g}^\zeta \right]  \exp \bigg(- \frac{1}{4 \pi \alpha^\prime} \int d^2 \sigma \sqrt{g} g^{a b} \partial_a X^\mu \partial_b X_\mu \bigg) \notag\\[5pt]
    &= \int\frac{ \mathcal{D} \zeta \mathcal{D}X }{V_{\text{diff}\times \text{Weyl}}} \Delta_\text{FP}[\hat{g}^\zeta] \exp \bigg(- \frac{1}{4 \pi \alpha^\prime} \int d^2 \sigma \sqrt{\hat{g}^\zeta} \, \hat{g}^{ \zeta a b} \partial_a X^\mu \partial_b X_\mu \bigg) \notag\\[5pt]
    &= \int \mathcal{D}X \Delta_\text{FP}[\hat{g}] \exp \bigg(- \frac{1}{4 \pi \alpha^\prime} \int d^2 \sigma \sqrt{\hat{g}} \, \hat{g}^{a b} \partial_a X^\mu \partial_b X_\mu \bigg).
    \label{FP computation}
\end{align}
In the last step, a gauge transformation with parameter $\zeta^{-1}$ was done to set all metrics to $\hat{g}$. This assumes that the integrand is gauge-invariant, which will be shown later. The integration over the gauge group $\int \mathcal{D} \zeta = V_{\text{diff}\times \text{Weyl}}$ then factorizes and cancels against the factor of $V_{\text{diff}\times \text{Weyl}}$ in the denominator, leaving behind just the integral over the embedding fields against the corrected measure $\mathcal{D}X \Delta_\text{FP}[\hat{g}]$. We then turn to the computation of
\begin{equation}
    \Delta_\text{FP}^{-1}[\hat{g}] = \int \mathcal{D} \zeta \delta \left[ \hat{g} - \hat{g}^\zeta \right].
\end{equation}
Since the delta functional is only nonvanishing for gauge transformations close to the identity, only the infinitesimal transformation
\begin{equation}
    \hat{g}^\zeta_{ab} =  \hat{g}_{ab} + 2 \omega \hat{g}_{ab} + \hat{\nabla}_a v_b + \hat{\nabla}_b v_a + \mathcal{O}\left( \zeta^2 \right) , \hspace{0.5cm} \zeta \sim (\omega,v_a)
\end{equation}
is needed, where the covariant derivatives are taken with respect to the $\hat{g}$ metric. Therefore
\begin{equation}
    \Delta_\text{FP}^{-1}[\hat{g}] = \int \mathcal{D} \omega \mathcal{D} v \, \delta \left[ 2 \omega \hat{g}_{ab} + \hat{\nabla}_a v_b + \hat{\nabla}_b v_a  \right].
\end{equation}
The delta can be exponentiated by using the functional analogue of $\delta(x) = \int dp \exp(2 \pi i p \cdot x)$,
\begin{equation}
    \Delta_\text{FP}^{-1}[\hat{g}] = \int \mathcal{D} \omega \mathcal{D} v \mathcal{D} \beta \exp \bigg[ 2 \pi i \int d^2 \sigma \sqrt{\hat{g}} \beta^{ab} \left( 2 \omega \hat{g}_{ab} + \hat{\nabla}_a v_b + \hat{\nabla}_b v_a  \right) \bigg].
\end{equation}
In this expression $\omega$ acts as a Lagrange multiplier that enforces the constraint $ \beta^{ab} \hat{g}_{ab} = \Tr \beta =0$, which we may integrate out to obtain
\begin{equation}
    \Delta_\text{FP}^{-1}[\hat{g}] = \int \mathcal{D} v \mathcal{D} \beta^\prime \exp \bigg( 4 \pi i \int d^2 \sigma \sqrt{\hat{g}} \, \beta^{\prime ab} \hat{\nabla}_a v_b \bigg),
    \label{Inverse FP det}
\end{equation}
where the integration over $\mathcal{D} \beta^{\prime}$ is over symmetric traceless tensors.

The inversion of \eqref{Inverse FP det} is done with the usual trick of replacing the bosonic fields with fermionic ones. We substitute every bosonic integration variable $x$ for $\theta/\sqrt{2 \pi}$, where $\theta$ is a Grassmann number, set $dx \to d \theta $, and trade the $i$'s for $-1$ on the exponents.\footnote{This simple one-dimensional example should be enough to motivate these substitutions:
\begin{align}
    &\int_{- \infty}^\infty dx \int_{- \infty}^\infty dy \, e^{2 \pi i \lambda xy} = \int_{- \infty}^\infty \frac{dx}{\sqrt{2 \pi}} \int_{- \infty}^\infty \frac{dy}{\sqrt{2 \pi}} \, e^{ i \lambda xy} = \int_{- \infty}^\infty dx \, \delta(\lambda x) = \frac{1}{|\lambda|} \notag\\[5pt]
    &\int d \theta \int d \varphi \, e^{- \lambda \theta \varphi} = \int d \theta \int d \varphi \, (1-\lambda \theta \varphi) = \lambda \int d \theta \int d \varphi \, \varphi \theta = \lambda
\end{align}
In higher dimensions one has $\exp( x_i M^{ij} x_j)$ for square matrix $M^{ij}$, and the bosonic integral gives $1/|\det M|$, whereas the fermionic one gives $\det M$.} 
\begin{equation}
    \Delta_\text{FP}[g]
    = \int \mathcal{D} b \mathcal{D} c \exp \bigg[  - 2 \int d^2 \sigma \sqrt{\hat{g}} \, b^{ab} \hat{\nabla}_a c_b  \bigg]
\end{equation}
The fermionic fields $b_{ab}$ and $c^a$, with $b_{ab}$ traceless, are the string's Faddeev-Popov ghosts. It is conventional to rescale $b_{ab} \to b_{ab}/4 \pi$. Writing the total exponential in the path integral as $-S=-(S_X +S_g )$, one then finds for the ghost action in conformal gauge
\begin{equation}
    S_g = \frac{1}{2 \pi} \int d^2 \sigma b_{ab} \partial^a c^b.
\end{equation}
Like the Polyakov action in conformal gauge, $S_g$ is conformally invariant. The ghosts and the physical fields do not couple to each other in the complete action, which remains noninteracting.

While it serves the purpose of producing the correct ghost action, which was the goal of this section, it 
should be mentioned that this derivation ignores some technical details that become relevant for string perturbation theory. We will examine these issues in Chapter \ref{ch6}, but, although they must be taken into account in order to compute amplitudes, they do not alter the fact that the total action of bosonic string theory in conformal gauge is given by
\begin{equation}
    S = \int d^2 \sigma \bigg( - \frac{1}{4 \pi \alpha^\prime} \partial^a X^\mu \partial_a X_\mu + \frac{1}{2 \pi} b_{ab} \partial^a c^b \bigg),
    \label{totalaction}
\end{equation}
which is all that will be necessary for the derivation of the spectrum. When the ghosts are present, it is customary to refer to the $X^\mu$ as the ``matter fields'', and call the part of the action that involves them the ``matter CFT''.

With the total action in our hands, we now turn to operator methods in order to derive the spectrum. The original gauge symmetry of the Polyakov action is still present in \eqref{totalaction} in the form of global BRST transformations, and one way to move forward would be to study the corresponding conserved BRST charge and its cohomology\footnote{The BRST charge is nilpotent ($Q_\text{BRST}^2=0$), and its cohomology is defined as
\begin{equation}
    \mathcal{H}_\text{BRST} = \frac{\mathcal{H}_\text{closed}}{\mathcal{H}_\text{exact}},
\end{equation}
where $\mathcal{H}_\text{closed}$ is the set of all states $\ket{\psi}$ such that $Q_\text{BRST} \ket{\psi}=0$ and $\mathcal{H}_\text{exact}$ is the set of all states $\ket{\varphi}$ that can be written as $\ket{\varphi}=Q_\text{BRST} \ket{\varphi^\prime}$ for some $\ket{\varphi^\prime}$. In BRST quantization, the physical Hilbert space is given by $\mathcal{H}_\text{BRST}$ \cite{Polchinskivol1:1998rq} .
}. Here, we will take the more pedestrian route of studying the matter and ghost CFTs separately, which can be done since they are not directly coupled to each other, and extract from each one the relevant information to derive the spectrum.

The necessary results for the canonical quantization of the matter CFT have already been derived in the classical analysis of last section. Both $X^\mu(\tau,\sigma)$ and its conjugate momentum density $\Pi^\mu(\tau,\sigma)$ are promoted to operators, whose equal time commutators are given by $i$ times the classical Poisson brackets. We therefore have
\begin{equation}
    \big[ X^\mu(\tau,\sigma) , \Pi^\mu(\tau,\sigma^\prime) \big] = i \eta^{\mu \nu} \delta(\sigma - \sigma^\prime),
\end{equation}
\begin{equation}
    \big[ X^\mu(\tau,\sigma) , X^\nu(\tau,\sigma^\prime) \big] = \big[ \Pi^\mu(\tau,\sigma) , \Pi^\nu(\tau,\sigma^\prime) \big]=0,
\end{equation}
for the fields and
\begin{equation}
    [x^\mu,p^\nu] = i \eta^{\mu \nu},
\end{equation}
\begin{equation}
    [\alpha^\mu_n,\alpha^\nu_m] = [\tilde{\alpha}^\mu_n,\tilde{\alpha}^\nu_m] = n \eta^{\mu \nu} \delta_{n+m,0},
    \label{canonical commutators}
\end{equation}
for the modes, with all others vanishing. The commutator of $x^\mu$ and $p^\nu$ is what should be expected for operators representing the position and momentum of the same object, while the ones for the modes describe two harmonic oscillators (one for the $\alpha^\mu_n$ and one for the $\tilde{\alpha}^\mu_n$), which can be put in the usual form by writing $\alpha^\mu_n = \sqrt{n} a^\mu_n $ and $\tilde{\alpha}^\mu_n = \sqrt{n} \tilde{a}^\mu_n $, with $a^\mu_n$ and $\tilde{a}^\mu_n$ having the standard interpretation of the mode's occupation number. The fact that the $\alpha^\mu_n$ were real in the classical case means that the corresponding operators are Hermitian, which leads to $\alpha^\mu_{-n}=\alpha^{\dagger \mu}_n$ and $\tilde{\alpha}^\mu_{-n}=\tilde{\alpha}^{\dagger \mu}_n$. We take the $\alpha^\mu_n$ for $n>0$ to be annihilation operators, and define the vacuum as the state which is annihilated by all of them
\begin{equation}
    \alpha^\mu_n \ket{k} = \tilde{\alpha}^\mu_n \ket{k}=0 \, , \hspace{0.5cm} n>0. 
\end{equation}
The notation $\ket{k}$ means that we have chosen to work in spacetime momentum space, and 
\begin{equation}
    p^\mu \ket{k} = k^\mu \ket{k},
\end{equation}
where $p^\mu$ is the center of mass momentum operator and $k^\mu$ the eigenvalue. Higher excited states are then obtained by acting on the vacuum with the creation operators $\alpha^\mu_{-n}$ and $\tilde{\alpha}^\mu_{-n}$ any number of times,
\begin{equation}
    (\alpha^{\mu_1}_{-1})^{n_{\mu_1}}(\alpha^{\mu_2}_{-2})^{n_{\mu_2}} \dots (\tilde{\alpha}^{\mu_1}_{-1})^{n_{\mu_1}}(\tilde{\alpha}^{\mu_2}_{-2})^{n_{\mu_2}} \dots \ket{k}.
    \label{states}
\end{equation}
The presence of the momentum quantum number on the vacuum is a reminder that, even though the Polyakov action taken at face value describes the dynamics of $D$ free scalar fields $X^\mu$, the physical system being modeled is that of one relativistic string traveling through spacetime, and the Hilbert space $\mathcal{H}$ of states such as \eqref{states} consists of all possible ways to excite the internal degrees of freedom of this one quantum string. The state $\ket{0}$ in particular does not represent empty space. It contains a string at rest in its ground state. In order to discuss the scattering of strings, it will be necessary to consider multiple string states. A general $n$-string state is taken to live on the Hilbert space\footnote{This is identical to the definition of the many-particle Hilbert space of quantum field theory, which us usually followed by the construction of the Fock space in order to set up the perturbative description of scattering amplitudes. Following the same route, we have that the full Hilbert space of string theory in the noninteracting limit is
\begin{equation}
    \mathcal{H}_{\text{total}} = \ket{\text{vacuum}} \oplus \mathcal{H} \oplus \mathcal{H}^2 \oplus \dots,
\end{equation}
where $\ket{\text{vacuum}}$ is the zero-string ground state. One may then proceed to do string theory directly in $\mathcal{H}_{\text{total}}$ in a way that parallels what is done in field theory. This second-quantized approach is called string field theory. In it, one has operators that create or annihilate entire strings. String field theory will not be used in this text. Instead, we will stick to the first-quantized worldsheet formalism, which features operators that move us up and down along the excitation levels of an individual string.\label{string field theory footnote}}
\begin{equation}
    \mathcal{H}^n = \underbrace{\mathcal{H} \otimes \mathcal{H} \otimes \dots \otimes \mathcal{H}}_{n \text{ times}}.
\end{equation}

When quantizing a classical conformal field theory, it is very common to find that conformal symmetry is broken at the quantum level. One of the most important cases in high energy physics is the Yang-Mills sector of QCD, where quantum corrections introduce the QCD scale. The appearance of such a characteristic energy scale is a sign that conformal symmetry has been broken.

In string theory conformal invariance is a worldsheet gauge symmetry. Its absence at the quantum level would cause the negative norm states, such as for instance $\alpha^0_{-1} \ket{k}$, to no longer decouple, breaking the usual probabilistic interpretation of quantum mechanics. It is therefore crucial to investigate whether or not a worldsheet conformal anomaly can occur. The conserved charges associated to classical conformal symmetry are the Virasoro generators
\begin{equation}
    L_n = \frac{1}{2} \sum_{m = -\infty}^\infty \alpha^\mu_{n-m} \alpha_{m \mu},
\end{equation}
and their antiholomorphic counterparts $\tilde{L}_n$ for closed strings. Since in the quantum theory the $\alpha^\mu_n$ modes are operators which do not always commute, ordering ambiguities may arise when writing down the quantum versions of $L_n$ and $\tilde{L}_n$. From the commutation relations $[\alpha^\mu_n,\alpha^\nu_m] = [\tilde{\alpha}^\mu_n,\tilde{\alpha}^\nu_m] = n \eta^{\mu \nu} \delta_{n+m,0}$, it is clear that one need only worry about the $n=0$ generator
\begin{equation}
    L_0 = \frac{1}{2} \sum_{m = -\infty}^\infty \alpha^\mu_{-m} \alpha_{m \mu},
\end{equation}
since it is the only one where $\alpha^\mu_{-m}$ and $\alpha^\mu_m$, which do not commute, appear together. Starting from an arbitrary ordering, one can use the commutation relation
\begin{equation}
    \alpha^\mu_m \alpha_{-m \mu} = \alpha^\mu_{-m} \alpha_{m \mu} + m D\hspace{0.5cm} (D = \delta^\mu_\mu)
    \label{commutation}
\end{equation}
multiple times to reorder the operators in any way desired. We therefore define the quantum $L_0$ to be normal-ordered, meaning that every annihilation operator is written on the right of its correspondent creation operator
\begin{equation}
    \normalord{\alpha^\mu_{m}  \alpha^\nu_{-m}} = \alpha^\nu_{-m} \alpha^\mu_{m}, \hspace{0.5cm} m>0
\end{equation}
and add an unknown normal ordering constant $a^X$ to account for the commutators:
\begin{equation}
    L_0 \to L_0 + a^X, \hspace{0.5cm} L_0 = \frac{1}{2} \sum_{m = -\infty}^\infty \normalord{\alpha^\mu_{-m} \alpha_{m \mu}} \, .
\end{equation}
This constant has physical meaning. To see it we take the particular case of a classical NN string and consider its Hamiltonian \eqref{Hamiltonians}
\begin{align}
    H &=  \frac{\pi \alpha^\prime}{l} p_\mu p^\mu + \frac{\pi}{2l} \sum_{n \neq 0} \alpha^\mu_{-n} \alpha_{n \mu}  \notag\\[5pt]
    &= \frac{\pi \alpha^\prime}{l}p_\mu p^\mu   + \frac{\pi}{2l} \sum_{n = 1}^\infty \bigg( \alpha^\mu_{n} \alpha_{-n \mu} + \alpha^\mu_{-n} \alpha_{n \mu} \bigg),
\end{align}
where we restored the dependence on $l$, the upper limit of $\sigma$.
Using the same expression for the quantum Hamiltonian, we see that the first set of modes inside the parenthesis is the one that needs reordering. The commutation relation gives
\begin{align}
    H &= \frac{\pi \alpha^\prime}{l} p_\mu p^\mu + \frac{\pi}{l} \sum_{n=1}^\infty \alpha^\mu_{-n} \alpha_{n \mu} + \frac{\pi D}{2l} \sum_{n=1}^\infty n \notag\\[5pt]
    &= \normalord{H} + \frac{\pi D}{2l} \sum_{n=1}^\infty n.
\end{align}
The divergent last term represents the sum of the zero-point energies of an infinite number of harmonic oscillators, leading to the same kind of vacuum energy that can be found when canonically quantizing any free field theory. Since the divergence comes from high energies, it makes sense to regularize the sum with a UV-cutoff $\Lambda$ \cite{TimoWeigandnotes}:
\begin{align}
    \frac{\pi D}{2l} \sum_{n=1}^\infty n \to \frac{\pi D}{2l} \sum_{n=0}^\infty n e^{-\frac{\pi n}{l \Lambda}} &= \frac{\pi D}{2l} \sum_{n=0}^\infty n e^{-q n} \bigg|_{q=\frac{\pi}{l \Lambda}} \notag\\[5pt]
    &= -\frac{\pi D}{2l} \frac{\partial}{\partial q} \frac{1}{1-e^{-q}} \bigg|_{q=\frac{\pi}{l \Lambda}} \notag\\[5pt]
    &= \frac{D}{2} \left( \frac{l}{\pi} \Lambda^2 - \frac{\pi}{l} \frac{1}{12} \right) + \mathcal{O}\left( \frac{1}{\Lambda} \right).
    \label{regulatedsum}
\end{align}
Not only does the first term diverge for large values of $\Lambda$, but also its explicit dependence on the cutoff signals a loss of conformal invariance. Fortunately, it is possible to renormalize away this vacuum energy by adding to the Polyakov action a ``cosmological constant'' counterterm
\begin{equation}
    S_\text{c} = -\frac{D \Lambda^2}{4 \pi^2} \int d\tau d\sigma \sqrt{-g}.
\end{equation}
$S_c$ explicitly breaks conformal symmetry since it is not Weyl invariant, but its only effect on the physics is the introduction of a constant energy density on the worldsheet given by $\mathcal{E}_0 = T^{\tau \tau}_c= -\frac{D \Lambda^2}{2 \pi}$ in conformal gauge, where $T^{ab}_c$ is the energy-momentum tensor associated to $S_c$. Integrating over the entire string, one finds the total contribution to the vacuum energy
\begin{equation}
    E_0 = -\frac{D}{2} \frac{l}{\pi} \Lambda^2,
\end{equation}
which cancels the offending term in the Hamiltonian, rendering it finite and independent of the cutoff, therefore saving conformal invariance. We are left with just the second term in \eqref{regulatedsum}, which is sometimes referred to as a Casimir energy, in analogy with the Casimir effect of QED. Since the NN Hamiltonian is given by $H=\frac{\pi}{l} L_0$, which in the quantum theory becomes $\frac{\pi}{l} ( L_0 +a^X ) $, we are led to identify the normal ordering constant with this Casimir energy:
\begin{equation}
    a^X = - \frac{D}{24},
\end{equation}
where we have restored $l=\pi$. Note that each spacetime dimension, for each of which there is one $X^\mu$ field, contributes $-1/24$ to the constant. The same reasoning applies to closed strings, $\tilde{L}_0$ also receives its constant $\tilde{a}^X$. In this case regularizing the zero point energy gives $a^X=\tilde{a}^X=-D/24$.

We then move on to the investigation of whether or not the quantum Virasoro generators satisfy the Witt algebra. The computation in Appendix \ref{A1} results in the so-called Virasoro algebra
\begin{equation}
    \left[ L_m , L_n \right] = (m-n)  L_{m+n} + \frac{c}{12} m \left( m^2-1  \right) \delta_{m+n,0},
\end{equation}
where the number $c = D$ is called the central charge. This shows that the classical conformal invariance is indeed broken. The fact that the extra term in $\left[ L_m , L_n \right]$ in relation to the classical Witt algebra comes from the normal ordering of the $L_n$ makes it clear that it is a purely quantum effect, an anomaly. 

We now move to the ghost CFT, whose action in Lorentzian signature is
\begin{equation}
    S_g = -\frac{i}{2 \pi} \int d\tau d \sigma \, b_{ab} \, \partial^a c^b.
\end{equation}
The equations of motion are particularly easy to find using the lightcone coordinates
\begin{equation}
    \sigma^\pm = \tau \pm \sigma,
\end{equation}
since in terms of them the tracelessness of $b_{ab}$ means that $b_{+ -} = b_{- +}=0$ and the action becomes simply
\begin{equation}
    S_g =  \frac{i}{2} \int d \sigma^+ d \sigma^- \big( b_{++} \partial_- c^+ + b_{--} \partial_+ c^- \big).
\end{equation}
Setting $\delta S_g=0$ and ignoring boundary terms gives $\partial_- b_{++} = \partial_+ b_{--} = \partial_-c^+ = \partial_+ c^-=0$. These are solved by
\begin{align}
    &b_{--}(\sigma^-) = \sum_{n= - \infty}^\infty b_n e^{-i n \sigma^-}, && b_{++}(\sigma^+) = \sum_{n= - \infty}^\infty \tilde{b}_n e^{-i n \sigma^+} \notag\\[5pt]
    &c^-(\sigma^-) = \sum_{n = -\infty}^\infty c_n e^{-in \sigma^-}, && c^+(\sigma^+) = \sum_{n = -\infty}^\infty \tilde{c}_n e^{-in \sigma^+}.
\end{align}
Reality of the action requires that both ghosts be real.\footnote{We adopt the convention where the complex conjugation of a product of Grassmann numbers mirrors the hermitian conjugation of operators, $(\epsilon_1 \epsilon_2)^\ast = \epsilon_2^\ast \epsilon_1^\ast = - \epsilon_1^\ast \epsilon_2^\ast$. This means that for $\epsilon_1$ and $\epsilon_2$ ``real'', in the sense of $\epsilon_1^\ast =\epsilon_1 $, the product $\epsilon_1 \epsilon_2$ is imaginary, so $i \epsilon_1 \epsilon_2$ is real.} The modes must then satisfy $b^\dagger_{n} = b_{-n}$, $c^\dagger_n = c_{-n}$, and likewise for the right-moving ones. In the classical theory the dagger should be taken to mean complex conjugation, but it will become hermitian conjugation once this system is quantized. In terms of the original worldsheet coordinates we have for closed strings
\begin{align}
    b_{\tau \tau}(\tau,\sigma) &= b_{\sigma \sigma}(\tau,\sigma)=  \sum_n \Big( b_n e^{-in(\tau-\sigma)} + \tilde{b}_n e^{-in(\tau+\sigma)} \Big)  \notag\\[5pt]
    b_{\tau \sigma} (\tau,\sigma) &= b_{\sigma \tau}(\tau,\sigma) = - \sum_n \Big( b_n e^{-in(\tau-\sigma)} - \tilde{b}_n e^{-in(\tau+\sigma)} \Big) \notag\\[5pt]
    c^\tau(\tau,\sigma) &= \frac{1}{2} \sum_n \Big( c_n e^{-in(\tau - \sigma)} + \tilde{c}_n e^{-in(\tau+\sigma)} \Big) \notag\\[5pt]
    c^\sigma(\tau,\sigma) &= -\frac{1}{2} \sum_n \Big( c_n e^{-in(\tau - \sigma)} - \tilde{c}_n e^{-in(\tau+\sigma)} \Big).
\end{align}
These solutions all satisfy the $\sigma \sim \sigma + 2 \pi$ periodicity of closed string worldsheets. If one considers instead a worldsheet with boundaries at $\sigma=0$ and $\sigma=\pi$, the variation of the action leads to the boundary term
\begin{equation}
    \int_{- \infty}^\infty d \tau \, b_{\sigma a} \delta c^a \Big|^{\sigma=\pi}_{\sigma=0} =\int_{-\infty}^\infty d \tau \Big( b_{\tau \sigma}(\tau,\pi) \delta c^\tau(\tau,\pi) - b_{\tau \sigma}(\tau,0) \delta c^\tau(\tau,0) \Big).
\end{equation}
There is some freedom in how this can be made to vanish. Recall however that the Faddeev-Popov determinant is the inverse of
\begin{equation}
    \int \mathcal{D} v \mathcal{D} \beta \exp \bigg[ 4 \pi i \int d^2 \sigma   \beta^{ab} \nabla_a v_{b} \bigg],
\end{equation}
where $v^a$ is a worldsheet diffeomorphism parameter. No diffeomorphism defined in a worldsheet with boundaries is allowed shift the boundaries themselves. This is guaranteed by imposing that the $v^\sigma$ component, the one normal to the boundaries, must vanish over them: $v^\sigma(\tau,0) = v^\sigma(\tau,\pi) = 0$. Given that $v^a$ is what becomes the $c^a$ ghost in the Faddeev-Popov procedure, these same boundary conditions are inherited by it
\begin{equation}
    c^\sigma(\tau,0) = c^\sigma(\tau,\pi) = 0.
\end{equation}
Taking this into account, the vanishing of the boundary term requires that
\begin{equation}
    b_{\tau \sigma}(\tau,\pi) = b_{\tau \sigma}(\tau,0) =0.
\end{equation}

Imposing these conditions on the mode expansions sets $b_n = \tilde{b}_n$, $c_n = \tilde{c}_n$, which leads to the following solutions for open boundary conditions
\begin{align}
    b_{\tau \tau}(\tau,\sigma) &= b_{\sigma \sigma}(\tau,\sigma)=  2 \sum_n b_n e^{-i n \tau} \cos (n \sigma)  \notag\\[5pt]
    b_{\tau \sigma} (\tau,\sigma) &= b_{\sigma \tau}(\tau,\sigma) = - 2i \sum_n b_n e^{-i n \tau} \sin(n \sigma) \notag\\[5pt]
    c^\tau(\tau,\sigma) &= \sum_n c_n e^{-i n \tau} \cos(n \sigma) \notag\\[5pt]
    c^\sigma(\tau,\sigma) &= -i \sum_n c_n e^{-i n \tau} \sin(n \sigma).
    \label{open ghost solutions}
\end{align}
Just like for the $X^\mu$ fields, a worldsheet with boundaries reduces by half the amount of independent modes.

The ghost energy-momentum tensor can be derived in the usual way from the form of the action in a curved background, although one must be careful in order to enforce the tracelessness of $b_{a b}$. This can be done via the introduction of a Lagrange multiplier term $\Omega \Tr b$ in the action before varying with respect to the metric. It will be useful to extract from it the ghost Virasoro generators. For that we Wick rotate to Euclidean signature and employ the complex coordinates defined in \eqref{complex coords}, in terms of which the mode expansions for the right-movers become the Laurent series $b_{zz}(z) \equiv b(z)=-\sum_n b_n z^{-n-2}$, $c^z(z) \equiv c(z)=i\sum_n c_n z^{-n+1}$. The holomorphic component of the energy-momentum tensor is
\begin{equation}
    T^g(z) = -i \big( 2 b \, \partial  c + \partial b \,  c \big) = \sum_{n= - \infty}^\infty \frac{L_n^g}{z^{n+2}},
\end{equation}
where 
\begin{equation}
    L^{g}_n = \sum_{m= - \infty}^\infty \left( 2n-m \right) b_m c_{n-m}.
\end{equation}
Identical expressions hold for the left-movers, with $\bar{z}$ in the place of $z$ and tildes over the modes. We will focus here on open strings, since all expressions involving the left-movers will be identical to the ones for the right-movers. The open string ghost classical Hamiltonian is
\begin{align}
    H^{g} &= L^{g}_0 \notag\\[5pt]
    &= - \sum_{m=-\infty}^\infty m b_m c_{-m} \notag\\[5pt]
    &= \sum_{m=1}^\infty m \big( b_{-m} c_m + c_{-m}  b_m \big).
\end{align}
The quantization of this system is done in the usual way for anticommuting variables. One imposes canonical anticommutation relations for the fields, which are equivalent to 
\begin{equation}
    \left\{ b_m, c_n \right\} = \delta_{m+n,0} \, ,
\end{equation}
\begin{equation}
    \left\{ b_m, b_n \right\} = \left\{ c_m, c_n \right\} = 0 ,
\end{equation}
for the modes, with identical relations holding for the $\tilde{b}_m$ and $\tilde{c}_n$.

The vacuum state is determined by requiring it to be annihilated by $H^g$. Like for the matter CFT we take the modes with $n>0$ to be lowering operators and the ones with $n<0$ to be raising operators. The vacuum should therefore be annihilated by all positive modes. The zero energy condition says nothing about the zero-modes $b_0$ and $c_0$, as these do not appear in the Hamiltonian. The anticommutation relation $b_0c_0=-c_0b_0+1$ makes it inconsistent to require the vacuum to be annihilated by both $b_0$ and $c_0$, so we actually have two different zero-energy states $\ket{\uparrow}$ and $\ket{\downarrow}$, such that
\begin{align}
    b_0 \ket{\downarrow} &= 0, \hspace{0.5cm} b_0 \ket{\uparrow} = \ket{\downarrow} \notag\\[5pt]
    c_0 \ket{\downarrow} &= \ket{\uparrow}, \hspace{0.5cm} c_0 \ket{\uparrow} =0,
\end{align}
and both are admissible vacua of the ghost CFT. The quantum Virasoro generators are taken to be normal ordered according to
\begin{equation}
    \normalord{b_m b_n} =
    \begin{cases}
        b_m b_n \, , \hspace{0.5cm} \text{if $m \leq n$} \\
        - b_n b_m  \, , \hspace{0.27cm} \text{if $m > n$}
    \end{cases},
\end{equation}
and similarly for $\normalord{c_n c_m}$ and $\normalord{b_m c_n}$, although the latter requires a choice of what to with $\normalord{b_0c_0}$. It is usual to set $\normalord{b_0 c_0} = - c_0 b_0$.\footnote{This choice can be motivated in a more thorough BRST treatment of the theory (see footnote \ref{b ghost footnote}), but nothing in this text will actually depend on it except for the computations in Appendix \ref{ApB}.} As in the matter CFT, only the $n=0$ generator suffers from an ordering ambiguity, so it receives a normal ordering constant $a^g$, which once again is related to a vacuum energy. The quantum Hamiltonian without normal ordering is
\begin{align}
    H^{(g)} &= - \sum_{m=-\infty}^\infty m b_m c_{-m} \notag\\[5pt]
    &= \sum_{m=1}^\infty m \big( b_{-m} c_m + c_{-m}  b_m \big) - \sum_{m=1}^\infty m \notag\\[5pt]
    &= \normalord{H^{(g)}} - \sum_{m=1}^\infty m.
\end{align}
The exact same regularization and renormalization procedure that was done for the matter CFT leads to
\begin{equation}
    a^g = \frac{1}{12}.
\end{equation}
As shown in Appendix \ref{B2}, the quantized ghost CFT also has a conformal anomaly. Its conformal generators satisfies a Virasoro algebra with central charge $c^g=-26$. The total generators for the matter plus ghost theory are 
\begin{equation}
    L^\text{total}_m = L_m + L^g_m + a \delta_{m0},
\end{equation}
where
\begin{equation}
    a = a^X + a^g = -\frac{D-2}{24}
\end{equation}
is the total normal ordering constant. Note that each of the $X^\mu$ fields adds $-1/24$ to it, but the contribution of the two nonphysical polarizations are removed by the ghosts. The total Virasoro algebra is given by the sum of the one for the matter and ghost CFTs:
\begin{equation}
    \left[ L^\text{total}_m, L^\text{total}_n \right] = (m-n)L^\text{total}_{m+n} + \frac{(D-26)}{12}m \left( m^2 - 1 \right)\delta_{m+n,0} -2m (a+1)\delta_{m+n,0},
\end{equation}
with total central charge of $D-26$. The vanishing of the conformal anomaly is required for the consistency of the quantum theory. This is what determines $D=26$ as the so-called critical dimension of bosonic string theory. This also sets $a=-1$, which eliminates the last term of the Virasoro algebra, leaving an unbroken Witt algebra at the quantum level.\footnote{This normal ordering constant is the Regge intercept mentioned in the previous chapter. It was originally believed to be free parameter that could be adjusted to match the Regge trajectories seen in experiments, but the choice $a = -1$ turns out to be required for the theory to be consistent. This value of $a$ renders the ``mesonic'' open string spectrum quite different from what is seen in the strong interactions, and this was one of the first major difficulties in applying dual models to the real world. Another one was of course the $D=26$ critical dimension.}

\section{The bosonic string spectrum}\label{sec34}

With the critical dimension determined, we now turn to the spectrum of the theory. Not all states of the form \eqref{states} correspond to physical excitations of the string, since we have not yet taken into account the Virasoro constraints from conformal symmetry. Only the states annihilated by the $L_n$ are gauge-invariant, and therefore physical. Conformal symmetry is only unbroken in the combined matter plus ghost theory, whose Hilbert space is expected to separate into a physical and a nonphysical sector, which must remain decoupled along the dynamics for the theory to be well-defined. Therefore, the natural requirement for a physical state is that it be annihilated by all $L^\text{total}_n$ generators of the combined theory. This leads to the BRST quantization method. However, for the purposes of simply deriving the spectrum, there is also available a more pedestrian route in which one sets the ghosts to their ground state and works only in terms of the $X^\mu$ fields. This method is referred to in string theory literature as ``old covariant quantization'', and a proof that the results obtained from it are equivalent to those of the BRST method (also called ``modern covariant quantization'') can be found in \cite{Polchinskivol1:1998rq}.
 
We proceed by imposing that the matter CFT's Virasoro generators have zero expectation value between physical states:
\begin{equation}
    \braket{\psi^\prime|(L_n + a \delta_{0n} )|\psi} = \braket{\psi^\prime|(\tilde{L}_n + \tilde{a} \delta_{0n} )|\psi} =0.
\end{equation}
This is analogous to the Gupta-Bleuler quantization of QED in Lorentz gauge, where the gauge-fixing condition $\partial_\mu A^\mu=0$ is imposed by requiring that the operator $\partial_\mu A^\mu$ have vanishing matrix elements between physical states \cite{Tong_qft}. This is a weaker requirement than asking that $L_n \ket{\psi} = 0$ for all $n$. The fact that $L^\dagger_n = L_{-n}$ means that it is enough to have
\begin{equation}
    (L_n +a \delta_{0n}) \ket{\psi} = (\tilde{L}_n +\tilde{a} \delta_{0n})  \ket{\psi} =0 \,, \hspace{0.5cm} n \geq 0.
    \label{physical}
\end{equation}
Any state that can be written as
\begin{equation}
    \ket{\chi} = \sum_{n=1}^\infty \left( L_{-n} \ket{\chi_n} + \tilde{L}_{-n} \ket{\tilde{\chi}_n} \right)
\end{equation}
for arbitrary $\ket{\chi_n}$ and $\ket{\tilde{\chi}_n}$ is orthogonal to all physical states:
\begin{equation}
    \braket{\chi|\psi} = \sum_{n=1}^\infty \left( \braket{\chi_n|L_n|\psi} + \braket{\tilde{\chi}_n|\tilde{L}_n|\psi} \right) =0.
\end{equation}
Such states are called spurious, and a state that is both spurious and physical is called null. One may always add to a physical state $\ket{\psi}$ any null state $\ket{\chi}$, since the inner products of $\ket{\psi}$ and $\ket{\psi}+\ket{\chi}$ with any other physical state are guaranteed to be equal. One should therefore identify
\begin{equation}
    \ket{\psi} \cong \ket{\psi} + \ket{\chi},
\end{equation}
which means that the Hilbert space is defined to be
\begin{equation}
    \mathcal{H} = \frac{\mathcal{H}_\text{phys}}{\mathcal{H}_\text{null}}.
\end{equation}
Starting with the closed string, recall that 
\begin{align}
    L_0 &= - \frac{\alpha^\prime M^2}{4} + N \notag\\[5pt]
    \tilde{L}_0 &= - \frac{\alpha^\prime M^2}{4} + \tilde{N}, 
\end{align}
where the level operators
\begin{align}
    N &= \sum_{m=0} \alpha^\mu_{-m} \alpha_{m \mu} = \sum_{m=0} m \, a^{\mu \dagger }_{m} a_{m \mu} \notag\\[5pt]
    \tilde{N} &= \sum_{m=0} \tilde{\alpha}^\mu_{-m} \tilde{\alpha}_{m \mu} = \sum_{m=0} m \, \tilde{a}^{\mu \dagger }_{m} \tilde{a}_{m \mu} 
\end{align}
count the amount of raising operators $\alpha^\mu_{-m}$ and $\tilde{\alpha}^\mu_{-m}$ that appear in each state, weighed by the mode number $m$. The only state with $N=\tilde{N}=0$ is the vacuum $\ket{k}$. It trivially satisfies the physical state condition \eqref{physical} for any $n>0$, while the $n=0$ case sets 
\begin{equation}
    M^2 \ket{k} = \frac{4a}{\alpha^\prime} \ket{k}= - \frac{4}{\alpha^\prime} \ket{k}.
\end{equation}
This tachyonic nature of the vacuum is a puzzling feature of the bosonic string that fortunately is not present in the superstring. 

Upon inserting $a=\tilde{a}$ into \eqref{physical} we recover the $N=\tilde{N}$ level matching condition for the closed string. The general state at level $N=\tilde{N}=1$ is of the form
\begin{equation}
    \xi_{\mu \nu}(k) \alpha^\mu_{-1} \tilde{\alpha}^\nu_{-1} \ket{k}
\end{equation}
where $\xi_{\mu \nu}(k)$ is a polarization tensor. The $n=0$ constraint gives $M^2=0$ for it. Note that any tensor with two Lorentz indices can be decomposed into a symmetric traceless part, an antisymmetric part and a scalar (trace) part, all of which are irreducible under Lorentz transformations, according to
\begin{equation}
    \xi_{\mu \nu}(k) = \left( \xi_{(\mu \nu)}(k) - \frac{1}{D}\xi^\rho{}_\rho(k) \eta_{\mu \nu} \right) + \xi_{[\mu \nu]}(k) + \frac{1}{D} \xi^\rho{}_\rho(k) \eta_{\mu \nu}.
\end{equation}
According to Wigner's classification scheme such irreducible representations correspond to particles, so we find that the closed bosonic string excited in its first level looks much like a graviton, when in the symmetric traceless state. $\xi_{\mu \nu}(k) \alpha^\mu_{-1} \tilde{\alpha}^\nu_{-1} \ket{k}$ satisfies trivially all physical state conditions for $n>1$, whereas the $n=1$ one leads to transversality of the polarization
\begin{equation}
    k^\mu \xi_{\mu \nu}(k)=0.
\end{equation}
At this level, the only spurious states are $ a_\mu \tilde{\alpha}^\mu_{-1} L_{-1} \ket{k}$ and $a_\mu \alpha^\mu_{-1} \tilde{L}_{-1} \ket{k}$, for some $a_\mu$. The third natural candidate $L_{-1}\tilde{L}_{-1} \ket{k}$ turns out to be proportional to either of these with $a_\mu = k_\mu$. Both of them are physical as long as $a_\mu k^\mu = 0$ and $k^2=0$. We should then identify
\begin{align}
    \xi_{\mu \nu}(k) \alpha^\mu_{-1} \tilde{\alpha}^\nu_{-1} \ket{k} &\cong \xi_{\mu \nu}(k) \alpha^\mu_{-1} \tilde{\alpha}^\nu_{-1} \ket{k}+ a_\mu \tilde{\alpha}^\mu_{-1} L_{-1} \ket{k} + b_\mu \alpha^\mu_{-1} \tilde{L}_{-1} \ket{k} \notag\\[5pt]
    &= \xi_{\mu \nu}(k) \alpha^\mu_{-1} \tilde{\alpha}^\nu_{-1} \ket{k} + \sqrt{\frac{\alpha^\prime}{2}} a_\mu \tilde{\alpha}^\mu_{-1} k_\nu \alpha^\nu_{-1} \ket{k} + \sqrt{\frac{\alpha^\prime}{2}} b_\mu \alpha^\mu_{-1} k_\nu \tilde{\alpha}^\nu_{-1} \ket{k} \notag\\[5pt]
    &= \left( \xi_{\mu \nu}(k) + a_\nu k_\mu + b_\mu k_\nu \right) \alpha^\mu_{-1} \tilde{\alpha}^\nu_{-1} \ket{k},
\end{align}
where in going from the first to the second line, we absorbed into $a_\mu$ and $b_\nu$ a factor of $\sqrt{\alpha^\prime / 2}$.

For the traceless symmetric case, we therefore have the equivalence relation
\begin{equation}
    \xi_{(\mu \nu)}(k) \cong \xi_{(\mu \nu)}(k) + a_\mu k_\nu + a_\nu k_\mu \, , \hspace{1cm} a \cdot k = 0.
    \label{equivalence relation}
\end{equation}
To see its consequences, note that a general such tensor has $(D+2)(D-1)/2$ independent components. Transversality $k^\mu \xi_{\mu \nu}(k)=0$ introduces $D$ constraints. As is expected for a massless particle, some polarizations lead to states with negative norm squared. These are the ones for which the nonzero components of $\xi_{\mu \nu}(k)$ are $\xi_{0i}(k)$, where $i = 1,\dots,D-1$:
\begin{align}
    \Big| \xi_{(0 i)}(k) \alpha^0_{-1} \tilde{\alpha}^i_{-1} \ket{k} \Big|^2 &= 2(\xi_{0,i}(k))^2 \braket{0;k| \alpha^0_1 \tilde{\alpha}^0_1 \alpha^0_{-1} \tilde{\alpha}^0_{-1} |0;k}  \notag\\[5pt]
    &= 2(\xi_{0,i}(k))^2 \eta^{00} \eta^{ii} (2 \pi)^D \delta^D(k-k)  \notag\\[5pt]
    &= -2(\xi_{0,i}(k))^2 (2 \pi)^D \delta^D(k-k),
\end{align}
where we normalized the vacuum state according to $\braket{0;k|0;k^\prime}=(2\pi)^D \delta^D(k-k^\prime)$. The identification \eqref{equivalence relation} is what eliminates these problematic states from the spectrum. This is easy to see by going to a frame where the momentum is given by
\begin{equation}
    k^\mu = (E,0,\dots,0,E),
\end{equation}
which is always possible for a massless particle. In this frame the transversality condition becomes $\xi_{(0 \mu)}(k) = - \xi_{(D \mu)}(k)$ and $a \cdot k=0 $ becomes $a_0 = -a_D$. The nontrivial equivalence relations are then
\begin{align}
    \xi_{00}(k) &\cong \xi_{00}(k) - 2 a_0 E \notag\\[5pt]
    \xi_{(0i)}(k) &\cong \xi_{(0i)}(k) - a_i E, \hspace{0.5cm} i=1,\dots D-1.
    \label{graviton equivalence relations}
\end{align}
The second line shows that the negative norm squared states are null and thus do not contribute to any amplitude. We can always pick $a_i$ to cancel the $\xi_{(0,i)}(k)$ part of any state. This is analogous to the elimination of the timelike polarization of the photon in the covariant quantization of QED. The $\xi_{00}(k)$ condition kills one more state, which in this case has positive norm squared. This is like the elimination of the photon's longitudinal momentum. These two conditions combined reduce the number of physical components of the polarization tensor by $1+ (D-2)$, leading to a total of
\begin{equation}
    \frac{(D+2)(D-1)}{2} - D - (1+(D-2)) =
    \frac{D(D-3)}{2}
\end{equation}
physical polarizations. This is what one expects for an on-shell graviton, giving for example $2$ in $D=4$. From the spacetime point of view, the unique noninteracting action that gives rise to spin two particles with these properties is the massless Fierz-Pauli action \cite{Schwartz_2013}
\begin{equation}
    S = \frac{1}{2 \kappa^2} \int d^Dx \bigg( \partial_\mu h^\rho{}_\rho \partial_\nu h^{\mu \nu} - \partial^\rho h^{\mu \nu} \partial_\mu h_{\rho \nu}  + \frac{1}{2} \partial_\rho h_{\mu \nu} \partial^\rho h^{\mu \nu} - \frac{1}{2} \partial_\mu h^\nu{}_\nu \partial^\mu h^\rho{}_\rho \bigg)
\end{equation}
for a symmetric traceless tensor field $h_{\mu \nu}$ satisfying the gauge symmetry
\begin{equation}
    h_{\mu \nu} \to h_{\mu \nu} + \partial_\mu \lambda_\nu + \partial_\nu \lambda_\mu.
    \label{graviton gauge symmetry}
\end{equation}
This happens to be the expansion of the Einstein-Hilbert action for a metric given by $G_{\mu \nu}(x) = \eta_{\mu \nu} + h_{\mu \nu}(x)$ to second order in the graviton field $h_{\mu \nu}(x)$, which is known to satisfy the gauge symmetry 
\begin{equation}
    g_{\mu \nu} \to g_{\mu \nu} + \nabla_\mu \lambda_\nu + \nabla_\nu \lambda_\mu,
\end{equation}
whose linearized form is \eqref{graviton gauge symmetry}. Although not a derivation, this is a strong sign that general relativity emerges from string theory. This will be shown later by an indirect method.

Besides the graviton $h_{\mu \nu}$, we also get from the irreducible decomposition of $\xi_{\mu \nu}(k)$ a massless antisymmetric tensor and a massless scalar. The correspondent spacetime fields are the so-called Kalb-Ramond field $B_{\mu \nu}(x)$, whose linearized gauge symmetry is
\begin{equation}
    B_{\mu \nu} \to B_{\mu \nu} + \partial_{\mu} \lambda_{\nu} - \partial_{\nu} \lambda_{\mu} ,
\end{equation}
and the scalar dilaton $\Phi(x)$. It will be useful later to know the number of physical degrees of freedom of the Kalb-Ramond field. A general antisymmetric tensor has $D(D-1)/2$ independent components. Transversality this time only adds $D-1$ constraints, since $\xi_{00}(k)=0$ by antisymmetry. For the same reason, the equivalence relation related to $\xi_{00}$ is absent, so one gets only the $D-2$ constraints from the lower line of \eqref{graviton equivalence relations}. The number of physical polarizations is therefore
\begin{equation}
    \frac{D(D-1)}{2} - (D-1) - (D-2) = \frac{(D-2)(D-3)}{2}.
\end{equation}

The higher excitation levels of the bosonic string form particle representations of higher spins due to the extra spacetime indices from acting on the vacuum with more mode operators. Writing the $n=0$ Virasoro condition as $M^2 = \frac{4}{\alpha^\prime}  \left( N-1 \right)$ shows that these are all massive, which is a necessary condition for the consistent quantization of fields of spin higher than 2 \cite{PhysRev.135.B1049,PhysRev.140.B516}. We will not need their explicit form.

We now move on to open strings, for which we have only one set of Virasoro conditions $(L_n - \delta_{n0}) \ket{\psi}=0$. Consider first a string with free endpoints. The vacuum $\ket{k}$ satisfies all Virasoro conditions for $n>1$, whereas the $n=0$ one, which for the open string can be written as
\begin{equation}
    M^2 = \frac{1}{\alpha^\prime} \Big( N - 1 \Big)
\end{equation}
sets the mass of the vacuum to $M^2=-1/\alpha^\prime$. Once again we find the ground state to be a tachyonic scalar particle. At level $N=1$ the general state is of the form
\begin{equation}
    e_\mu(k) \alpha^\mu_{-1} \ket{k}.
\end{equation}
Similarly to the first level of the closed string, the $n=0$ constraint requires this state to be massless and the $n=1$ constraint requires it to be transverse, $e_\mu(k) k^\mu=0$. The only spurious state at this level is $L_{-1} \ket{k}$, which is physical if $k^2=0$. Therefore we identify
\begin{align}
    e_\mu(k) \alpha^\mu_{-1} \ket{k} &\cong e_\mu(k) \alpha^\mu_{-1} \ket{k} + a L_{-1} \ket{k} \notag\\[5pt]
    &=e_\mu(k) \alpha^\mu_{-1} \ket{k} + \sqrt{2 \alpha^\prime} a k_\mu \alpha^\mu_{-1} \ket{k} \notag\\[5pt]
    &= \left( e_\mu(k) +a k_\mu \right) \alpha^\mu_{-1} \ket{k},
    \label{open bosonic equivalence}
\end{align}
where in the last line we absorbed $\sqrt{2 \alpha^\prime}$ into the constant $a$. The equivalence relation
\begin{equation}
    e_\mu(k) \cong e_\mu(k) + ak_\mu
\end{equation}
corresponds precisely the gauge invariance of a $U(1)$ gauge field. An NN open string excited in its first level therefore behaves like a 26-dimensional photon $A_\mu(x)$, with gauge symmetry given by
\begin{equation}
    A_\mu \to A_\mu + \partial_\mu \lambda.
\end{equation}
Once again we have also an infinite tower of higher excited states, which are massive and whose form will not be needed. 

Lastly we discuss open strings with endpoints attached to D-branes, which means that DD boundary conditions are imposed for some of the spacetime directions. The general one string vacuum state in this case is given by $\ket{k;ij}$, with the indices $i$, $j$ denoting on which brane the left and right endpoints are attached, respectively (see figure \ref{D-brane figure}). The $L_0$ condition gives the mass relation\footnote{Recall from Section \ref{sec32} that $y^I_i$ is position of one brane and $y_j^I$ that of the other, in the directions perpendicular to them, and $\Delta y^2 = (y_2 - y_1)^I(y_2 - y_1)^I$.}
\begin{equation}
    M^2 = \frac{\Delta y^2}{(2 \pi \alpha^{\prime})^2} +  \frac{1}{\alpha^\prime} \Big( N - 1 \Big).
    \label{open string DD mass relation}
\end{equation}
As mentioned after equation \eqref{Dirichlet sol}, the components of the spacetime momentum in the DD directions vanish, so all particle states found in the spectrum only propagate inside the branes.

There are two qualitatively different types of vacuum: $\ket{k;ii}$, the ground state of a string whose endpoints lie on the same brane, and $\ket{k;ij}$ with $i\neq j$, the ground state of a string stretched between two different branes. $\ket{0;k;ii}$ has exactly the same tachyonic mass the vacuum of the NN string, whereas $\ket{k;ij}$ may or may not be tachyonic depending on how separated the branes are. At level $N=1$, two kinds of states can be built,
\begin{equation}
    e_\mu(k) \alpha^\mu_{-1} \ket{k;ij} \hspace{0.5cm} \text{or} \hspace{0.5cm} \alpha^I_{-1} \ket{k;ij},
\end{equation}
Their mass is given by
\begin{equation}
    M^2 = \frac{\Delta y^2}{(2 \pi \alpha^\prime)^2}.
\end{equation}
From the point of view of the reduced Lorentz group inside the worldvolume of the branes, the states of the first kind form a vector, while the states of the second kind are scalars. We have
\begin{equation}
    L_1 = \sqrt{2 \alpha^\prime} \alpha^\mu_{1} p_\mu + \frac{(y_j^I - y_i^I)}{\pi \sqrt{2 \alpha^\prime}} \alpha^I_1 + \dots,
\end{equation}
where the dots contain operators that annihilate all level $1$ states. For the vector, the $L_1$ condition sets $e_\mu(k) k^\mu=0$. For the scalars it instead gives $y^I_i = y^I_j$, which means that they are only physical if both endpoints lie on the same brane. The only $N=1$ spurious state is 
\begin{equation}
    L_{-1} \ket{k;ij} = \sqrt{2 \alpha^\prime} k_\mu \alpha^\mu_{-1} \ket{k;ij} + \frac{(y_j^I - y_i^I)}{\pi \sqrt{2 \alpha^\prime}} \alpha^I_{-1} \ket{k;ij}.
\end{equation}
From the $n=1$ Virasoro condition we know that for $y^I_i \neq y^I_j$ this is not physical, so there are no null states at level 1 for a string going from one brane to another. If both endpoints are on the same brane, this state becomes physical and only the term with $k_\mu$ remains, giving precisely the same structure as the NN case, since for $y^I_i=y^I_j$ the vector also becomes massless.

Overall one finds on the first two levels of an open string stretched between branes one massive vector and one possibly tachyonic scalar. For a string with both endpoints on the same brane, we get a scalar field for each direction perpendicular to the brane plus a lower-dimensional copy of what was found for the NN string: a tachyon and a massless gauge boson. In both cases there are of course also the higher spin massive particles from the other levels.

A particularly important concept for AdS/CFT is that of a stack of branes. One may understand this construction as the result of starting with $N$ branes of the same dimension at different positions and then taking the limit in which all branes approach each other.

\begin{figure}[h]
\begin{center}
\includegraphics[width= 0.95\textwidth]{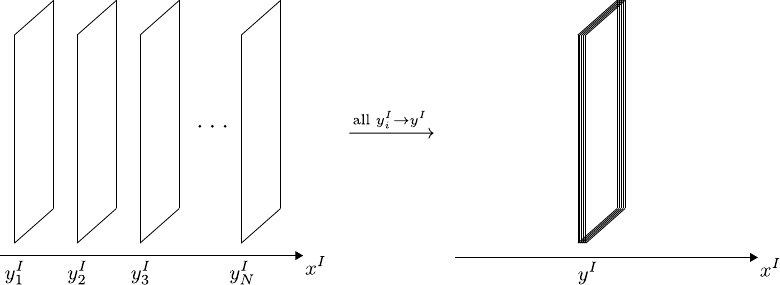}
\caption[Brane stack limit]{Brane stack limit.}
\end{center}
\end{figure}

\begin{figure}[h]
\begin{center}
\includegraphics[width= 0.3\textwidth]{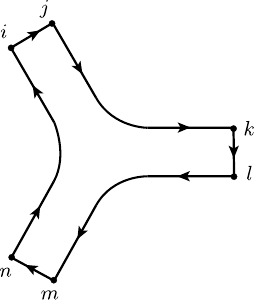}
\caption[Open string three-point function with Chan-Paton factors]{Open string three-point function with Chan-Paton factors. The arrows point
from one endpoint to the other. The similarity with the gauge theory three point function
in double line notation (figure \ref{double line 3pt function}) is noteworthy.}
\label{fig37}
\end{center}
\end{figure}

This makes all factors of $y^I_j - y^I_i$ disappear from the previous expressions, without setting $i=j$ on the states themselves. Of course, if all branes lie on top of each other there is no sense in which one can be distinguished from another, so the $i$, $j$ indices lose their geometric meaning and become simply a new pair of discrete degrees of freedom associated to the string endpoints. In this context they are called Chan-Paton indices. All that can be said about the endpoints of some open string state in a stack of branes is that both lie in the worldvolume of the branes. There is no way to differentiate a particular state $ \alpha^{\mu_1}_{n_1} \alpha^{\mu_2}_{n_2} \dots \ket{k;12}$ from $\alpha^{\mu_1}_{n_1} \alpha^{\mu_2}_{n_2} \dots\ket{k;13}$ because their masses are equal. To this $N^2$-degeneracy in the spectrum is associated a symmetry: the overlap of any two states is invariant under the ``rotation'' of the Chan-Paton indices
\begin{equation}
    \ket{N;k;ij} \to U^1_{i r} U^2_{js} \ket{N;k;rs},
    \label{Chan-Paton sym}
\end{equation}
where $U^1_{ir}$ and $U^2_{js}$ are $U(n)$ matrices. Dynamical considerations reduce this apparent $U(N) \times U(N)$ symmetry down to $U(N)$. To see why, it is useful to go back to considering a system of $n$ separated branes that do not intersect at any points. A particular open string in a state $\alpha^{\mu_1}_{n_1} \alpha^{\mu_2}_{n_2} \dots\ket{k;11}$ can only move inside the first brane. It is not allowed, for instance, to decay into strings propagating along different branes, at least at leading order in perturbation theory\footnote{At higher order, meaning with more than one insertion of the string's three point function, there can be exchanges of strings between different branes. One may for instance have an open string propagating along one brane such that its endpoints meet and form a closed loop, turning it into a closed string that detaches from the brane. It may then decay into a pair o closed strings each with some nonzero momentum in the directions transverse to the branes, such that one of them eventually reaches another brane and is absorbed by it, turning again into open string excitations. One such process where branes emit and absorbed closed strings will be studied in Chapter \ref{ch6}.}. The same holds for a string stretched between two distinct branes, there is no way for it to change to what branes it is attached to if the different branes never intersect. This means that the index structure of the string propagator is simply a pair of Kronecker deltas enforcing the conservation of the index along each endpoint. This trivial dependence on the indices is insensitive to the spatial arrangement of the branes, and it survives the limit in which all are on top of each other. With this in mind, consider the open string three-point vertex in a stack of $N$ branes shown in figure \ref{fig37}.
The fact that the states satisfy the symmetry \eqref{Chan-Paton sym} means that each endpoint is charged with respect to its $U(N)$. Assume time flows from left to right. The endpoints with indices $i$ and $n$ meet and become a regular bulk point of the worldsheet, which carries no Chan-Paton indices. If one views each point of the string as a propagating particle, this corresponds to two $U(N)$-charged particles annihilating into a singlet, which is only consistent if the initial particles had opposite charges. Consistency of the three point function therefore requires that if the left endpoint transforms in the fundamental representation of $U(n)$, the right one should transform in the antifundamental
\begin{equation}
    \ket{k;ij} \to U_{i r} \ket{k;rs} U^{\dagger}_{s j}.
\end{equation}
Each state of the spectrum in a stack of $N$ branes therefore transforms in the adjoint representation of $U(N)$. The adjoint of $U(1)$ is trivial and $U(N) =U(1) \times SU(N)$, so all states are neutral under the $U(1)$ factor and transform nontrivially only under the $SU(N)$ part. At the lowest level we get the tachyonic vacuum with mass $M^2 = -1/\alpha^\prime$. On the next level we get the same vector and set of scalars that were found for the general brane setup, but this time, since $y^I_i=y^I_j$, they are all massless and physical, and there are $N^2$ copies of each. 

It is an interesting fact that Chan-Paton factors actually predate the entire concept of D-branes by two decades.\footnote{D-branes were introduced in 1989, when string theory was already regarded as theory of quantum gravity \cite{Dai:1989ua}, whereas the paper by Chan and Paton \cite{Paton:1969je} was published in 1969.} They were proposed not long after the discovery of the Veneziano amplitude, as a way of introducing flavor degrees of freedom at the endpoints of the mesonic string \cite{Paton:1969je}. The unitary symmetry seen in the spectrum was then interpreted as $SU(2)_\text{flavor}$ or $SU(3)_\text{flavor}$, depending on the number of light quarks included. It turns out that in the spacetime dynamics of open strings this rigid $SU(N)$ symmetry of the states becomes a gauge symmetry. This is in fact required for the theory to be consistent, since the fact that all excitations transform in the adjoint of $SU(N)$ means that, as long as $N>1$, the massless vector found in the spectrum is a nonabelian gauge boson. The field theory action that gives rise to such particles as excitations is necessarily Yang-Mills, possibly with higher dimension gauge-invariant operators added. The appearance of nonabelian gauge dynamics in string theory is fundamental for AdS/CFT, and will be considered in more detail in Section \ref{sec52}.

%\end{document}

%\documentclass[a4paper,12pt]{memoir}
%\usepackage{graphicx}
%\usepackage[utf8]{inputenc}
%\usepackage{indentfirst}
%\usepackage{braket}
%\usepackage{setspace}
%\usepackage{amsmath, amsthm, amssymb, amsfonts,bm}
%\usepackage[multiple]{footmisc}
%\usepackage{mathtools, changepage, slashed}
%\usepackage{tikz-feynman}
%\usepackage{bm, mathrsfs}
%\usepackage{gensymb}
%\usepackage[a4paper,top=3cm,left=3cm,right=2cm,bottom=2cm]{geometry}
%\usepackage{epstopdf}
%\usepackage{hyperref}
%\usepackage{pgfplots}
%\pgfplotsset{compat=1.18} 
%\usepackage[sorting=none]{biblatex}
%\addbibresource{refs.bib}
%\numberwithin{equation}{section}
%\usepackage[inkscapelatex=false]{svg}
%\usepackage[super]{natbib}
%\usepackage{doi}
%\hypersetup{
%  colorlinks   = true, %Colours links instead of ugly boxes
%  urlcolor     = black, %Colour for external hyperlinks
%  linkcolor    = black, %Colour of internal links
%  citecolor   = black %Colour of citations
%}
%\DeclareMathOperator{\Tr}{Tr}

%\newcommand{\normalord}[1]{%
%  {:\mathrel{\mspace{1mu}#1\mspace{1mu}}:}%
%}

%\OnehalfSpacing
%\usepackage{newtx}
%\usepackage{newtxtext}
%\usepackage{lmodern}

%\title{T-duality chapter}
%\author{pedrobairrao}

%\begin{document}

\chapter{Compactification and T-duality}\label{ch4}

\section{T-duality for closed strings}\label{sec41}

The starting point for constructing the Polyakov action was the description of a relativistic string propagating in Minkowski space, and consistency of the quantum theory requires that this Minkowski space be 26-dimensional. We will see in Chapter \ref{ch7} that the same situation happens with the superstring, although with a different critical dimension. This is a challenge that must be overcome in any attempt to describe lower-dimensional physics with string theory.\footnote{There is an additional complication due to the tachyon, since its presence in the spectrum means that the Minkowski vacuum we expanded around is actually unstable and should decay into some other target space configuration. This process is referred to as tachyon condensation in string theory literature. Since we are using the bosonic string only as a toy model of the superstring, which has no tachyon, we will ignore this issue. \label{tachyon condensation footonote} } One possible strategy, called compactification, is to study string theory over 26-dimensional manifolds of the form
\begin{equation}
    M^{26} = \mathbb{R}^{1,3} \times K^{22},
\end{equation}
where $\mathbb{R}^{1,3}$ is flat four-dimensional Minkowski space and $K^{22}$ some compact space. The idea is that as long as $K^{22}$ is made small enough, attaching a copy of it to each point of a four-dimensional space allows one to still avoid the conformal anomaly while at the same time describing the dynamics of strings that, at length scales large enough to make the entirety of $K^{22}$ shrink to a point, appears four-dimensional.

\begin{figure}[h]
\begin{center}
\includegraphics[width= 0.55\textwidth]{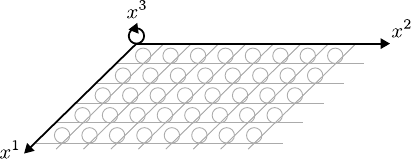}
\caption[Representation of $\mathbb{R}^2 \times S^1$]{Representation of $\mathbb{R}^2 \times S^1$. Over each point of the $(x^1,x^2)$ plane there is a small circle spanned by the coordinate $x^3$. At large enough length scales these circles become indistinguishable from points, and the total space appears to be just $\mathbb{R}^2$.}
\end{center}
\end{figure}

Any symmetries of $M ^{26}$ coming from transformations of the compact space $K^{22}$ would, in this large distance limit, appear to be internal symmetries of the system. This idea is the basic content of Kaluza-Klein theory, which started with papers by Kaluza and Klein showing that the low-energy (large distance) limit of general relativity in five dimensions, with one dimension assumed to be a circle of small radius, was given by four-dimensional gravity coupled to electromagnetism and an additional scalar field, the Kaluza-Klein dilaton \cite{Kaluza:1921tu, Klein:1926tv}.\footnote{See \cite{Appelquist:1987nr} for translations to English.} 

For our purposes it will be enough to study only the simplest possible compactification, where $K^{22}$ is taken to be a 22-torus 
\begin{equation}
    \mathbb{T}^{22} = \underbrace{S^1 \times \dots \times S^1}_{\text{22 times}}
\end{equation}
with each circle $S^1$ having radius $R$. The line element in $\mathbb{R}^4 \times \mathbb{T}^{22}$ is identical to that of $\mathbb{R}^{26}$, so the worldsheet action, energy-momentum tensor and equations of motion are the same as before. The classical solutions in the compact directions must now respect the target space periodicity
\begin{equation}
    X^m \cong X^m + 2 \pi R \, , \hspace{0.5cm} m=4, \dots, 26.
\end{equation}
The operator that translates $X^m$ to $X^m+ 2 \pi R$ is $\exp(2 \pi i R p^m)$, where $p^m$ is the component of the center of mass momentum in the $m$-th direction. Asking that it leaves the states invariant requires the momenta to be quantized according to
\begin{equation}
    k^m = \frac{n^m}{R}, \hspace{0.5cm} n^m \in \mathbb{Z}.
\end{equation}
If a particular state satisfies
\begin{equation}
    k^M k_M =k^\mu k_\mu + k^m k^m = - M^2_0, \hspace{0.5cm}  M=(\mu,m),
\end{equation}
with the $k^m$ components of the momentum quantized, its energy $k^0 \equiv E$ satisfies
\begin{equation}
    E^2 = M_0^2 + k^i k^i + \frac{n^m n^m}{R^2},
\end{equation}
where $i=1,2,3$ and all repeated indices are summed over. In terms of the motion through the four-dimensional extended directions, the momenta along the compact dimensions act as a contribution to the mass. At energy scales small compared to $1/R$, all states with $n^m \neq 0$ cannot be excited and effectively decouple from the dynamics. The ones that are left have $k^m=0$ and therefore no $x^m$-dependence on their position-space wave functions, so their dynamics is effectively four-dimensional. 
This forms the basis of Kaluza-Klein theory.

Another consequence of having some directions curled up into a circle is that a closed string can wind around a compact dimension an integer number of times before returning to its starting point. This possibility is expressed by the boundary condition

\begin{figure}[t]
\begin{center}
\includegraphics[width= 0.8\textwidth]{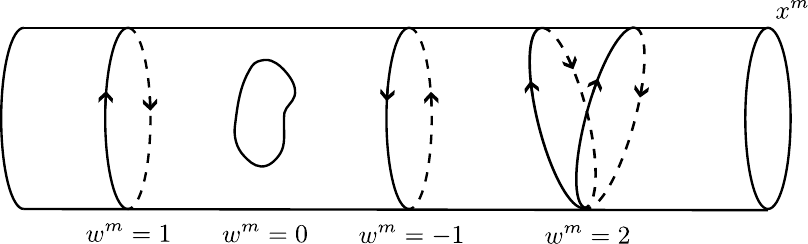}
\caption[Closed strings with different winding numbers]{Closed strings with different winding numbers. The arrows represent in what direction the strings wind around the compact dimension.}
\end{center}
\end{figure}
\begin{figure}[b]
\begin{center}
\includegraphics[width= 0.8\textwidth]{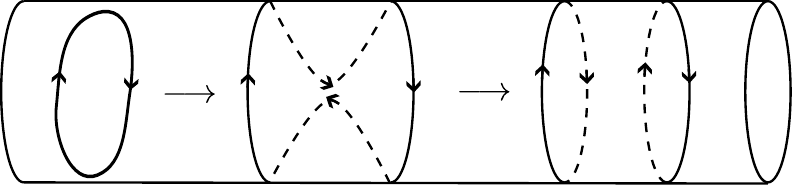}
\caption[Dynamical process exemplifying winding conservation]{Dynamical process exemplifying winding conservation. A closed string with $w^m=0$ splits off into two strings, one with $w^m = 1$ and another with $w^m=-1$. }
\label{winding conservation}
\end{center}
\end{figure}


\begin{equation}
    X^m(\tau,\sigma + 2 \pi) = X^m(\tau,\sigma) + 2 \pi R w^m, \hspace{0.5cm} w^m \in \mathbb{Z}.
    \label{winding BD}
\end{equation}
The integers $w^m$ that count the amount of times the strings wind around each compact direction are called winding numbers. One may convince oneself that these are conserved quantities by picturing a closed string that splits into a pair, as in figure \ref{winding conservation}.
For the splitting to occur, the points of the $w^m=0$ string must approach each other until they touch, resulting in the string pinching off at that point. If this happens like in figure \ref{winding conservation}, the arrows make it clear that at the end one of the strings has $w^m=+1$ while the other one has $w^m=-1$, adding up to the $w^m=0$ of the original configuration.

The solution to the equations of motion compatible with \eqref{winding BD} is
\begin{equation}
    X^m(\tau,\sigma) = x^m + \alpha^\prime \left( \frac{ n^m}{R} \right) \tau + w^m R \sigma + i \sqrt{\frac{\alpha^\prime}{2}} \sum_{n \neq 0} \frac{1}{n} \Big( \alpha^m_n e^{-in(\tau-\sigma)} + \tilde{\alpha}^m_n e^{-in(\tau + \sigma)}\Big).
\end{equation}
Since these boundary conditions only differ from the noncompact closed string ones by a constant, the derivatives of $X^m$ still satisfy the usual periodicity and thus can still be written in the form \eqref{derivatives X}. Comparing with the expansion above fixes
\begin{equation}
    \alpha^m_0 = \sqrt{\frac{\alpha^\prime}{2}} \bigg( \frac{n^m}{R} - \frac{w^m R}{\alpha^\prime} \bigg), \hspace{0.5cm} \tilde{\alpha}^m_0 = \sqrt{\frac{\alpha^\prime}{2}} \bigg( \frac{n^m}{R} + \frac{w^m R}{\alpha^\prime} \bigg).
\end{equation}
After plugging these results into the formula for $L_0$ and $\tilde{L}_0$, the $(L_0-1)\ket{\psi} = (\tilde{L}_0 -1)\ket{\psi}=0$ physical state condition is found to be equivalent to
\begin{align}
    M^2 &= \frac{n^m n^m}{R^2} + \frac{w^m w^m R^2}{\alpha^{\prime 2}} + \frac{2}{\alpha^\prime} \Big( N + \tilde{N} - 2 \Big) \notag\\[5pt]
    N &= \tilde{N} + n^m w^m,
    \label{compact mass spectrum}
\end{align}
where $M^2 = - p_\mu p^\mu$ involves only the continuous momenta. The contribution of the compact momenta to the mass is the same that is found for point particles, but the contribution from the winding number is exclusive to strings. The closed spectrum can be constructed along the lines of Section \ref{sec34}, but its detailed form will not be needed. Note that as the radius is decreased, the first term in $M^2$ gets increasingly large, leading ultimately to the decoupling of the $n^m \neq 0$ states that was mentioned before. On the other hand, the second term gets smaller, making states with high winding number increasingly accessible. This is reasonable because the string's energy grows with its length, so it should cost very little energy for a string to wind around a small circular dimension. Also, as the radius is reduced the energy spacing between each winding eigenstate gets smaller, and in the $R \to 0$ limit the winding spectrum tends to a continuum. In the limit of large radius the opposite happens: the momenta $p^m = n^m/R$ approach the continuous spectrum seen in the uncompactified theory, while the winding states become extremely massive, expressing the fact that it costs a lot of energy to wrap a string around a very large dimension. This similarity of both limits is captured by the fact that the spectrum \eqref{compact mass spectrum} is invariant under the simultaneous inversion of the radius and swapping of compact momentum and winding
\begin{equation}
    R \to R^\prime = \frac{\alpha^\prime}{R}, \hspace{0.5cm} n^m \leftrightarrow w^m,
    \label{Tduality on spectrum}
\end{equation}
which is equivalent to doing $\alpha^m_0 \to - \alpha^m_0$, while $\tilde{\alpha}^m_0$ stays the same. 

This points to the existence of an equivalence between string theory defined on a background of compactification radius $R$ and the theory on a different background with radius $R^\prime = \alpha^\prime /R$, although an analysis of just the zero-modes $\alpha^m_0$ and $\tilde{\alpha}^m_0$ is not enough to assert this. As is required by the equation of motion $\partial^2 X^M =0$, the fields $X^M$ all split into the sum of a right-moving part $X^M_R(\tau-\sigma)$ and a left-moving part $X^M_L(\tau+\sigma)$. For a closed string coordinate in a compact direction one has
\begin{align}
    X^m_R(\tau-\sigma) &= \frac{1}{2} ( x^m -c^m ) + \frac{ \alpha^\prime}{2} \Big( \frac{n^m}{R} - \frac{w^m R}{\alpha^\prime} \Big) ( \tau - \sigma ) + i \sqrt{\frac{\alpha^\prime}{2}} \sum_{n \neq 0} \frac{\alpha^m_n}{n} e^{-in(\tau-\sigma)} \notag\\[5pt]
    X^m_L(\tau+\sigma) &= \frac{1}{2} ( x^m + c^m ) + \frac{ \alpha^\prime}{2} \Big( \frac{n^m}{R} + \frac{w^m R}{\alpha^\prime} \Big) ( \tau + \sigma ) + i \sqrt{\frac{\alpha^\prime}{2}} \sum_{n \neq 0} \frac{\tilde{\alpha}^m_n}{n}e^{-in(\tau + \sigma)},
\end{align}
where $c^m$ is a constant. One way to implement the transformation \eqref{Tduality on spectrum} on the full form of the worldsheet fields is by taking $X^m(\tau,\sigma) \to X^{\prime m}(\tau,\sigma)$, where
\begin{align}
    X^{\prime m}(\tau,\sigma) &\equiv - X_R^m(\tau - \sigma) + X_L^m(\tau + \sigma) \notag\\[5pt]
    &= c^m + \alpha^\prime \left( \frac{ w^m}{R^\prime} \right) \tau + n^m R^\prime \sigma + i \sqrt{\frac{\alpha^\prime}{2}} \sum_{n \neq 0} \frac{1}{n} \Big( -\alpha^m_n e^{-in(\tau-\sigma)} + \tilde{\alpha}^m_n e^{-in(\tau + \sigma)}\Big).
\end{align}
In terms of the modes, taking $X^m \to X^{\prime m}$ is equivalent to
\begin{equation}
    x^m \to c^m, \hspace{0.5cm} \alpha^m_n \to - \alpha^m_n, \hspace{0.5cm} \tilde{\alpha}^m_n \to  \tilde{\alpha}^m_n,
\end{equation}
whose $n=0$ term is precisely \eqref{Tduality on spectrum}. Clearly $X^{\prime m}$ satisfies the same equations of motion as $X^m$, and the Hamiltonians built from $X^m$ and $X^{\prime m}$ are identical, since they are quadratic in the modes and do not involve the center of mass position. Using the canonical commutation relations of the modes it is straightforward to verify that $X^{\prime m}$ and its momentum density $\Pi^{\prime m} = - \partial^\tau X^{\prime m} / (2 \pi \alpha^\prime)$ satisfy
\begin{equation}
    \big[ X^{\prime m}(\tau,\sigma) , \Pi^{\prime m^\prime}(\tau,\sigma^\prime) \big] = i \delta^{m m^\prime} \delta(\sigma - \sigma^\prime),
\end{equation}
as long as one sets
\begin{equation}
\left[ c^m , \frac{w^{m^\prime}}{R^\prime} \right] = i \delta^{m m^\prime},
\end{equation}
which is nothing but the image of $[x^m,p^{m^\prime}] = [x^m, n^{m^\prime} /R ]$ under the $X^m \to X^{\prime m}$ map. This, together with the equality of the Hamiltonians, means that either the $X^m$ or the $X^{\prime m}$ are equally valid degrees of freedom for the same theory. The difference between using $X^m$ and $X^{\prime m}$ is that the former describes strings on a compactified background of radius $R$, while the latter describes strings on a background of radius $\alpha^\prime / R$. The fact these two setups result in the same physics is a nontrivial property of string theory called T-duality.\footnote{The T in T-duality is usually said to stand for target space, or toroidal, but the original reason for using the letter T was the fact that the Kaluza-Klein dilaton, a scalar field which emerges in the spacetime action of a compactified field theory and acts as a dynamical compactification radius, was commonly called $T$ \cite{Font:1990gx}.} $X^{m}$ and $X^{\prime m}$ are said to be T-duals of each other.

\section{Open strings and D-branes}\label{sec42}

T-duality means that in the limit of vanishing compactification radius the closed string spectrum becomes identical to that of the uncompactified theory, with compact momentum swapped for winding number. Utilization of the T-dual coordinates $X^{\prime m}$ in place of the original $X^m$ simply undoes this swapping, giving back a closed string theory in $26$ flat dimensions. This is very different from what happens in quantum field theory, where compactifying some dimensions and sending their radius to zero is a way to eliminate them from the theory entirely. 

It is then natural to wonder what happens to open strings in toroidal compactification. Since open strings have no winding number (they can always be unwound from a compact dimension), the spectrum for NN boundary conditions is given simply by 
\begin{equation}
    M^2 = \frac{n^m n^m}{R^2} + \frac{1}{\alpha^\prime} \Big( N - 1 \Big).
\end{equation}
There is no new degree of freedom to approach a continuum as the radius gets smaller, the $R \to 0$ limit simply gives open strings in a lower-dimensional space, just like what happens in field theory. One might then think that open strings would be incompatible with T-duality, but this is not the case. To see why, split the open string solution
\begin{equation}
        X^m(\tau,\sigma) = x^m +  2  \alpha^{\prime} \left( \frac{n^m}{R} \right) \tau + i \sqrt{2 \alpha^\prime} \sum_{n \neq 0} \frac{\alpha^m_n}{n} e^{-i n \tau} \cos \left( n \sigma \right).
    \end{equation}
into $X^m_R(\tau-\sigma) + X^m_L(\tau + \sigma)$, where
\begin{align}
    X^m_R(\tau - \sigma) &= \frac{1}{2} ( x^m - c^m ) +  \alpha^\prime \left( \frac{n^m}{R} \right) (\tau - \sigma) + i \sqrt{\frac{\alpha^\prime}{2}} \sum_{n \neq 0} \frac{\alpha^m_n}{n} e^{-in(\tau - \sigma)} \notag\\[5pt]
    X^m_L(\tau + \sigma) &= \frac{1}{2} ( x^m + c^m ) +  \alpha^\prime \left( \frac{n^m}{R} \right) (\tau + \sigma) + i \sqrt{\frac{\alpha^\prime}{2}} \sum_{n \neq 0} \frac{\alpha^m_n}{n} e^{-in(\tau + \sigma)},
\end{align}
and form the T-dual coordinate
\begin{align}
    X^{\prime m}(\tau, \sigma) &= - X_R^m(\tau - \sigma) + X_L^m(\tau + \sigma) \notag\\[5pt]
    &= c^m + 2 n^m R^\prime \sigma + \sqrt{2 \alpha^\prime} \sum_{n \neq 0} \frac{\alpha^m_n}{n} e^{-in \tau} \sin (n \sigma).
\end{align}
Once again one finds that $X^{\prime m}$ and its momentum density satisfy the canonical commutation relations, and lead to the same Hamiltonian as the original $X^m$. T-duality should therefore still hold for open strings. Comparison with \eqref{Dirichlet sol} reveals that $X^{\prime m}$ describes an open string with Dirichlet boundary conditions, with endpoints fixed at positions
\begin{equation}
    x^{\prime m}_1 = c^m, \hspace{0.5cm} x^{\prime m}_2 = c^m + 2 \pi n^m R^\prime.
\end{equation}
D-branes thus naturally appear in the T-dual description of open strings with compact dimensions. One also recovers in the T-dual picture a sense in which the discrete momentum becomes a winding number, since open strings with fixed endpoints cannot unwind themselves from the circular dimension, and $n^m$ counts the amount of times a particular string circles around it before attaching back to the brane.

\begin{figure}[t]
\begin{center}
\includegraphics[width= 0.85\textwidth]{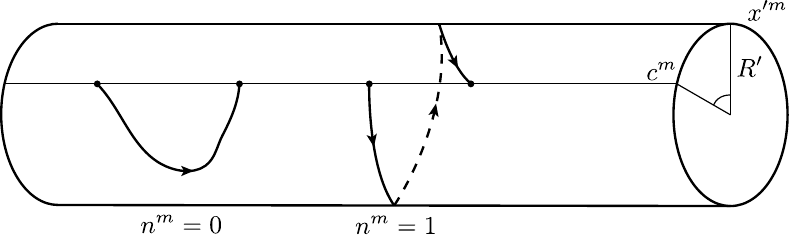}
\caption[T-dual picture of open strings]{T-dual picture of open strings with compact momenta $n^m=0$ and $n^m=1$. In the T-dual spacetime the strings have their endpoints attached to a brane at $x^{\prime m}_ 1=c^m$. The winding momenta count how many times the string winds around the compact direction before going back to the brane.}
\end{center}
\end{figure}

The $R \to 0$ limit, in which the $p^m \neq 0$ strings decouple, corresponds to $R^\prime \to \infty$ in the T-dual picture, unwrapping the compact dimension into a flat one. This makes all $n^m \neq 0$ strings, the ones that wind around the compact dimension before coming back to the brane, infinitely long and therefore infinitely massive. One is left with a state containing one brane at position $x^{\prime m} = c^m$ in a noncompact dimension and open strings attached to it. 

While this construction allows us to construct single brane states, it is not capable of producing multiple branes. Both endpoints of the string along a compact direction sit in the same spacetime point, due to the periodicity: $(x^\prime_2 - x^\prime_1)^m = 2 \pi n^m R^\prime \cong 0 \text{ (mod $2 \pi R^\prime$)}$. Relative to the parameters of the original theory (before T-duality), we have
\begin{equation}
    (x^\prime_2 - x^\prime_1)^m = 2 \pi \alpha^\prime \left( \frac{n^m}{R} \right) = 2 \pi \alpha^\prime p^m.
\end{equation}
If we find a way to shift the momentum $p^m$ by a constant amount, this should, by the formula above, correspond in the T-dual theory to separated branes. Note that if the string happens to be charged under a $U(1)$ gauge field $A_M$, a nonzero value for its components in the compact directions is expected to shift the momenta in the Hamiltonian according to
\begin{equation}
    p^m \to p^m - q A^m,
\end{equation}
where $q$ is the string's charge. At the massless level of the open string spectrum there is such a gauge field, under which we saw that the string's endpoints are oppositely charged. Assume that one adds Chan-Paton indices going over $N$ values, so that the gauge group is $SU(N)$. We now consider a state with open strings in a coherent background of the gauge field such that in the compact directions $A_m$ is a constant, and in the flat directions $A_\mu =0$. Under a gauge transformation $\Omega \in SU(N)$,
\begin{equation}
    A_m \to \Omega A_m \Omega^\dagger + i \Omega \partial_m \Omega^\dagger, 
\end{equation}
with the charge normalized to 1. Choosing a constant $\Omega$, so that the last term drops out, we are left with just the action of a unitary transformation on the hermitian matrix $A_m$. Any hermitian matrix can be diagonalized by a unitary matrix, so without loss of generality we can choose $\Omega$ so that the gauge field takes the form
\begin{equation}
    \hspace{0.5cm} A_m = - \frac{1}{2 \pi R} \text{diag}\left( \theta^m_1,\theta^m_2, \dots , \theta^m_N \right), \hspace{0.5cm} \theta_i \in \mathbb{R}.
\end{equation}
This sits in the abelian $U(1)^N$ subgroup of $SU(N)$. An open string in a state built over the $\ket{k;ij}$ vacuum has one endpoint coupled to the $i$-th $U(1)$, with charge $+1$, and the other coupled to the $j$-th $U(1)$, with charge $-1$. As stated in Section \ref{sec31}, any particle charged under a $U(1)$ gauge field feels this charge via a $q A_M(x) dx^M$ term in its worldline action. This should hold in particular for the open string's endpoints, so the correct modification of the Polyakov action to account for the $A_m$ background is
\begin{align}
    S &= - \frac{1}{4 \pi \alpha^\prime} \int d \tau d \sigma \partial^a X^M \partial_a X_M + \int d \tau A_{m,ii} \partial_\tau X^m (\tau,0) - \int d \tau A_{m,jj} \partial_\tau X^m (\tau,\pi) \notag\\[5pt]
    &= - \frac{1}{2 \pi \alpha^\prime} \int d\tau d\sigma \Big[ \frac{1}{2} \partial^a X^M \partial_a X_M + \frac{\alpha^\prime}{R} \Big( \theta^m_i \delta(\sigma) - \theta^m_j \delta(\sigma - \pi) \Big)\partial_\tau X^m  \Big].
    \label{open string coupling}
\end{align}
This boundary term does not affect the equations of motion, so the general form of the solution remains the same. However, the momentum density in the compact directions becomes
\begin{equation}
    \Pi^m = - \frac{1}{2 \pi \alpha^\prime} \partial^\tau X^m + \frac{1}{2 \pi R} \Big( \theta^m_i \delta(\sigma) - \theta^m_j \delta(\sigma - \pi) \Big),
\end{equation}
from which one obtains the total momentum
\begin{equation}
    p^m = \int_0^\pi d\sigma \Pi^m = \frac{n^m}{R} + \frac{\theta^m_i - \theta^m_j}{2 \pi R}.
\end{equation}
The new boundary conditions after T-duality are
\begin{equation}
    x^{\prime m}_1 = c^m, \hspace{0.5cm} x^{\prime m}_2 = c^m + 2 \pi n^m R^\prime + \theta^m_i R^\prime - \theta^m_j R^\prime,
\end{equation}
which describe two branes separated by a distance of $(\theta^m_i - \theta_j^m)R^\prime$ in the $m$-th direction. $\theta^m_i$ and $\theta^m_j$ are the angular position of the branes. All winding states once again decouple in the $R^\prime \to \infty$ limit, but the strings that stretch from one brane to the other without going around the cylinder survive. Setting $c^m=0$, we see that in this limit the constant values
\begin{equation}
    \theta_i^m R^\prime = - 2 \pi \alpha^\prime A_{m,ii} \equiv y_i^m , \hspace{0.5cm} \theta_j^m R^\prime = - 2 \pi \alpha^\prime A_{m,jj} \equiv y^m_j \hspace{0.5cm}
    \label{T-duality identification}
\end{equation}
become the position of the branes along the $X^{\prime m}$ axis, which is no longer compact.

\begin{figure}[t]
\begin{center}
\includegraphics[width= 0.75\textwidth]{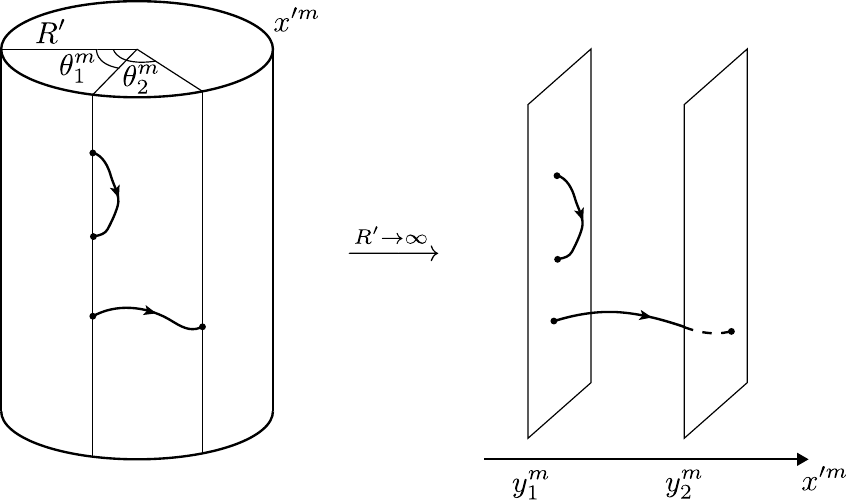}
\caption[Decompactification limit]{Decompactification limit.}
\label{decompactification}
\vspace{16cm}
\end{center}
\end{figure}

%\end{document}

%\documentclass[a4paper,12pt]{memoir}
%\usepackage{graphicx}
%\usepackage[utf8]{inputenc}
%\usepackage{indentfirst}
%\usepackage{braket}
%\usepackage{setspace}
%\usepackage{amsmath, amsthm, amssymb, amsfonts,bm}
%\usepackage[multiple]{footmisc}
%\usepackage{mathtools, changepage, slashed}
%\usepackage{tikz-feynman}
%\usepackage{bm, mathrsfs}
%\usepackage{gensymb}
%\usepackage[a4paper,top=3cm,left=3cm,right=2cm,bottom=2cm]{geometry}
%\usepackage{epstopdf}
%\usepackage{hyperref}
%\usepackage{pgfplots}
%\pgfplotsset{compat=1.18} 
%\usepackage[sorting=none]{biblatex}
%\addbibresource{refs.bib}
%\numberwithin{equation}{section}
%\usepackage[inkscapelatex=false]{svg}
%\usepackage[super]{natbib}
%\usepackage{doi}
%\hypersetup{
%  colorlinks   = true, %Colours links instead of ugly boxes
%  urlcolor     = black, %Colour for external hyperlinks
%  linkcolor    = black, %Colour of internal links
%  citecolor   = black %Colour of citations
%}
%\DeclareMathOperator{\Tr}{Tr}
%
%\newcommand{\normalord}[1]{%
%  {:\mathrel{\mspace{1mu}#1\mspace{1mu}}:}%
%}

%\OnehalfSpacing
%\usepackage{newtx}
%\usepackage{newtxtext}
%\usepackage{lmodern}

%\title{low-energy actions chapter}
%\author{pedrobairrao}

%\begin{document}

\chapter{Effective actions}\label{ch5}

\section{The low-energy action for closed strings}\label{sec51}

In order to compute string scattering amplitudes, one must build the correspondent worldsheet and integrate the Polyakov action over it. For our purposes it will not be necessary to develop the details of how this is done, although some aspects of it will investigated in the next chapter. It will, however, be important to understand what string dynamics looks like in the low-energy limit. At low energies one can only probe distances larger than the string length $l_\text{s} = \sqrt{\alpha^\prime}$, so in this limit the extended nature of strings is expected to become undetectable, making them indistinguishable from point particles. 

\begin{figure}[h]
\begin{center}
\includegraphics[width= 0.6\textwidth]{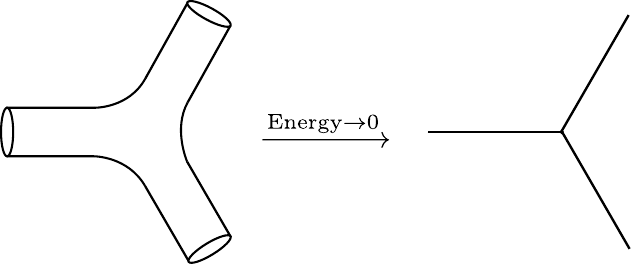}
\caption[Low-energy limit of the closed string three-point function]{Low-energy limit of the closed string three-point function, in which it becomes a regular field theory vertex.}
\end{center}
\end{figure}

Indeed, if one does not have enough spatial resolution to see the spacelike direction of the worldsheet, it is natural to average over it, so that the basic degree of freedom becomes
\begin{equation}
    x^\mu(\tau) \equiv \frac{1}{l} \int_0^{l} d \sigma  X^\mu(\tau,\sigma). 
\end{equation}
Upon plugging in the closed string solution \eqref{closed} for $X^\mu(\tau,\sigma)$, one finds that the integral kills all the terms with the modes, leading to the free point particle solution
\begin{equation}
    x^\mu(\tau) = x^\mu + \frac{2 \pi}{l} \alpha^\prime p^\mu \tau.
\end{equation}

From the point of view of the spectrum, in the low-energy limit all massive excitations are expected to decouple, effectively restricting the dynamics to the massless level. String theory should therefore reduce to some quantum field theory of interacting massless particles.\footnote{The tachyon completely spoils this argument, since in its presence the actual low-energy limit of bosonic string theory is expected to be the endpoint of tachyon condensation, which may look very different from simply truncating the theory at the massless level \cite{Adams_2001} (see footnote \ref{tachyon condensation footonote}). We once again ignore this issue because it does not occur in the supersymmetric theory. \label{second tachyon footnote} } The goal of this chapter is to understand what this field theory is. The most direct way to do this would be to compute string scattering amplitudes involving the massless states, take the low-energy limit and try to identify what spacetime action reproduces them. We shall instead use an indirect method that relies solely on consistency of the quantum string theory. 

The Polyakov action was introduced to describe the dynamics of a string in flat space, which is why it features explicitly the spacetime Minkowski metric $\eta_{\mu \nu}$. If one considers instead a string propagating in some other gravitational background of metric $G_{\mu \nu}$, the natural generalization of the Polyakov action is the so-called nonlinear sigma model action
\begin{equation}
    S_\sigma= \frac{1}{4 \pi \alpha^\prime} \int d^2 \sigma  \sqrt{g} \,  G_{\mu \nu}(X) \partial_a X^\mu \partial_b X^\nu g^{a b},
    \label{nonlinear sigma model}
\end{equation}
written here in Euclidean signature. Just as in the flat case, we can use worldsheet reparametrization and Weyl invariance to go to conformal gauge $g_{ab} = \delta_{ab}$. Varying $X^\mu$ leads to
\begin{equation}
    \delta S_\sigma = - \frac{1}{2 \pi \alpha^\prime} \int d^2 \sigma \Big( \partial_a \partial_a X^\mu + \Gamma^\mu_{\rho \sigma}(X) \, \partial_a X^\rho \partial_a X^\sigma  \Big) \delta X_\mu,
\end{equation}
which reduces the geodesic equation for the center of mass $x^\mu(\tau)$ upon setting $X^\mu(\tau,\sigma) \sim x^\mu(\tau)$.

If the graviton were not part of the string's spectrum, this coupling of the string to an external curved background would generate a different quantum theory from the flat space one we have been working with so far. However, we know that the graviton is part of the spectrum, which means that the background described by $G_{\mu \nu}$ can be found in the Hilbert space of the original theory, as a coherent state of gravitons. One may in fact explicitly construct the sigma model action by considering string scattering in the presence of such a coherent state (see Section 3.4.1 of \cite{GreenSchwarzWitten_vol1}). If the curved and flat space actions correspond to the same theory expanded around different states, they must have the same gauge symmetries. Requiring conformal invariance of $S_\sigma$ puts restrictions on $G_{\mu \nu}$ which can be interpreted as its equations of motion \cite{CALLAN1985593}.

The most striking difference between the flat space Polyakov action and the curved space one is that in the former the $X^\mu$ fields are free, whereas in the latter they become interacting, due to the $X$-dependence of the spacetime metric $G_{\mu \nu}(X)$. In any field theory with anomalous conformal symmetry, the origin of this anomaly can be traced to the need to renormalize. We saw this for the free theory, where the need to choose a particular ordering for the Virasoro generators was shown to be equivalent to adding a cosmological constant counterterm to the worldsheet. This introduces an energy scale into the model and therefore generically breaks conformal symmetry. The same general reasoning applies to interacting theories, although the actual renormalization procedure becomes much more involved. Conformal symmetry of an interacting theory with a set of coupling constants $g_i$ is equivalent to the vanishing of all the beta functions
\begin{equation}
    \beta_i(g_i) = \mu \frac{d g_i(\mu)}{d \mu},
\end{equation}
where $g_i(\mu)$ is the running coupling at renormalization scale $\mu$. To identify more clearly the form of the interactions in $S_\sigma$, we expand the embedding fields around some point $x^\mu_0$,
\begin{equation}
    X^\mu(\sigma) = x^\mu_0 + \sqrt{\alpha^\prime} \, Y^\mu(\sigma),
\end{equation}
where $Y^\mu$ is dimensionless and has zero vacuum expectation value. After Taylor expanding the metric, the action becomes
\begin{align}
    \frac{1}{4 \pi} \int d^2 \sigma \Big[ G_{\mu \nu}(x_0) \partial_a Y^\mu \partial_a Y^\nu &+ \sqrt{\alpha^\prime} \partial_\rho G_{\mu \nu}(x_0) Y^\rho  \partial_a Y^\mu \partial_a Y^\nu \notag\\[5pt]
    &+ \frac{\alpha^\prime}{2} \partial_\rho \partial_\sigma G_{\mu \nu}(x_0) Y^\rho Y^\sigma \partial_a Y^\mu \partial_a Y^\nu + \mathcal{O} \left( \alpha^{\prime 3/2} \right) \Big].
    \label{sigma expansion}
\end{align}
We thus find an infinite amount of interaction vertices, with coupling constants given by derivatives of the metric. If the target space determined by $G_{\mu \nu}$ has a characteristic radius of curvature $R_c$, the derivatives of the metric are generically of order $1/R_c$, so the expansion in $\sqrt{\alpha^\prime}$ that appears in $S_\sigma$ should be interpreted as a shorthand for an expansion in powers of the dimensionless parameter $\sqrt{\alpha^\prime} / R_c$.\footnote{$R_c$ may be determined for instance by the scale set by the inverse of the Ricci scalar associated to $G_{\mu \nu}$.} Since $\sqrt{\alpha^\prime} = l_\text{s}$ is what sets the length scale of the string, $\sqrt{\alpha^\prime} / R_c \ll 1 $ means that the strings are much smaller than the scale set by the background, making the point particle approximation applicable. This is the regime in which this $\alpha^\prime$ expansion is perturbative.

The condition for conformal invariance is that the beta function of all the interactions in the $\alpha^\prime$ expansion, each of which is a term in the Taylor expansion of the metric, vanish. This is usually restated as the vanishing of the beta functional
\begin{equation}
    \beta_{\mu \nu} \left( G \right) = \mu \frac{d G_{\mu \nu} \left( X;\mu \right)}{d \mu},
\end{equation}
where the renormalized metric $G_{\mu \nu} \left( X ; \mu \right)$ is the formal result of renormalizing each vertex and resumming the Taylor series for the metric. To compute $\beta_{\mu \nu}(G)$ we exploit the fact that the integrand $G_{\mu \nu}(X) \partial_a X^\mu \partial_b X^\nu$ in $S_\sigma$ is invariant under spacetime coordinate changes
\begin{equation}
    X^\mu \to \tilde{X}^\mu \left( X \right)
\end{equation}
(as long as the metric is covariantly transformed), to pick locally inertial coordinates at the point $x^\mu_0$, meaning that
\begin{equation}
    G_{\mu \nu} \left( X \right) \Big|_{X = x_0} = \eta_{\mu \nu},
\end{equation}
and that the first derivatives $\partial_\rho G_{\mu \nu} \left( X \right)$ vanish at $X^\mu = x_0^\mu$. A convenient choice are the Riemann normal coordinates, in terms of which
\begin{equation}
    G_{\mu \nu} \left( X \right) = \eta_{\mu \nu} - \frac{\alpha^\prime}{3} R_{\mu \rho \nu \sigma} \left( x_0 \right) Y^\rho Y^\sigma - \frac{\alpha^{\prime 3/2}}{6} \nabla_\lambda R_{\mu \rho \nu \sigma} \left( x_0 \right) Y^\lambda Y^\rho Y^\sigma + \mathcal{O} \left( \alpha^{\prime 2} \right),
\end{equation}
where $R_{\mu \rho \nu \sigma} \left( x_0 \right)$ is the spacetime Riemann tensor at $x^\mu_0$ and $\nabla_\lambda$ the spacetime covariant derivative. The action up to order $\alpha^\prime$ becomes
\begin{equation}
    \frac{1}{2 \pi} \int d^2 \sigma \Big( \frac{1}{2} \eta_{\mu \nu} \partial_a Y^\mu \partial_a Y^\nu - \frac{\alpha^\prime}{6} R_{\mu \rho \nu \sigma} \left( x_0 \right) Y^\rho Y^\sigma \partial_a Y^\mu \partial_a Y^\nu \Big).
\end{equation}
This is a two-dimensional field theory with a four-point vertex 
\begin{equation}
    \includegraphics[scale=1.1,valign=c]{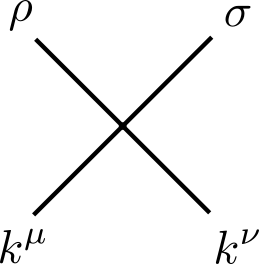} \sim  \alpha^\prime R_{\mu \rho \nu \sigma} \left( x_0 \right) k^\mu \cdot k^\nu,
\end{equation}
where $k^\mu_a$ is the two-momentum carried by $\partial_a Y^\mu$ and $k^\mu \cdot k^\nu = k^{\mu}_a k^{\nu}_a$. One source of possible divergences in this theory is the one-loop correction to the kinetic term
\begin{equation}
    \includegraphics[scale=1.1,valign=c]{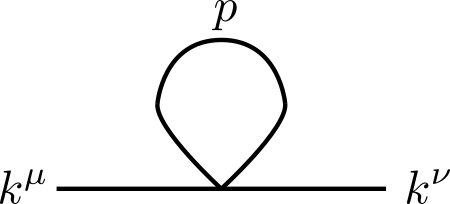} \sim -\alpha^\prime R_{\mu \rho \nu \sigma }(x_0) k^\mu \cdot k^\nu \int \frac{d^2p}{(2 \pi)} \frac{\eta^{\rho \sigma}}{p^2}.
\end{equation}
This diagram diverges both in the IR and the UV, but only the latter divergence is relevant for the beta function. In order to separate them, we first regulate the IR by adding a small fictitious mass $m^2$ for the $Y$ fields, which should be taken to zero at the end of the calculation. This of course explicitly breaks conformal symmetry, but it does so in a controlled manner, in the sense that it is restored upon taking the $m^2 \to 0$ limit. The UV divergence will be treated with dimensional regularization, which leads to
\begin{equation}
    -\alpha^\prime R_{\mu \rho \nu \sigma }(x_0)k^\mu \cdot k^\nu 
    \mu^\epsilon \int \frac{d^{2 - \epsilon}p}{(2 \pi)^{2 - \epsilon}} \frac{\eta^{\rho \sigma}} {p^2+m^2} = -\alpha^\prime R_{\mu \nu}(x_0)k^\mu \cdot k^\nu 
    \mu^\epsilon \int \frac{d^{2 - \epsilon}p}{(2 \pi)^{2 - \epsilon}} \frac{1} {p^2+m^2}
\end{equation}
to first order in $\alpha^\prime$, where $\mu^\epsilon$ is an arbitrary parameter with dimensions of energy that will serve as the renormalization scale. For small $\epsilon$ this evaluates to
\begin{equation}
   -\frac{\alpha^\prime}{4 \pi} R_{\mu \nu}(x_0)k^\mu \cdot k^\nu (\mu^2)^{\epsilon/2} (m^2)^{- \epsilon/2} \Gamma(\epsilon/2) = - \frac{\alpha^\prime}{4 \pi} R_{\mu \nu}(x_0)k^\mu \cdot k^\nu \bigg( \frac{2}{\epsilon}- \gamma_\text{E} + \ln \left( \frac{\mu^2}{m^2} \right) \bigg) + \mathcal{O} \left( \epsilon \right),
\end{equation}
where $\gamma_\text{E}$ is the Euler-Mascheroni constant. This diagram therefore leads to a divergent contribution to the one-loop effective action
\begin{equation}
    S_\text{div} \sim -\frac{\alpha^\prime}{2} \int d^2 \sigma \, R_{\mu \nu}(x_0) \partial_a Y^\mu \partial_a Y^\nu \bigg( \frac{2}{\epsilon}- \gamma_\text{E} + \ln \left( \frac{\mu^2}{m^2} \right) \bigg),
\end{equation}
up to numerical factors. We subtract it by adding a counterterm for the spacetime metric
\begin{equation}
    \delta G_{\mu \nu}(X;\mu) =  \frac{\alpha^\prime}{2} R_{\mu \nu}(X) \bigg( \frac{2}{\epsilon}- \gamma_\text{E} + \ln \left( \frac{\mu^2}{m^2} \right) \bigg) .
\end{equation}
From the renormalized metric $G_{\mu \nu}(X;\mu) = G_{\mu \nu}(X) + \delta G_{\mu \nu}(X;\mu)$ we extract the beta function
\begin{equation}
    \beta_{\mu \nu}(G) = \frac{d \delta G_{\mu \nu}(X;\mu)}{d \ln \mu} =  \alpha^\prime R_{\mu \nu}.
\end{equation}
As advertised, this result is insensitive to the IR regulator $m^2$, which we may now set to zero. Conformal symmetry of the nonlinear sigma model therefore requires\footnote{There is a second divergent diagram made from contracting into a loop the two legs carrying the momenta from $\partial_a Y^\mu \partial_a Y^\nu$. This would generate a quadratically divergent quantum correction to $Y$'s mass. This diagram does not contribute in dimensional regularization \cite{GreenSchwarzWitten_vol1}.}
\begin{equation}
    \beta_{\mu \nu} \left( G \right) = \alpha^\prime R_{\mu \nu} =0,
\end{equation}
which are nothing but the vacuum Einstein equations. This means that, in the low-energy limit, physics involving only gravitons is governed by the Einstein-Hilbert action
\begin{equation}
    S = \frac{1}{2 \kappa_0^2} \int d^{26}x \sqrt{-G} \, \mathcal{R}, \hspace{1cm} \mathcal{R} = R^\mu{}_\mu.
\end{equation}
The field equations alone do not determine $\kappa_0$. Going beyond one loop and $\mathcal{O} \left( \alpha^\prime \right)$ in $S_\sigma$, one may compute quantum corrections to the Einstein equations. At two loops for instance one finds \cite{FRIEDAN1985318}
\begin{equation}
    \beta_{\mu \nu} \left( G \right) \sim \alpha^\prime R_{\mu \nu} + \frac{\alpha^{\prime 2}}{2} R_{\mu \rho \sigma \lambda} R_\nu{}^{\rho \sigma \lambda}.
\end{equation}
Modifications of the Einstein equations of this kind are expected to appear in any quantum gravity theory \cite{Donoghue:1995cz}. They can be derived from an effective field theory expansion
\begin{equation}
    S_\text{EFT} = \int d^{26}x \sqrt{-G} \Big(  \frac{\mathcal{R}}{2 \kappa_0^2} + c_1 \mathcal{R}^2 + c_2 R^{\mu \nu} R_{\mu \nu} + c_3 R^{\mu \nu \rho \sigma} R_{\mu \nu \rho \sigma} \Big) + \mathcal{O} \left( \mathcal{R}^3 \right), 
\end{equation}
which is the most general local action for $G_{\mu \nu}$ that has the same symmetries as general relativity and reproduces it at low energies, expanded up to second order in the curvature. The part of this expression that actually depends on the details of string theory are the Wilson coefficients $c_1$, $c_2$ and $c_3$, which should be determined by comparing the equations of motion to the sigma model beta function. $S_\text{EFT}$ is the spacetime form of the $\alpha^\prime$ expansion: dimensional analysis shows that the $c_i$ must have one more power of $\alpha^\prime$ than the Einstein-Hilbert term. The order $\mathcal{R}^3$ terms must have one more power of $\alpha^\prime$ than the quadratic ones, and so on. For strong gravitational fields, meaning large curvature and thus small curvature radius, this expansion of course breaks down. This is the regime in which the string's length is no longer negligible. 

This shows that general relativity is a prediction of string theory, at least if only gravitons are present. If a background for all three massless fields of the closed string $(G_{\mu \nu}, B_{\mu \nu}, \Phi)$ is turned on, it can be shown that the sigma model action takes the form \cite{CALLAN1985593}
\begin{equation}
    S_\sigma = \frac{1}{4 \pi \alpha^\prime} \int \sqrt{g} \, d^2 \sigma  \Big[  \Big( g^{a b}  G_{\mu \nu}(X) + i \varepsilon^{ab} B_{\mu \nu}(X) \Big) \partial_a X^\mu \partial_b X^\nu + \alpha^\prime \Phi(X) R \Big],
\end{equation}
where $R$ is the worldsheet Ricci scalar and $\varepsilon^{ab}$ is the Levi-Civita tensor, related to the usual Levi-Civita symbol $\epsilon^{ab}$ by $\varepsilon^{ab} = \epsilon^{ab} /\sqrt{g}$. In this case the beta functions are \cite{Callan:1989nz}
\begin{align}
    \beta_{\mu \nu}(G) &= \alpha^\prime R_{\mu \nu} + 2 \alpha^\prime \nabla_\mu \nabla_\nu \Phi - \frac{\alpha^\prime}{4} H_{\mu \rho \sigma} H_\nu{}^{\rho \sigma} + \mathcal{O}\left(\alpha^{\prime 2} \right) \notag\\[5pt]
    \beta_{\mu \nu}(B) &= - \frac{\alpha^\prime}{2} \nabla^\rho H_{\rho \mu \nu} + \alpha^\prime H_{\rho \mu \nu}  \nabla^\rho \Phi + \mathcal{O}\left(\alpha^{\prime 2} \right) \notag\\[5pt]
    \beta(\Phi) &= \frac{D-26}{6} - \frac{\alpha^\prime}{2} \nabla^2 \Phi + \alpha^\prime \nabla_\rho \Phi \nabla^\rho \Phi - \frac{\alpha^\prime}{24} H_{\mu \nu \rho} H^{\mu \nu \rho} + \mathcal{O}\left(\alpha^{\prime 2} \right),
\end{align}
where $D$ is the spacetime dimension and 
\begin{equation}
    H_{\mu \nu \rho} = \partial_\mu B_{\nu \rho} + \partial_\rho B_{\mu \nu} + \partial_\nu B_{\rho \mu}
\end{equation}
is a field strength tensor for $B_{\mu \nu}$, just like $F_{\mu \nu}$ is for $A_\mu$ in a regular abelian gauge theory. The equations that follow from setting all beta functions to zero are obtained as equations of motion from the spacetime action
\begin{equation}
    S = \frac{1}{2 \kappa^2_0} \int d^Dx \sqrt{-G}\, e^{- 2 \Phi} \bigg[ - \frac{2(D-26)}{3 \alpha^\prime} + \mathcal{R} - \frac{1}{12} H_{\mu \nu \rho} H^{\mu \nu \rho} + 4 \nabla_\mu \Phi \nabla^\mu \Phi + \mathcal{O} \left( \alpha^{\prime} \right) \bigg].
\end{equation}
This effective action is said to be in ``string frame'', since it is written directly in terms of the fields that couple to the string in the worldsheet action. It is sometimes useful to split the dilaton into a constant part $\Phi_0$ and a dynamical part with zero expectation value $\phi$,
\begin{equation}
    \Phi(x) = \Phi_0 + \phi(x),
\end{equation}
and rewrite the spacetime action in terms of the modified metric
\begin{equation}
    \tilde{G}_{\mu \nu}(x) = e^{ -4 \phi/(D-2) } G_{\mu \nu}(x).
\end{equation}
This leads to \cite{Polchinskivol1:1998rq}
\begin{align}
    S = \frac{1}{2 \kappa^2} \int d^Dx \sqrt{-\tilde{G}} \bigg[& - \frac{2(D-26)}{3 \alpha^\prime} e^{4 \phi/(D-2)} + \tilde{\mathcal{R}} - \frac{1}{12} e^{-8 \phi/(D-2)} H_{\mu \nu \rho} \tilde{H}^{\mu \nu\rho} \notag\\[5pt]
    &\hspace{4cm} - \frac{4}{D-2} \nabla_\mu \phi \tilde{\nabla}^{\mu} \phi + \mathcal{O} \left( \alpha^{\prime} \right) \bigg],
\end{align}
where tildes have been placed in objects with upper indices to signal that these are raised with $\tilde{G}_{\mu \nu}$. We have also defined 
\begin{equation}
    \kappa = \kappa_0 e^{\Phi_0},
\end{equation}
which is the physical value of the gravitational constant, in the sense that it is $\kappa$ and not $\kappa_0$ that appears in Newton's law of gravitation in the nonrelativistic limit (this is also true in string frame). Written in this form the action is said to be in ``Einstein frame'', in which the gravitational action takes the standard Einstein-Hilbert form (note the absence of the overall factor of $e^{-2 \Phi}$).

\section{Low-energy action for open strings}\label{sec52}

As mentioned in Section \ref{sec42}, the fact that only the endpoints of open strings are charged under the massless gauge boson means that the coupling to a coherent background of it is done via a boundary term:
\begin{equation}
    S= \frac{1}{4 \pi \alpha^\prime} \int d^2 \sigma  \sqrt{g} \,  \eta_{\mu \nu} \partial_a X^\mu \partial_b X^\nu g^{a b} - i \int d \sigma^2 A_\mu (X) \partial_2 X^\mu.
    \label{boundary action}
\end{equation}
This is the Euclidean version of \eqref{open string coupling} for an open string with free endpoints. We set $g^{ab}=\delta^{ab}$, vary the $X^\mu$ field and integrate by parts to find
\begin{equation}
    \delta S = -\frac{1}{2 \pi \alpha^\prime} \int d^2 \sigma \, \partial^2 X^\mu \delta X_\mu + \frac{1}{2 \pi \alpha^\prime} \int d \sigma^2 \Big( \partial_1 X^\mu - 2 \pi i \alpha^\prime  \big( \partial^\mu A^\nu - \partial^\nu A^\mu \big) \partial_2 X_\nu  \Big) \delta X_\mu.
\end{equation}
The coupling to the gauge field does not alter the equation of motion of $X^\mu$ in the bulk of the worldsheet, but the Neumann boundary conditions become
\begin{equation}
    \partial_1 X^\mu - 2 \pi i \alpha^\prime F^{\mu \nu} \partial_2 X_\nu =0  \hspace{0.5cm} \text{at $\sigma^1 =0,\pi$},
\end{equation}
where $F^{\mu \nu} = \partial^\mu A^\nu - \partial^\nu A^\mu$.

The beta function can be derived using a similar method as before. We expand
\begin{equation}
    X^\mu(\sigma) = x^\mu(\sigma) + \sqrt{\alpha^\prime} \, Y^\mu(\sigma),
\end{equation}
where $x^\mu(\sigma)$ this time is a dynamical solution of the bulk equations of motion and modified boundary conditions. After some algebra one finds that the action becomes
\begin{align}
    S[x + \sqrt{\alpha^\prime} \, Y] =& S[x] + \frac{1}{4 \pi} \int_W d^2 \sigma \, \partial_a Y^\mu \partial_a Y_\mu \notag\\[5pt]
    &- \frac{i \alpha^\prime}{2} \int_{\partial W} d \sigma^2 \Big( F_{\mu \nu} Y^\mu \partial_2 Y^\nu + \partial_\mu F_{\nu \rho} Y^\mu Y^\nu \partial_2 x^\rho + \mathcal{O} \left( \partial^2 F \right) \Big),
\end{align}
where the field strengths $F_{\mu \nu}$ are all evaluated at $x^\mu(\sigma)$. In the approximation of slowly varying $F_{\mu \nu}$, so that one may ignore terms with more than one derivative acting on it, the vanishing of the beta function is found to be equivalent to \cite{Tong_string,Abouelsaood:1986gd}
\begin{equation}
    \partial_\nu F_{\mu \rho} \bigg[ \frac{1}{1 - 4 \pi \alpha^{\prime 2} F^2 } \bigg]^{\rho \nu} = 0,
    \label{open eq of motion}
\end{equation}
which are therefore the equations of motion of the gauge field. These only hold at first order in the derivatives of $F_{\mu \nu}$, but they are exact with respect to $\alpha^\prime$. At leading order in $\alpha^\prime$ these are the Maxwell equations
\begin{equation}
    \partial_\nu F_{\mu \rho} \eta^{\rho \nu} = \partial_\nu F_\mu{}^\nu = 0 + \mathcal{O} \left( \alpha^{\prime 2} \right).
\end{equation}
The action whose equations of motion are \eqref{open eq of motion} is the Born-Infeld action
\begin{equation}
    S_\text{BI} = - T_{26} \int d^{26} x \sqrt{-\det \left( \eta_{\mu \nu} + 2 \pi \alpha^\prime F_{\mu \nu} \right)},
\end{equation}
which was originally introduced in 1934 as a generalization of Maxwell electrodynamics \cite{Born:1934gh}. The dimensional constant $T_{26}$ plays a role analogous to the string's tension, and its precise meaning will become clearer later.

We have considered only the case of a string with free endpoints. T-duality allows us to easily obtain from $S_\text{BI}$ an action for strings on a single D$p$-brane.\footnote{The notation D$p$-brane means a D-brane with $(p+1)$-dimensional worldvolume.} To do this we take $25-p$ dimensions to be compact with radius $R$, and assume that the gauge fields depend only on the remaining $p+1$ noncompact directions. With $\mu,\nu = 0,\dots,p$ running over the noncompact directions, $I,J=p+1,\dots,26$ running over the compact ones, and $M,N$ running over all 26 dimensions, we have
\begin{equation}
    \eta_{MN} + 2 \pi \alpha^\prime  F_{MN}=
    \begin{bmatrix}
        \eta_{\mu \nu} + 2 \pi \alpha^\prime F_{\mu \nu} && 2 \pi \alpha^\prime  \partial_\mu A_I \\
        - 2 \pi \alpha^\prime  \partial_\nu A_J && \delta_{IJ}
    \end{bmatrix}
    \label{matrix split}
\end{equation}
The block matrix determinant formula \cite{Abadir_Magnus_2005}
\begin{equation}
    \det \begin{pmatrix}
    A &&  B \\
    C && D
    \end{pmatrix}
    =
    \det(A - B D^{-1} C) \det(D),
\end{equation}
leads to
\begin{equation}
    \det \left( \eta_{MN} + 2 \pi \alpha^\prime  F_{MN} \right) = \det \left( \eta_{\mu \nu} + 2 \pi \alpha^\prime F_{\mu \nu} +(2 \pi \alpha^\prime)^2 \partial_\mu A_I \partial_\nu A_I \right).
\end{equation}
In Section \ref{sec42} we saw that the components of the gauge field on the compact directions become the position coordinates of a D-brane under T-duality. We therefore follow \eqref{T-duality identification} and set
\begin{equation}
    - 2 \pi \alpha^\prime A_I(x^\mu) \equiv \phi^I(x^\mu),
    \label{dynamical T-duality}
\end{equation}
interpreting the $\phi^I$ scalar fields as coordinate positions for a D$p$-brane in the directions perpendicular to it's worldvolume. Plugging this into $S_\text{BI}$ leads to
\begin{equation}
    S_\text{single brane} = - T_p \int d^{p+1} x \sqrt{-\det \left( \eta_{\mu \nu} + \partial_\mu \phi^I \partial_\nu \phi^I + 2 \pi \alpha^\prime F_{\mu \nu} \right)},
    \label{single brane action}
\end{equation}
where $T_p = T_{26} (2 \pi R)^{25-p}$ comes from the integral over the compact dimensions. Changing the constant part of $A_I$ in the compactified theory was shown in Section \ref{sec42} to be T-dual to a rigid shift of the whole D-brane. The constant mode of $\phi^I$ is therefore interpreted as the brane's center of mass coordinate in the $25-p$ transverse directions. By taking into account the spacetime dependence of $A_I(x^\mu)$, the more general identification \eqref{dynamical T-duality} allows for the shape of the D-brane to fluctuate. This becomes very clear once one recognizes the geometric nature of $S_\text{single brane}$. Let the 26 functions $\phi^M$ be coordinates on spacetime:
\begin{equation}
    \phi^M : \text{spacetime} \to \mathbb{R} \, , \hspace{0.5cm} M =0,\dots 25.
\end{equation}
A D$p$-brane is a $p+1$-dimensional submanifold which we parameterize by the worldvolume coordinates $x^\mu$, with $\mu = 0,\dots p$. Every value of $x^\mu$ on the brane corresponds to a spacetime point 
\begin{equation}
    \phi^M = \phi^M(x)
\end{equation}
lying inside the brane. This is a higher-dimensional version of the relation between the string's worldsheet coordinates $\sigma$ and the embedding fields $X^\mu(\sigma)$. If spacetime is described by some metric $G_{M N}(\phi)$, the induced metric on the brane is the pullback of $G_{M N}$ to the worldvolume:
\begin{equation}
    \mathcal{G}_{\mu \nu}(x) = G_{M N} ( \phi(x) ) \frac{\partial \phi^M}{\partial x^\mu} \frac{\partial \phi^N}{\partial x^\nu}.
\end{equation}
For a brane sitting inside flat space we have $G_{MN} = \eta_{MN}$, and may for simplicity take the spacetime and brane coordinate systems to be aligned along the directions inside the brane, i.e. $\phi^\mu(x) = x^\mu$. The remaining coordinates $\phi^I(x)$ parameterize the shape taken by the brane in the directions perpendicular to it. The induced metric is then
\begin{equation}
    \mathcal{G}_{\mu \nu} = \eta_{\mu \nu} + \delta_{I J}  \frac{\partial \phi^I}{\partial x^\mu} \frac{\partial \phi^J}{\partial x^\nu} =\eta_{\mu \nu} +  \partial_\mu \phi^I \partial_\nu \phi^I.
\end{equation}
For $F_{\mu \nu}=0$, \eqref{single brane action} can therefore be written as
\begin{equation}
    S_\text{Dirac} = -T_p \int d^{p+1} x \sqrt{-\det ( \mathcal{G}_{\mu \nu} )},
\end{equation}
an action that was first proposed proposed with $p=2$ by Dirac in 1962 in an attempt to model electrons as charged surfaces \cite{Dirac:1962iy}. It is the higher-dimensional analogue of the Nambu-Goto action: the value of $S_\text{Dirac}$ in Euclidean signature is $-T_p$ times the brane's invariant volume.

The formula $\det M = e^{\text{Tr} \ln M}$ allows $S_\text{single brane}$ to written as
\begin{align}
    S_\text{single brane} &= - T_p  \int d^{p+1}x \exp \bigg[ \frac{1}{2} {\text{Tr} \ln \left( \delta^\mu_\nu + \partial^\mu \phi^I \partial_\nu \phi^I + 2 \pi \alpha^\prime F^\mu{}_\nu \right)} \bigg] \notag\\[5pt]
    &= \int d^{p+1}x \bigg( - \frac{T_p}{2} - \frac{T_p}{2} \partial^\mu \phi^I \partial_\mu \phi^I - T_p (\pi \alpha^\prime)^2 F_{\mu \nu} F^{\mu \nu} + \dots \bigg),
    \label{single abelian brane action}
\end{align}
where $\dots$ stands for higher dimensional operators. The Klein-Gordon action for $\phi^I$ means that small deviations from flatness in the brane's shape propagate as transverse waves. 

\begin{figure}[t]
\begin{center}
\includegraphics[width= 0.7\textwidth]{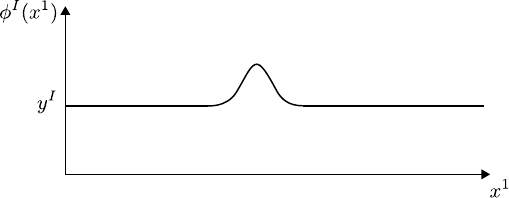}
\caption[Perturbation of a brane's shape]{Perturbation of a brane's shape in the $I$-th direction away from $\braket{\phi^I}=y^I$, localized in the $x^1$ worldvolume direction. }
\end{center}
\end{figure}

By having the displacement fields depend only on the time coordinate $x^0=t$, one describes the dynamics of a brane that is completely flat, but may still move as a whole in space. The action for $\phi^I$ in this case becomes the kinetic energy of this center of mass motion:
\begin{equation}
   \frac{T_p}{2} \bigg( \int d^px \bigg) \int dt \, \left( \partial_t \phi^I \right)^2 = \int dt \, \frac{1}{2} \big(V_p T_p\big) \left( \partial_t \phi^I \right)^2,
\end{equation}
where $V_p$ is the volume occupied by the brane, whose mass is therefore $M_p=V_p T_p$, from which one sees that $T_p$ should be understood as the brane's tension, or equivalently its mass density\footnote{This argument is taken from \cite{Liu2014StringTheory}}
\begin{equation}
    T_p = \frac{M_p}{V_p}.
\end{equation}
This interpretation extends also to the $T_{26}$ factor of the Born-Infeld action if one thinks of strings with all endpoints free as being attached to a $26$-dimensional brane that fills all of spacetime. 

One generalization that will be particularly important is that of a stack of branes, or equivalently that of a brane with Chan-Paton factors, since the $U(N)$ symmetry that was found on the spectrum in this case (see Section \ref{sec34}) is what allows for the appearance of nonabelian gauge theories in the low-energy limit. In a stack of $N$ branes one finds $N^2$ massless fields $(A_\mu)^i{}_j$ that transform in the $N \times \bar{N}$ representation of $U(N)$, which is isomorphic to the adjoint. The effect of such a modification on the coupling to the worldsheet \eqref{boundary action} is not trivial, due to the fact a matrix-valued $A_\mu(x)$ field does not in general commute with itself at different values of the argument $x$. It is possible to generalize the analysis of the abelian case to this one, and it leads to the spacetime action \cite{Tseytlin:1997csa,Dorn:1996an,Dorn:1996xk}
\begin{equation}
    S_\text{BI, nonabelian} = - T_{26} \int d^{26} x \, \text{STr} \sqrt{-\det \left( \eta_{\mu \nu} + 2 \pi \alpha^\prime F_{\mu \nu} \right)},
\end{equation}
with $F_{\mu \nu}$ being the nonabelian field strength
\begin{equation}
    F_{\mu \nu} = \partial_\mu A_\nu - \partial_\nu A_\mu - i [A_\mu, A_\nu ],
\end{equation}
and with the determinant under the square root taken only with respect to the Lorentz indices. STr is a symmetrized trace over the gauge group, meaning that
\begin{equation}
    \text{STr} (A_1 \dots A_n) = \frac{1}{n!} \big(\text{Tr} (A_1 \dots A_n) + \text{all permutations} \big).
\end{equation}
In contrast to the abelian case, here the separation between powers of $F_{\mu \nu}$ and its derivatives is not unambiguous, due to the fact that $[D_\mu, D_\nu] F_{\rho \sigma} = [F_{\mu \nu},F_{\rho \sigma}]$, where $D_\mu$ is the gauge covariant derivative. This allows one to trade derivative terms for commutator terms and vice-versa. One way to deal with this is to treat all commutators as higher order and keep them out of the action, which is what the symmetrized trace accomplishes. We will only actually need the leading order terms in the expansion of the square root, for which the symmetrized trace and the regular trace coincide. 

T-duality proceeds as before, but extra terms appear due to the commutator in the definition of $F_{\mu \nu}$. One has for instance
\begin{align}
   2 \pi \alpha^\prime F_{\mu I} =2 \pi \alpha^\prime \big( \partial_\mu A_I - i [A_\mu,A_I] \big) = -\partial_\mu \phi^I + i[A_\mu,\phi^I] = - D_\mu \phi^I.
\end{align} 
Instead of \eqref{matrix split} we find
\begin{equation}
    \eta_{MN} + 2 \pi \alpha^\prime  F_{MN}=
    \begin{bmatrix}
        \eta_{\mu \nu} + 2 \pi \alpha^\prime F_{\mu \nu} && - D_\mu \phi^I  \\
        D_\nu \phi^J && \delta_{IJ} - i(2 \pi \alpha^\prime)^{-1} [\phi^I,\phi^J]
    \end{bmatrix},
\end{equation}
which leads to the action
\begin{align}
    S_\text{$n$ branes}  &= - T_p \int d^{p+1} x \, \text{STr} \bigg[ \sqrt{-\det \left( \eta_{\mu \nu} + D_\mu \phi^I \left( \delta^{IJ} + i (2 \pi \alpha^\prime)^{-1} [\phi^I,\phi^J]  \right)^{-1}  D_\nu \phi^I + 2 \pi \alpha^\prime F_{\mu \nu} \right)} \notag\\[5pt]
    & \hspace{9cm} \times \sqrt{ \det \big( \delta^{IJ} + i(2 \pi \alpha^\prime)^{-1} [\phi^I,\phi^J] \big) } \bigg] \label{full T-dual action} \\[5pt]
    &= \int d^{p+1}x \, \text{Tr} \bigg( -T_p - (\pi \alpha^\prime)^2 T_p F_{\mu \nu} F^{\mu \nu} - \frac{T_p}{2} D_\mu \phi^I D^\mu \phi^I + \frac{T_p}{4} [\phi^I,\phi^J][\phi^I,\phi^J] + \dots \bigg),
    \label{expanded T-dual action}
\end{align}
where in the second line we have discarded higher order terms in field strengths, covariant derivatives and commutators. This confirms what was already suspected from the discussion of the spectrum: at low energies the dynamics of open strings in a stack of branes reduces to Yang-Mills coupled to scalars that transform in the adjoint. By comparing the kinetic term of the gauge fields with the usual $(1/4g^2_\text{YM}) \text{Tr}F^2$, one finds the relation between the brane tension and the Yang-Mills coupling constant,
\begin{equation}
    g_\text{YM} = \frac{1}{2 \pi \alpha^\prime \sqrt{T_p}}.
\end{equation}

It is known since the early days of string theory that any theory of open strings contains closed strings as well, and therefore gravity \cite{CREMMER1972222}. An example of a closed-open string interaction is a worldsheet in the shape of a closed tube that splits open into two sheets, a process which at low energies could take the form of a graviton decaying into two photons. Consistency thus requires that the massless excitations of the open string couple to gravity. Even if we start from a state with $A_\mu=\phi^I=0$, meaning a completely flat brane with the lowest possible energy, there is still a rest energy
\begin{equation}
    \int d^{p+1}x \,  T_p \text{Tr}(\delta^i{}_j)  = N T_p V_p = NM_p
\end{equation}
from the constant term in $S_\text{$n$ branes}$, which is nonzero as long as the brane occupies some volume in spacetime. In a gravitational theory such an extended massive object must itself generate a gravitational field that distorts the spacetime around it away from the starting flat configuration, driving the embedding fields $\phi^I$ away from zero. The coupling of a D-brane to closed strings can be found by turning on backgrounds for both closed and open strings in the Polyakov action and computing the beta functions. For nonabelian gauge group and fully Neumann boundary conditions the result is the Dirac-Born-Infeld action \cite{Leigh:1989jq,Myers:1999ps}
\begin{equation}
    S_\text{DBI} = -T_{26} \int d^{26}x \, \text{STr} \left( e^{-\Phi} \sqrt{-\det (G_{\mu \nu} + B_{\mu \nu} + 2 \pi \alpha^\prime F_{\mu \nu} )} \right).
\end{equation}
It is still possible to obtain the D-brane action from this by T-duality, although the process is more complicated in a curved background. One finds an action similar to \eqref{full T-dual action} (see \cite{Myers:1999ps} for the explicit form), which can also be expanded as a series in $[\phi^I,\phi^J]$. The zeroth order term of this series is 
\begin{equation}
    \tilde{S}_\text{DBI} = -T_{p} \int d^{p+1}x \, \text{STr} \left( e^{-\Phi} \sqrt{-\det (\mathcal{G}_{\mu \nu} + \mathcal{B}_{\mu \nu} + 2 \pi \alpha^\prime F_{\mu \nu} )} \right),
\end{equation}
where
\begin{align}
    \mathcal{G}_{\mu \nu} &= G_{MN} D_\mu \phi^M D_\nu \phi^N \notag\\[5pt]
    &= G_{MN} \partial_\mu \phi^M \partial_\nu \phi^N + \mathcal{O} \left( [A,\phi] \right) \notag\\[5pt]
    &= G_{\mu \nu} + 2 G_{I(\mu} \partial_{\nu)} \phi^I + G_{IJ} \partial_\mu \phi^I \partial_\nu \phi^J + \mathcal{O} \left( [A,\phi] \right)
\end{align}
is the pullback of the spacetime metric to the brane together with $\mathcal{O} \left( [A,\phi] \right)$ terms that come from the T-dualizing $F_{\mu I}$. These extra terms in relation to the abelian case introduce the necessary interactions to minimally couple the scalars to the gauge group. $\mathcal{B}_{\mu \nu}$ is the equivalent for the Kalb-Ramond field. At higher orders in $[\phi^I,\phi^J]$ one finds a wealth of couplings, among them $G_{KJ}G_{LI}[\phi^I,\phi^K][\phi^J,\phi^L]$, the curved space version of the quartic interaction found in \eqref{expanded T-dual action}.

%\end{document}

%\documentclass[a4paper,12pt]{memoir}
%\usepackage{graphicx}
%\usepackage[utf8]{inputenc}
%\usepackage{indentfirst}
%\usepackage{braket}
%\usepackage{setspace}
%\usepackage{amsmath, amsthm, amssymb, amsfonts,bm}
%\usepackage[multiple]{footmisc}
%\usepackage{mathtools, changepage, slashed}
%\usepackage{tikz-feynman}
%\usepackage{bm, mathrsfs}
%\usepackage{gensymb}
%\usepackage[a4paper,top=3cm,left=3cm,right=2cm,bottom=2cm]{geometry}
%\usepackage{epstopdf}
%\usepackage{hyperref}
%\usepackage{pgfplots}
%\pgfplotsset{compat=1.18} 
%\usepackage[sorting=none]{biblatex}
%\addbibresource{refs.bib}
%\numberwithin{equation}{section}
%\usepackage[inkscapelatex=false]{svg}
%\usepackage[super]{natbib}
%\usepackage{doi}
%\hypersetup{
%  colorlinks   = true, %Colours links instead of ugly boxes
%  urlcolor     = black, %Colour for external hyperlinks
%  linkcolor    = black, %Colour of internal links
%  citecolor   = black %Colour of citations
%}
%\DeclareMathOperator{\Tr}{Tr}

%\newcommand{\normalord}[1]{%
%  {:\mathrel{\mspace{1mu}#1\mspace{1mu}}:}%
%}

%\OnehalfSpacing
%\usepackage{newtx}
%\usepackage{newtxtext}
%\usepackage{lmodern}

%\title{Open-closed duality chapter chapter}
%\author{pedrobairrao}

%\begin{document}

\chapter{Open-closed duality}\label{ch6}

\section{The possibility of an open-closed duality}\label{sec61}

At low energies the dynamics of closed and open bosonic strings looks radically different. For closed strings one has a gravitational theory in 26 dimensions coupled to a scalar and an antisymmetric tensor. Open strings on the other hand lead to Yang-Mills theories in any spacetime dimension up to 26, possibly with some adjoint scalars. Both sectors of course couple to each other, but separately they seem to describe very different kinds of physics. 

\begin{figure}[b]
\begin{center}
\includegraphics[width= 0.4\textwidth]{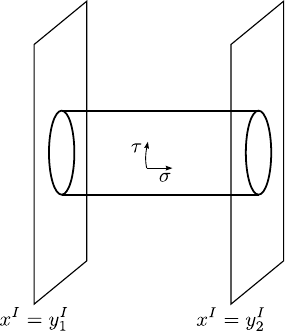}
\caption[Brane interaction]{A vacuum fluctuation of an open string whose endpoints are each attached to a different branes. }
\label{brane interaction}
\end{center}
\end{figure}

However, if one thinks not in terms of the field theory limit but in terms of worldsheet diagrams, the distinction in some cases appears much milder. Consider for instance the physical process described by figure \ref{brane interaction}.

The two planes represent flat D-branes of the same dimension at positions $y^I_1$ and $y^I_2$ in transverse space. Between them there is a one-loop vacuum fluctuation: a pair of open strings appears from the vacuum, each one propagates for some distance until they finally annihilate each other. Since an open string tends to shrink under the force of its tension, one would expect such a fluctuation to pull the branes toward each other. Stated differently, the fact that the energy of a stretched string is given by the tension times its length means that the occurrence of this vacuum process should contribute to a higher expectation value for the energy density in the region between the branes than in the rest of spacetime, leading to an attractive Casimir-like force between them. 

The same diagram can also be given a different interpretation if one assumes time flows from left to right instead of upwards. By treating $\sigma$ as the the timelike direction and $\tau$ as the spacelike one, one of the branes seems to emit a closed string, which travels until it is absorbed by the other one. Indeed, we have seen on Section \ref{sec52} that D-branes do couple to closed string modes, so this is a reasonable process to consider. If one takes the closed string to be in a graviton state, figure \ref{brane interaction} describes the branes feeling the gravitational pull of each other via graviton exchange, which should, as in the open string case, lead to an attractive interaction.

Wick-rotating to an Euclidean worldsheet makes the difference between the two interpretations seem even less relevant. In both cases the action is given by the area of the worldsheet, a quantity that does not depend on whether one views the intermediate states as that of a one-loop open string diagram or a tree level closed string one.

The purpose of this chapter is to verify if the open and closed interpretations of figure \ref{brane interaction} are in fact equivalent, by computing both amplitudes. We will find that they are indeed the same, making this an example of open-closed string duality. One might then imagine that taking some sort of low-energy limit of both descriptions would lead to a duality between what is found in the low-energy spectrum of closed and open strings: gravity and gauge theories. Although this turns out to be too simplistic an idea to provide a direct derivation of AdS/CFT, it does capture the essence of the correspondence. By carefully setting up the closed string computation we will also gain some intuition on an alternative description of D-branes that will prove fundamental to holography.

\section{The cylinder Faddeev-Popov measure}\label{sec62}

In the open string description what must be computed is a one-loop vacuum diagram with $25-p$ coordinates having Dirichlet boundary conditions and $p$ coordinates having Neumann conditions. Vacuum diagrams are zero-point functions, which are usually divided out of any physical correlation function in field theory. In this case what we want is the actual value of the path integral, which we shall compute in Euclidean signature:
\begin{equation}
    \mathcal{A} = \int\frac{\mathcal{D}g \mathcal{D}X }{V_{\text{diff}\times \text{Weyl}}(C^2)}\exp \bigg(- \frac{1}{4 \pi \alpha^\prime} \int_{C^2} d^2 \sigma \sqrt{g} g^{a b} \partial_a X^\mu \partial_b X_\mu  \bigg).
    \label{non gauge fixed amp}
\end{equation}
The cylinder $C^2$ has two boundaries and no holes, so its Euler number is zero. In Section \ref{sec32} we used the fact that all components of the worldsheet metric can be locally fixed with a diff$\times$Weyl gauge transformation to impose conformal gauge. In doing this, two important assumptions were made. First, it was assumed that after going to conformal gauge no further gauge fixing was necessary. We expect this not to be true, since much of the usefulness of this gauge lies in the very fact that one still has the freedom to do conformal transformations. The $g_{ab} = \delta_{ab}$ condition does not affect the integral over $\mathcal{D}X$, which will naturally run over all forms the $X^\mu(\sigma_1,\sigma_2)$ functions may take, 
including gauge-equivalent ones related to each other by a conformal transformation. This leads to an overcounting in the path integral. The second assumption was that every metric on the worldsheet can be made flat globally with a diff$\times$Weyl transformation. We will investigate these two points in detail for the particular case of the cylinder, starting with the latter.

As usual for open strings, we take the spatial coordinate $\sigma^1$ that parametrizes the length of the cylinder to go from $0$ to $\pi$. The Euclidean time coordinate $\sigma^2$ then takes us around the cylinder's circumference, and we choose it to vary from $0$ to $2 \pi$. The path integral in \eqref{non gauge fixed amp} therefore runs over every possible metric one may define on $[0,\pi] \times [0,2 \pi]$. The vast majority of these metrics of course do not describe actual cylinders, which are usually defined as being flat, but general curved surfaces of cylindrical topology. All we ask is that $\sigma^1=0,\pi$ describe boundaries and that $\sigma^2$ be periodically identified
\begin{equation}
    \sigma^2 \cong \sigma^2 + 2 \pi.
\end{equation}
Starting from one such metric $ 
g_{ab}(\sigma)$, we use worldsheet diffeomorphism invariance to move to a new metric
\begin{equation}
    \tilde{g}_{ab}(\tilde{\sigma}) = \frac{\partial \tilde{\sigma}^c}{\partial \sigma^a} \frac{\partial \tilde{\sigma}^d}{\partial \sigma^b} g_{cd}(\sigma).
    \label{metric transformation}
\end{equation}
Setting $\tilde{g}_{12}(\tilde{\sigma})=0$ and $\tilde{g}_{11}(\tilde{\sigma})=\tilde{g}_{22}(\tilde{\sigma})$ leads to two differential equations for $\tilde{\sigma}^1(\sigma)$ and $\tilde{\sigma}^2(\sigma)$, or equivalently for the inverse functions $\sigma^1(\tilde{\sigma})$ and $\sigma^2(\tilde{\sigma})$. The solution of this system will naturally involve integration constants, whose value we fix by imposing the boundary conditions
\begin{align}
    &\sigma^1(0,\tilde{\sigma}^2) =0, \hspace{0.5cm} \sigma^1( \pi l^1,\tilde{\sigma}^2) = \pi, \\[5pt]
    &\sigma^2(\tilde{\sigma}^1,\tilde{\sigma}^2+ 2 \pi l^2) = \sigma^2(\tilde{\sigma}^1,\tilde{\sigma}^2) + 2 \pi \cong \sigma^2(\tilde{\sigma}^1,\tilde{\sigma}^2). 
\end{align}
By varying $\tilde{\sigma}^1$ from 0 to $l^1$ one moves along the cylinder, from one boundary to the other, whereas varying $\tilde{\sigma}^2$ corresponds to going around cylinder's circumference, ultimately reaching the starting point at $\tilde{\sigma}^2 + 2 \pi l^2$. These requirements determine the values taken by $l^1$ and $l^2$ for each metric, and they mean that the region in the $\tilde{\sigma}$-plane that spans the cylinder is $[0,\pi l^1]\times [0,2 \pi l^2]$. 

To see that it is in general inconsistent to demand that the coordinate region be $[0,\pi] \times [0, 2 \pi]$ in the $\tilde{\sigma}$ coordinates, consider the particular case in which the original metric already satisfies $g_{12}(\sigma)=0$, with the remaining diagonal components arbitrary functions of $\sigma$. The area of this cylinder, a diffeomorphism-invariant quantity, is given by
\begin{equation}
    A = \int d^2 \sigma \sqrt{\det g} = \int d^2 \sigma \sqrt{g_{11} \, g_{22}}.
\end{equation}
In this case the two functions from our diffeomorphism are enough to make the metric completely flat. In the new coordinates the area is simply
\begin{equation}
    A = \int d^2 \tilde{\sigma} =\Delta \tilde{\sigma}^1 \Delta \tilde{\sigma}^2, 
\end{equation}
where $\Delta \tilde{\sigma}^a$ is the range of $\tilde{\sigma}^a$. If the domain in the $\tilde{\sigma}$-plane could always be chosen to be $[0,\pi] \times [0, 2 \pi]$, we would conclude that $A = 2 \pi^2$ for every diagonal metric. By working instead in $[0,\pi l^1]\times [0,2 \pi l^2]$ we allow the dependence of $A$ on the original metric components $g_{ab}(\sigma)$ to be contained in the value of the product $l^1 l^2$, which makes sense since both $l^1$ and $l^2$ are determined by the map $\sigma \to \tilde{\sigma}$, itself determined by the form of $g_{ab}(\sigma)$. 

Going back to the general case, the $\tilde{\sigma}$ coordinates reduce the metric to $\tilde{g}_{ab} = \text{diag}(\tilde{g}_{11},\tilde{g}_{11})$. We then do a Weyl transformation with $e^{- 2 \omega} = 1/\tilde{g}_{11}$ to make the metric flat, resulting in an ordinary straight cylinder, whose length and circumference are given by the parameters
\begin{align}
    \pi l^1 &= \underset{d \tilde{\sigma}^2=0}{\int} \sqrt{\tilde{g}_{ab} \, d \tilde{\sigma}^a d \tilde{\sigma}^b} =\int  d \tilde{\sigma}^1, \notag\\[5pt]
     2 \pi l^2 &=\underset{d \tilde{\sigma}^1=0}{\int} \sqrt{\tilde{g}_{ab} \, d \tilde{\sigma}^a d \tilde{\sigma}^b} = \int d \tilde{\sigma}^2,
\end{align}
respectively.

\begin{figure}[b]
\begin{center}
\includegraphics[width= 0.6\textwidth]{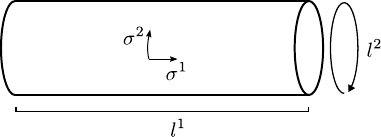}
\caption[Euclidean cylinder 1]{Euclidean cylinder with length $l^1$ and circumference $l^2$.}
\label{sigma cylinder}
\end{center}
\end{figure}

Having used up most of the gauge symmetry to make the metric flat, not much more can be done to simplify things. To maintain the conventional range of $\pi$ for $\tilde{\sigma}^1$, we do a global rescaling of both coordinates $\tilde{\sigma}^a \to l^1 \tilde{\sigma}^a$ along with a compensating constant Weyl transformation to keep the metric normalized. The coordinate region then becomes
\begin{equation}
    [0,\pi] \times [0, 2\pi  t],
\end{equation}
where $t = l^2 / l^1$. No further gauge transformation can eliminate $t$ from the theory. The original interval of $[0,\pi] \times [0,2 \pi ]$ can of course be restored by rescaling $\tilde{\sigma}^2 \to \tilde{\sigma}^2/t $, but this turns the metric into\footnote{From now on we drop the tildes on the flat coordinates.}
\begin{equation}
    \delta_{a b}(t) =
    \begin{bmatrix}
        1 && 0 \\
        0 && t^2
    \end{bmatrix}.
    \label{modulus dependent metric}
\end{equation}
Each worldsheet metric in the path integral is gauge-equivalent to $\delta_{ab}(t)$ for some value of $t \in (0,\infty)$, but for $t \neq t^\prime$ the metrics $\delta_{ab}(t)$ and $\delta_{ab}(t^\prime)$ with the same coordinate region cannot be obtained from each other via conformal transformations. Parameters such as $t$, which classify metrics on a surface modulo diffeomorphisms and Weyl transformations, are called moduli, or Teichmüller parameters. The set of such parameters for a given topology is called its moduli space. What remains of the path integral over all metrics after fixing conformal gauge is an integral over the moduli space:
\begin{equation}
    \int \mathcal{D}g \, F[g]\to \int_0^\infty dt \, F[\delta(t)].
\end{equation}

We now consider the issue of conformal transformations, the
residual gauge transformations that are not fixed by choosing a form for the metric. Although the two-dimensional conformal algebra is infinite-dimensional, most of the conformal Killing vectors do not in general exponentiate to well-defined, invertible transformations on the worldsheet. In the case of the cylinder, the only continuous conformal transformation that respects the $\sigma^2 \cong \sigma^2 + 2 \pi t$ periodicity and the boundaries is a rigid translation along the periodic direction
\begin{equation}
    \sigma^2 \to \sigma^2 + c^2.
\end{equation}
The set of all conformal Killing vectors that exponentiate to global transformations on a given worldsheet generates its conformal Killing group (CKG). On the cylinder we also have the discrete inversion $\sigma^2 \to - \sigma^2$, which is not the exponential of any infinitesimal transformation. Whenever there are residual global conformal transformations after imposing conformal gauge, one must divide the path integral by the volume of the conformal Killing group and also by possible symmetry factors from discrete symmetries in order to avoid overcounting. 

Both the issues of moduli and the conformal Killing group can be dealt with via a modification of the Faddeev-Popov procedure described in Section \ref{sec33}. Taking moduli into account is simple. Starting from $1=\int \mathcal{D} h \, \delta [g-h]$, we write $\int \mathcal{D}h=\int dt \int \mathcal{D}h(t)$, where $\mathcal{D}h(t)$ is a measure over the metrics of modulus $t$. Each $h_{ab}(t)$ can then be written as $\hat{g}_{ab}^\zeta(t)$, the image of some metric $\hat{g}_{ab}(t)$ in the same conformal class as $h_{ab}(t)$ under a gauge transformation. The result is
\begin{equation}
    1 = \int_0^\infty dt \int \mathcal{D}h(t) \, \delta [g-h(t)] = \Delta_\text{FP}[g]  \int^\infty_0 dt \, \int \mathcal{D}\zeta \, \delta \big[ g- \hat{g}^\zeta(t) \big].
\end{equation}
The problem with this expression is that the $\int \mathcal{D} \zeta$ integral runs over the entire gauge group, in particular the CKG, a blind spot of the gauge-fixing delta functional. If we split the measure schematically as $\mathcal{D} \zeta = \mathcal{D} \zeta^\prime \mathcal{D} v_0$, where $v_0$ denotes a transformation in the CKG of $\hat{g}_{ab}(t)$ and $\zeta^\prime$ contains all other ones, we find
\begin{equation}
    \int \mathcal{D} \zeta \, \delta \big[ g - \hat{g}(t)^{\zeta} \big] F[g] = \underset{\text{CKG}}{\int} \mathcal{D}v_0 \int \mathcal{D} \zeta^\prime \, \delta \big[ g - \hat{g}(t)^{\zeta^\prime} \big] F[g]
    = V(\text{CKG}) F[\hat{g}(t)],
\end{equation}
where $V(\text{CKG})= \int_{\text{CKG}} \mathcal{D}v_0$ is the volume of the conformal Killing group. To better understand how one may modify $\Delta_{\text{FP}}[g]$ in order to eliminate this factor of $V(\text{CKG})$, we proceed by writing $\mathcal{D} \zeta = \mathcal{D} \omega \mathcal{D}v$, and focus on the $\mathcal{D}v$ part, where we expect to find the conformal Killing vectors, since for the rigid translation $\sigma^2 \to \sigma^2 + c^2$ the Weyl part of the transformation is trivial. In order to put the integral over diffeomorphisms on slightly less heuristic (but by no means mathematically rigorous) grounds, we should look for a metric on the functional space they inhabit, which can be derived from a norm on the tangent space.\footnote{A good example of this procedure is the construction of the area element on a two-dimensional surface with coordinates $(\sigma^1,\sigma^2) \equiv \sigma$ and metric $g_{ab}(\sigma)$. Fix some arbitrary point $\sigma_0= (\sigma_0^1,\sigma_0^2)$ and consider the vectors $ \mathrm{v}= \mathrm{v}^a \partial_a$ that live in the tangent space above it. We require that the norm $|\mathrm{v}|^2$ be a quadratic function of the vector components, so that it can reduce to the usual Euclidean norm on a flat surface, and also that it be a scalar with respect to reparametrizations (the size of the tangent vectors should not depend on how the base space is described).  The only possible choice is then $|\mathrm{v}|^2=g_{ab}(\sigma_0) \mathrm{v}^a \mathrm{v}^b$. The tangent space therefore inherits the base space's metric, evaluated at $\sigma_0$. This of course also defines the scalar product $\mathrm{v} \cdot \mathrm{u} = g_{ab}(\sigma_0) \mathrm{v}^a \mathrm{u}^b$. The tangent space is by definition flat, so the area of the parallelogram enclosed by two vectors $\mathrm{v}$ and $\mathrm{u}$ is given by the elementary formula of base times height, $ A=\sqrt{|\mathrm{v}|^2 |\mathrm{u}|^2 - (\mathrm{v} \cdot \mathrm{u})^2} $ in terms of the vector components. The area of a small region of the surface $[\sigma^1_0,\sigma^1_0 + \delta \sigma^1 ] \times [\sigma^2_0,\sigma^2_0 + \delta \sigma^2 ]$ can, for small $\delta \sigma^i$, be approximated by the area of the parallelogram of sides $\delta \sigma^1 \partial_1$ and $\delta \sigma^2 \partial_2$ in the tangent space of $\sigma_0$. The base times height formula for these vectors gives $ \delta A = \sqrt{\det g(\sigma_0)}  \, \delta \sigma^1 \delta \sigma^2 $. Since the area of a finite region of the surface can be broken into a sum of the areas of these infinitesimal parallelograms, each one anchored over some base point, the surface's area element is the familiar $dA = \sqrt{g(\sigma)} \, d \sigma^1 d \sigma^2$. We know that this formula generalizes to any finite number of dimensions, the volume element will in general be given the determinant of the metric times the product of all coordinate differentials. If it is the norm on the tangent space that is known instead of the base space metric, this argument can be inverted so that the metric is derived from the tangent space norm. Path integrals can always in principle be formulated in terms of some finite lattice spacing for the coordinates, so we expect some version of this construction to also hold in the (suitably regularized) infinite-dimensional case, where some of the coordinate labels are continuous.} Each point of the space of diffeomorphisms is a function $v=v^a(\sigma)$ from the worldsheet to itself. We fix one such function $v_0$ and consider a variation $v_0 \to v_0 + \delta v $, where $ \delta v =\delta v^a(\sigma)$ represents a some linearized diffeomorphism around $v_0$. These will play the role of vectors in the tangent space around $v$. The norm $|\delta v|^2$ must contract all indices such that the result is a scalar with respect to the base space symmetry transformations, which in this case are the worldsheet diffeomorphisms themselves. The only admissible form is \cite{Polyakov:1987hqn}
\begin{equation}
    |\delta v|^2 = \int d^2 \sigma \sqrt{g(\sigma)} \, g_{ab}(\sigma) \, \delta v^a(\sigma) \, \delta v^b(\sigma) \equiv \sum_{\substack{a,b \\ \sigma_1,\sigma_2} } G_{ab}(\sigma_1,
    \sigma_2 ) \, \delta v^a(\sigma_1) \, \delta v^b(\sigma_2),
\end{equation}
where\footnote{In these expressions $\sigma_1$ is shorthand for $(\sigma^1_1,\sigma^2_1)$. It should not be confused with $g_{1a} \,\sigma^a$.}
\begin{equation}
    G_{ab}(\sigma_1,
    \sigma_2 ) = \delta^2(\sigma_1-\sigma_2) \sqrt{g(\sigma_1)} \, g_{ab}(\sigma_1)
\end{equation}
is the tangent space metric. In a general gauge the integration measure on the space of diffeomorphisms is thus $ \mathcal{D}v= \sqrt{\det G} \prod_{a,\sigma} d v^a(\sigma)$, but if this hits a gauge-fixing delta functional that sets $g_{ab}=\delta_{ab}$ (with no explicit modulus dependence on $\delta_{ab}$), the metric in diff-space becomes the identity and we get just
\begin{equation}
    \int_{C^2} \mathcal{D}v = \int \prod_{a,\sigma} dv^a(\sigma).
    \label{naive diff measure}
\end{equation}
For each pair $(\sigma^1,\sigma^2)$, the values taken by $v^1(\sigma^1,\sigma^2)$ and $v^2(\sigma^1,\sigma^2)$ are the possible worldsheet points to which $(\sigma^1,\sigma^2)$ may be mapped,
\begin{equation}
    (\sigma^1,\sigma^2) \to (v^1(\sigma^1,\sigma^2),v^2(\sigma^1,\sigma^2)),
\end{equation}
which is what fixes the integration limits in
\begin{equation}
    \int_{C^2} \mathcal{D}v = \prod_{\sigma^1,\sigma^2} \int_0^\pi dv^1(\sigma^1,\sigma^2) \int_0^{2 \pi t} dv^2(\sigma^1,\sigma^2).
\end{equation}
Among these integrals is the one that corresponds to the constant shift 
\begin{equation}
    (\sigma^1, \sigma^2) \to (\sigma^1,v^2) =  (\sigma^1,\sigma^2+c^2).
\end{equation}
The sum over all values the shift can take is the volume of the conformal Killing group
\begin{equation}
    \int dc^2 =\int^{2 \pi t}_0  dv^2 = 2 \pi t = V(\text{CKG}).
\end{equation}
This integral is of course identical to any one of the other $\int dv^2(\sigma)$ factors in $\int \mathcal{D}v$. Omitting a single one from the product should therefore take care of the overcounting problem. We do this by adding a delta function (not functional) $\delta(v^2(\hat{\sigma}))$ to the definition  of the Faddeev-Popov measure:
\begin{equation}
    1 = \Delta_\text{FP}[g]  \int^\infty_0 dt  \int \mathcal{D}\zeta \, \delta \big[ g- \hat{g}^\zeta(t) \big] \delta(v^2(\hat{\sigma})).
    \label{FP measure with moduli}
\end{equation}
The point $\hat{\sigma}$ is arbitrary, what matters is that this delta function eliminates precisely the contribution of a single diffeomorphism parallel to the boundary. Inserting this factor of $1$ into the amplitude \eqref{non gauge fixed amp} and proceeding as in \eqref{FP computation}, we find
\begin{equation}
    \mathcal{A} = \int_0^\infty \frac{dt}{2 \pi t}\int \mathcal{D}X \, \Delta_\text{FP}[\hat{g}(t)] e^{-S[X,\hat{g}(t)]},
\end{equation}
where we used the fact that the delta function removes the integration over the CKG \cite{ReidEdwardsnotes}:
\begin{equation}
    \int \mathcal{D} \zeta \, \delta(v^2(\sigma)) = \frac{V_{\text{diff}\times \text{Weyl}}(C^2)}{V(\text{CKG})} = \frac{V_{\text{diff}\times \text{Weyl}}(C^2)}{2 \pi t}.
\end{equation}
To compute 
\begin{equation}
    \Delta^{-1}_\text{FP}[\hat{g}(t)] = \int^\infty_0 dt^\prime  \int \mathcal{D}\zeta \, \delta \big[ \hat{g}(t)- \hat{g}^\zeta(t^\prime) \big] \delta(v^2(\sigma))
\end{equation}
we once again use the fact that the integrals are only nonvanishing in a neighborhood of $\hat{g}_{ab}(t)$ to linearize the variation inside the delta:
\begin{equation}
    \delta \big[ \hat{g}(t)- \hat{g}^\zeta(t^\prime) \big] = \delta \left[ 2 \omega \hat{g}_{ab} + \hat{\nabla}_a v_b + \hat{\nabla}_b v_a + \delta t \, \partial_{t} \hat{g}_{ab}(t) \right].
\end{equation}
The last term corresponds to the physical variation of the metric due to a change in the modulus. It is simpler to compute it with an explicitly modulus-dependent metric, and after doing so one may transform back to coordinates such that the metric components are $t$-independent, which are the ones we will use for the remainder of this chapter. The inverse of the Faddeev-Popov determinant is given by
\begin{align}
    &\Delta^{-1}_\text{FP}[\hat{g}(t)] = \int^\infty_0 d \delta t \int \mathcal{D} \omega \, \mathcal{D} v \, \delta \left[ 2 \omega \hat{g}_{ab}(t) + \hat{\nabla}_a v_b + \hat{\nabla}_b v_a + \delta t \, \partial_{t} \hat{g}_{ab}(t) \right] \delta (v^2(\hat{\sigma})) \notag\\[5pt]
    &= \int_0^\infty d \delta t \int \mathcal{D} \omega \, \mathcal{D} v \, \mathcal{D} \beta \exp \bigg[ 2 \pi i \int d^2 \sigma \sqrt{\hat{g}(t)} \beta^{ab} \left( 2 \omega \hat{g}_{ab}(t) + \hat{\nabla}_a v_b + \hat{\nabla}_b v_a  + \delta t \partial_t \hat{g}_{ab}(t) \right) \bigg] \notag\\[5pt]
    & \hspace{11cm} \times \int_{-\infty}^\infty d \xi \exp{ \Big( 2 \pi i \xi v^2(\hat{\sigma}) \Big) } \notag\\[5pt]
    &= \int_0^\infty d \delta t \int_{- \infty}^\infty d \xi \int \mathcal{D}v \, \mathcal{D}\beta^\prime \exp{ \Big( 4 \pi i  \big( \hspace{-4pt} \big(\beta^\prime , \nabla v  \big) \hspace{-4pt} \big)
    \Big)} \exp{ \Big( 2 \pi i \delta t  \big( \hspace{-4pt} \big( \beta^\prime , \partial_t \hat{g}(t)  \big) \hspace{-4pt} \big)
    \Big)}  \exp{ \Big( 2 \pi i \xi v^2(\hat{\sigma})
    \Big)} \notag\\[5pt]
\end{align}
where
\begin{equation}
    \big( \hspace{-4pt} \big( A, B \big) \hspace{-4pt}  \big) \equiv \int d^2 \sigma \sqrt{\hat{g}(t)} \, A^{ab} B_{ab}.
\end{equation}
In going from the first to the second line both the delta functional and the regular delta function were exponentiated. We invert this expression by substituting all bosonic integration variables with fermionic ones, including $\delta t$ and $\xi$, which become the Grassmann numbers $\theta$ and $\varphi$. The result is\footnote{This formula generalizes in a somewhat straightforward way to the general case of a worldsheet with any number of moduli and conformal Killing vectors. See \cite{ReidEdwardsnotes} for a derivation similar to the one here.}
\begin{align}
    \Delta_\text{FP}[\hat{g}] &= \int d \theta \int d \varphi \int \mathcal{D}b \, \mathcal{D}c \, \exp{ \Big( -2   \big( \hspace{-4pt} \big(b , \nabla c  \big) \hspace{-4pt} \big)
    \Big)} \exp{ \Big( -  \theta \,  \big( \hspace{-4pt} \big(b , \partial_t \hat{g}(t)  \big) \hspace{-4pt} \big)
    \Big)} \exp{ \Big( - \varphi c^2(\hat{\sigma})
    \Big)} \notag\\[5pt]
    &=\int \mathcal{D}b \, \mathcal{D}c \exp{ \Big( -\frac{1}{2 \pi}  \big( \hspace{-4pt} \big(b , \nabla c  \big) \hspace{-4pt} \big) \Big)} \frac{1}{4 \pi} \big( \hspace{-4pt} \big( b, \partial_t \hat{g}(t) \big) \hspace{-4pt} \big) \, c^2(\hat{\sigma}),
    \label{final FP measure}
\end{align}
where in the second line we rescaled $b_{ab} \to b_{ab} / 4 \pi$ and did the integrals over $\theta$ and $\varphi$. The first exponent in \eqref{final FP measure} is the ghost action
\begin{equation}
     \frac{1}{2 \pi}  \big( \hspace{-4pt} \big(b , \nabla c  \big) \hspace{-4pt} \big) =  \frac{1}{2 \pi} \int d^2 \sigma \sqrt{\hat{g}(t)} \, b_{ab} \hat{\nabla}^a c^b = S_g.
\end{equation}

\section{The open string interpretation}\label{sec63}

Taking into account the Faddeev-Popov determinant, the final form of the open string cylinder amplitude is
\begin{equation}
    \mathcal{A}_\text{open} = 2 \frac{1}{2} \int_0^\infty \frac{dt}{2 \pi t} \int \mathcal{D}X \, \mathcal{D}b \, \mathcal{D}c \, e^{-S_\text{P}[X] - S_g[b,c]  } \, \frac{1}{4 \pi} \big( \hspace{-4pt} \big( b,\partial_t \hat{g}(t) \big) \hspace{-4pt} \big) \, c^2(\hat{\sigma})
\end{equation}
The factor of 1/2 is due to the discrete conformal transformation $\sigma^2 \to - \sigma^2$. The factor of 2 next to it is present due to the fact that the open string can attach to the branes with either orientation. Both give the same amplitude, so to get the total force felt by the branes we should multiply by 2. In conformal gauge the $b$-ghost insertion is found to be $\big( \hspace{-3.5pt} \big( b,\partial_t \delta(t) \big) \hspace{-3.5pt} \big) = \frac{2}{t} \int d^2 \sigma \, b_{22}(\sigma)$, so 
\begin{align}
    \mathcal{A}_\text{open} &= 2 \int_0^\infty \frac{dt}{2 t} \int \mathcal{D}X \, e^{-S_\text{P}[X]} \, \frac{1}{4 \pi^2 t}\int d^2\sigma \int \mathcal{D}b \, \mathcal{D}c \, e^{- S_g[b,c]  } \,  b_{22}(\sigma) c^2(\hat{\sigma}) \notag\\[5pt]
    &\equiv 2\int_0^\infty \frac{dt}{2t} \, \mathcal{A}^X_\text{open}(t) \, \mathcal{A}^g_\text{open}(t),
    \label{cylinder amplitude formula}
\end{align}
where in the last step we used the fact that the matter and ghost actions do not couple to each other to separate their respective path integrals:
\begin{equation}
    \mathcal{A}^X_\text{open}(t) =  \int \mathcal{D}X \, e^{-S_\text{P}[X]} \hspace{0.2cm} , \hspace{1cm} \mathcal{A}^g_\text{open}(t) = \frac{1}{4 \pi^2 t} \int d^2\sigma \int \mathcal{D}b \, \mathcal{D}c \, e^{- S_g[b,c]  } \,  b_{22}(\sigma) c^2(\hat{\sigma}).
\end{equation}
The distribution of the $1/(2 \pi t)^2$ prefactor between the two amplitudes is of course arbitrary. Our choice, whose usefulness will become clear later, is to factor out the $2$ from summing over the two orientations and to turn the worldsheet integral $\int d^2 \sigma$ in the ghost amplitude into an average by dividing it by the twice the cylinder's area of $ \pi (2 \pi t) =2 \pi^2 t$.

Instead of directly computing these, our strategy will be to relate them to thermal partition functions. Given the generating functional of a quantum field theory describing particles in $d+1$-dimensional Minkowski space, the general procedure to obtain from it the partition function for a gas of such particles in $d$ dimensions and temperature $T = 1/k_\text{B} \beta $ is to Wick rotate the action to Euclidean signature and compactify the Euclidean time direction into a circle of radius $\beta$ \cite{Altland_Simons_2010}. This is precisely what we have in
\begin{equation}
    \int \mathcal{D}X \, e^{-S_\text{P}[X]} = \int \mathcal{D}X \exp \bigg( -\int_0^{\beta} d \sigma^2 \int^{\pi}_0 d \sigma^1 \mathcal{L}_\text{P}(X) \bigg),
\end{equation}
with $\beta = 2 \pi t$. The Wick rotation to Euclidean space has been done from the start, and the $X$ fields are periodic with respect to the Euclidean time $\sigma^2$ since this is the direction that goes around the cylinder's circumference. We exploit this correspondence to write the amplitude $\mathcal{A}^X_\text{open}(t)$ as a trace over the matter CFT's spectrum,
\begin{equation}
    \int \mathcal{D}X \, e^{-S_\text{P}[X]} = \text{Tr} \, e^{- 2 \pi t H} = \text{Tr} \, e^{- 2 \pi t (L_0 + a^X)}.
    \label{trace open}
\end{equation}
where $H$ is the open string Polyakov Hamiltonian, which we know is given by the normal ordering constant $a$ plus the DD open string Virasoro generator found in Section \ref{sec32}
\begin{equation}
    L_0 = \alpha^\prime p_\mu p^\mu + \frac{\Delta y^2}{4 \pi^2 \alpha^\prime} + \sum_{n=1}^\infty \big( \alpha^\mu_{-n} \alpha_{n \, \mu} + \alpha^I_{-n} \alpha^I_n \big) .
\end{equation}
For this computation the careful distinction between NN and DD directions will not necessary, so we will use the spacetime indices $M=0,\dots25$. After inserting $L_0$ into \eqref{trace open}, the trace splits into a sum over the noncompact momenta and the modes,
\begin{equation}
    \text{Tr} \, e^{- 2 \pi t (L_0 + a^X)} = e^{- 2 \pi a^X t - \frac{t \Delta y^2}{2 \pi \alpha^\prime}} \sum_k e^{-2 \pi \alpha^\prime t k^2} \sum_i \exp{ \bigg( -2 \pi t \sum_{n=1}^\infty \alpha^M_{-n} \alpha_{n \, M} \bigg) },
\end{equation}
where the sum $\sum_i$ goes over the spectrum. As it stands, this quantity is divergent due to the wrong sign Gaussian over the energy $k^0$:
\begin{equation}
    \sum_k  e^{-2 \pi \alpha^\prime t k^2} = V_{p+1} \int \frac{d^{p+1 }k}{(2 \pi)^{p+1}}  \, e^{-2 \pi \alpha^\prime t k^2} = V_{p+1} \int_{- \infty}^{\infty} \frac{dk^0}{2 \pi} e^{2 \pi \alpha^\prime t (k^0)^2} \int \frac{d^{p}k}{(2 \pi)^{p}}  \, e^{-2 \pi \alpha^\prime t k^i k^i}.
\end{equation}
This is dealt with in string theory in the same way it is usually done in field theory: one Wick-rotates to Euclidean spacetime (which in this case means using an Euclidean metric from the start), where the integral converges. After some momentum space correlation function has been computed to the desired order in perturbation theory, one should then analytically continue the external Euclidean momenta back to Lorentzian signature.\footnote{See \cite{Witten:2013pra} for a clear discussion of analytical continuation between Euclidean and Lorentzian spacetime signatures in string theory.} Vacuum diagrams such as the one we are computing of course involve no external momenta, but they are still part of the disconnected contributions to any correlation function, and therefore should also be computed in Euclidean spacetime if this is the philosophy adopted. Note also that vacuum fluctuations do in general contribute to a theory's vacuum energy density, and therefore can only be neglected if the same can be done for the zero-point energy. This is usually the case in field theory, but not in string theory, due to the presence of gravity.\footnote{See Section 7.3 of \cite{Polchinskivol1:1998rq} for more on this.} We therefore have
\begin{equation}
    \sum_k  e^{-2 \pi \alpha^\prime t k^2} = i V_{p+1} \bigg( \int_{- \infty}^{\infty} \frac{dk}{2 \pi} \, e^{-2 \pi \alpha^\prime t k^2} \bigg)^{p+1} = i V_{p+1} \left( \frac{1}{8 \pi^2 \alpha^\prime t} \right)^{(p+1)/2}.
\end{equation}
To compute the contribution of the modes to the trace it is convenient to write the level operators for each spacetime index $M$ and mode number $n$,  $ N_{M,\, n} = \alpha^M_{-n} \alpha^M_{n}$ (no sum over $M$ or $n$),\footnote{Since from now on we use Euclidean spacetime signature, for the remainder of this chapter we will write all spacetime indices up.} in terms of number operators $\mathbf{N}_{n,\, M} = a^M_{-n} a^{\dagger M}_{n}$ (see the paragraph below \eqref{canonical commutators}),
\begin{equation}
    \alpha^M_{-n} \alpha^M_{n} = n \, a^{\dagger M}_{n} a^M_{n}  =n \, \mathbf{N}_{n, \,  M} \, ,\hspace{1cm} \text{(no sum)}.
\end{equation}
For each value of $M$ and $n$ the bosonic number operator can taken any positive value, so 
\begin{align}
     \sum_i \exp{ \bigg( -2 \pi t \sum_{n=1}^\infty \alpha^M_{-n} \alpha^M_{n} \bigg) } &= \prod_{M=1}^{26} \, \prod_{n=1}^{\infty} \, \sum_{\mathbf{N}_{n, M}=0}^\infty e^{- 2 \pi t n \mathbf{N}_{n, M}} \notag\\[5pt]
     &= \left( \prod_{n=1}^{\infty} \frac{1}{1 - e^{- 2 \pi t n}}  \right)^{26} \notag\\[5pt]
     &= e^{- \frac{26 \pi t}{12}} \, \eta(it)^{-26},
\end{align}
where
\begin{equation}
    \eta(\tau) = e^{\frac{\pi i \tau}{12}} \prod_{n=1}^\infty \Big( 1 - e^{2 \pi i n \tau} \Big)
    \label{eta function}
\end{equation}
is the Dedekind eta function. Putting these results back into the formula for the trace gives 
\begin{equation}
    \mathcal{A}^X_\text{open}(t) = i V_{p+1} \left( \frac{1}{8 \pi^2 \alpha^\prime t} \right)^{(p+1)/2} e^{- \frac{t \Delta y^2}{2 \pi \alpha^\prime} - 2 \pi t \big( a^X + \frac{26}{24} \big)} \eta(it)^{-26}.
\end{equation}
Each of the $X^M$ fields contributes with one factor of  $\eta(it)^{-1}$, regardless of the boundary conditions. Note that the part involving the normal ordering constant vanishes upon setting $a^X=-\frac{26}{24}$, which is the value found in Section \ref{sec33}. 

The same method can be used to obtain the ghost contribution
\begin{equation}
    \mathcal{A}^g_\text{open}(t) = \frac{1}{4 \pi^2 t}  \int d^2 \sigma \, e^{-2 \pi t\big(a^g - \frac{1}{12} \big)} \, \eta(it)^2 = e^{-2 \pi t \big( a^g - \frac{1}{12} \big)} \, \eta(it)^2.
\end{equation}
See Appendix \ref{B1} for the details. The path integral is found to be independent of the fixed point $\hat{\sigma}$, causing the worldsheet integral to cancel the factor of $1/(2 \pi^2 t)$, while the additional factor of $\frac{1}{2}$ is canceled by the factor of $2$ in the mode expansion of $b_{22}$ in \eqref{open ghost solutions}. Once again the dependence on the normal ordering constant $a^g$ vanishes for the physical value $a^g=1/12$. The total amplitude is therefore
\begin{equation}
    \mathcal{A}_\text{open} = 2i \, V_{p+1} \int_0^\infty \frac{dt}{2t}   \left( \frac{1}{8 \pi^2 \alpha^\prime t} \right)^{(p+1)/2} e^{ -\frac{t \Delta y^2}{2 \pi \alpha^\prime} } \,\eta(it)^{-24}.
    \label{full open amplitude}
\end{equation}
The integrand in this expression only differs from the matter CFT amplitude $\mathcal{A}^X_\text{loop}(t)$ by the fact that the exponent of $\eta(it)$ is shifted from $-26$ to $-24$. As always, the ghosts are responsible for canceling the contribution of the two unphysical polarizations of $X^M$. Since this is their only effect, we would have obtained the right result by simply ignoring the ghosts and writing
\begin{equation}
    \mathcal{A}_\text{open} = 2 \int_0^\infty \frac{dt}{2t} \, \mathcal{A}^{\perp}_X(t) = 2 \int_0^\infty \frac{dt}{2t} \text{Tr}_{\perp} \, e^{- 2 \pi t (L_0 + a)},
    \label{Physical trace formula}
\end{equation}
where $\text{Tr}_{\perp}$ is a trace over only the momenta and the transverse modes of $X^M$ and $a = a^X + a^g=-1$ is the physical normal ordering constant. When we discuss the superstring, the worldsheet CFT we have been using so far will need to be modified, but its path integral over the cylinder can still be written as the right-hand side of \eqref{Physical trace formula}, up to the inclusion of $(-1)^F$ to account for the fact that bosonic and fermionic fields contribute with opposite signs to the amplitude. This point is elaborated on in Appendix \ref{B1}.

For small $t$ the cylinder becomes a very long tube, the spacetime interpretation being that of a pair of strings that only travel an extremely short distance before annihilating. One would be correct to guess that the $t \to 0$ region of moduli space corresponds to the ultraviolet, in the same way that the high loop momenta region does in usual field theory. In the opposite limit of large $t$, one has a stubby, short cylinder. The spacetime picture is that of two strings that travel for a long distance relative to their lengths before annihilating each other. This corresponds to the infrared. Indeed, for $t \to \infty$ the length of the strings becomes negligible compared to the distance traveled and one expects their dynamics to reduce to that of point particles, in the sense discussed in Chapter \ref{ch5}.

\begin{figure}[h]
     \centering
     \begin{subfigure}[h]{0.37\textwidth}
         \centering
         \includegraphics[width=\textwidth]{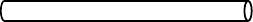}
         \caption{small $t$}
     \end{subfigure}
     \hspace{3cm}
     \begin{subfigure}[h]{0.13\textwidth}
         \centering
         \includegraphics[width=\textwidth]{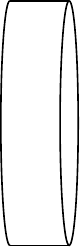}
         \caption{large $t$}
         \label{large t figure}
     \end{subfigure}
     \caption[Cylinders in opposite limits of moduli space]{Cylinders in opposite limits of moduli space. For small $t$ we have a thin, long cylinder, while for large $t$ we get a short, stubby cylinder. }
\end{figure}

It is instructive to confirm this intuitive picture by taking the large $t$ limit of $\mathcal{A}_\text{closed}$. Before getting our hands dirty we should know what to expect. If the low-energy dynamics of the string reduces to that of particles, the diagram on figure \ref{large t figure} should in this limit be well described by a sum of one loop vacuum bubbles of the string's lightest particles, all propagating only along the directions spanned by the branes. Let us then for a moment forget about the details of string theory and consider what such a sum of bosonic particle loops looks like. Take some field theory with bosonic fields $\phi_i$, assumed to be all scalars for simplicity, each with mass $m^2_i$. The connected partition of this theory with sources turned off, $W[0]=- \log Z[0]$, is given by the sum of all vacuum bubbles. The sum of all one-loop bubbles can be obtained by considering the theory in its noninteracting limit:
\begin{equation}
    Z_\text{1-loop}[0] = \prod_i \int \mathcal{D}\phi_i \exp{ \bigg[ -\frac{1}{2} \int d^{p+1}x \, \phi_i \big( -\partial^2  + m_i^2 \big)\phi_i \bigg] } = \prod_i \frac{1}{\sqrt{\det (- \partial^2 +m_i^2 ) }}.
\end{equation}
The quantity we expect to find is thus of the form
\begin{align}
   W_\text{1-loop}[0] &= \frac{1}{2}\sum_i \log \det \big( - \partial^2 + m_i^2 \big) 
    = \frac{1}{2}\sum_i \text{Tr} \log \big( - \partial^2 + m_i^2 \big) \notag\\[5pt]
    &= \frac{i V_{p+1}}{2} \sum_i \int \frac{d^{p+1}k}{(2 \pi)^{p+1}} \log \big( k^2 + m^2_i \big).
\end{align}
Using the representation of the logarithm \cite{Donoghue:2017pgk}
\begin{equation}
    \log \left( \frac{a}{b} \right) = \int_0^\infty \frac{dT}{T} \Big( e^{-bT} - e^{-aT} \Big)
\end{equation}
we obtain
\begin{align}
    W_\text{1-loop}[0] &= i V_{p+1} \sum_i \int_0^\infty \frac{dt}{2t} \int  \frac{d^{p+1}k}{(2 \pi)^{p+1}} e^{-2 \pi \alpha^\prime t(k^2 + m^2_i )} \notag\\[5pt]
    &= i  V_{p+1} \int_0^\infty \frac{dt}{2t} \left( \frac{1}{8 \pi^2 \alpha^\prime t} \right)^{(p+1)/2} \sum_i e^{- 2 \pi \alpha^\prime t m_i^2} \, ,
    \label{Coleman-Weinberg formula}
\end{align}
where we have set $T = 2 \pi \alpha^\prime t$ and discarded a constant term from the logarithm. This way of writing the effective action puts it in the form that comes out of the so-called Schwinger proper time formalism, which is a particular representation of the first-quantized worldline formalism mentioned below figure \ref{closed string perturbation series figure}. It is in this language that field theory amplitudes can be directly compared to those of string theory. 

Now we expand the $\eta(it)^{-24}$ factor found in $\mathcal{A}_\text{open}$ around $t = \infty$ to find
\begin{align}
    \eta(it)^{-24} &= e^{2 \pi t} \prod_{n=1}^\infty \left( 1 - e^{-2 \pi n t} \right)^{-24} \notag\\[5pt]
    &= e^{2 \pi t} \prod_{n=1}^\infty \left( 1 - 24 \,e^{-2 \pi n t} + \mathcal{O} \left( e^{- 4 \pi n t} \right) \right) \notag\\[5pt]
    &= e^{2 \pi t} + 24 + \mathcal{O} \left( e^{- 2 \pi n t} \right).
\end{align}
Plugging this back into the amplitude, we obtain
\begin{align}
     \mathcal{A}_\text{open} \xrightarrow{ t \to \infty } &2i V_{p+1} \int^\infty \frac{dt}{2t}  \left( \frac{1}{8 \pi^2 \alpha^\prime t} \right)^{(p+1)/2} \exp \left( -\frac{t \Delta y^2}{2 \pi \alpha^\prime}  \right) \bigg( e^{2 \pi t} + 24 + \dots \bigg)  \notag\\[5pt]
     &= 2i V_{p+1} \int^\infty \frac{dt}{2t}  \left( \frac{1}{8 \pi^2 \alpha^\prime t} \right)^{(p+1)/2}\bigg\{ \exp{ \bigg[ -2 \pi \alpha^\prime t \bigg( \frac{\Delta y^2}{(2 \pi \alpha^\prime)^2} - \frac{1}{\alpha^\prime} \bigg) \bigg] } \notag\\[5pt]
     & \hspace{5.33cm} + 24 \exp{ \bigg( - 2 \pi \alpha^\prime t \frac{\Delta y^2}{(2 \pi \alpha^\prime)^2} \bigg) } + \dots \bigg\}.
\end{align}
Comparison with the DD open string mass relation \eqref{open string DD mass relation} shows that this expansion is precisely of the form \eqref{Coleman-Weinberg formula}, with the inclusion of the factor of 2 from the string's two orientations. The first term corresponds to the ground state and the second one comes from the contribution of the 24 polarization states of the $N=1$ level. The higher the level (and therefore mass), the smaller is the contribution to the amplitude for large $t$, as expected.

\section{The closed string interpretation}\label{sec64}

If one considers \ref{brane interaction} from a closed string perspective, the interpretation of which region of moduli space corresponds to low and high energies turns out to be inverted in relation to the open string case. The long, thin tube found at small $t$ represents a closed string that travels a large distance in relation to its circumference. In the $t \to 0$ limit, in which the radius collapses to zero, its dynamics should reduce to that of point particles, so the small $t$ region clearly corresponds to the infrared. At large $t$ we have an ultraviolet process where a closed string travels a very short distance compared to its circumference.

Looking at \ref{sigma cylinder} as a closed string worldsheet requires us to interpret the $\sigma^1$ coordinate as the temporal one. To facilitate this we define new coordinates
\begin{equation}
    \xi^1 = \frac{\sigma^2}{t} \, , \hspace{0.5cm} \xi^2 = \frac{\sigma^1}{t},
\end{equation}
so that we recover the usual $\xi^1 \in [0,2 \pi]$ for the closed string compact coordinate. The new Euclidean time coordinate $\xi^2$ goes from $0$ to $\pi/t \equiv s$. A constant Weyl transformation can then be done to make the metric flat again, setting the cylinder's area to $2 \pi s$.

\begin{figure}[b]
\begin{center}
\includegraphics[width= 0.6\textwidth]{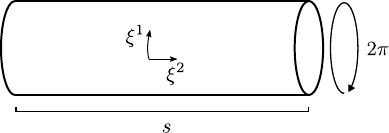}
\caption[Euclidean cylinder 2]{Euclidean cylinder with length $s$ and circumference $2 \pi$.}
\end{center}
\end{figure}

If the force felt by the branes for small $t$, or large $s$, can be understood as coming from an exchange of particles between them, this should be visible in the appropriate limit of $\mathcal{A}_\text{open}$. To see what such an interaction looks like in the Schwinger proper time formalism we once again consider the simple case of scalars. The starting point is the identity \cite{Schwartz_2013}
\begin{equation}
    \frac{1}{A} = \int_0^\infty dT \, e^{-T A},
    \label{Schwinger trick}
\end{equation}
which allows the Feynman propagator for a scalar particle in 26 dimensions to be written as
\begin{align}
   D^i_F(x,x^\prime) &= \int \frac{d^{26}k}{(2 \pi)^{26}} \frac{e^{i k \cdot (x-x^\prime)}}{k^2 + m_i^2} = \frac{\alpha^\prime}{2} \int_0^{\infty} ds \int \frac{d^{26}k}{(2 \pi)^{26}} e^{- \frac{\alpha^\prime s}{2} ( k^2 + m_i^2 ) + i k \cdot (x-x^\prime) } \notag\\[5pt]
   &= \frac{\alpha^\prime}{2 (2 \pi)^{26}} \left( \frac{2 \pi}{\alpha^\prime} \right)^{13} \int_0^\infty \frac{ds}{s^{13}} \, e^{  - \frac{(x-x^\prime)^2}{2 \alpha^\prime s} } e^{- \frac{\alpha^\prime s}{2}m^2_i}, 
\end{align}
where we have used \eqref{Schwinger trick} with $T= \alpha^\prime s/2$ and done the Gaussian integrals over the momenta. If $\Gamma_i$ is the amplitude for this particle to be emitted or absorbed by a brane, which we assume is constant throughout the brane's worldvolume, the amplitude for a particle to be emitted at some point $x=(x_1^\mu,y^I_1)$ on the first brane and absorbed at a point $x^\prime=(x_2^\mu,y^I_2)$ on the second brane is $\Gamma_i \, D^i_F(x,x^\prime) \, \Gamma_i$. The total interaction amplitude is then given by the integral of this object over both brane's worldvolumes, summed over each species of particle exchanged:
\begin{align}
    \mathcal{A} &= \sum_i \int d^{p+1}x \int d^{p+1}x^\prime \, D^i_F(x,x^\prime) \notag\\[5pt]
    &= \sum_i \frac{i V_{p+1} \, \alpha^\prime \, \Gamma^2_i }{2 (2 \pi \alpha^\prime )^{13}} \int_0^\infty \frac{ds}{s^{(25-p)/2}} \, e^{- \frac{\Delta y^2}{2 \alpha^\prime s}} \, e^{- \frac{\alpha^\prime s}{2}m^2_i},
    \label{D-brane interaction general reasoning}
\end{align}
where as before $\Delta y^2 \equiv (y^I_2 - y^I_1 )(y^I_2 - y^I_1 )$ contains only the directions orthogonal to the branes. In order to rewrite the amplitude $\mathcal{A}_\text{open}$ in terms of our new parametrization of the cylinder we change integration variables to $s = \pi/ t$ and use the fact that the eta function satisfies the property 
\begin{equation}
\eta(it) = \frac{\eta(i/t)}{\sqrt{t}} = \sqrt{\frac{s}{\pi}} \, \eta(is/\pi),
\label{eta function S transf}
\end{equation}
a proof of which is available in \cite{Siegel_1954}. This gives
\begin{equation}
    \mathcal{A}_\text{open} = \frac{i V_{p+1} \pi^{24}}{(8 \pi^3 \alpha^\prime)^{(p+1)/2}} \int_0^\infty \frac{ds}{s^{(25-p)/2}} \, e^{- \frac{\Delta y^2}{2 \alpha^\prime s}} \, \eta (i s/\pi)^{-24}.
\end{equation}
The expansion of $\eta(is/ \pi)$ for large $s$ is done as before, leading to
\begin{equation}
    \mathcal{A}_\text{open} \xrightarrow{ s \to \infty }  \frac{i V_{p+1} \pi^{24}}{(8 \pi^3 \alpha^\prime)^{(p+1)/2}} \int^\infty \frac{ds}{s^{(25-p)/2}} \, e^{- \frac{\Delta y^2}{2 \alpha^\prime s}} \, \Big( e^{2s} + 24 + \dots \Big),
\end{equation}
which is of the exact form we expected, with the sum going over the spectrum of the closed string.

Encouraged by this apparent miracle, we now examine whether the full interaction amplitude can be obtained from a first principles closed string computation. The form of the cylinder Faddeev-Popov measure is the same as before, as it makes no reference to what boundary conditions are chosen for the $X^\mu$ fields. In terms of the new $\xi^a$ coordinates the $c$ ghost insertion is $c^1(\hat{\xi})$, because the translation parallel to the boundary is now a shift of $\xi^1$. This also means that the volume of the conformal Killing group is $2 \pi$. The modulus-dependent version of the metric in the new coordinates is $\text{diag}(1,s^2)$, so the $b$ ghost insertion is $\big( \hspace{-3.5pt} \big( b,\partial_s \delta(s) \big) \hspace{-3.5pt} \big) = \frac{2}{s} \int d^2 \xi \, b_{22}(\xi)$. Therefore we have
\begin{equation}
    \mathcal{A}_\text{closed} = \frac{1}{2} \int_0^\infty \frac{ds}{2 \pi} \int \mathcal{D} X \,e^{-S_\text{P}[X]} \, \frac{1}{2 \pi s} \int d^2 \xi \int \mathcal{D}b \,  \mathcal{D}c \, e^{-S_g[b,c]}  \, b_{22}(\xi) \, c^1(\hat{\xi})
\end{equation}
for the amplitude. The fundamental difference with respect to the open case is hidden in the path integration limits. Whereas the open string path integral over the matter CFT was a zero-point function, or a partition function, in this case we integrate over worldsheets that interpolate between two different states, an initial one where we have a closed string being emitted from the brane at $X^I=y_1^I$, and a final one where it is absorbed by the brane at $X^I=y_2^I$. The interpretation is that of a transition amplitude between these states, like the general matrix element \eqref{amplitude}. We once again separate the path integrals, this time as
\begin{equation}
    \mathcal{A}_\text{closed} = \frac{1}{2} \int_0^\infty \frac{ds}{2 \pi} \, \mathcal{A}^X_\text{closed}(s) \, \mathcal{A}^g_\text{closed}(s),
\end{equation}
with
\begin{equation}
    \mathcal{A}^X_\text{closed}(s) = \int \mathcal{D}X \, e^{-S_\text{P}[X]} \hspace{0.2cm} , \hspace{1cm} \mathcal{A}^g_\text{closed}(s) = \frac{1}{2 \pi s} \int d^2\xi \int \mathcal{D}b \, \mathcal{D}c \, e^{- S_g[b,c]  } \,  b_{22}(\xi)  c^1(\hat{\xi}).
\end{equation}
As for the open string computation, a useful strategy is to translate these into operator language. Starting with the matter CFT, we have \cite{Blumenhagen:2013fgp}
\begin{align}
    \mathcal{A}^X_\text{closed}(s) &=   \braket{Dp,y_2|e^{-s H}|Dp,y_1} \notag\\[5pt]
    &= \int_0^{2 \pi} \frac{d \theta}{2 \pi} \,  \braket{Dp,y_2|e^{-s (L_0 +\tilde{L}_0+2a^X )} \, e^{i \theta(L_0 -\tilde{L}_0)} |Dp,y_1},
\end{align}
where $H=L_0+\tilde{L}_0 + 2a^X$ is the closed string Hamiltonian, responsible for propagating a closed string from the first D-brane to the second one. The states $\ket{Dp,y_1}$ and $\ket{Dp,y_2}$ are called boundary states. They are responsible for imposing on the worldsheet the boundary conditions that attach it to the branes. The integral
\begin{equation}
    \int_0^{2 \pi} \frac{d \theta}{2 \pi} \, e^{i \theta(L_0 - \tilde{L}_0)} = \delta_{L_0 \, \tilde{L}_0}
\end{equation}
is added to impose level matching. 

In the open string description, the Dirichlet conditions imposed on $25-p$ directions fix the string's endpoints to the worldvolume of the branes, whereas the Neumann conditions on the remaining $p+1$ directions leave the string's endpoints free to move inside the branes. The boundary states are determined by requiring these same physical conditions to hold for the closed string worldsheet:
\begin{align}
    &\partial_2 X^\mu(\xi^1,0)  \ket{Dp,y_1} = 0 \\[5pt]
    &X^I(\xi^1,0) \ket{Dp,y_1} = y^I_1,
\end{align}
with similar conditions for $\ket{Dp,y_2}$. Upon expanding these in terms of the mode expansion \eqref{closed}, one obtains 
\begin{equation}
    (\alpha^\mu_n + \tilde{\alpha}^\mu_{-n})\ket{Dp,y_1} = (\alpha^I_n - \tilde{\alpha}^I_{-n})\ket{Dp,y_1} = 0.
\end{equation}
for the modes and
\begin{align}
    &p^\mu \ket{Dp,y_1} = 0 \notag\\[5pt]
    &y^I \ket{Dp,y_1} = y^I_1 \ket{Dp,y_1}
    \label{conditions on spatial state}
\end{align}
for the center of mass variables, where on the left-hand side of the second line we have the center of mass operator $y^I$ that appears on the mode expansion and on the right-hand side we have the position of the first brane $y^I_1$, which appears here as an eigenvalue of $y^I$. We can split $\ket{Dp,y_1} = \ket{Dp} \ket{y_1}$ and determine each state separately. The two conditions \eqref{conditions on spatial state} mean that $\ket{y_1}$ is a center of mass eigenstate on the transverse directions and is translationally invariant on the worldvolume directions, so
\begin{equation}
    \ket{y_1} = \int \frac{d^{25-p}k}{(2 \pi)^{25-p}} e^{-i k^I y^I_1} \ket{k^\perp},
\end{equation}
where $\ket{k^\perp} = \ket{0,\dots,0,k^{p+ 1},\dots,k^{26}}$ is the momentum sector of a closed string ground state with $k^\mu=0$. The conditions related to the modes are solved by
\begin{equation}
    \ket{Dp} = N_p \exp{ \left( - \sum_{n=1}^\infty \frac{ \alpha^\mu_{-n} \, \tilde{\alpha}_{-n \, \mu} - \alpha^I_{-n} \, \tilde{\alpha}^I_{-n} }{n}  \right) } \ket{0}
    \label{bosonic boundary state}
\end{equation}
where $\ket{0}$ is the mode part of the closed string ground state and $N_p$ is some normalization that is not fixed by the boundary conditions.\footnote{This becomes easy to see with the following trick. The mode operators $\alpha^\mu_n$ can be traded for occupation number operators $a^\mu_n = \alpha^\mu_n /\sqrt{n}$, $a^\mu_{-n} = \alpha^{\dagger \mu}_n /\sqrt{n}$ which satisfy the usual harmonic oscillator commutation relation 
\begin{equation}
    a_{n \mu} \, a^{\dagger \nu}_n \ket{Dp} = a^{\dagger \nu}_n \, a_{n \mu} \ket{Dp} + \delta^\nu_\mu \ket{Dp}.
\end{equation}
The algebraic structure is identical to the one found in 
\begin{equation}
    \partial_\mu ( x^\nu f(x) ) = x^\nu \partial_\mu f(x) + \delta^\nu_\mu\, f(x).
\end{equation}
This motivates the identifications
\begin{align}
     a^{\dagger \mu}_n \to x^\mu , && a_{n \, \mu} \to \partial_{x^\mu}, && \tilde{a}^{\dagger \mu}_n \to y^\mu , && \tilde{a}_{n \, \mu} \to \partial_{y^\mu}, && \ket{Dp} \to f(x,y),
\end{align}
in terms of which the $(a^\mu_n + \tilde{a}^\mu_{-n})\ket{Dp}=0$ condition becomes $\partial_{x^\mu} f(x,y) = - y^\mu f(x,y) $. Taking $n \to -n$ produces $\partial_{y^\mu} f(x,y) = - x^\mu f(x,y)$. The solution of this simple system is $f(x,y) = K e^{- x_\mu y^\mu }$, where $K$ is some normalization constant. For the Dirichlet directions one finds $f(x,y) = K e^{ x^I y^I }$ instead. Doing this for every positive $n$ and undoing the identifications leads to the form shown for $\ket{Dp}$. \label{boundary state trick footnote} } The other boundary state is identical except for the $y^I_1 \to y^I_2$ substitution. These states already satisfy $L_0 = \tilde{L}_0$, so the $\theta$ integral is trivial and we have
\begin{equation}
    \mathcal{A}^X_\text{closed}(s) = e^{-2 a^X s} \braket{y_2 | e^{- \frac{\alpha^\prime s}{2} p^2 } | y_1} \braket{Dp| e^{-s(N+\tilde{N})} |Dp}.
\end{equation}
With the plane wave decomposition of the position eigenstates and the normalization $\braket{k^\perp|q^\perp} = i V_{p+1} (2 \pi)^{25-p} \delta^{25-p}(k^\perp - q^\perp)$, the factor of $i$ coming from the Euclidean signature, the first matrix element is easily found to be
\begin{align}
    \braket{y_2 | e^{- \frac{\alpha^\prime s}{2} p^2 } | y_1} &= i V_{p+1} \int \frac{d^{25-p}k}{(2 \pi)^{25-p}} e^{- \frac{\alpha^\prime s}{2}k^I k^I + i k^I(y_2-y_1)^I} \notag\\[5pt]
    &= i V_{p+1} \left( \frac{1}{2 \pi \alpha^\prime s} \right)^{(25-p)/2} \, e^{- \frac{\Delta y^2}{2 \alpha^\prime s}}.
\end{align}
For the second one, we start by computing just
\begin{align}
    &e^{-s(N+\tilde{N})} \ket{Dp} = N_p \prod_{\mu,I} \, \prod_{n=1}^\infty e^{-s(N+\tilde{N})} \, e^{-\frac{1}{n} (\alpha^\mu_{-n} \, \tilde{\alpha}^\mu_{-n} - \alpha^I_{-n} \, \tilde{\alpha}^I_{-n}) } \ket{0} \notag\\[5pt]
    &= N_p \prod_{\mu,I} \, \prod_{n=1}^\infty \sum_{m,k} \frac{(-1)^m}{m!} \frac{1}{k!} \frac{1}{n^m} \frac{1}{n^k} e^{-s(N+\tilde{N})} \, (\alpha^\mu_{-n})^m \, (\tilde{\alpha}^\mu_{-n })^m \, (\alpha^I_{-n})^k \, (\tilde{\alpha}^I_{-n})^k \ket{0},
\end{align}
with the Einstein summation convention temporarily suspended. This string of creation operators acting on the vacuum has level $N = \tilde{N} = n(m+k)$, so we find
\begin{equation}
    e^{-s(N+\tilde{N})} \ket{Dp} = N_p \prod_{\mu,I, n} \, \sum_{m,k} \frac{1}{m!} \left( -\frac{e^{-2sn}}{n} \right)^m \frac{1}{k!} \left( \frac{e^{-2sn}}{n} \right)^k (\alpha^\mu_{-n})^m \, (\tilde{\alpha}^\mu_{-n })^m \, (\alpha^I_{-n})^k \, (\tilde{\alpha}^I_{-n})^k \ket{0}.
\end{equation}
We similarly expand the bra
\begin{equation}
    \bra{Dp} = \bar{N}_p \prod_{\mu,I,n} \sum_{l,r} \frac{(-1)^l}{l!} \frac{1}{r!} \bra{0}  (\alpha^\mu_n)^l \, (\tilde{\alpha}^\mu_n)^l \, (\alpha^I_n)^r \, (\tilde{\alpha}^I_n)^r.
\end{equation}
The contraction of these two expressions leads to products and sums of the matrix element
\begin{equation}
    \braket{0|(\alpha^\mu_n)^l \, (\tilde{\alpha}^\mu_n)^l \, (\alpha^I_n)^r \, (\tilde{\alpha}^I_n)^r \, (\alpha^\mu_{-n})^m \, (\tilde{\alpha}^\mu_{-n })^m \, (\alpha^I_{-n})^k \, (\tilde{\alpha}^I_{-n})^k |0}.
\end{equation}
For each conjugate pair we use
\begin{equation}
    \braket{0|(\alpha^M_n)^a \, (\alpha^M_{-n})^b|0} = n^{(a+b)/2}  \braket{0|(a^M_n)^a \, (a^{\dagger M}_{n})^b|0} = n^a a! \, \delta_{ab},
\end{equation}
the last equality being a consequence of the usual orthogonality relation of number eigenstates $\braket{n|n^\prime} = \delta_{n \, n^\prime}$, where $\ket{n} = \frac{(a^{\dagger})^n}{\sqrt{n!}} \ket{0}$. After a wealth of cancellations one finally obtains
\begin{align}
    \braket{Dp| e^{-s(N+\tilde{N})} |Dp} &= |N_p|^2  \prod_{n=1}^\infty \left( \prod_{\mu=0}^p \sum_{m=0}^\infty e^{-2sn m} \right) \left(  \prod_{I=p+1}^{26} \sum_{l=0}^\infty e^{-2sn l} \right) \notag\\[5pt]
    &= |N_p|^2   \prod_{n=1}^\infty  \left( \frac{1}{1-e^{-2sn}} \right)^{p+1} \left( \frac{1}{1-e^{-2sn}} \right)^{25-p} \notag\\[5pt]
    &= |N_p|^2 e^{-\frac{26s}{12}} \,  \eta(is/\pi)^{-26}.
\end{align}
Therefore the total contribution from the matter CFT is
\begin{equation}
    \mathcal{A}^X_\text{closed}(s) = \frac{i V_{p+1} |N_p|^2}{(2 \pi \alpha^\prime s)^{(25-p)/2}} \,e^{-2s \big(a^X+\frac{26}{24}\big)} \, e^{- \frac{\Delta y^2}{2 \alpha^\prime s}} \, \eta(i s /\pi)^{-26}.
\end{equation}
Once again, the term involving $a^X$ goes away once we set $a^X = -26/24$.

The ghost contribution
\begin{equation}
    \mathcal{A}^g_\text{closed}(s) = 2 e^{-2s ( a^g - \frac{1}{12} \big)} \eta ( is/\pi )^2
\end{equation}
is computed in Appendix \ref{B2}. As usual, it cancels the contribution from the two unphysical matter fields. The final form of the amplitude is then
\begin{equation}
    \mathcal{A}_\text{closed} = \frac{i V_{p+1} |N_p|^2}{2 \pi (2 \pi \alpha^\prime)^{(25-p)/2}} \int_0^\infty \frac{ds}{s^{(25-p)/2}} \, e^{ -\frac{\Delta y^2}{2 \alpha^\prime s} } \, \eta(i s /\pi)^{-24},
\end{equation}
which is equal to $\mathcal{A}_\text{loop}$ for $N_p = (2 \pi^3 \alpha^\prime)^6/(2 \pi \sqrt{\alpha^\prime})^p$. This concludes our verification of open-closed duality, for arguably the simplest process where it can be seen. Although surprising from the point of view of the field theory limit, this duality has an almost tautological nature from the worldsheet point of view. The path integral of the worldsheet CFT over a cylindrical surface stretched between two branes is a well defined quantity in itself. It is only when one makes a decision of how to cut open the cylinder into one-dimensional slices that the distinction between open and closed strings makes sense. At each fixed $\tau$, the path integration measure over the string's shape is given by
\begin{equation}
    \mathcal{D} X \big|_{\text{fixed }\tau} = \prod_{\mu,\sigma} dX^\mu(\tau,\sigma).
\end{equation}
What we have called the open string interpretation corresponds heuristically to writing the path integral over the worldsheet as
\begin{equation}
    \int \mathcal{D}X = \int \prod_{\mu,\tau,\sigma} dX^\mu(\tau,\sigma) = \prod_\tau \int  \mathcal{D} X \big|_{\text{fixed }\tau}
\end{equation}
By doing the integrals at fixed $\tau$ before taking the product, the intermediary configurations that the integral sums over are necessarily open strings. If one instead fixes $\sigma$ and integrates first with respect to all shapes the string can take as one moves along the cylinder's diameter, one finds
\begin{equation}
    \int \mathcal{D}X  = \prod_{\sigma} \int  \mathcal{D} X \big|_{\text{fixed }\sigma},
\end{equation}
and the intermediary states are closed strings. Either way, the result must be the same, no matter how distinct these intermediary states can seem to be at low energies. 

As mentioned in the beginning of this chapter, this duality is intimately related to AdS/CFT. One relevant fact is that the boundary states
\begin{equation}
    \ket{Dp,y} = N_p \exp{ \left( - \sum_{n=1}^\infty \frac{ \alpha^\mu_{-n} \, \tilde{\alpha}_{-n \, \mu} - \alpha^I_{-n} \, \tilde{\alpha}^I_{-n} }{n}  \right) } \ket{y},
\end{equation}
which are how D-branes are incorporated into the closed string interpretation, have the form of coherent states in the closed string Hilbert-space.\footnote{Coherent states can always be written as exponentials of creation operators acting on the vacuum \cite{Altland_Simons_2010}.} Since the coherent states of a theory are precisely those that correspond to classical solutions, one may take the success of our computation as an indication that perhaps D-branes, which up this point have been discussed exclusively from an open string point of view, also admit a dual description as closed string classical backgrounds.

%\end{document}

%\documentclass[a4paper,12pt]{memoir}
%\usepackage{graphicx}
%\usepackage[utf8]{inputenc}
%\usepackage{indentfirst}
%\usepackage{braket}
%\usepackage{setspace}
%\usepackage{amsmath, amsthm, amssymb, amsfonts,bm}
%\usepackage[multiple]{footmisc}
%\usepackage{mathtools, changepage, slashed}
%\usepackage{tikz-feynman}
%\usepackage{bm, mathrsfs}
%\usepackage{gensymb}
%\usepackage[a4paper,top=3cm,left=3cm,right=2cm,bottom=2cm]{geometry}
%\usepackage{epstopdf}
%\usepackage{hyperref}
%\usepackage{pgfplots}
%\pgfplotsset{compat=1.18} 
%\usepackage[sorting=none]{biblatex}
%\addbibresource{refs.bib}
%\numberwithin{equation}{section}
%\usepackage[inkscapelatex=false]{svg}
%\usepackage[super]{natbib}
%\usepackage{doi}
%\hypersetup{
%  colorlinks   = true, %Colours links instead of ugly boxes
%  urlcolor     = black, %Colour for external hyperlinks
%  linkcolor    = black, %Colour of internal links
%  citecolor   = black %Colour of citations
%}
%\DeclareMathOperator{\Tr}{Tr}

%\newcommand{\normalord}[1]{%
%  {:\mathrel{\mspace{1mu}#1\mspace{1mu}}:}%
%}

%\OnehalfSpacing
%\usepackage{newtx}
%\usepackage{newtxtext}
%\usepackage{lmodern}

%\title{Superstrings chapter}
%\author{pedrobairrao}

%\begin{document}

\chapter{Superstrings}\label{ch7}

\section{The RNS formalism}\label{sec71}

The bosonic string we have discussed so far has two undesirable features. One is the absence of spacetime fermions. Each propagating particle that comes out of string theory is in direct correspondence with some way the string can be excited, that is, with some state
\begin{equation}
    \alpha^{\mu_1}_{-n_1}  \alpha^{\mu_2}_{-n_2} \dots \ket{k}  
\end{equation}
of the worldsheet CFT's spectrum. Since the $X^\mu$ fields are bosonic, so are all of its modes and therefore all states of the above form. The second is the tachyon, which we have so far conveniently ignored. Although its presence does not automatically mean that the theory is inconsistent (see footnotes \ref{tachyon condensation footonote} and \ref{second tachyon footnote}), it does mean its actual low-energy limit is not given by the truncation to the massless level, making all of the previous chapters' statements related to low-energy dynamics unreliable. As will become clear later in this chapter, the AdS/CFT correspondence emerges from taking certain low-energy limits of string theory, so having a good handle on the theory in this regime is fundamental for understanding it. 

Our strategy will be to focus at first on the elimination of the tachyon, since this will naturally lead to spacetime fermions. Recall that the ground state's mass is determined by the normal ordering constant, which was shown in Section \ref{sec33} to be the proportional to the worldsheet's regularized zero-point energy. It is a well known property of field theory that bosonic and fermionic degrees of freedom contribute oppositely to the zero-point energy. This raises the possibility of adding one fermion for each boson on the worldsheet so that their contributions exactly cancel. Worldsheet Lorentz symmetry requires that these fermions be packaged into worldsheet spinors
\begin{equation}
    \psi(\sigma) =
    \begin{pmatrix}
        \psi_+(\sigma) \\
        \psi_-(\sigma)
    \end{pmatrix},
\end{equation}
out of which one builds the scalars $\overline{\psi} \rho^a \partial_a \psi$, where
\begin{equation}
    \rho^0 = 
    \begin{pmatrix}
        0 && 1 \\
        -1 && 0
    \end{pmatrix} \, , \hspace{1cm} \rho^1 = 
    \begin{pmatrix}
        0 && 1 \\
        1 && 0
    \end{pmatrix}
    \label{2d Dirac matrices}
\end{equation}
are two-dimensional Dirac matrices satisfying $\{ \rho^a,\rho^b \} = 2 \eta^{ab}$ and $\overline{\psi} = \psi^\dagger \rho^0$. The $+$ or $-$ signs on the components of $\psi$ refer to the chirality, i.e. the eigenvalue with respect to the chiral matrix
\begin{equation}
    \overline{\rho} = \rho^0 \rho^1 =
    \begin{pmatrix}
        1 && 0 \\
        0 && -1
    \end{pmatrix}.
\end{equation}
If $\psi$ is a Dirac spinor, its components are complex, giving a total of four off-shell independent fermionic degrees of freedom per spinor. The equations of motion reduce this to two on-shell degrees of freedom. In two dimensions one has Majorana spinors, which are real, so for these the number of on-shell independent degrees of freedom per spinor is 1. From two real spinors we can always build a complex one, so for simplicity we take the $\psi$ to be Majorana. Since the central charge of a CFT is tied to its field content, we expect our modifications to alter the the critical dimension, so we go back to calling it $D$. We must therefore add $D$ Majorana spinors to the Polyakov action in order to cancel the zero-point energy. Our choice, whose utility will become clear later, is to make each fermion a spacetime vector and add $\overline{\psi}^\mu \rho^a \partial_a \psi_\mu$ to the action. The result is the Ramond-Neveu-Schwarz action
\begin{equation}
    S_\text{RNS} = - \frac{1}{4 \pi} \int  d \tau d \sigma \bigg( \frac{1}{\alpha^\prime} \partial^a X^\mu \partial_a X_\mu + i\overline{\psi}^\mu \rho^a \partial_a \psi_\mu \bigg).
    \label{RNS superconformal gauge}
\end{equation}
This action possesses the rigid $N=1$ on-shell supersymmetry\footnote{On-shell supersymmetry means that the supersymmetry algebra only closes upon using the equations of motion. This is to expected for $S_\text{RNS}$ because each Majorana spinor has two off-shell degrees of freedom, the Majorana equation of motion is necessary to reduce this to one. Therefore the matching of bosonic and fermionic degrees of freedom only works on-shell. In order to have an off-shell supersymmetric version of the RNS action is is necessary to add $D$ new bosonic auxiliary fields $F^\mu$:
\begin{equation}
    S_\text{RNS} = - \frac{1}{4 \pi} \int d \tau  d\sigma \bigg( \frac{1}{\alpha^\prime} \partial^a X^\mu \partial_a X_\mu + i\overline{\psi}^\mu \rho^a \partial_a \psi_\mu + F^\mu F_\mu \bigg).
\end{equation}
The $N=1$ off-shell supersymmetry is \cite{GreenSchwarzWitten_vol1}
\begin{align}
    \sqrt{\frac{2}{\alpha^\prime}} \delta_\epsilon X^\mu = i \overline{\epsilon} \psi^\mu, && \delta_\epsilon \psi^\mu = \bigg( \frac{1}{2} \sqrt{\frac{2}{\alpha^\prime}} \rho^a \partial_a X^\mu + F^\mu \bigg) \epsilon \, , && \delta_\epsilon F^\mu = - i \overline{\epsilon} \rho^a \partial_a \psi^\mu.
\end{align}
While necessary for the off-shell closure of the algebra, $F^\mu$ plays no dynamical role in the theory, as its equation of motion is $F^\mu=0$. One may therefore set it to zero, thus arriving at the action \eqref{RNS superconformal gauge}. \label{off shell SUSY footnote} }
\begin{equation}
    \sqrt{\frac{2}{\alpha^\prime}} \delta_\epsilon X^\mu = i \overline{\epsilon} \psi^\mu, \hspace{1cm} \delta_\epsilon \psi^\mu =  \frac{1}{2} \sqrt{\frac{2}{\alpha^\prime}} \rho^a \epsilon \, \partial_a X^\mu  ,
    \label{SUSY of RNS}
\end{equation}
where $\epsilon$ is a constant Majorana fermion parameter. This supersymmetry is expected to guarantee the cancellation of the zero-point energy.

The fact that the $\psi^\mu$ are worldsheet spinors and spacetime vectors might appear strange, but such constructions are commonplace in the worldline formalism for point particles. Consider for instance the equivalent of $S_\text{RNS}$ in one lower dimension, commonly called the spinning particle action
\begin{equation}
    S_\text{spin part} =  \frac{1}{2} \int d \tau\left( \dot{x}^\mu \dot{x}_\mu - i \psi^\mu  \dot{\psi}_\mu \right).
\end{equation}
This is nothing but the point particle action \eqref{Sprime} in ``conformal gauge'' $g_{\tau \tau}=1$, with worldline fermions included. The equations of motion are $\ddot{x}^\mu(\tau)=0$, solved by $x^\mu(\tau)=x^\mu_0 + p^\mu \tau$, and $\dot{\psi}^\mu=0$, solved by constant $\psi^\mu$. By varying $\delta x^\mu = \omega^{\mu}{}_\nu x^\nu$ and $\delta \psi^\mu = \omega^\mu{}_\nu \psi^\nu$ with $\omega^{\mu \nu}$ antisymmetric, we obtain the Lorentz generators
\begin{equation}
    M^{\mu \nu} = x^\mu P^\nu - x^\nu P^\mu + S^{\mu \nu},
\end{equation}
where $P^\mu = \dot{x}^\mu$ are the translation generators and
\begin{equation}
    S^{\mu \nu} = - \frac{i}{2}[\psi^\mu,\psi^\nu].
\end{equation}
The conjugate momentum of $\psi^\mu$ is $ p^\mu_\psi = \frac{\delta S_\text{spin part}}{\delta \dot{\psi}_\mu} = \frac{i}{2}  \psi^\mu$. The canonical equal time anticommutator one must impose to quantize this theory is therefore equivalent to
\begin{equation}
    \{ \psi^\mu , \psi^\nu \} = 2  \eta^{\mu \nu}.
\end{equation}
There is no $\tau$ dependence because the $\psi^\mu$ are constant by their equations of motion. This means that upon quantization these variables effectively become spacetime Dirac matrices,
and $S^{\mu \nu} = -\frac{i}{2}[\psi^\mu , \psi^\nu]$ becomes the spin generator for a Dirac spinor.\footnote{This is an example of a more general procedure of using fermionic variables to add internal degrees of freedom to a system. For a succinct but quite physical discussion of this, see \cite{Barducci:1977gy}. A detailed introduction to this method can be found in \cite{BEREZIN1977336}. } One might be tempted to say that $S_\text{spin part}$ describes the dynamics of a single spacetime fermion, but there is still one element missing: the spacetime Dirac equation. In the case of the scalar particle we managed to get the Klein-Gordon equation by interpreting $S_\text{scalar part} = \frac{1}{2}\int d \tau \, ( \dot{x}^\mu \dot{x}_\mu - m^2 )$ as a particular gauge-fixing of the reparametrization-invariant action
\begin{equation}
    S_\text{scalar part} =  \frac{1}{2} \int d\tau \, e \Big( e^{-2} \dot{x}^\mu \dot{x}_\mu - m^2 \Big),
\end{equation}
written here in terms of the einbein $e = \sqrt{|g_{\tau \tau}|}$. One may understand the process of obtaining this covariant form of the action from the flat one as coupling the scalars $x^\mu$ to one-dimensional gravity, or, equivalently, taking the rigid proper time translation symmetry of $S_\text{scalar part}$ and gauging it. As shown in Section \ref{sec31}, the equation of motion from varying the einbein becomes the $\dot{x}^\mu \dot{x}_\mu +m^2=0$ constraint after eliminating $e$ with a gauge transformation. This trick by itself does not work for the spinning particle, as minimally coupling its action to worldsheet gravity merely changes it to
\begin{equation}
    \frac{1}{2} \int d\tau \, e \big( e^{-2}\dot{x}^\mu  \dot{x}_\mu - i e^{-1} \psi^\mu \dot{\psi}_\mu \big)
\end{equation}
and the variation with respect to $e$ is the same as for the scalar particle. To obtain an additional constraint, a new gauge symmetry is required. Note that in addition to shifting $\tau$, the spinning particle action is also invariant under the global worldline supersymmetry transformation
\begin{equation}
    \delta_\epsilon x^\mu = i \epsilon \psi^\mu , \hspace{0.5cm} \delta_\epsilon \psi^\mu =  \epsilon \dot{x}^\mu,
\end{equation}
with $\epsilon$ being a Grassmann number. The $\tau$ shift is actually part of this symmetry algebra, as it is easily verified that
\begin{equation}
    \delta_{\epsilon_2} \delta_{\epsilon_1} = \delta_a\big|_{a = i \epsilon_1 \epsilon_2}
\end{equation}
where $\delta_a = a \partial_\tau$ is the operator that implements the transformation $\tau \to \tau - a$ for infinitesimal $a$.  We can then see what comes out of gauging this supersymmetry, or, equivalently, of coupling the spinning particle to $N=1$ worldline supergravity. The result is \cite{Ortin_2015}
\begin{equation} 
    S_\text{SUGRA part} =  \frac{1}{2} \int d \tau \, e \big( e^{-2}\dot{x}^\mu  \dot{x}_\mu - ie^{-1} \psi^\mu \dot{\psi}_\mu - ie^{-2} \chi \psi^\mu \dot{x}_\mu  \big),
\end{equation}
where the fermionic field $\chi$ is the one-dimensional gravitino, the supersymmetric partner of the einbein. The local supersymmetries satisfied by this action are
\begin{align}
    &\delta_\epsilon x^\mu = i\epsilon \psi^\mu, \hspace{0.5cm} \delta_\epsilon \psi^\mu = \epsilon \left( \dot{x}^\mu - \frac{i}{2} \chi \psi^\mu \right) e^{-1}, \notag\\[5pt]
    &\delta_\epsilon e = i\epsilon \chi \, , \hspace{0.9cm} \delta_\epsilon \chi = 2 \dot{\epsilon},
\end{align}
where now $\epsilon = \epsilon(\tau)$ is a local Grassmann parameter. By doing a SUSY-gauge transformation with $\dot{\epsilon} = -\frac{1}{2} \chi$ we eliminate the gravitino, and a worldline reparametrization can then be done to set $e=1$, taking us back to the spinning particle action. The equations of motion from varying with respect to $e$ and $\chi$, respectively $\dot{x}^\mu \dot{x}_\mu - i\chi \psi^\mu \dot{x}_\mu=0$ and $\psi^\mu \dot{x}_\mu=0$, then become the off-shell constraints
\begin{equation}
    p^2 =0  \, , \hspace{0.5cm} \psi^\mu p_\mu =0,
\end{equation}
where $p^\mu = \dot{x}^\mu$ is the particle's momentum. Upon quantization the second one becomes the massless Dirac equation. It is the dynamics defined by the spinning particle action with the addition of these two constraints that is identified with that of a spacetime fermion.\footnote{Many more details on the spinning particle and some generalizations of it can be found in \cite{BRINK197776}, where the authors show that the theory has also a bosonic sector in addition to the fermionic one alluded to here. This also parallels what happens for the RNS string. } The inclusion of the Dirac constraint from worldline supergravity is not really a choice, it is required for the theory to be well defined, since it is what renders unphysical the fermionic state that follows from acting with $\psi^0(\tau)$ on the vacuum. This state has negative norm as a consequence of the minus sign in $\{ \psi^0,\psi^0 \}=-2$. This mirrors how the Klein-Gordon constraint from worldline diffeomorphism symmetry renders unphysical the bosonic negative norm state created by $x^0(\tau)$. 

The RNS string is a direct worldsheet generalization of the spinning particle, and for this reason it is commonly called the spinning string, or fermionic string. Its gauge symmetries and constraints parallel those of the particle very closely, just like what happens for the bosonic string and particle. In that context the Virasoro constraints were a consequence of the worldsheet diffeomorphism symmetry of the Poyakov action. We will see that local worldsheet supersymmetry likewise leads for the RNS string to a second set of crucial constraints. We therefore declare the gauge-invariant form of the RNS action as the coupling of it to $N=1$ worldsheet supergravity \cite{Blumenhagen:2013fgp}\footnote{A detailed construction of it via the Noether procedure can be found in \cite{Bailin:1994qt}}:
\begin{equation}
    S = - \frac{1}{4 \pi} \int d \tau d \sigma \, e  \bigg[ \frac{1}{\alpha^\prime} g^{ab} \partial_a X^\mu \partial_b X_\mu + i\overline{\psi}^\mu \rho^a \partial_a \psi_\mu - i \overline{\chi}_a \rho^b \rho^a\psi^\mu \bigg( (2 \alpha^\prime)^{-1/2}  \partial_b X_\mu - \frac{i}{8} \overline{\chi}_b \psi_\mu \bigg)  \bigg],
\end{equation}
with auxiliary fields required for off-shell closure of the supersymmetry algebra omitted (see footnote \ref{off shell SUSY footnote}). We also omit all spinor indices and use a tilde to differentiate coordinate from tangent space indices, so the zweibein is written as $e^{\tilde{a}}_a$ and satisfies
\begin{align}
    e^{\tilde{a}}_a e^{\tilde{b}}_b \, g^{ab} &= \eta^{\tilde{a} \tilde{b}} \notag\\[5pt]
    e^{\tilde{a}}_a e^{\tilde{b}}_b \, \eta_{\tilde{a} \tilde{b}} &= g_{ab},
\end{align}
where $\eta_{\tilde{a} \tilde{b}}$ is the flat tangent space metric. The worldsheet gravitino $\chi_a$ has, in addition to the vector index shown, one hidden Majorana spinor index, as is appropriate for the spin $\frac{3}{2}$ superpartner of the graviton. The factor of $e$ next to $d \tau d \sigma$ is shorthand for the metric determinant
\begin{equation}
    e \equiv |\det(e^{\tilde{a}}_a)| = \sqrt{-g}.
\end{equation}
The Dirac matrices with coordinate indices that appear in the action are defined as $\rho^a = e^a_{\tilde{a}} \rho^{\tilde{a}}$, with the $\rho^{\tilde{a}}$ given by \eqref{2d Dirac matrices}, and satisfy
\begin{equation}
    \{ \rho^a , \rho^b \} = 2 g^{ab}.
\end{equation}
In addition to local supersymmetry, this action is also invariant under the super-Weyl transformation
\begin{equation}
    \delta_\eta \chi_a = \rho_a \eta \, , \hspace{0.5cm} \delta_\eta (\text{others})=0,
\end{equation}
where $\eta(\tau,\sigma)$ is a Majorana spinor parameter. See \cite{Blumenhagen:2013fgp} for the explicit form of all the gauge symmetries, as well as how to fix them to restore the CFT form \eqref{RNS superconformal gauge}, called ``superconformal gauge''. The procedure is analogous to the particle, one uses diffeomorphisms, local supersymmetry, Weyl and super-Weyl transformations to gauge-fix the gravitino to zero and the metric to $\eta_{ab}$. 

Just like the equation of motion from varying with respect to the zweibein means the vanishing of the energy-momentum tensor
\begin{equation}
    T_{ab} = \frac{4 \pi}{\sqrt{-g}} \frac{\delta S}{\delta g^{ab}} = \frac{2 \pi}{e} e_{a \tilde{a}} \frac{\delta S}{\delta e^b_{\tilde{a}}} =0 ,
\end{equation}
the equation of motion from varying with respect to the gravitino guarantees the vanishing of the supercurrent
\begin{equation}
    J_a = - \frac{2 \pi i}{e} \frac{\delta S}{\delta \overline{\chi}^a}=0.
\end{equation}
In superconformal gauge these become the constraints
\begin{align}
    T_{ab} &= - \frac{1}{\alpha^\prime} \bigg( \partial_a X^\mu \partial_b X_\mu - \frac{1}{2} g_{ab} \partial_c X^\mu \partial^c X_\mu \bigg) - \frac{i}{4} \bigg( \overline{\psi}^\mu \rho_a \partial_b \psi_\mu + \overline{\psi}^\mu \rho_b \partial_b \psi_\mu  \bigg) = 0, \\[5pt]
    J_a &= - \frac{1}{4} \sqrt{\frac{2}{\alpha^\prime}} \rho^b \rho_a \psi^\mu \partial_b X_\mu =0.
\end{align}
$T_{ab}$ is the conserved current associated to worldsheet translation symmetry and $J_a$ is the conserved current associated to worldsheet supersymmetry \eqref{SUSY of RNS}. The equations of motion derived from this action are
\begin{equation}
    \partial^2 X^\mu =0, \hspace{0.5cm} \rho^a \partial_a \psi^\mu=0.
\end{equation}
We know how the story goes for the bosonic fields, so here we focus on the fermionic ones. After separating the $\psi^\mu$ into chirality the eigenvectors $\psi^\mu_{\pm}$ (also known as Weyl spinors), the equations of motion read
\begin{equation}
    (\partial_\tau - \partial_\sigma)\psi^\mu_+ = (\partial_\tau + \partial_\sigma)\psi^\mu_- =0,
\end{equation}
which are solved by
\begin{equation}
    \psi^\mu_+ = \psi^\mu_+(\tau + \sigma) \, , \hspace{0.5cm} \psi^\mu_- = \psi^\mu_-(\tau - \sigma).
\end{equation}
The variation of the action also leads to the boundary term
\begin{equation}
    \int d \tau \big( \bar{\psi}^\mu \rho^1 \delta \psi_\mu \big)\Big|^{\sigma=l}_{\sigma=0} =
    \int d \tau \Big( \psi^\mu_+ \delta \psi_{+ \, \mu} - \psi^\mu_- \delta \psi_{- \, \mu} \Big)\Big|^{\sigma=l}_{\sigma=0}=0.
    \label{fermion boundary term}
\end{equation}
The choice of boundary conditions for the solutions involves a subtlety that is not present in the bosonic case. Assume we want to describe closed strings. The $\psi^\mu(\sigma)$ are then functions defined on a closed worldsheet, so it would be natural to expect them to satisfy the same $\sigma \sim \sigma + 2 \pi$ periodicity as the $X^\mu(\sigma)$. However, one could also satisfy \eqref{fermion boundary term} by choosing them to be antiperiodic instead, $\psi^\mu_{\pm}(\tau,0) = - \psi^\mu_{\pm}(\tau,2 \pi)$, as the minus signs in each product would always cancel. Every observable of the theory that involves the fermions can only do so via fermion bilinears, and both the periodic and antiperiodic choices guarantee that all such bilinears satisfy the worldsheet's periodicity.\footnote{Note that it is not consistent to have $\psi^\mu_{\pm}$ be periodic for some values of $\mu$ and antiperiodic for other values, because then bosonic fields such as $\psi_+^\mu \psi^\nu_+$ can still be antiperiodic for some $\nu \neq \mu$. Another argument is that the worldsheet supercurrent, which generates residual gauge symmetries and therefore must be well defined, involves the contraction $\psi^\mu X_\mu$. Since the $X_\mu$ are all periodic, all the $\psi^\mu$ must be either periodic or antiperiodic for the supercurrent to have a well defined periodicity \cite{bachas2024dbranes}.} The periodic and antiperiodic choices are referred to as Ramond (R) and Neveu-Schwarz (NS) boundary conditions, respectively. Writing the general periodicity as $\psi^\mu(\tau,\sigma + 2 \pi) = e^{2 \pi i \nu} \psi^\mu(\tau,\sigma)$, the solutions are given by
\begin{equation}
   \psi^\mu_-(\tau,\sigma) = \sum_{r \in\mathbb{Z} + \nu} \psi^\mu_r \, e^{- i r (\tau - \sigma)} , \hspace{0.5cm}  \psi^\mu_+(\tau,\sigma) = \sum_{r \in\mathbb{Z} + \tilde{\nu}} \tilde{\psi}^\mu_r \, e^{- i r (\tau + \sigma)} ,
\end{equation}
where $\nu = 0$ for Ramond and $\nu = \frac{1}{2}$ for Neveu-Schwarz. Reality of the fermions translates into $(\psi^{\mu}_r)^\ast = \psi^\mu_{-r}$ and $(\tilde{\psi}^{\mu}_r)^\ast = \tilde{\psi}^\mu_{-r}$ for the modes. The values of $\nu$ and $\tilde{\nu}$ can be chosen independently for each chirality, so the closed string sector splits into four: R-R, where $(\nu,\tilde{\nu}) = (0,0)$, R-NS and NS-R, with $(\nu,\tilde{\nu}) = (0,\frac{1}{2})$ or $(\frac{1}{2},0)$, and NS-NS, with $(\nu,\tilde{\nu}) = (\frac{1}{2},\frac{1}{2})$. 

The analysis of the constraints is simpler if we do a conformal transformation to the complex coordinates defined in \eqref{complex coords}. The scaling dimension of the worldsheet fermions is $\frac{1}{2}$, so the transformations are
\begin{align}
    \psi_-^\mu(\sigma) &\to \Psi^\mu(z) = \sqrt{\frac{\partial(\sigma^1 + i \sigma^2)}{\partial z}} \psi^\mu_-(z) = \sqrt{\frac{i}{z}}\psi^\mu_-(z) = \sqrt{i} \sum_{r \in\mathbb{Z} + \nu} \frac{\psi^\mu_r}{z^{r+1/2}} \\[5pt]
    \psi_+^\mu(\sigma) &\to \tilde{\Psi}^\mu(\bar{z}) = \sqrt{\frac{\partial(\sigma^1 - i \sigma^2)}{\partial \bar{z}}} \psi^\mu_+(\bar{z}) = \sqrt{\frac{1}{i \bar{z} }}\psi^\mu_+(\bar{z}) = \frac{1}{\sqrt{i}} \sum_{r \in\mathbb{Z} + \tilde{\nu}} \frac{\tilde{\psi}^\mu_r}{\bar{z}^{r+1/2}}.
\end{align}
The nonzero components of the energy-momentum tensor become
\begin{align}
    T(z) &= \frac{1}{\alpha^\prime} \partial X^\mu(z) \partial X_\mu (z) - \frac{1}{2} \Psi^\mu(z) \partial \Psi_\mu(z), \\[5pt]
    \tilde{T}(\bar{z}) &= \frac{1}{\alpha^\prime} \bar{\partial} X^\mu(\bar{z}) \bar{\partial} X_\mu (\bar{z}) - \frac{1}{2} \tilde{\Psi}^\mu(\bar{z}) \bar{\partial} \Psi_\mu(\bar{z}),
\end{align}
in terms of the notation defined in \eqref{complex T}. The mode expansion leads to the Laurent series
\begin{equation}
    T(z) = \sum_{n = - \infty}^\infty \frac{L_n}{z^{n+2}} = \sum_{n = - \infty}^\infty \frac{L^X_n + L^\psi_n}{z^{n+2}},
\end{equation}
where $L_n^X$ are the bosonic Virasoro generators studied in Section \ref{sec32} and 
\begin{equation}
    L^\psi_n = \frac{1}{4} \sum_{r \in \mathbb{Z} + \nu} (2r - n) \psi^\mu_{n-r} \psi_{r \, \mu}.
\end{equation}
The expressions for $\tilde{T}(\bar{z})$ are completely analogous. For the supercurrent we have
\begin{align}
    J(z) &= i \sqrt{\frac{2}{\alpha^\prime}} \Psi^\mu(z) \partial X_\mu(z) = \sum_{r \in \mathbb{Z} + \nu} \frac{G_r}{z^{r+3/2}} , \\[5pt]
    \tilde{J}(\bar{z}) &= i \sqrt{\frac{2}{\alpha^\prime}} \tilde{\Psi}^\mu(\bar{z}) \bar{\partial} X_\mu(\bar{z}) = \sum_{r \in \mathbb{Z} + \tilde{\nu}} \frac{\tilde{G}_r}{\bar{z}^{r+3/2}},
\end{align}
where
\begin{equation}
    G_r = \sum_{n= - \infty}^\infty \alpha^\mu_n \, \psi_{r-n \, \mu} \, , \hspace{0.5cm} \tilde{G}_r = \sum_{n= - \infty}^\infty \tilde{\alpha}^\mu_n \, \tilde{\psi}_{r-n \, \mu}.
\end{equation}
The $G_r$ together with the $L_n$ are referred to as super Virasoro generators, since they form a closed algebra which is in a sense the supersymmetric extension of the Virasoro algebra. It is called the superconformal algebra, or Ramond/Neveu-Schwarz algebra, depending on the value of $\nu$. We will not need its explicit form. The antiholomorphic modes form a second copy of this algebra. 

For open strings, we require the $\sigma=0$ and $\sigma=\pi$ terms of \eqref{fermion boundary term} to vanish separately. This binds one chirality to the other,
\begin{align}
    \psi_+^\mu(\tau,0) &= \pm  \psi_-^\mu(\tau,0), \notag\\[5pt]
    \psi_+^\mu(\tau,\pi) &= \pm  \psi_-^\mu(\tau,\pi).
\end{align}

The redefinition $\psi^\mu_{\pm} \to - \psi^\mu_{\pm}$ is a symmetry of the action, so only the relative sign between the boundary conditions at $\sigma=0$ and $\sigma=\pi$ is relevant. We can therefore choose $\psi_+^\mu(\tau,0) =   \psi_-^\mu(\tau,0)$ for an NN open string and have the relevant boundary condition be specific to the other endpoint:
\begin{equation}
    \psi^\mu_+(\tau,\pi) = \pm \psi^\mu_-(\tau,\pi) =e^{2 \pi i \nu} \psi^\mu_-(\tau,\pi) , \hspace{0.5cm} \text{(NN)}.
    \label{open RNS fermions NN}
\end{equation}
The supersymmetry transformation of the worldsheet spinors in terms of the components is
\begin{equation}
    \delta_\epsilon \psi^M_\pm =  \frac{1}{2} \sqrt{\frac{2}{\alpha^\prime}} \epsilon^\pm (\partial_\tau \pm \partial_\sigma ) X^M,
\end{equation}
where $\epsilon^A = (\epsilon^+,\epsilon^-)=(\epsilon_-,-\epsilon_+)$. Evaluating this at the boundaries gives $\delta_\epsilon \psi^\mu_\pm =  \frac{1}{2} \sqrt{\frac{2}{\alpha^\prime}} \epsilon^\pm \partial_\tau X^\mu$ for NN conditions, which is only compatible with \eqref{open RNS fermions NN} if the supersymmetry parameter satisfies
\begin{equation}
    \epsilon^+(\tau,0) = \epsilon^-(\tau,0) \, , \hspace{0.5cm} \epsilon^+(\tau,\pi) = e^{2 \pi i \nu} \epsilon^-(\tau,\pi).
\end{equation}
For a string with Dirichlet boundary conditions at both endpoints, one has $\delta_\epsilon \psi^I_\pm = \pm \frac{1}{2} \sqrt{\frac{2}{\alpha^\prime}} \epsilon^\pm \partial_\sigma X^I$ at the boundaries. Taking into account the boundary conditions of $\epsilon^\pm$, this determines $\psi_+^I(\tau,0) = -\psi_-^I(\tau,0)$ and $\psi_+^I(\tau,\pi) = - e^{2 \pi i \nu} \psi_-^I(\tau,\pi)$. As for the bosonic fields, both NN and DD boundary conditions lead to the two fermionic fields sharing the same set of modes:
\begin{equation}
    \psi^M_{\pm}(\tau,\sigma) =
    \begin{dcases}
        \hspace{0.36cm} \sum_{r \in\mathbb{Z} + \nu} \psi^\mu_r \, e^{- i r (\tau \pm \sigma)}, \hspace{0.5cm} \text{(NN)} \\[0.3cm]
        \pm 
        \sum_{r \in\mathbb{Z} + \nu} \psi^I_r \, e^{- i r (\tau \pm \sigma)}, \hspace{0.5cm} \text{(DD)}
    \end{dcases}
\end{equation}
In either case the open string splits into one R and one NS sector, depending on the value of $\nu$. The energy-momentum tensor and supercurrent lead to only one set of superconformal generators. Recall from Section \ref{sec42} that for the worldsheet bosons Dirichlet boundary conditions could be obtained from Neumann ones by changing the sign of the right-moving fields. We see that the same also holds for the worldsheet fermions.

Canonical quantization of the fermionic variables is done as usual. One promotes the fields to operators and imposes equal time canonical anticommutation relations, which in terms of the modes read
\begin{equation}
    \big\{ \psi^\mu_r , \psi^\nu_s \big\} = \big\{ \tilde{\psi}^\mu_r , \tilde{\psi}^\nu_s \big\} = \eta^{\mu \nu} \delta_{r+s,0},
    \label{RNS anticommutators}
\end{equation}
with all others vanishing. For $r=0$ we get the spacetime Clifford algebra
\begin{equation}
   \, \hspace{0.5cm} \big\{ \psi^\mu_0 , \psi^\nu_0 \big\} = \eta^{\mu \nu} ,
\end{equation}
or $\{ \Gamma^\mu, \Gamma^\nu \} = 2 \eta^{\mu \nu}$ where $\Gamma^\mu \equiv \sqrt{2} \psi^\mu_0$ (we will use $\Gamma^\mu$ for the ten-dimensional Dirac matrices and $\gamma^\mu$ for the ten-dimensional Pauli matrices. See Appendix \ref{C1} for more details on the ten-dimensional Clifford algebra.)
The spacetime Dirac equation emerges from the $G_0$ constraint
\begin{equation}
    G_0 = \sum_{n= -\infty}^\infty \alpha_n^\mu \psi_{-n \, \mu} \sim \Gamma^\mu p_\mu + \dots =0,
\end{equation}
just like the $L_0$ constraint leads to the spacetime Klein-Gordon equation. The $n \neq 0 $ terms in the ellipsis allow for the mass term in the equation to take many values, depending on which modes are excited.

The vacuum is defined as the state which is annihilated by all positive modes. The canonical anticommutator is expected to lead to a conformal anomaly and an ordering ambiguity in the Virasoro generators, both of which must be studied in order to obtain the spectrum. Starting with the latter, as before only the $n=0$ generators require a normal ordering constant. We derive it by normal ordering their classical expressions and regulating the zero-point energy:
\begin{equation}
    L_0^\psi = \frac{1}{2} \sum_{r \in \mathbb{Z} + \nu} r \, \psi^\mu_{-r} \psi_{r \, \mu}
    = \frac{1}{2} \underset{r \geq 0}{\sum_{r \in \mathbb{Z} + \nu}} r \big( 2 \psi^\mu_{-r} \psi_{r \, \mu} - \delta^\mu_\mu \big) = \normalord{L_0^\psi} - \frac{D}{2} \underset{r \geq 0}{\sum_{r \in \mathbb{Z} + \nu}}  r.
\end{equation}
The divergent sum can be dealt with as in \eqref{regulatedsum}:
\begin{align}
    - \frac{D}{2} \underset{r \geq 0}{\sum_{r \in \mathbb{Z} + \nu}}  r &= - \frac{D}{2} \sum_{n= 0}^\infty (n+\nu) \notag\\[5pt]
    &\to -\frac{D}{2} \sum_{n= 0}^\infty (n+\nu)e^{-(n+\nu)/\Lambda} \notag\\[5pt]
    &= \frac{D}{2} \frac{\partial }{\partial q}  \frac{e^{- \nu q}}{1-e^{-q}} \Bigg|_{q = 1/\Lambda} \notag\\[5pt]
    &= \frac{D}{2} \bigg( - \Lambda^2 + \frac{1}{12} - \frac{\nu}{2} + \frac{\nu^2}{2}
    \bigg) + \mathcal{O} \left( \frac{1}{\Lambda} \right).
\end{align}
When we add to this the contribution from the bosonic fields \eqref{regulatedsum}, the divergent piece cancels exactly, no renormalization required. Upon setting $\nu=0$ we find that each periodic fermion adds $\frac{1}{24}$ to the normal ordering constant, the exact opposite of the contribution of a periodic boson. Unfortunately for $\nu = \frac{1}{2}$, we get that each antiperiodic fermion adds $-\frac{1}{48}$, so taking both bosons and fermions into account,
\begin{equation}
    a^{X+ \psi}_\text{R} = - \frac{D}{24} + \frac{D}{24} = 0 \, , \hspace{0.5cm} a^{X + \psi}_\text{NS} = - \frac{D}{24} - \frac{D}{48} = - \frac{D}{16}.
\end{equation}
A tachyon is thus expected to appear in the NS sector. While this undermines part of our motivation for the construction of the RNS action, a tachyon-free theory can still be obtained from it via the so-called GSO projection, to be introduced in the next section. For now we simply press on with the analysis. The computation of $[ L^\psi_m, L^\psi_n ]$ done in Appendix \ref{A3} shows that each fermion adds $1/2$ to the central charge, so the total central charge from both bosons and fermions is
\begin{equation}
    c = D+\frac{D}{2} = \frac{3D}{2}.
\end{equation}

The direct canonical quantization of the RNS action that we have discussed up this point is bound to lead to negative norm states due to the wrong sign anticommutation relations \eqref{RNS anticommutators} for $\eta^{\mu \nu}= \eta^{00}$. The correct way to fix superconformal gauge is via the Faddeev-Popov method, which we studied in detail for the bosonic string. For the superstring, it leads to the ghost action $S_g = S_g^{(bc)} + S_g^{(\beta \gamma)}$, where $S_g^{(bc)}$ is the same fermionic ghost CFT that appears in the bosonic case, from fixing two bosonic gauge symmetries (diff $+$ Weyl), and \cite{Blumenhagen:2013fgp}
\begin{equation}
    S^{(\beta \gamma)}_g = \frac{1}{2 \pi} \int d^2 z \Big( \beta \bar{\partial} \gamma + \tilde{\beta} \partial \tilde{\gamma} \Big),
\end{equation}
is the superconformal gauge action for two new bosonic ghosts that come from gauge-fixing two fermionic gauge symmetries (SUSY $+$ super-Weyl), written in complex coordinates. The energy-momentum tensor is
\begin{align}
    T^{(\beta \gamma)} &= - \frac{1}{2} \partial \beta  \gamma - \frac{3}{2} \beta \partial \gamma \notag\\[5pt]
    \tilde{T}^{(\beta \gamma)} &=  -\frac{1}{2} \bar{\partial} \tilde{\beta}  \tilde{\gamma} - \frac{3}{2} \tilde{\beta}  \bar{\partial}  \tilde{\gamma}.
\end{align}
One should then quantize this CFT and extract its central charge and contribution to the normal ordering constants. The details can be found in \cite{Polchinskivol1:1998rq,Polchinskivol2:1998rr}. The central charge is found to be $c^{(\beta \gamma)}=11$, which together with $c^{(bc)}=-26$ from the other ghosts sets
\begin{equation}
    c^{g} = -15
\end{equation}
for the total ghost central charge. For the normal ordering constants, we guess correctly that these ghosts are responsible for canceling the contribution of two unphysical worldsheet fermions. The ghost system as a whole thus sets the physical normal ordering constants to
\begin{equation}
    a_\text{R} = - \frac{D-2}{24} + \frac{D-2}{24} = 0 \, , \hspace{0.5cm} a_\text{NS} = - \frac{D-2}{24} - \frac{D-2}{48} = - \frac{D-2}{16}.
\end{equation}
Unbroken conformal invariance of the quantized theory requires
\begin{equation}
    c^\text{total} = \frac{3D}{2}  -15 =0,
\end{equation}
which determines $D=10$ as the critical dimension of the RNS string. The physical normal ordering constants are then
\begin{equation}
    a_\text{R} = 0 \, , \hspace{0.5cm} a_\text{NS}= - \frac{1}{2}.
\end{equation}

We now have everything we need to derive the spectrum, following the same procedure as in Section \ref{sec34}. The open string physical state conditions are now
\begin{equation}
    (L_n +a  \delta_{0n}) \ket{\psi} = G_r \ket{\psi} =0  \,, \hspace{0.5cm} n,r \geq 0,
\end{equation}
and the general spurious state is of the form
\begin{equation}
    \ket{\chi} = \sum_{n=1}^\infty L_{-n} \ket{\chi_n} + \sum_{r \in \mathbb{N} + \nu >0} G_{-r} \ket{\chi^\prime_r}.
\end{equation}
The $n=0$ condition is equivalent to the mass relation
\begin{equation}
    M^2 = \frac{1}{\alpha^\prime} \Big( N + a \Big),
\end{equation}
where the level $N$ is now given by 
\begin{equation}
    N = \sum_{n=1}^\infty \alpha^\mu_{-n} \alpha_{n \, \mu} + \sum_{r \in \mathbb{N} + \nu} r \, \psi^\mu_{-r} \psi_{r \, \mu}. 
\end{equation}
Starting with the Neveu-Schwarz sector, for $N=0$ we have a tachyonic vacuum $\ket{\text{NS};k}$ with mass $M^2 = - 1/(2 \alpha^\prime) $. The next level is $N=\frac{1}{2}$, consisting of massless states of the form
\begin{equation}
    e_\mu(k) \psi^\mu_{- \frac{1}{2}} \ket{\text{NS};k}.
\end{equation}
Clearly all $L_n$ for $n>0$ annihilate it, as well as all $G_r$ for $r > \frac{1}{2}$. The $r= \frac{1}{2}$ case gives
\begin{equation}
    G_{\frac{1}{2}} \,  e_\mu(k) \psi^\mu_{-\frac{1}{2}} \ket{\text{NS};k} = e_\mu(k) \alpha_0^\nu \psi_{\frac{1}{2} \, \nu} \, \psi^\mu_{-\frac{1}{2}} \ket{\text{NS};k} = \sqrt{2 \alpha^\prime} \, k^\mu e_\mu(k) \ket{\text{NS};k},
\end{equation}
so this state is physical if the polarization is transverse. The only null (spurious and physical) state at this level is
\begin{equation}
    \ket{\chi} = G_{- \frac{1}{2}} \ket{\text{NS};k} = \sqrt{2 \alpha^\prime} \, k^\mu \psi_{-\frac{1}{2} \, \mu} \ket{\text{NS};k}
\end{equation}
with $k^2=0$, so we find the familiar $ e_\mu(k) \cong e_\mu(k) + ak_\mu$ equivalence relation of a massless gauge boson in spacetime. The first two levels of the open RNS string in the NS sector are thus essentially identical to those of the open bosonic string, one tachyon and one $U(1)$ gauge boson, the only difference being the tachyon's mass and of course the spacetime dimension.

Since $a_\text{R}=0$, in the Ramond sector there are no tachyons and the only massless state is the vacuum itself $\ket{\text{R};k}$. An important feature of the Ramond sector is that the vacuum is actually degenerate. We can act with the fermionic zero-modes on $\ket{\text{R};k}$ to reach many other states with zero energy:
\begin{equation}
     \psi^\mu_0 \, \ket{\text{R};k}, \hspace{0.3cm} \psi^\mu_0 \psi^\nu_0 \, \ket{\text{R};k}, \hspace{0.1cm} \dots
    \label{R vacua}
\end{equation}
Since the zero-modes are nothing but rescaled spacetime Dirac matrices, this means that these Ramond vacua furnish a representation of the spacetime Clifford algebra. The Ramond vacuum is therefore a spacetime spinor.\footnote{For an explicit construction, we follow \cite{Polchinskivol2:1998rr} and build from the $10$-dimensional Dirac matrices $\Gamma^\mu = \sqrt{2} \psi^\mu_0$ the operators 
\begin{equation}
    \Gamma^{0  \pm} = \frac{1}{2}\Big( \pm \Gamma^0 + \Gamma^1 \Big), \hspace{0,5cm}
    \Gamma^{a \pm} = \frac{1}{2}\Big( \Gamma^{2a} \pm i\Gamma^{2a+1} \Big),
\end{equation}
where $a=1,\dots, 4$. These play the role of raising and lowering operators for the Clifford algebra. Taking any one of the R vacua, we can act on it with the lowering operators $\Gamma^{0-} = \frac{1}{\sqrt{2}}(-\psi^0_0 + \psi^1_0 )$ and $\Gamma^{a-}=\frac{1}{\sqrt{2}}(\psi^{2a}_0 - \psi^{2a+1}_0 )$ until we reach a state that is annihilated by all of them, which we call $\big| -\frac{1}{2},-\frac{1}{2},-\frac{1}{2},-\frac{1}{2},-\frac{1}{2};k\big\rangle$. A set of mutually commuting spin projections are given by $S_a = i^{\delta_{a0}} \Sigma^{2a,2a+1} = \Gamma^{a+} \Gamma^{a-} - \frac{1}{2}$, where $\Sigma^{\mu \nu} = - \frac{i}{4}[\Gamma^\mu,\Gamma^\nu]$ are the spin generators. Since $\big| -\frac{1}{2},-\frac{1}{2},-\frac{1}{2},-\frac{1}{2},-\frac{1}{2};k\big\rangle$ is annihilated by all the $\Gamma^{a-}$, it has all spin projections $s_a$ equal to $-\frac{1}{2}$, which explains the notation. By acting on this state with the raising operators $\Gamma^{a+}$ one builds states with all possible spin projections $\ket{\text{R};k}_A \equiv \ket{s_0,s_1,s_2,s_3,s_4;k}$, $s_a=\pm \frac{1}{2}$, $A = (s_0,\dots,s_4)$, which form the components of a spacetime spinor. It is straightforward to show that the chiral matrix $\Gamma = \Gamma^0 \dots \Gamma^{9}$ can be written as $\Gamma = 2^{5} S_0 \dots S_4$, so the $\ket{\text{R};k}_A$ states have $\Gamma=+1$ if they contain an even number of $-\frac{1}{2}$'s and $\Gamma=-1$ if there is an odd number. Since it diagonalizes the chirality matrix, this is a Weyl basis for the 10-dimensional spinors.}  We have kept worldsheet spinors indices hidden, but it will be useful to write spacetime spinor indices explicitly. Our conventions are that $\lambda_A$ are the components of a 10-dimensional Majorana spinor $\lambda$, and $\lambda^A$ are the components of the adjoint spinor $\overline{\lambda} = \lambda^\text{T} C$, where $C$ is the charge conjugation matrix. More details on 10-dimensional Majorana spinors can be found in Appendix \ref{C1}. 

The most general form of the Ramond vacuum is a contraction of the spinor components $\ket{R;k}_A$ with an adjoint polarization spinor $u^A(k)$:
\begin{equation}
    \ket{\overline{u},k}= u^A(k) \ket{\text{R};k}_A.
\end{equation}
Defining $\psi_0^\mu \ket{\text{R};k}_A = \frac{1}{\sqrt{2}} (\Gamma^\mu)_A{}^B \ket{\text{R};k}_B$, the $n=0$ physical state condition becomes
\begin{equation}
    G_0 \ket{\overline{u},k} = \sqrt{\alpha^\prime} k^\mu u^A(k) (\Gamma_\mu)_A{}^B \ket{\text{R};k}_B =0,
\end{equation}
which is the massless adjoint Dirac equation for the polarization:
\begin{equation}
    \overline{u}(k) \Gamma_\mu k^\mu =0.
\end{equation}
Using the rules for the raising and lowering of spinor indices explained in Appendix \ref{C1} and the fact that $( \Gamma^\mu)^{AB} = (\Gamma^\mu)^{BA}$, this constraint can be manipulated into 
\begin{align}
    G_0 \ket{\overline{u},k} &= -\sqrt{\alpha^\prime} k^\mu u_A(k) (\Gamma_\mu)^{AB} \ket{\text{R};k}_B \notag\\[5pt]
    &= -\sqrt{\alpha^\prime} k^\mu \ket{\text{R};k}_B (\Gamma_\mu)^{BA} u_A(k) \notag\\[5pt]
    &= \sqrt{\alpha^\prime} \ket{\text{R};k}^B  k^\mu (\Gamma_\mu)_B{}^A u_A(k)
\end{align}
which gives the usual Dirac equation for $u_A(k)$:
\begin{equation}
    k^\mu \Gamma_\mu u(k) =0.
\end{equation}

It is convenient to classify these low-lying states in terms of their $(-1)^F$ eigenvalue, where $F$ is the worldsheet fermion number. For the NS vacuum we choose
\begin{equation}
    (-1)^F \ket{\text{NS};k} = - \ket{\text{NS};k},
\end{equation}
after which $(-1)^F$ is determined for every state in this sector, as $(-1)^F$ is defined to anticommute with all fermionic modes. In particular, the states forming the spacetime gauge boson have $(-1)^F=+1$. In the R sector the 10-dimensional chiral matrix
\begin{equation}
    \Gamma = \Gamma^0 \dots \Gamma^{9} = 2^5 \psi^0_0 \dots \psi^9_0
\end{equation}
is a natural candidate for an operator that anticommutes with all zero-modes. We will only be interested in the massless level of the R sector, so for our purposes we can set $(-1)^F = \Gamma$.\footnote{The generalization of this to the full theory is \cite{Blumenhagen:2013fgp}
\begin{equation}
    (-1)^F = \Gamma (-1)^{\sum_{n>0} \psi^\mu_{-n} \psi_{n \, \mu} }.
\end{equation}}
The notation NS$\pm$ and R$\pm$ will be used to refer to the states with $(-1)^F=\pm 1$ in each sector. We use a Majorana-Weyl representation for the Dirac matrices, so that they are all of the form
\begin{equation}
    (\Gamma^\mu)_A{}^B = 
    \begin{pmatrix}
        0 && (\gamma^\mu)_{\alpha}{}^{\dot{\beta}} \\
         (\gamma^\mu)_{\dot{\alpha}}{}^{\beta} && 0
    \end{pmatrix}.
\end{equation}
The matrices $\gamma^\mu$ are the 10-dimensional analogue of the Pauli matrices. The R vacuum thus splits into a right-handed Majorana-Weyl spinor $\ket{\text{R};k}_{\alpha}$ and a left-handed one $\ket{\text{R};k}_{\dot{\alpha}}$:
\begin{equation}
    \ket{\text{R};k}_A = 
    \begin{pmatrix}
        \ket{\text{R};k}_{\alpha} \\
        \ket{\text{R};k}_{\dot{\alpha}}
    \end{pmatrix},
\end{equation}
which have opposite chirality ($\Gamma$ eigenvalue) and therefore belong to different sectors. The open spectrum is summarized in table \ref{open RNS table}.

    \begin{table}
        \centering
        \begin{tabular}{|c|c|c|c|}
             \hline
        Sector & SO(9,1) representation & Spacetime field & $(\text{Mass})^2$ \\[0pt]  \hline
        NS$+$ & Vector & $A_\mu(x)$ & 0 \\[5pt] 
        NS$-$ & Scalar & $T(x)$ & $-\frac{1}{2 \alpha^\prime}$ \\[5pt]  
        R$+$ & Right-handed Majorana-Weyl spinor & $\chi_\alpha(x)$ & 0 \\[5pt]  
        R$-$ & Left-handed Majorana-Weyl spinor & $\chi_{\dot{\alpha}}(x)$ & 0  \\ \hline
        \end{tabular}
        \caption[Low-energy states of the open RNS string]{Low-energy states of the open RNS string. We split the Dirac spinor index $s=1,\dots 32$ into two Weyl spinor indices $\alpha$ and $\dot{\alpha}$, each one going from $1$ to $16$.}
        \label{open RNS table}
    \end{table}

To obtain the closed spectrum we consider also the left-moving modes $\tilde{\alpha}^\mu_n$ and $\tilde{\psi}^\mu_r$, as well as the physical state conditions that refer to them, $(\tilde{L}_n - \tilde{a}\delta_{n0} )\ket{\psi} = \tilde{G}_r \ket{\psi}=0$. Like for the bosonic string, the structure of the closed spectrum is that of a product of two copies of the open one, subject to level matching. Using that in both NS and R sectors the physical normal ordering constant is given by $\nu$, the closed string mass relation that takes level-matching into account is
\begin{equation}
    M^2 = \frac{4}{\alpha^\prime}\Big( N - \nu \Big) =  \frac{4}{\alpha^\prime}\Big( \tilde{N} - \tilde{\nu} \Big).
\end{equation}
The choice of boundary condition for the fermions splits the closed spectrum into four sectors, $(\text{R},\text{R})$, $(\text{NS},\text{R})$, $(\text{R},\text{NS})$ and $(\text{NS},\text{NS})$, each of which further splits into four if one considers also the $(-1)^F$ eigenvalue of each side. We will only need the massless or possibly tachyonic states, and therefore will restrict each sector to what is shown in table \ref{open RNS table}. The combinations $(\text{R}+,\text{R}+)$, $(\text{R}+,\text{R}-)$, $(\text{R}-,\text{R}+)$ and $(\text{R}-,\text{R}-)$ all have $N=\tilde{N}=0$ and $\nu=\tilde{\nu}=0$, and therefore are all allowed and lead to massless particles. $(\text{NS}+,\text{R}+)$ and $(\text{NS}+,\text{R}-)$ have $N-\nu= \frac{1}{2}-\frac{1}{2}=0=\tilde{N}=\tilde{\nu}$ and so do the other two from swapping the right and left-movers, so they are also allowed and massless. $(\text{NS}-,\text{NS}-)$ has $N-\nu = -\frac{1}{2}=\tilde{N}-\tilde{\nu}$, so it is allowed. From the mass formula we see that this sector contains a tachyon of mass $M^2 = - 2/\alpha^\prime$. The $\text{NS}-$ sector only combines with itself, no other choice satisfies level matching. For instance $(\text{NS}-,\text{NS}+)$ has $N-\nu=- \frac{1}{2}$ and $\tilde{N}-\tilde{\nu}=0$. Finally we have $(\text{NS}+,\text{NS}+)$, which has $N-\nu= \frac{1}{2} - \frac{1}{2} =0= \tilde{N} -\tilde{\nu}$ and is therefore allowed. The particle content is obtained by decomposing each of these products into irreducible representations of the Lorentz group. Of particular importance are the mixed sectors, for instance $(\text{NS}+,\text{R}+)$, whose massless level consists of states of the form
\begin{equation}
    \xi^\alpha_{\mu}(k)  \Big(  \psi^\mu_{-\frac{1}{2}}\ket{\text{NS};k}\, \ket{\widetilde{\text{R}};k}_\alpha \Big),
    \label{NSR general state}
\end{equation}
where $\xi^\alpha_{\mu}(k)$ is a polarization with one vector and one Weyl spinor index. The $G_0=G_{\frac{1}{2}}=0$ physical state conditions require that the polarization be transverse and satisfy the momentum space massless Weyl equation:
\begin{equation}
    k^\mu \xi^\alpha_\mu(k) =0 , \hspace{0.5cm} k^\nu \gamma_\nu \xi_\mu(k) =0.
\end{equation}
In this sector and level the null states are of the form
\begin{equation}
    \ket{\chi} = \epsilon^\alpha \, G_{-\frac{1}{2}} \, \ket{\text{NS};k} \,  \ket{\widetilde{\text{R}};k}_\alpha = \sqrt{\frac{\alpha^\prime}{2}} \epsilon^\alpha k_\mu \psi^\mu_{-\frac{1}{2}}  \, \ket{\text{NS};k} \,  \ket{\widetilde{\text{R}};k}_\alpha
\end{equation}
where $k^2=0$ and $\epsilon^\alpha$ is a constant spinor that satisfies the massless Weyl equation $k^\mu \gamma_\mu \epsilon=0$. In close analogy with \eqref{open bosonic equivalence}, one obtains the equivalence relation 
\begin{equation}
    \xi_{\mu}^\alpha(k) \cong \xi_{\mu}^\alpha(k) +  k_\mu \epsilon^\alpha
\end{equation}
for the polarization. This is the momentum space manifestation of the spacetime gauge symmetry
\begin{equation}
    \varphi_{\mu \alpha}(x) \to  \varphi_{\mu \alpha}(x) + \partial_\mu \epsilon_\alpha(x).
    \label{gravitino gauge symmetry}
\end{equation}

From basic angular momentum addition one expects a spin $\frac{3}{2}$ product state such as \eqref{NSR general state} to decompose into irreducible states with helicities $h=\frac{1}{2}$ and $h=\frac{3}{2}$, depending on whether the spins of the two subsystems are aligned parallel to each other, giving $h=1+\frac{1}{2}= \frac{3}{2} $, or antiparallel, in which case $h=1- \frac{1}{2} = \frac{1}{2}$. Indeed, one can produce a simple Weyl spinor from $\varphi_{\mu \alpha}$ by contracting with the 10-dimensional Pauli matrices:
\begin{equation}
    \hspace{1cm} \lambda_{\dot{\alpha}} =( \gamma^\mu)_{\dot{\alpha}}{}^\alpha \varphi_{\mu \alpha} .
    \label{dilatino decomposition}
\end{equation}
The irreducible decomposition is obtained by writing
\begin{align}
    \varphi_{\mu \alpha} &= \varphi_{\mu \alpha} - \frac{1}{10} (\gamma_\mu)_\alpha{}^{\dot{\alpha}}  (\gamma^\nu)_{\dot{\alpha}}{}^\beta  \varphi_{\nu \beta} +  \frac{1}{10} (\gamma_\mu)_\alpha{}^{\dot{\alpha}}  (\gamma^\nu)_{\dot{\alpha}}{}^\beta  \varphi_{\nu \beta}  \notag\\[5pt]
    &=\chi_{\mu \alpha } +\frac{1}{10} (\gamma_\mu)_\alpha{}^{\dot{\alpha}}  \lambda_{\dot{\alpha}} \, ,
\end{align}
where the helicity $\frac{1}{2}$ part is $\lambda_{\dot{\alpha}}$ and the helicity $\frac{3}{2}$ part is
\begin{equation}
    \chi_{\mu \alpha} = \Big( \delta^\nu_\mu \delta^\beta_\alpha - \frac{1}{10} (\gamma_\mu)_\alpha{}^{\dot{\alpha}}  (\gamma^\nu)_{\dot{\alpha}}{}^\beta   \Big) \varphi_{\nu \beta}.
\end{equation}
The factors of $10$ are related to the fact that the 10-dimensional Pauli matrices satisfy a Clifford algebra, which can have its Lorentz indices contracted to give
\begin{equation}
    \frac{1}{2}
    \big\{ \gamma^\mu , \gamma_\mu \big\}_\alpha{}^\beta= \delta^\mu_\mu \delta^\beta_\alpha = 10 \delta_\alpha^\beta.
\end{equation}
They guarantee that the spin $\frac{1}{2}$ part of $\chi_{\mu \alpha}$ vanishes:
\begin{equation}
    (\gamma^\mu)_{\dot{\alpha}}{}^\alpha  \chi_{\mu \alpha} =0.
\end{equation}
For the next section it will be useful to count the physical degrees of freedom associated to these particles. A ten-dimensional Majorana spinor has $2^{10/2}=32$ real components, so each Weyl spinor has $16$. The total number of components of $\chi_{\mu \alpha}$ is therefore $10 \times 16 = 160$. The Weyl equation $ k^\mu \gamma_\nu \xi_\mu(k) =0 $ reduces the amount of independent degrees of freedom by half. In the frame where $k^\mu=(E, \dots,E)$, the transversality and gauge-invariance conditions for the polarization become
\begin{equation}
    \xi_0^\alpha(k) + \xi^\alpha_9(k) =0, \hspace{0.5cm} \xi^\alpha_0(k) \cong \xi^\alpha_0(k)  + E \epsilon^\alpha.
\end{equation}
The first condition leads to 16 constraints, one for each value of $\alpha$. The second one gives 8 more constraints, one for each independent component of $\epsilon^\alpha$, which is required to satisfy $k^\mu \gamma_\mu \epsilon=0$. The total number degrees of freedom in $\chi_{\mu \alpha}$ is thus $80-16-8=56$. The spin $\frac{1}{2}$ particle $\lambda_{\dot{\alpha}}$ possesses no gauge-invariance and therefore has the $8$ on-shell degrees of freedom one would expect from a 10-dimensional Weyl-fermion. 

It is the presence of this helicity $\frac{3}{2}$ massless particle on the spectrum that allows us to state with certainty that the RNS string cannot be given consistent interactions. The reason for this can be easily seen in the field theory limit. The general form of an interaction one can write involving a spacetime spin $\frac{3}{2}$ field $\chi_{\mu \alpha}(x)$ is a linear coupling $\mathcal{L}_\text{int} \sim iJ^{\mu \alpha}(x)   \chi_{\mu \alpha}(x) \, $, where $J^{\mu \alpha}(x)$ is some fermionic operator built from the fields. The gauge transformation $\delta \chi_{\mu \alpha} = \partial_\mu \epsilon_\alpha$ changes the Lagrangian by 
\begin{equation}
    \delta \mathcal{L}_\text{int} \sim i J^{\mu \alpha} \partial_\mu \epsilon_\alpha \to -i \partial_\mu J^{\mu \alpha} \epsilon_\alpha,
\end{equation}
where as usual we assume the gauge transformation parameter goes to zero at infinity and integrate by parts. Therefore, the action is only gauge-invariant if $J^{\mu \alpha}$ is a conserved current, and this can only hold in a supersymmetric theory. It can be seen in a variety of ways that the low-energy limit of RNS does not have spacetime supersymmetry. There is, for instance, the fact that the tachyons found in both open and closed sectors have no fermionic counterparts and are themselves a sign that the state with zero energy, empty Minkowski space, is not the theory's lowest energy state, as supersymmetry would require. In light of this, it is clear that the RNS string by itself is not the way to obtain the superstring theories alluded to at the beginning of this chapter. There is a way to, in a sense, modify it, so that these issues are no longer present. This will be the first subject of next section, and we delay the discussion of the rest of the closed string particle content until then.

\section{Type II superstring theories}\label{sec72}

The main issue we found in our investigation of the RNS string is that it seems to produce too many particles, some of which are known to be impossible to simultaneously embed into an interacting theory. In spite of this, someone obstinate enough might still attempt to use the RNS action to compute scattering amplitudes. This means doing a path integral of $e^{-S_\text{RNS}}$ over worldsheets with external legs that each approach asymptotically one of the states we found in the RNS string's spectrum. Since we have not developed the formalism required to do such a computation, we will content ourselves with a simple account of what would happen. Focusing on closed strings, at tree level there exist different groupings of the RNS sectors such that, if one computes scattering amplitudes involving only states inside each one, no apparent inconsistencies arise. One such group is given by 
\begin{equation}
\big\{ (\text{NS}+,\text{NS}+), \, (\text{R}+,\text{NS}+), \, (\text{NS}+,\text{R}+), \, (\text{R}+,\text{R}+) \big\}.
\label{type IIB group example}
\end{equation}
If every asymptotic state chosen for the worldsheet is from one of these sectors, the path integral can be computed using known CFT methods, and gives some finite result for the amplitude. It also happens that any amount of particles from one of these groups can only scatter into a final state consisting of particles of this same group. Each group is disconnected in this sense with respect to the S-matrix. If one, however, attempts to set up the computation of a scattering amplitude involving states from different groups, one runs into ill-defined expressions that cannot be evaluated. This state of affairs begs the question of whether it would be consistent to simply take each of these groups of sectors to define the spectrum of a separate, independent theory from the spacetime point of view, all of which happen to share $S_\text{RNS}$ as the worldsheet action. The answer is affirmative, and the way this is done is via the so-called Gliozzi-Scherk-Olive (GSO) projection \cite{Gliozzi:1976qd}.

The central object in the GSO projection is the operator $(-1)^F$. Note that all states from the group \eqref{type IIB group example} have 
\begin{equation}
    (-1)^F=(-1)^{\tilde{F}}=1,
    \label{type IIB projection}
\end{equation}
where we take $F$ to be the right-moving fermion number and $\tilde{F}$ the left-moving one. Truncating the RNS spectrum to only the sectors in \eqref{type IIB group example} is therefore equivalent to projecting the total RNS Hilbert space into the subspace satisfying \eqref{type IIB projection}. This is one possible GSO projection, and the resulting theory is called the type IIB superstring. Another possibility is to keep the sectors with
\begin{equation}
    (-1)^F = 1 \, , \hspace{0.5cm} (-1)^{\tilde{F}} = -(-1)^{2\tilde{\nu}},
\end{equation}
leading to
\begin{equation}
\big\{ (\text{NS}+,\text{NS}+), \, (\text{R}+,\text{NS}+), \, (\text{NS}+,\text{R}-), \, (\text{R}+,\text{R}-) \big\}.
\label{type IIA group}
\end{equation} 
This also leads to consistent interactions, and defines the type IIA superstring.\footnote{One could also try $(-1)^F=-(-1)^{2\nu}$ $ (-1)^{\tilde{F}} = -(-1)^{2\tilde{\nu}}$ or $(-1)^F= -(-1)^{2 \nu}$ $ (-1)^{\tilde{F}} = 1$,
leading to
\begin{equation}
\big\{ (\text{NS}+,\text{NS}+), \, (\text{R}-,\text{NS}+), \, (\text{NS}+,\text{R}-), \, (\text{R}-,\text{R}-) \big\}
\end{equation} 
and
\begin{equation}
\big\{ (\text{NS}+,\text{NS}+), \, (\text{R}-,\text{NS}+), \, (\text{NS}+,\text{R}+), \, (\text{R}-,\text{R}+) \big\}.
\end{equation} 
These represent the same physical theories as \eqref{type IIB group example} and \eqref{type IIA group}, since they only differ from those by an inversion of the chirality of the R sector. This amounts to switching right- and left-moving Weyl representations and therefore to a spacetime parity transformation.} There are other consistent projections, leading to the type I and type 0 theories, which will not be discussed. Neither type IIA nor IIB theories contain the $(\text{NS}-,\text{NS}-)$ sector, so both are tachyon-free. We now discuss their low-energy spectrum. Both contain the sector $(\text{NS}+,\text{NS}+)$, whose massless states are of the form
\begin{equation}
    \xi_{\mu \nu}(k) \psi^\mu_{-\frac{1}{2}} \tilde{\psi}^\nu_{-\frac{1}{2}} \ket{\text{NS};k} \, \ket{\widetilde{\text{NS}};k}.
\end{equation}
 The polarization tensor can be decomposed into traceless symmetric, antisymmetric, and scalar parts, giving a 10-dimensional copy of the bosonic string massless states: a graviton, a Kalb-Ramond field, and a dilaton. The physical state conditions produce the same equations of motion and gauge invariances that were found for these fields in Section \ref{sec34} We know that a ten-dimensional on-shell graviton has $10(10-3)/2=35$ physical degrees of freedom, while a ten-dimensional on-shell Kalb-Ramond field has $(10-2)(10-3)/2=28$ degrees of freedom. Together with the dilaton, we thus far have $64$ bosonic degrees of freedom on-shell.

The sector $(\text{R}+,\text{NS}+)$ is also shared by the two theories. Its field content was worked out in the last section: a massless left-handed Weyl fermion $\lambda_{\dot{\alpha}}$ and a massless right-handed Rarita-Schwinger field $\chi_{\mu \alpha}$. They have the interpretation of a dilatino and a gravitino, the supersymmetric partners of the dilaton and graviton, and the physical state conditions take the form of the equations of motion expected for such fields. In the type IIB string we have also $(\text{NS}+,\text{R}+)$, which gives another copy of these same two fields. In the type IIA string we have instead $(\text{NS}+,\text{R}-)$, which differs only by the chirality of the Weyl fermions, leading to $\lambda_\alpha$ and $\chi_{\mu \dot{\alpha}}$. In the previous section we showed that these fields each have $8$ and $56$ on-shell physical degrees of freedom. Adding the two copies we have that in either type II theory there are $128$ fermionic degrees of freedom on-shell.

The analysis of the R-R sectors is a bit more intricate and must be done separately for type IIA and IIB. Starting with the former, the general $(\text{R}+,\text{R}-)$ state is given by
\begin{equation}
    f^{\alpha \dot{\beta}}(k) \ket{\text{R};k}_\alpha \ket{\widetilde{\text{R}};k}_{\dot{\beta}}.
\end{equation}
At level $N=\tilde{N}=0$ there are no null states and the only nontrivial physical state conditions are $G_0 = \tilde{G}_0=0$. These lead to one Weyl equation for each index,
\begin{equation}
    k_\mu (\gamma^\mu)_{\alpha}{}^{\dot{\alpha}} f_{\dot{\alpha} \beta}(k)  =k_\mu (\gamma^\mu)_{\dot{\alpha}}{}^{\alpha} f_{\dot{\beta} \alpha}(k) = 0.
\end{equation}
The polarizations $f_{\dot{\alpha}\beta}(k)$ and $f_{ \alpha \dot{\beta}}(k)$ are each elements of the tensor product of two spin $\frac{1}{2}$ representations, so one would expect them to decompose into irreducible tensors with integer spin. This is the content of Fierz identities, which reexpress uncontracted spinor bilinears as sums of antisymmetric tensors,
\begin{equation}
    \psi \bar{\varphi} \sim (\bar{\varphi} \psi) + (\bar{\varphi} \Gamma_\mu \psi) \Gamma^\mu + (\bar{\varphi} \Gamma_{\mu \nu} \psi) \Gamma^{\mu \nu} + \dots \, .
\end{equation}
The derivation of the exact form these decompositions take in ten dimensions is worked out in Appendix \ref{C1}. One finds that $f_{\dot{\alpha} \beta}(k)$ contains a scalar, a 2-form and a 4-form:
\begin{equation}
    f_{\dot{\alpha} \beta}(k) = \frac{1}{16} \bigg[ F(k) \mathcal{C}_{\dot{\alpha} \beta } - \frac{1}{2!} F_{\mu_1 \mu_2}(k) (\gamma^{\mu_1 \mu_2} )_{\dot{\alpha} \beta } + \frac{1}{4!}  F_{\mu_1 \mu_2 \mu_3 \mu_4 }(k) (\gamma^{\mu_1 \mu_2 \mu_3 \mu_4} )_{\dot{\alpha} \beta }  \bigg],
\end{equation}
where $F_{\mu_1 \dots \mu_n}(k) = \text{Tr}(\mathcal{C} \gamma_{\mu_1 \dots \mu_n} f(k) )$ with 
\begin{equation}
     C^{AB} = \begin{pmatrix}
        0 && \mathcal{C}^{\alpha \dot{\beta}} \\
         \mathcal{C}^{\dot{\alpha} \beta } && 0
    \end{pmatrix}.
\end{equation}
The general state $(\text{R}+,\text{R}+)$ of type IIB theory is
\begin{equation}
    f^{\alpha \beta}(k) \ket{\text{R};k}_\alpha \ket{\widetilde{\text{R}};k}_{\beta}.
\end{equation}
The physical state conditions once again produce one Weyl equation for each index, but this time the decomposition contains forms of degree $1$, $3$ and $5$:
\begin{equation}
    f_{\dot{\alpha} \dot{\beta} }(k) = \frac{1}{16} \bigg[ F_{\mu_1}(k) ( \gamma^{\mu_1} )_{\dot{\alpha} \dot{\beta}} - \frac{1}{3!} F_{\mu_1 \mu_2 \mu_3}(k) ( \gamma^{\mu_1 \mu_2 \mu_3})_{\dot{\alpha} \dot{\beta}}
    + \frac{1}{5!} F_{\mu_1 \mu_2 \mu_3 \mu_4 \mu_5}^{+}(k) ( \gamma^{\mu_1 \mu_2 \mu_3 \mu_4 \mu_5} )_{\dot{\alpha} \dot{\beta}}
    \bigg]
\end{equation}
with the 5-form satisfying a self-duality condition
\begin{equation}
    (\ast F^+)_{\mu_1 \mu_2 \mu_3 \mu_4 \mu_5}(k) = \frac{1}{5!} \epsilon_{ \mu_{1} \dots \mu_{5} \nu_1 \dots \nu_5 } F^{+ \nu_1 \dots \nu_5}(k) =  F^+_{\mu_1 \mu_2 \mu_3 \mu_4 \mu_5}(k).
\end{equation}
Contracting the left Weyl index of either the type IIA or type IIB expansion with $k_\mu \gamma^\mu$ and using the identity
\begin{equation}
    \gamma^\mu \gamma^{\nu_1 \dots \nu_n} = \gamma^{\mu \nu_1 \dots \nu_n} + n \eta^{\mu [ \nu_1} \gamma^{\nu_2 \dots \nu_n]},
\end{equation}
proven in Appendix \ref{C2}, leads to 
\begin{equation}
    k_{[ \mu} F_{\nu_1 \dots \nu_n]}(k) = k^{\nu_1} F_{\nu_1 \dots \nu_n}(k) =0
\end{equation}
as the physical state conditions for each tensor. The position space version
\begin{equation}
    \partial_{[\mu} F_{\nu_1 \dots \nu_n]}(x) = \partial^{\nu_1} F_{\nu_1 \dots \nu_n}(x)=0
\end{equation}
can perhaps be more easily be recognized as the Bianchi identity and equations of motion of abelian $n$-form gauge field strengths. No additional information comes from the physical state condition for which the contraction is done with the second Weyl index.

The number of independent components of a completely antisymmetric rank $n$ tensor in ten dimensions is equal to the number of different ways one may choose $n$ different values for the indices out of 10 possibilities,
\begin{equation}
    \binom{10}{n} = \frac{10!}{n!(10-n)!}.
\end{equation}
It is complicated to calculate the amount of constraints brought by the Bianchi identity and the equations of motion separately. One way to simplify this analysis is to relate the field strength to a gauge field,
\begin{equation}
    F_{\mu_1 \dots \mu_n} = n \, \partial_{[\mu_n} C_{\mu_1 \dots \mu_{n-1}]},
\end{equation}
which trivializes the Bianchi identity.\footnote{This is a consequence of the fact that in flat space every form $F$ that is closed ($dF=0$) is also exact ($F=dC$). } The fact that $C_{\mu_1 \dots \mu_{n-1}}$ has one index less than $F_{\mu_1 \dots \mu_n}$ brings the total number of independent components down to $\binom{10}{n-1}$. Not all of them count as physical degrees of freedom because, as usual, one finds that there is a freedom to do gauge transformations
\begin{equation}
    C_{\mu_1 \dots \mu_{n-1}} \to C_{\mu_1 \dots \mu_{n-1}} + \partial_{[ \mu_{n-1} } \Lambda_{\mu_1 \dots \mu_{n-2}]}
\end{equation}
that do not affect $F_{\mu_1 \dots \mu_n}$. Since $\Lambda_{\mu_1 \dots \mu_{n-2}}$ has $\binom{10}{n-2}$ independent components, one then might be tempted to say that the number of degrees of freedom is $\binom{10}{n-1}-\binom{10}{n-2}$, and for the most familiar case of Maxwell electrodynamics this is true, but for $n>2$ there is subtlety that must be taken into consideration: The gauge parameter is itself invariant under the so-called ``gauge-for-gauge'' transformations
\begin{equation}
    \Lambda_{\mu_1 \dots \mu_{n-2}} \to \Lambda_{\mu_1 \dots \mu_{n-2}} + \partial_{[\mu_{n-2}} \Omega_{\mu_1 \dots \mu_{n-3}]}.
\end{equation}
This means that not every independent degree of freedom contained in $\Lambda_{\mu_1 \dots \mu_{n-2}}$ actually contributes to the gauge transformation, so the number of degrees of freedom removed by the gauge invariance is smaller than $\binom{10}{n-2}$. This number is not $\binom{10}{n-2}-\binom{10}{n-3}$ either because $\Omega_{\mu_1 \dots \mu_{n-3}}$ has itself a gauge-for-gauge symmetry related to a $(n-4)$-form. This goes on until we arrive at a gauge parameter with $n-n=0$ indices, at which point there are no further transformations. The number of physical degrees of freedom in $C_{\mu_1 \dots \mu_{n-1}}$ is therefore
\begin{align}
    \binom{10}{n-1} - \bigg[  \binom{10}{n-2} - \bigg[ \binom{10}{n-3} - \dots \bigg] \bigg] &= \binom{10}{n-1} - \binom{10}{n-2} + \binom{10}{n-3} - \dots  \notag\\[5pt]
    &= \binom{9}{n-1},
\end{align}
where we used the identity
\begin{equation}
    \sum_{i=0}^k (-1)^i \binom{D}{k-i} = (-1)^k \binom{D-1}{k}.
\end{equation}
This effective reduction from ten to nine dimensions reflects the fact that, as in electrodynamics, one may use gauge invariance to eliminate all the timelike components of the gauge field, which are the ones associated with wrong sign canonical commutators when quantizing. The constraints from the equations of motion are most easily counted in momentum space in the $k^\mu=(E,0,\dots0,E)$ frame, where we have
\begin{equation}
    k^\mu F_{\mu \mu_1 \dots \mu_{n-1}}(k) =n \,  k^\mu k_{[\mu} C_{\mu_1 \dots \mu_{n-1}]}(k) =n E^2 \big( \delta^0_{[0} C_{\mu_1 \dots \mu_{n-1}]}(k) + \delta^9_{[9} C_{\mu_1 \dots \mu_{n-1}]}(k) \big) =0.
\end{equation}
After going to the $C_{0 \mu_2 \dots \mu_{n-1}}=0$ gauge this becomes $C_{\mu_1 \dots \mu_{n-1}}(k) = -\delta^9_{[9} C_{\mu_1 \dots \mu_{n-1}]}(k)$, which means that the gauge field vanishes if any index is set to $9$. This effectively removes one more dimension, the one associated to longitudinal polarizations, making the number of on-shell physical degrees of freedom
\begin{equation}
    \binom{8}{n-1}.
\end{equation}
The number of bosonic on-shell degrees of freedom from the R-R sector of type IIA theory is therefore
\begin{equation}
    \binom{8}{1} + \binom{8}{3} = 64.
\end{equation}
The scalar $F$ does not contribute because it has no on-shell propagating degrees of freedom (its equation of motion $\partial_\mu F=0$ sets it to a constant). In type IIB theory we have
\begin{equation}
    \binom{8}{0} + \binom{8}{2} + \frac{1}{2}\binom{8}{4} = 64,
\end{equation}
with the $\frac{1}{2}$ coming from the self-duality of the 5-form. In both type II superstring theories we therefore have $64+64=128$ bosonic degrees of freedom on-shell and the same number of fermionic ones, which must happen for the theory to be supersymmetric. The massless level of the type II theories is summarized in table \ref{Closed type II table}. 

\begin{table}
        \centering
        \begin{tabular}{|c|c|c|c|}
             \hline
        Sector & Theory & SO(9,1) representation & Spacetime field  \\[0pt]  \hline
        (NS$+$,NS$+$) & Both & $(35) \oplus [28] \oplus 1 $ & $G_{\mu \nu}(x)$, $B_{\mu \nu}(x)$, $\Phi(x)$  \\[5pt] 
        (R$+$,NS$+$) & Both & $56  \oplus \bar{8} $ & $\chi^1_{\mu \alpha}(x)$, $\lambda^1_{\dot{\alpha}}(x)$  \\[5pt] 
        (NS$+$,R$+$) & IIB & $56  \oplus \overline{8} $ & $\chi^2_{\mu \alpha}(x)$, $\lambda^2_{\dot{\alpha}}(x)$ \\[5pt]
        (NS$+$,R$-$) & IIA & $\overline{56}  \oplus 8 $  & $\chi^2_{\mu \dot{\alpha}}(x)$, $\lambda^2_{\alpha}(x)$ \\[5pt]
        (R$+$,R$+$) & IIB & $1 \oplus [28] \oplus [35]^+ $ & $F_{\mu_1}(x)$, $F_{\mu_1 \mu_2 \mu_3}(x)$, $F_{\mu_1 \dots \mu_5}^+(x)$  \\[5pt]
        (R$+$,R$-$) & IIA & $[8] \oplus [56] $ & $F_{\mu_1 \mu_2}(x)$, $F_{\mu_1 \dots \mu_4}(x)$  \\ \hline
        \end{tabular}
        \caption[Massless level of the closed type II superstring theories]{Massless level of the closed type II superstring theories. We label the SO(9,1) representations according to its dimension, and for the integer spin ones use $(n)$ for a traceless symmetric tensor, $[n]$ for an antisymmetric tensor and $[n]^+$ for a self-dual antisymmetric tensor. For the half-integer spin representations we use a number with no brackets for right-handed spinors and put a bar over it for left-handed spinors.}
        \label{Closed type II table}
    \end{table}

The symmetries and field content of the type II theories uniquely determine the form of their low-energy actions. They are the type IIA and IIB $\mathcal{N}=2$ ten-dimensional supergravities, the latter being the most relevant for our discussion of AdS/CFT. It is not known how to write an action such that the 5-form field strength comes out automatically self-dual, but for a classical treatment it is enough to write an action for a general 5-form and impose the self duality condition on the solutions. This action is most easily expressed in differential form notation:\footnote{Our notation for differential forms is such that
\begin{align}
    &A_p = \frac{1}{p!} A_{\mu_1 \dots \mu_p} dx^{\mu_1} \wedge \dots \wedge dx^{\mu_p}, \\[5pt]
    &(A_p \wedge B_q)_{\mu_1 \dots \mu_p \nu_1 \dots \nu_q} = \frac{(p+q)!}{p! q!} A_{[ \mu_1 \dots \mu_p} B_{\nu_1 \dots \nu_q]} , \\[5pt]
    &|A_p|^2 = \frac{1}{p!} A_{\mu_1 \dots \mu_p}A^{\mu_1 \dots \mu_p}, \\[5pt]
    &\int A_{10} = \frac{1}{10!} \int d^{10}x \sqrt{-G} A_{012345679},
\end{align}}
\begin{align}
    S_{\text{IIB}} &= \frac{1}{2 \kappa_0^2} \int d^{10}x \sqrt{-G} \bigg[ e^{-2 \Phi } \Big( \mathcal{R} + 4 \nabla_\mu \Phi  \nabla^\mu \Phi  -\frac{1}{2} |H_3|^2 \Big) - \frac{1}{2} |F_1|^2 - \frac{1}{2} |\widetilde{F}_3|^2  \notag\\[5pt]
    &\hspace{3.5cm}  - \frac{1}{4}|\widetilde{F}_5|^2 \bigg] - \frac{1}{2 \kappa_0^2} \int C_4 \wedge H_3 \wedge F_3 + \text{fermionic terms},
    \label{type IIB SUGRA action}
\end{align}
where $F_{n} = dC_{n-1}$, $H_3 = dB_2$, and
\begin{align}
    \widetilde{F}_3 &= F_3 - C_0 \wedge H_3, \\[5pt]
    \widetilde{F}_5 &= F_5 - \frac{1}{2} C_2 \wedge H_3 + \frac{1}{2} B_2 \wedge F_3.
\end{align}
The fermionic terms are uniquely determined from the bosonic part by supersymmetry. One may verify that this is indeed the action that governs the low-energy dynamics of the type IIB superstring by comparing the scattering amplitudes it generates (supplemented with the $\ast F_5=F_5$ constraint) with the low-energy limit of the string's scattering amplitudes, computed via worldsheet methods. It should also be possible to derive the type II supergravity equations of motion from the vanishing of the worldsheet beta-functions in a general background, as was done for the bosonic string in Section \ref{sec51}, and one can do this for (NS+,NS+) backgrounds without too much trouble \cite{Callan:1989nz}. The same unfortunately does not hold for backgrounds involving fields from the Ramond sector, since the inclusion of backgrounds for them in the RNS action is much more complicated than that of NS backgrounds.\footnote{See \cite{Berenstein:1999jq} for details on the difficulties that arise and an attempt to overcome them.} There exist alternative worldsheet formalisms such as the Green-Schwarz or the pure spinor formalism that allow for a simpler treatment of Ramond backgrounds, at the cost of working with a more complicated worldsheet action  \cite{Berkovits:2017ldz}.

\section{Supersymmetric D-branes}\label{sec73}

In our discussion of the type II theories there has been, so far, no mention of open strings. Since open string boundary conditions effectively set the right and left-moving modes equal to each other, the GSO-projected open sector of the RNS string is given by the states of table \ref{open RNS table} satisfying $(-1)^F=1$, which are a gauge boson-gaugino pair $\{ A_\mu, \chi_{\alpha}\}$. Adding open strings to either type II theory is therefore expected to result in a theory that at low energies reduces to the corresponding supergravity coupled to ten-dimensional supersymmetric Yang-Mills. The issue with this is that such a theory would only admit the $\mathcal{N}=1$ supersymmetry that rotates $A_\mu$ and $\chi_{\alpha}$ into each other, while the type II supergravities have $\mathcal{N}=2$. Directly coupling a Yang-Mills sector to them would explicitly break half of their fermionic gauge symmetries, which would not result in a consistent theory.

It is useful to know that this breaking can also be understood directly from the open string boundary conditions for the worldsheet fields, independently of the low-energy field theory. Only a sketch of the argument will be provided here, the details are left for the references. First, note that a quantity which is conserved in spacetime should also be conserved on the worldsheet. After all, one may always choose to go to physical gauge by identifying the timelike worldsheet coordinate $\tau$ with the spacetime time coordinate $X^0$. Consider for instance the spacetime momentum carried by the string $p^\mu \sim \int_0^l d \sigma \,  \partial^\tau X^\mu$. This is conserved in the spacetime dynamics due to target space Poincaré symmetry, and is conserved on the worldsheet because it is the integral over the string of the timelike component of the worldsheet current $j^a = \partial^a X^\mu$, whose conservation follows directly from the equations of motion $\partial_a \partial^a X^\mu=0$. The spacetime supercharges $Q_A^1$ and $Q_A^2$ of the type II strings are similarly also conserved on the worldsheet, and can therefore be written as 
\begin{equation}
    Q^i_A = \int_0^l d \sigma \, j^{i\tau}_A(\tau,\sigma),
\end{equation}
where $j^{ia}(\tau, \sigma)$ is a current defined over the string and $i=1,2$ goes over the two independent supercharges of each theory. The actual construction of these densities is somewhat involved, due to the lack of manifest spacetime supersymmetry of the RNS formalism. For our purposes it will only be necessary to know that they are built out of the worldsheet fields, with the fermions $\psi^\mu_\pm$ satisfying Ramond boundary conditions, and that $Q^1_A$ differs from $Q^2_A$ only by swapping $\psi^\mu_+ \leftrightarrow \psi^\mu_- $. Worldsheet conservation of the supercharges means that 
\begin{equation}
    \partial_\tau Q^i_A = \int_0^l d \sigma \partial_\tau j^{i \tau}_A = - \int_0^l d \sigma \partial_\sigma j^{i \sigma}_A = - j^{i \sigma}_A(\tau,l) + j^{i \sigma}_A(\tau,0).
    \label{supercharge conservation}
\end{equation}
For closed strings this is automatic, since all fields are periodic in the Ramond sector. Both $Q^1_A$ and $Q^2_A$ are then separately conserved, leading to $\mathcal{N}=2$ spacetime supersymmetry. Each Majorana-Weyl spinor in ten dimensions has $16$ components, so this means a total of $32$ individual supercharges. An open string worldsheet however has boundaries, over which the Ramond fermions must satisfy $\psi^\mu_+ = \psi^\mu_-$ for a string with free endpoints. In this case we have
\begin{align}
    \partial_\tau Q^1_A = - j^{1 \sigma}_A(\tau,l) + j^{1 \sigma}_A(\tau,0) = - j^{2 \sigma}_A(\tau,l) + j^{2 \sigma}_A(\tau,0) = \partial_\tau Q^2_A.
\end{align}
It follows that, for fully NN boundary conditions, the only supercharge which is conserved on the worldsheet is $Q_A = Q^1_A - Q^2_A$. If there are some directions in which DD conditions are imposed, the correspondent fermions satisfy $\psi^I_\pm = - \psi^I_\mp$ on the boundaries. One then finds the more general linear relation $\partial_\tau Q^1_A = \partial_\tau (P Q^2)_A$, where
\begin{equation}
    P = \prod_I \big( \Gamma^I \Gamma \big),
\end{equation}
with the product running over all DD directions \cite{bachas2024dbranes}. The conserved supercharges in the presence of open string boundary conditions are therefore
\begin{equation}
    Q_A = Q^1_A - ( P Q^2 )_A.
\end{equation}
The number of Dirichlet conditions affects $P$, and therefore changes the particular combinations of $Q^1_A$ and $Q^2_A$ that are preserved, but does not change the fact that there are always only 16 conserved supercharges, half of the original amount. This is in contrast with the number of Poincaré symmetries broken, which of course grows with the number of DD conditions. 

One should not, however, take this to mean that open strings do not exist in the type II superstring theories. All it means is that they do not arise as low-energy excitations of the trivial vacuum, where the expectation value of all fields is set to zero. If one quantizes the theory around some other classical background that happens to only preserve 16 supersymmetries, it is perfectly reasonable to assume that open strings do emerge as quantum excitations of this different, nontrivial vacuum. 

The existence of open strings is equivalent to the existence of D-branes, which in the context of the bosonic string were shown to admit a representation as coherent states in the closed string Hilbert space, the boundary states. Since so far we have only established the existence of closed strings in the type II theories, investigating whether or not they admit boundary states can provide a definitive answer as to the existence of open strings. The procedure is the same as what was done for the bosonic string: one looks for a state $\ket{Dp,y}$ in the closed string Hilbert space that is annihilated by the boundary conditions appropriate for a $Dp$-brane, imposed at $\xi^2=0,s$:
\begin{align}
    \partial_2 X^\mu(\xi^1,0) = X^I(\xi^1,0) = 0, && \psi^\mu_+(\xi^1,0) = i \eta \psi^\mu_-(\xi^1,0) , && \psi^I_+(\xi^1,0) = -i \eta \psi^I_-(\xi^1,0) ,
\end{align}
with $\eta=\pm 1$ (recall that we used $(\xi^1,\xi^2)$ for the worldsheet coordinates when discussing closed string exchange by branes). The conditions for the fermions differ from the ones defined in Section \ref{sec71} by the factor of $i \eta$. The $i$ comes from the fact that the boundary sits at fixed worldsheet time instead of space. The boundary term one obtains when varying the action in this case is
\begin{equation}
    \int d \xi^1 \big( \bar{\psi}^\mu \rho^2 \delta \psi_\mu \big) \Big|^{\xi^2=s}_{\xi^2=0} =i
    \int d \xi^1 \Big( \psi^\mu_+ \delta \psi_{+ \, \mu} + \psi^\mu_- \delta \psi_{- \, \mu} \Big) \Big|^{\xi^2=s}_{\xi^2=0}=0, \hspace{0.5cm} (\rho^2 = i \rho^0),
\end{equation}
so one must set $\psi^\mu_+ = \pm i \psi^\mu_-$. Whereas before we chose for simplicity the positive sign on $\xi^2=0$ for all NN directions, in this context keeping a general overall sign $\eta=\pm 1$ makes it easier to build a boundary state that respects the GSO projection. In an analogous manner to our construction of bosonic boundary states in Section \ref{sec64}, the fermionic boundary conditions can be solved for the $\eta$-dependent boundary states $\ket{B,\eta}$. We refer the reader to \cite{Blumenhagen:2013fgp,Callan:1987px} for the detailed computations. The main point of interest for us is that these states are of the general form
\begin{equation}
    \ket{B,\eta} \sim  \exp{ \bigg( i \eta \sum_{r>\nu} \big( \psi^\mu_{-r} \tilde{\psi}_{\mu \, -r} - \psi^I_{-r} \tilde{\psi}^I_{-r} \big) \bigg) } \ket{\text{vac}}, 
\end{equation}
where $\ket{\text{vac}}$ contains no excited modes. They are therefore coherent states. One finally obtains $\ket{Dp,y}$ by joining the $\ket{B,\eta}$ with different values of $\eta$ and different fermion periodicities with the bosonic boundary states found before into GSO-invariant combinations.\footnote{In this discussion we are sweeping under the rug the issue of building the part of the boundary states related to the superconformal ghosts. We instead eliminate by hand the two unphysical components of the matter fields and ignore the ghosts. This was shown in Section \ref{sec64} to produce the correct results for the bosonic string. } 

The interaction amplitude between two parallel branes, one at $x^I=y^I_1$ and another one at $x^I=y^I_2$, is given by \cite{Blumenhagen:2013fgp}
\begin{align}
    \mathcal{A}_\text{closed} &= \frac{1}{2} \int_0^\infty \frac{ds}{2 \pi} \braket{Dp,y_2|e^{-s(L_0+\tilde{L}_0 + 2a )} \delta_{L_0 \tilde{L}_0} |Dp,y_1} \notag\\[5pt]
    &= \frac{i V_{p+1} |N_p|^2}{2 \pi (2 \pi \alpha^\prime)^{(9-p)/2}} \int_0^\infty \frac{ds}{s^{(9-p)/2}} \, e^{ -\frac{\Delta y^2}{2 \alpha^\prime s} } \left( \frac{ \vartheta_{00}(0|is/\pi)^4 - \vartheta_{01}(0|is/\pi)^4 - \vartheta_{10}(0|is/\pi)^4 }{\eta(is/\pi)^{12}} \right),
    \label{D-brane interaction amplitude}
\end{align}
where the $\vartheta_{\alpha \beta}(0|is/\pi)$ are the Jacobi theta-functions
\begin{equation}
    \vartheta_{\alpha \beta}(\nu|\tau) = \sum_{n = - \infty}^\infty \exp \bigg[ i \pi \Big( n+\frac{\alpha}{2} \Big)^2 \tau + 2 \pi i \Big( n+ \frac{\alpha}{2} \Big) \Big(\nu + \frac{\beta}{2} \Big) \bigg].
\end{equation}

This amplitude should be related by worldsheet duality to the open string one-loop vacuum amplitude. Indeed, equation \eqref{Physical trace formula} in the case of the type II superstring leads to \cite{Blumenhagen:2013fgp}\footnote{Is is necessary to add to the trace also the GSO projection operator $\frac{1}{2} (1 + (-1)^F)$.}
\begin{align}
    \mathcal{A}_\text{open} = 2 i V_{p+1} \int_0^\infty \frac{dt}{2t} \left( \frac{1}{8 \pi^2 \alpha^\prime t} \right)^{(p+1)/2} e^{ -\frac{t \Delta y^2}{2 \pi \alpha^\prime} } \frac{1}{2}  \left( \frac{ \vartheta_{00}(0|it)^4 - \vartheta_{10}(0|it)^4 - \vartheta_{01}(0|it)^4 }{\eta(it)^{12}} \right).
\end{align}
For $N_p= \frac{1}{\sqrt{32}} (4 \pi \alpha^\prime)^{(4-p)/2} $ the two amplitudes are indeed the same. This can be shown using the fact that the theta-functions satisfy
\begin{align}
    \vartheta_{00}(0|it) = \sqrt{\frac{s}{\pi}} \vartheta_{00}(0|is/\pi), && \vartheta_{01}(0|it) = \sqrt{\frac{s}{\pi}} \vartheta_{10}(0|is/\pi), && \vartheta_{10}(0|it) = \sqrt{\frac{s}{\pi}} \vartheta_{01}(0|is/\pi),
\end{align}
for $s=\pi / t$, together with the transformation of the eta function \eqref{eta function S transf}.\footnote{Another argument for the existence of open strings in the type II theories starts from the type I superstring, which has only $\mathcal{N}=1$ spacetime supersymmetry. This theory has both closed and open strings as excitations of the vacuum. T-dualizing one dimension maps a state in the type I theory to one in the type IIA theory. T-duality in more directions then moves us back and forth between type IIA and IIB. In this way it is possible to directly obtain states in the type II theories containing open strings. For a detailed account of this argument see Chapter 13 of \cite{Polchinskivol2:1998rr}. } 

Having established the existence of D-branes in the type II theories, we now turn to the investigation of some of their properties. Since the excitations of a D-brane are open strings propagating inside their worldvolume, in the low-energy limit we expect them to be well described by the field theory that governs the massless level of the GSO-projected open superstring. From our analysis of the spectrum we know this theory is a supersymmetric extension of Maxwell theory, which can be promoted to an $SU(N)$ Yang-Mills theory by stacking $N$ branes on top of each other, or equivalently, by adding $U(N)$ Chan-Paton factors to the open strings. The massless level of the NS-NS sector of the superstring was found to be a copy of the bosonic string's massless level. In light of this it should be no surprise that, at low energies, the bosonic part of a single D-brane's coupling to the NS-NS closed strings is given by the Dirac-Born-Infeld action discussed in Section \ref{sec52}:
\begin{equation}
    S_\text{DBI} = -T_{p} \int d^{p+1}x \, e^{-\Phi} \sqrt{-\det (\mathcal{G}_{\mu \nu} + \mathcal{B}_{\mu \nu} + 2 \pi \alpha^\prime F_{\mu \nu} )}.
\end{equation}
From this we extract an important fact: since a shift of the dilaton expectation value $\braket{\Phi} \equiv \Phi_0$ changes the effective value of $T_p$, it is actually the combination 
\begin{equation}
    \tau_p =T_p e^{- \Phi_0} =\frac{T_p}{g_s}
\end{equation}
that represents the physical tension of the brane. By similarly separating the dilaton expectation value in the supergravity action \eqref{type IIB SUGRA action}, we find that the physical gravitational coupling is 
\begin{equation}
    \kappa = e^{\Phi_0} \kappa_0 = g_s \kappa_0.
\end{equation}
This ties nicely with our previous discussion of the apparent absence of open strings in the type II theories. The starting point for the derivation of the spectrum was the quadratic RNS action, which describes strings propagating in flat spacetime. The particles obtained in this way, among them gravitons, are those contained in the perturbative limit of supergravity $\kappa \sim g_s =0$. In this region of parameter space the tension $\tau_p \sim 1/g_s$ diverges, making the brane an infinitely massive and therefore invisible object in perturbation theory. The rigid, symmetry breaking Dirichlet boundary conditions are a manifestation of this. They imply that momentum flowing into the brane is not conserved, and only an object of infinite mass can absorb momentum without changing its shape or position. As we raise the dilaton expectation value the tension becomes smaller and the brane less rigid. Its degrees of freedom become accessible. The coupling of the worldvolume fields to the closed string modes means that momentum flowing into the brane is actually conserved, and it can change the brane's shape by exciting the $\phi^I$ scalars that describe its embedding in spacetime. Of course, by raising $g_s$ we are also moving the theory away from the perturbative supergravity limit, so one should expect the gravitational dynamics to no longer resemble that of free gravitons in flat space. This is an important point, to which we will return.

The coupling of the brane's worldvolume fermions to the closed string fermionic fields from the NS-R and R-NS sector can be found in the full supersymmetric version of the DBI action. We will not need its detailed form. For a comprehensive account of such actions the reader is referred to \cite{caltechthesis4867}. There are many ways in which D-branes couple to the massless R-R fields. The simplest one is given by integrating the gauge field of rank $p+1$ over the brane's $(p+1)$-dimensional worldvolume:
\begin{equation}
    S_{\mu_p} = \mu_p \int C_{p+1} = \mu_p \int \frac{1}{(p+1)!} C_{\mu_1 
    \dots \mu_{p+1}} dx^{\mu_1} \wedge \dots \wedge dx^{\mu_{p+1}},
    \label{brane RR coupling}
\end{equation}
where for simplicity we assumed the brane to be sitting inside a flat ambient spacetime on the second equality. This is the natural higher-dimensional generalization of the coupling of a charged particle to an electromagnetic field, $ q\int A= q \int A_\mu dx^\mu$. The constant $\mu_p$ is the brane's R-R charge. Interaction terms of higher dimension can be built by taking wedge products of the other forms available. For instance, the forms $F_2 \wedge C_{p-1}$ and $B_2 \wedge C_{p-1}$, where $F_2$ is the gauge field intrinsic to the brane and $B_2$ is the Kalb-Ramond field pulled back to the brane's worldvolume, also have the correct rank to be integrated over the brane. The same goes for $F_2 \wedge F_2 \wedge C_{p-3}$ and a variety of other possible combinations. These are all present in the spacetime action, and come with increasing powers of $\alpha^\prime$ according to their dimension, just like the higher powers of $F_{\mu \nu}$ found in the expansion of the DBI action. We will only be concerned with the basic coupling $S_{\mu_p}$, which is the most important one at low energies.

In the $s \to \infty$ low-energy limit the interaction amplitude between the two branes $\mathcal{A}_\text{closed}$ should reduce to the amplitude for exchanging the massless, long range closed string modes. The effective actions for these modes provide us with a second way of finding the interaction amplitude, and comparing the result of both computations allows us to fix the value of the brane's tension and charge.

The first thing to note about $\mathcal{A}_\text{closed}$ is that it actually vanishes, due to Jacobi's ``abstruse identity''
\begin{equation}
    \vartheta_{00}(0|\tau)^4 - \vartheta_{01}(0|\tau)^4 - \vartheta_{10}(0|\tau)^4=0.
\end{equation}
Physically this can be understood from supersymmetry. In the open string interpretation we have at all mass levels an equal amount of bosonic and fermionic particles going around the loop. Their contributions cancel each other exactly, leaving no net energy density between the branes capable of producing a Casimir force between them. In the closed string interpretation one should instead speak of particles being exchanged by the branes. At the massless level we have that the attractive interaction from exchanging NS-NS gravitons and dilatons cancels against the repulsive one from the exchange of R-R gauge fields, leading once again to zero net force. This cancellation of course occurs also at the higher mass levels. In the $s \to \infty$ limit \eqref{D-brane interaction amplitude} becomes
\begin{equation}
    \mathcal{A}_{\text{closed}} = \left( \frac{\pi}{2} \right)^{5/2} \frac{i V_{p+1}}{(2 \pi)^{3p/2} {\alpha^{\prime}}^{(p+1)/2}} \int^\infty \frac{ds}{s^{(9-p)/2}} \, e^{ -\frac{\Delta y^2}{2 \alpha^\prime s} } \left( 16 -16 + \mathcal{O}\left( e^{-s} \right)  \right).
    \label{low-energy D-brane amplitude}
\end{equation}
The first factor of 16 comes from the exchange of gravitons and dilatons, and the second one with the minus sign comes from the exchange of $C_{p+1}$ quanta. The interaction amplitude from only the exchange of the p-form particles is therefore
\begin{equation}
    \mathcal{A}^{C_{p+1}} = - \left( \frac{\pi}{2} \right)^{5/2} \frac{16i V_{p+1}}{(2 \pi)^{3p/2} {\alpha^{\prime}}^{(p+1)/2}} \int^\infty \frac{ds}{s^{(9-p)/2}} \, e^{ -\frac{\Delta y^2}{2 \alpha^\prime s} }.
    \label{worldsheet D-brane interaction}
\end{equation}

We now consider how the same interaction amplitude is set up in the language of the low-energy effective field theory. The general reasoning is the same one that led to equation \eqref{D-brane interaction general reasoning}, but we shall review it in more detail here, taking at first a bosonic scalar of action $ -\frac{1}{2} \int d^{10}x (\partial_\mu \phi \partial^\mu \phi + m^2 \phi^2 )$ to play the role of the force-mediating particle. If this field is found in its vacuum state at $t = - \infty$, the amplitude for it to be found in the same state at $t = + \infty$ is given by
\begin{equation}
     \braket{0,+ \infty|0,-\infty} \equiv  Z[0] = \int \mathcal{D} \phi \,
     e^{i S[\phi]},
\end{equation}
where the boundary conditions of the path integral are such that $\phi=0$ at $t \to \pm \infty$. For this reason $Z[0]$ is also commonly called the vacuum persistence amplitude. Now let there be some system capable of emitting or absorbing quanta of this field. We ignore the actual dynamics of this system, so that its only effect is the introduction of a source term $J(x)$ in the equation of motion for the scalar:
\begin{equation}
    \partial^2 \phi + m^2 \phi = J(x).
\end{equation}
This is obtained as an equation of motion if one adds to the action the linear coupling term $S_{\text{int}} = \int d^{10}x J(x) \phi(x)$. In the presence of this interaction the vacuum persistence amplitude is given by the generating functional
\begin{equation}
    Z[J] = Z[0] \exp \bigg( - \frac{1}{2} \int d^4x_1 \, d^4x_2 J(x_1) \Delta(x_1-x_2) J(x_2) \bigg),
\end{equation}
where $\Delta(x_1-x_2)$ is the propagator of $\phi$. Take $J(x) = J_a(x) + J_e(x)$, where $J_e(x)$ is a function localized around an emission point $x_e=(t_e,\mathbf{x}_e)$. In the limit of infinite localization we would have $J_e(x) \sim \delta^{10}(x-x_e)$. $J_a(x)$ is defined similarly, but for a point $x_a$, the absorption point. In this case
\begin{align}
    Z[J] = Z[0] \exp \bigg( - \frac{1}{2} \int d^{10}x_1 \, d^{10}x_2  \Big( J_a(x_1) \Delta(x_1-x_2) J_a(x_2) + & 2 J_a(x_1) \Delta(x_1-x_2) J_e(x_2) \notag\\[5pt]
    +& J_e(x_1) \Delta(x_1-x_2) J_e(x_2)  \Big) \bigg).
\end{align}
The first and last terms represent self-interaction of the sources, which we are not interested in. We therefore discard them and keep only the middle term:
\begin{align}
    \frac{Z[J]}{Z[0]} &= \exp \bigg( - \int d^{10}x_1 \, d^{10}x_2   J_a(x_1) \Delta(x_1-x_2) J_e(x_2) \bigg) \notag\\[5pt]
    &= 1 - \int d^{10}x_1 \, d^{10}x_2   J_a(x_1) \Delta(x_1-x_2) J_e(x_2) \notag\\[5pt]
    &\hspace{0.65cm}+ \frac{1}{2!} \int d^{10}x_1 \, d^{10}x_2 \, d^{10}x_3 \, d^{10}x_4   J_a(x_1) \Delta(x_1-x_2) J_e(x_2) J_a(x_3) \Delta(x_3-x_4) J_e(x_4) + \dots
\end{align}
According to our earlier interpretation, this is an amplitude for the system to be found in the vacuum in the far past and future. By the usual paradigm of summing over histories, the amplitude is given by the sum of the amplitudes of all possible ways this may happen. One possibility is that no particles are emitted or absorbed. This corresponds to the factor of $1$ in the expansion above. This guarantees that for $J_e=J_a=0$ we recover $Z[J]=Z[0]$. If one particle is emitted by $J_e$, the only way for the system be in the vacuum state at $t=+\infty$ is for this particle to be absorbed $J_a$ at time $t_a > t_e$. The amplitude for this is the second term in the expansion
\begin{equation}
    \mathcal{A}_\text{one particle exchange} = -\int d^{10}x_1 \, d^{10}x_2   J_a(x_1) \Delta(x_1-x_2) J_e(x_2).
    \label{one particle exchange amplitude}
\end{equation}
Another possibility is to have two particles being emitted at $x_e$ and absorbed at $x_a$. This corresponds to the second term, quartic in the sources. We can go on indefinitely and end up resumming the entire exponential. Of course, one could also consider processes in which each source emits and itself reabsorbs any number of particles. These correspond to the self-interaction terms that we ignored. 

For the determination of the D-brane's charge, the lowest order interaction amplitude \eqref{one particle exchange amplitude} is enough. The source term is easily determined by rewriting the coupling of the brane to the R-R field as
\begin{equation}
    S_{\mu_p} = \int d^{10}x J^{\mu_1 \dots \mu_{p+1}}(x) C_{\mu_1 \dots \mu_{p+1}}(x).
\end{equation}
Comparison with \eqref{brane RR coupling} leads to 
\begin{equation}
    J^{01 \dots p}(x^\mu,y^I) = \frac{\mu_p \delta^{9-p}(y - y_0)}{(p+1)!} 
\end{equation}
for a flat, static brane at fixed position $y^I = y_0^I$ in transverse space, with all other components not related to this one by antisymmetry vanishing. The factor of $1/(p+1)!$ compensates for the overcounting due to summing over all Lorentz indices, given that both $C_{p+1}$ and the current and totally antisymmetric. The propagator for $C_{p+1}$ is extracted from the kinetic term
\begin{equation}
    -\frac{1}{2 \kappa_0^2} \int d^{10}x \frac{1}{2(p+2)!} F_{\mu_1 \dots \mu_{p+2}} F^{\mu_1 \dots \mu_{p+2}}
\end{equation}
in the supergravity actions in flat space. Upon setting $F_{\mu_1 \dots \mu_{p+2}} =(p+2) \partial_{[ \mu_1} C_{\mu_2 \dots \mu_{p+2}]}$ and imposing the Lorenz-like gauge $\partial^\nu C_{\nu \mu_1 \dots \mu_{p}}=0$, this becomes
\begin{equation}
    \frac{1}{2 \kappa_0^2} \int d^{10}x \frac{1}{2(p+1)!} C_{\mu_1 \dots \mu_{p+1}} I^{\mu_1 \dots \mu_{p+1} \nu_1 \dots \nu_{p+1}} \partial^2 C_{\nu_1 \dots \nu_{p+1}},
\end{equation}
where $I^{\mu_1 \dots \mu_{p+1}}{}_{\nu_1 \dots \nu_{p+1}} = \delta^{[\mu_1}_{\nu_1} \dots \delta^{\mu_{p+1}]}_{\nu_{p+1}}$. The propagator is then easily found by inverting the kinetic operator:
\begin{equation}
    \Delta^{\mu_1 \dots \mu_{p+1} \nu_1 \dots \nu_{p+1}}(x_1 - x_2) = 2i \kappa_0^2 (p+1)! \int \frac{d^{10}k}{(2 \pi)^{10}} \frac{I^{\mu_1 \dots \mu_{p+1} \nu_1 \dots \nu_{p+1}}}{k^2} e^{i k \cdot (x_1 - x_2)}.
\end{equation}
The interaction amplitude is therefore given by
\begin{align}
    \mathcal{A}^{C_{p+1}} &= - \int d^{10} x \, d^{10}x^\prime \, J_{\mu_1 \dots \mu_{p+1}}(x) \Delta^{\mu_1 \dots \mu_{p+1} \nu_1 \dots \nu_{p+1}}(x - x^\prime) J_{\nu_1 \dots \nu_{p+1}}(x^\prime) \notag\\[5pt]
    &=- (p+1)! \int d^{10} x \, d^{10}x^\prime \, J_{0 \dots p}(x) \Delta^{0 \dots p\,  0 \dots p }(x - x^\prime) J_{0 \dots p}(x^\prime) \notag\\[5pt]
    &= - 2i \kappa_0^2  \mu_p^2 \int \frac{d^{10}k}{(2 \pi)^{10}} e^{i k^I (y_1^I - y_2^I)} \int d^{p+1}x \, d^{p+1}x^\prime  \frac{e^{ik^\mu(x^\mu - x^{\prime \mu})}}{k^2}.
\end{align}
After using \eqref{Schwinger trick} to write the propagator in exponential form, all spacetime and momentum integrals are Gaussian, leading to
\begin{equation}
    \mathcal{A}^{C_{p+1}} = - \frac{i V_{p+1} \alpha^\prime \mu_p^2 \kappa_0^2}{(2 \pi \alpha^\prime)^{(9-p)/2}} \int_0^\infty \frac{ds}{s^{(9-p)/2}} \, e^{ -\frac{\Delta y^2}{2 \alpha^\prime s} }.
\end{equation}
Comparing this with \eqref{worldsheet D-brane interaction} gives
\begin{equation}
    \mu_p = \frac{\sqrt{\pi}}{\kappa_0} (4 \pi^2 \alpha^\prime)^{(3-p)/2}
    \label{brane charge}
\end{equation}
for the charge. The brane's tension can be similarly fixed by comparing \eqref{low-energy D-brane amplitude} with the field theory amplitude for the exchange of gravitons and dilatons. The currents in this case come from the DBI action. The computation is slightly more involved, due to the fact that the graviton and dilaton kinetic terms are mixed in the string frame supergravity actions. The best strategy is to go to Einstein frame, where each propagator separates. This is done in detail in \cite{Blumenhagen:2013fgp}, and leads to 
\begin{equation}
    T_p = \frac{\sqrt{\pi}}{\kappa_0} (4 \pi^2 \alpha^\prime)^{(3-p)/2} = \mu_p.
\end{equation}
It is conventional to relate the gravitational constant $\kappa_0$ and $\alpha^\prime$ by setting the tension of the fundamental string $T=1/(2 \pi \alpha^\prime)$ equal to that of the $D1$-brane, which is itself a string:
\begin{equation}
    T = T_1.
\end{equation}
This determines $\kappa_0 = 8 \pi^{7/2} \alpha^{\prime 2}$. The physical D-brane tension is thus given by
\begin{equation}
    \tau_p = \frac{T_p}{g_s} = \frac{1}{g_s (2 \pi)^p \alpha^{\prime (p+1)/2}},
\end{equation}
and the physical gravitational constant is
\begin{equation}
    \kappa = g_s \kappa_0 = 8 g_s \pi^{7/2} \alpha^{\prime 2}.
\end{equation}

\section{D-branes as solitons}\label{sec74}

We have gathered enough information on D-branes to allow for a discussion of what is perhaps their most important property for holography: they possess a dual description in terms of closed string solitons. By ``soliton'' we mean a stable, localized solution of the equations of motion, which therefore has particle-like properties without being itself among the theory's elementary excitations. They arise instead as nontrivial collective excitations that depend on the interactions of the theory to preserve their properties. Their mass usually is inversely proportional to some power of the coupling constant, making them inherently nonperturbative objects. This is of course the case for D-branes, because $\tau_p \sim 1/g_s$. That branes must be in some sense made out of closed strings can be argued simply from the fact that the only way to move the type II theories away from the vacuum is to add closed strings to it. There are no other kinds of excitations to work with, at least perturbatively. Also in favor of this is the fact that the worldsheet CFT's boundary states, which are our only explicit construction of a D-brane state so far, are given precisely by closed string coherent states. 

With this in mind, we look for supergravity solutions with the expected properties of a D-brane, which should give the low-energy description of such a soliton. These solutions should have the symmetry group of a flat $(p+1)$-dimensional hypersurface inside ten-dimensional spacetime:
\begin{equation}
    \mathbb{R}^{p+1} \times SO(1,p) \times SO(9-p).
\end{equation}
Let $x^\mu$ with $\mu=0,\dots,p$ be coordinates that cover the brane's worldvolume, and $y^I$ with $I= 1, \dots, 9-p$ be coordinates on the directions perpendicular to the brane.
The factor of $\mathbb{R}^{p+1} \times SO(1,p)$ is the Poincaré group inside the brane's worldvolume. This forces the metric in the directions parallel to the brane to be at most a rescaling of the Minkowski metric. The $SO(9-p)$ consists of rotations on the perpendicular directions, and it forces the metric in these directions to be a rescaling of the Euclidean metric. Worldvolume translation symmetry means that the metric coefficients cannot depend on the $x^\mu$ coordinates, whereas transverse rotation symmetry means that they can only depend on the transverse radius
\begin{equation}
    r = \sqrt{y^I y^I}.
\end{equation}
The general ansatz for the metric is therefore
\begin{equation}
    ds^2 = f(r) \eta_{\mu \nu} dx^\mu dx^\nu + g(r) \delta_{IJ} dy^I dy^J,
\end{equation}
where the functions $f(r)$ and $g(r)$ should approach 1 for large $r$, since far away from the brane one should recover Minkowski space. We also set $e^{\Phi} =e^{\Phi_0} e^{\phi(r)} = g_s e^{\phi(r)}$ for the dilaton and require that $C_{p+1} = C_{p+1}(r)$ be nontrivial, since a D$p$-brane is a source for this field. The rest of the supergravity fields do not couple directly to the brane at low energies and can all be set to zero. The reader is referred to \cite{West_2012} for the details on how to solve the supergravity equations of motion subjected to these assumptions. The result for $p \leq 6$ in string frame is the so-called extremal black $p$-brane solution \cite{Blumenhagen:2013fgp}
\begin{align}
    ds^2 &= H(r)^{-1/2} \eta_{\mu \nu} dx^\mu dx^\nu + H(r)^{1/2} \delta_{IJ} dy^I dy^J \notag\\[5pt]
    e^{ \Phi} &= g_s H(r)^{(3-p)/4} \notag\\[5pt]
    F_{p+2} &= g_s^{-1} dx^0 \wedge \dots \wedge dx^{p+1} \wedge dH(r)^{-1}, \hspace{0.5cm} p \neq 3.
\end{align}
where
\begin{equation}
     H(r) = 1 + \frac{L^{7-p}}{r^{7-p}},
\end{equation}
with $L$ an integration constant with dimension of length. For $p=3$ there is the additional condition that $F_{5}$ be self-dual, in which case the solution is
\begin{equation}
    F_5 = g_s^{-1}\big(1+ \ast \big) dx^0 \wedge dx^1 \wedge dx^2 \wedge dx^3 \wedge dH(r)^{-1}.
\end{equation}
In all cases we have $F_{0\dots p I} = g_s^{-1} \partial_I(H(r)^{-1})$. It can be verified that this solution breaks half of the supersymmetries of the correspondent ten-dimensional supergravity theory. Since for $L =0$ we obtain flat spacetime with no Ramond-Ramond flux, $L$ must depend on both the brane's tension and charge (which are, after all, equal to each other), so either quantity can be used to fix its value. The simpler choice is the charge, which is given by the total flux of $\ast F_{p+2}$ through a ($8-p$)-dimensional hypersurface that encloses the brane:\footnote{This formula is the higher-dimensional version of computing the total electric charge of an object by integrating the flux of the electric field through a sphere surrounding it. In the simplest case of a static point charge in the origin of four-dimensional spacetime, we have in spherical coordinates
\begin{equation}
    F=\frac{1}{2} F_{\mu \nu} \, dx^\mu \wedge dx^\nu = E_{r}(r) \,  dr \wedge dt,
\end{equation}
with $E_r(r)=q/(4 \pi r^2)$ and all other components vanishing. Taking $\epsilon_{t r\theta \phi}= +1$, the spherical coordinate metric $g_{\mu \nu} = \text{diag}(-1,1,r^2,r^2 \sin^2 \theta)$ leads to $\ast(dr \wedge dr) = r^2 \sin \theta \, d \theta \wedge d \phi$. One then readily computes the charge via
\begin{equation}
    \int_{S^2} \ast F = \int \left( \frac{q}{4 \pi r^2}  \right) r^2 \sin \theta \, d \theta \wedge d \phi = \frac{q}{4 \pi} \int_0^\pi \sin \theta  \, d \theta \int_0^{2 \pi} d \phi = q.
\end{equation}
}
\begin{equation}
    \mu_p = \frac{1}{2 \kappa_0^2}\int_{S^{8-p}} \ast F_{p+2}.
\end{equation}
The hypersurface is taken for simplicity to be $(8-p)$-sphere at infinity in transverse space, $r \to \infty$, and the factor of $1/2 \kappa_0^2$ comes from the noncanonical normalization of the gauge fields in the supergravity action. For this computation it is convenient to parametrize the transverse directions with a hyperspherical coordinate system
\begin{equation}
    \big( x^0, \dots, x^p , y^1, \dots, y^{9-p} \big) \longrightarrow \big( x^0, \dots, x^p , r,\theta^1, \dots, \theta^{8-p} \big)
\end{equation}
which is defined by
\begin{align}
    y^I &= r \sin \theta^1 \dots\sin \theta^{I-1} \cos \theta^I, \hspace{0.5cm} I=1,\dots, 7-p \notag\\[5pt]
    y^{8-p} &=r \sin \theta^1 \dots \sin \theta^{7-p} \,  \sin \theta^{{8-p}},
\end{align}
where $\theta^{I} \in[0, \pi] $ for $I= 1, \dots7-p$ and $\theta^{8-p} \in [0,2 \pi]$. Naturally we have $r^2 = y^I y^I$. In these coordinates the surface element of the $(8-p)$-sphere is \cite{West_2012}
\begin{equation}
    d \Omega_{8-p} = r^{8-p} d \theta^1 \wedge \sin \theta^1 d \theta^2 \wedge \sin \theta^1 \sin \theta^2 d \theta^3 \wedge \dots \wedge \sin \theta^1 \dots \sin \theta^{7-p} d \theta^{8-p}.
\end{equation}
From this it is clear that the component of $\ast F_{p+2}$ that occurs in the integral is $\ast F_{\theta^1 \dots \theta^{8-p}}$:
\begin{equation}
    \mu_p = \frac{1}{2 \kappa_0^2} \int (\ast F)_{\theta^1 \dots \theta^{8-p}} \, d \Omega_{8-p}.
    \label{charge integral}
\end{equation}
This component is given by
\begin{align}
    (\ast F)_{\theta^1 \dots \theta^{8-p}} &= \frac{\sqrt{-G}}{(p+1)!} \epsilon_{\theta^1 \dots \theta^{8-p} \mu_1 \dots \mu_{p+1} r } F^{r \mu_1 \dots \mu_{p+1}} \notag\\[5pt]
    &= \sqrt{-G}\,  \epsilon_{\theta^1 \dots \theta^{8-p} x^0 \dots x^p r} F^{0 1 \dots p  r} \notag\\[5pt]
    &= - g_s^{-1} (-1)^p H(r)^{(4+p)/2} \partial_r (H(r)^{-1})
\end{align}
where we set $\epsilon_{x^0 \dots x^p r \theta^1 \dots \theta^{8-p}}=+1$, used 
 that $\det G = \det(G_{\mu \nu}) \det(G_{IJ}) = - H(r)^{4-p}$ and also that
 \begin{equation}
     F^{0 1 \dots p r} = - H(r)^p F_{0 1 \dots p r} = -g_s^{-1} H(r)^p \partial_r( H(r) ^{-1}).
 \end{equation}
At large $r$ the field strength tends to the Coulomb-like form
\begin{equation}
    (\ast F)_{\theta^1 \dots \theta^{8-p}} \sim -\frac{(-1)^p g_s^{-1} (7-p)L^{7-p}}{r^{8-p}}.
\end{equation}
When this is plugged into \eqref{charge integral}, the $1/r^{8-p}$ cancels against the factor of $r^{8-p}$ in $d \Omega_{8-p}$, so the integral is simply the surface area of the $(8-p)$-sphere of unit radius,
\begin{equation}
    \text{Area}\big(S^{8-p}\big) = \frac{ \pi^{(9-p)/2}}{(7-p)\Gamma\big(\frac{7-p}{2}\big)}.
\end{equation}
Using $\kappa_0 = 8 \pi^{7/2} \alpha^{\prime 2}$ leads to 
\begin{equation}
    \mu_p = - \frac{4(-1)^p  L^{7-p} \pi^{(9-p)/2} }{ (2 \pi)^7 \alpha^{\prime 4} \Gamma \big( \frac{7-p}{2} \big)},
\end{equation}
which, when set equal to \eqref{brane charge}, fixes
\begin{equation}
    L^{7-p} = \alpha^{\prime (7-p)/2} g_s (4 \pi)^{(5-p)/2} \Gamma \Big( \frac{7-p}{2} \Big).
\end{equation}
The fact that this constant is proportional to the string coupling is important. It means that for $g_s \sim 0$, where the string theory is perturbative, the D-brane spacetime 
\begin{equation}
    ds^2 = \bigg( 1+ \frac{L^{7-p}}{r^{7-p}} \bigg)^{-1/2} \eta_{\mu \nu} dx^\mu dx^\nu + \bigg( 1+ \frac{L^{7-p}}{r^{7-p}} \bigg)^{1/2}  \delta_{IJ} dy^I dy^J
    \label{D-brane metric}
\end{equation}
becomes ten-dimensional Minkowski space everywhere except on top of the brane, $r=0$. At any nonzero distance from the brane, the only dynamics seen are that of weakly interacting closed strings in flat space. This is very reminiscent of the open string picture of D-branes, where they take the form of static defect-like structures inside of Minkowski space. The $g_s =0$ case is of course not very physical in the presence of a brane, since we know that this makes the tension diverge. Raising the value of $g_s \sim L^{7-p}$ smoothens out the metric and makes the gravitational distortion caused by the brane visible. The apparent divergence one obtains from setting $r=0$ in the metric corresponds to a coordinate singularity, not a physical one, just like the event horizon of a Schwarzschild black hole. For this reason the $r =0$ region is called the horizon, and the near-horizon limit $r \to 0$ plays an important role in holography.

\section{AdS/CFT}\label{sec75}

We have finally identified the two sides of the holographic duality, each corresponding to a different way to describe a D-brane. The intrinsic description in terms of the worldvolume excitations (open strings) is where we expect to find gauge theories. The ten-dimensional solitonic description is the gravitational side. The only ingredient still missing is $N$, the rank of the gauge group on the brane's worldvolume. We know that this parameter is the number of branes stacked on top of each other, so we should look for a generalization of the supergravity solution that describes more than one brane. This is achieved by setting
\begin{equation}
    H(r) \to H(y) = 1 + \sum_{i=1}^N \frac{L^{7-p}}{|y-y_i|^{7-p}},
\end{equation}
where $y_i^I$ is the position of the $i$-th brane in transverse space. Setting $y^I_i =0$ for all $i$ collapses all $N$ branes on top of each other at $r=0$, and we recover the same solution as before, but with an additional factor of $N$ inside the constant $L^{7-p}$:
\begin{equation}
    L^{7-p} = \alpha^{\prime (7-p)/2} g_s N (4 \pi)^{(5-p)/2} \Gamma \Big( \frac{7-p}{2} \Big).
\end{equation}
This is equivalent to simply multiplying the tension and charge of a single brane by $N$.

In open string language the introduction of multiple branes mirrors exactly what was done for the bosonic string in Section \ref{sec34}. One adds to each state of the spectrum Chan-Paton factors to identify to what brane each of the string's endpoints are fixed. Upon letting the branes coincide, one obtains a new symmetry in the spectrum that corresponds to $U(N)$ rotations of the Chan-Paton factors. The consequence in spacetime is that all particles transform in the adjoint representation of an $SU(N)$ gauge group. The supersymmetrized DBI action must therefore be substituted by its nonabelian version, which at low energies, meaning to lowest order on the $\alpha^\prime$ expansion, should reduce to a supersymmetric Yang-Mills theory. From now on we specialize to $p=3$, so that the brane's worldvolume is four-dimensional. In this case the low-energy limit can be understood as truncating the nonabelian version of the supersymmetric DBI action to operators of energy dimension up to four, as interactions with dimension above this are non-renormalizable in four dimensions. The result is the unique $\mathcal{N}=4$ four-dimensional super Yang-Mills theory \cite{dhoker2002supersymmetricgaugetheoriesadscft}:
\begin{align}
    S_{\mathcal{N}=4} = \tau_3(\pi \alpha^\prime)^2 \int d^4x \, \text{Tr} \bigg( -  F_{\mu \nu} F^{\mu \nu} &- \frac{1}{2} D_\mu \phi^I D^\mu \phi^I  - \frac{i}{2} \bar{\lambda}^a
    \gamma^\mu D_\mu \lambda^a \notag\\[5pt]
    & - \frac{1}{2} C^I_{ab} \bar{\lambda}^a \gamma^5 [ \phi^I , \lambda^b ] + \frac{1}{4} [\phi^I,\phi^J][\phi^I,\phi^J] \bigg),
\end{align}
which we will call the $\mathcal{N}=4$ theory for short. The $a,b$ indices go from 1 to 8, and $\lambda^a$ are four-dimensional Majorana fermions. The $C^I_{ab}$ are constants related to the six-dimensional Gamma matrices, with $I = 1, \dots, 6$. The simplest way to derive this action is to start from the unique $\mathcal{N}=1$ ten-dimensional super Yang-Mills theory and use T-duality, meaning that one dimensionally reduces down to four spacetime dimensions and then sets $A^I(x) = -\phi^I(x)$.\footnote{In Section \ref{sec42} we chose $2 \pi \alpha^\prime A^I(x) = -\phi^I(x)$ , but in this context it is more convenient to not include the dimensional factor.} The fact that this theory has $\mathcal{N}=4$ four-dimensional supersymmetry follows directly from the fact that toroidal compactification breaks no supersymmetry. One therefore expects all of the 16 supercharges of $\mathcal{N}=1$ ten-dimensional super Yang-Mills to still be present. This is in accordance with our previous discussion on how open string boundary conditions always break half of the 32 supersymmetries of the type II theories, regardless of the dimension of the brane. A four-dimensional Majorana spinor has four independent real components, so one needs $\mathcal{N}=4$ Majorana fermions to accommodate a total of 16 supercharges. Like the action for regular four-dimensional Yang-Mills, $S_{\mathcal{N}=4}$ contains no dimensionful parameters, and therefore describes a theory that is classically conformally invariant. What is surprising is that, unlike nonsupersymmetric Yang-Mills, the $\mathcal{N}=4$ theory retains its conformal symmetry also at the quantum level. It is the CFT that appears in AdS/CFT.

All interactions that couple the worldvolume fields to the closed string excitations, such as gravitons and R-R gauge fields, have energy dimension of at least six, so they all decouple at low energies, leading to a gauge theory in flat spacetime. The Yang-Mills coupling is given by
\begin{equation}
    g_\text{YM} = \frac{1}{2 \pi \alpha^\prime \sqrt{\tau_3}} = \sqrt{2 \pi g_s},
\end{equation}
which is dimensionless, as expected. The 't Hooft coupling is
\begin{equation}
    \lambda = g_\text{YM}^2 N = 2 \pi g_s N.
\end{equation}

We now discuss the low-energy limit in the gravitational description. For this we first review how this works in a general context, and then specialize to the brane metric.

Conservation of energy in a curved background depends on the existence of a timelike Killing vector. Recall that a Killing vector is, by definition, the generator of an isometry of the spacetime. Since under an infinitesimal diffeomorphism generated by a vector $K^\mu$, the metric changes as $\delta g_{\mu \nu} = \nabla_\mu K_\nu + \nabla_\nu K_\mu$, the metric is only left invariant by diffeomorphisms that satisfy the Killing equation $\nabla_{(\mu} K_{\nu )}=0$.\footnote{For this general argument we use $x^\mu$ for all coordinates of an arbitrary spacetime.} If $x^\mu(\tau)$ parametrizes a trajectory in spacetime, the covariant derivative along this trajectory is given by $\nabla_\tau = U^\mu \nabla_\mu$, where $U^\mu = dx^\mu / d \tau$ is the velocity. The covariant derivative of the velocity itself is
\begin{equation}
    \nabla_\tau U^\mu = U^\nu \nabla_\nu U^\mu = U^\nu \partial_\nu U^\mu + U^\nu \Gamma^\mu_{\nu \rho} U^\rho = \frac{dU^\mu}{d \tau} + \Gamma^\mu_{\nu \rho} U^\nu U^\rho.
\end{equation}
Setting this to zero gives the geodesic equation, which therefore can be written as $U^\nu \nabla_\nu U^\mu=0$, or, if the geodesic describes the motion of a particle of mass $m$, 
\begin{equation}
    p^\nu \nabla_\nu p^\mu=0,
\end{equation}
where $p^\mu = mU^\mu$. This, together with the Killing equation, guarantee that the quantity $K_\nu p^\nu$ is conserved along geodesics:
\begin{equation}
    \frac{d}{d \tau} \big( K_\nu p^\nu \big) = \frac{1}{m}p^\mu \nabla_\mu \big( K_\nu p^\nu \big) = \frac{1}{m}p^\mu p^\nu \nabla_{(\mu} K_{\nu )} + \frac{1}{m}K_\nu p^\mu \nabla_\mu p^\nu =0.
\end{equation}
This scalar is the conserved charge that corresponds to the symmetry generated by the Killing vector $K^\mu$ \cite{Carroll:2004st}. For timelike $K^\mu$, the conserved quantity is called ``energy at infinity''
\begin{equation}
    E_\text{infinity} = - g_{\mu \nu} K^\mu p^\nu,
\end{equation}
with the minus sign added for later convenience. The reason for the name will become clear shortly. Consider now an observer situated at some point of this spacetime, making measurements with respect to his own local inertial frame. This frame is a flat space, tangent to his location on the ambient curved space, whose basis vectors are the vielbeins $e_{\tilde{\mu}} =e_{\tilde{\mu}}{}^\mu \partial_\mu $ satisfying $e_{\tilde{\mu}} \cdot e_{\tilde{{\nu}}} = e_{\tilde{\mu}}{}^\mu \,  e_{\tilde{{\nu}} \mu} = \eta_{\tilde{\mu} \tilde{\nu}}$. Assume this observer measures the motion of some particle and assigns to it the momentum $p^{\tilde{\mu}}$ and energy $E_\text{local} = p^{\tilde{0}}$. Since in his frame his own velocity is $U^{\tilde{0}} = 1$, with all other components vanishing, this energy can be written in Lorentz invariant form as $E_\text{local} = -\eta_{\tilde{\mu} \tilde{\nu}} U^{\tilde{\mu}} p^{\tilde{\nu}}$. If the spacetime happens to approach Minkowski space at spatial infinity, the energy measured by an observer at infinity is $p^{\tilde{0}} = p^0$, where we used the fact that at infinity $g_{\mu \nu} \to \eta_{\mu \nu}$, and therefore $e_{\tilde{\mu}}{}^\mu \to \delta_{\tilde{\mu}}^{\mu}$, to convert the local frame index into a global coordinate one. This shows that the conserved quantity $E_\text{infinity}$ has the interpretation of the energy an observer at infinity would measure. Whenever there is a timelike Killing vector, it is possible and useful to use a coordinate system where this vector is $\partial_0$ ($K^\mu = \delta^\mu_0$), for a suitably chosen time coordinate $x^0$. The symmetry under $x^0$-translations is expressed by the fact that no metric component depends on $x^0$. It is then easy to relate $E_\text{infinity}$ and $E_\text{local}$. Using the local observer's frame, one has 
\begin{equation}
    E_\text{infinity} = - \eta_{\tilde{\mu} \tilde{\nu} } K^{\tilde{\mu}} p^{\tilde{\nu}} = - \eta_{\tilde{\mu} \tilde{\nu}} \, e^{\tilde{\mu}}{}_\mu K^\mu p^{\tilde{\nu}} =  e^{\tilde{0}}{}_0 \, p^{\tilde{0}} = e^{\tilde{0}}{}_0 E_\text{local}.
\end{equation}
For a $D3$-brane the dilaton is constant and the metric is
\begin{equation}
    ds^2 = \bigg( 1+ \frac{L^{4}}{r^{4}} \bigg)^{-1/2} \eta_{\mu \nu} dx^\mu dx^\nu + \bigg( 1+ \frac{L^{4}}{r^{4}} \bigg)^{1/2}  \big( dr^2 + r^2 d \Omega_{S^5} \big)
    \label{D3 brane metric}
\end{equation}
with $L^4 = 4 \pi g_s N \alpha^{\prime2}$,
which is the $p=3$ case of \eqref{D-brane metric} with hyperspherical coordinates used for the six transverse directions. This has the timelike Killing vector $\partial_0$ and approaches flat space as $r \to \infty$, so the previous discussion applies. For a diagonal metric $e^{\tilde{0}}{}_0 = \sqrt{-g_{00}}$, so one has
\begin{equation}
    E_\text{infinity} = \bigg( 1+ \frac{L^{4}}{r^{4}} \bigg)^{-1/4} E_\text{local}. 
\end{equation}
The low-energy limit corresponds to $E_\text{infinity} \to 0$. At any finite distance $r$ from the brane, this also results in $E_\text{local} \to 0$. Since $E_\text{local}$ is the locally measured energy, around any point in the bulk of spacetime the only dynamics seen in this limit are that which can still occur at infinitesimally small energies: free propagation of gravitons and other massless closed string modes. Note, however, that as we move closer to the brane by lowering the value of $r$, the prefactor that relates $E_\text{infinity}$ and $E_\text{local}$ gets smaller. Upon taking the $r \to 0$ limit we have that the $E_\text{infinity} \to 0$ limit is reached independently of the value of the locally measured energy. Close to the brane a local observer would still be able to find interacting closed strings of arbitrarily high energies. For small $r$ we can approximate $H(r) = 1 + L^4/r^4 \approx L^4/r^4$, which turns the D3-brane metric into 
\begin{equation}
    ds^2 \approx \frac{r^2}{L^2} \Big( -dt^2 + d \vec{x}^2 \Big) + \frac{L^2}{r^2} dr^2 + L^2 d \Omega_{S^5},
\end{equation}
with $\eta_{\mu \nu} dx^\mu dx^\nu \equiv -dt^2 + d \vec{x}^2 $. The first two factors form the metric of five-dimensional Anti-de Sitter space. This is a maximally symmetric solution of Einstein's equations with a cosmological constant $\Lambda = -20 / L^2$, and is the AdS$_5$ in AdS/CFT. The second factor describes a five-sphere of radius $L$, so the total space is AdS$_5 \times S^5$.

Consider a process in which some supergravity particle comes toward the brane from the bulk of spacetime. It can be shown that the cross-section for the the brane to absorb such a particle goes like $\sigma \sim \kappa^2 E^3$, where $E$ is the particle's energy at infinity \cite{Aharony:1999ti}.\footnote{The cross-sections for many specific supergravity modes were computed in \cite{Klebanov:1997kc,GUBSER1997217}, and shown to agree with the cross-sections computed in the open string language, where one extracts the interaction vertex from the DBI action.} This of course vanishes in the low-energy limit. A particle that is instead traveling away from the brane, starting at some radius $r$ with locally measured energy $E_r$ and moving toward $r \to \infty$, has an energy at infinity dampened by the redshift factor $E_\text{infinity} = (1+L^4 / r^4)^{-1/4} E_r$, which grows smaller as $r$ is brought closer to $0$. Intuitively this ``energy loss'' can be attributed to the particle having the climb out of the gravitational well produced by the brane, which is steeper the closer one is to the horizon. For an excitation emitted around $r \sim 0$ the redshift goes to zero, and the particle cannot travel to the bulk. We therefore have that the near-horizon region of the D3-brane metric and the rest of ten-dimensional spacetime are decoupled, just like what was found earlier in the open string language.

Having studied the low-energy limit of the same brane system in both the open and closed string descriptions, we now compare the results. In both cases one finds two decoupled systems: low-energy, noninteracting supergravity modes away from the branes, and the branes themselves, described in one language as the usual open string D-brane, reducing to super Yang-Mills at low energies, and in the other as the $p$-brane supergravity solution, which reduces to the full theory of type IIB superstrings over an AdS$_5 \times S^5$ background. By declaring these to indeed be two descriptions of the same system, we arrive at the original AdS/CFT correspondence:

\begin{equation*}
    \fbox{\begin{tabular}{@{}c@{}}
        Type IIB superstring theory \\ on an AdS$_5 \times S^5$ background 
  \end{tabular}}
  =
  \fbox{\begin{tabular}{@{}c@{}}
        $\mathcal{N}=4$ super Yang-Mills theory in \\ 
        four-dimensional Minkowski space
  \end{tabular}}
\end{equation*}

\vspace{0.3cm}

The equality sign should be understood as a full duality, meaning that each side consists of a different language to describe the exact same physics. In order to investigate this duality further, it is convenient to understand the low-energy limit on the gravitational side in a slightly different way. Recall from Chapter \ref{ch5} that, for both open and closed strings, this limit is obtained by formally setting $\alpha^\prime \to 0$ on the spacetime actions. This is because $1/\sqrt{\alpha^\prime} = 1/l_s$, where $l_s$ is the string scale, so one can only probe distances of order $l_s$ with energies $E_\text{UV} \sim 1/\sqrt{\alpha^\prime}$, meaning that the dimensionless ratio $E / E_\text{UV}$, where $E$ is the characteristic energy of the process under consideration, should not be too small. The low-energy limit
\begin{equation}
    \frac{E}{E_\text{UV}} = \sqrt{\alpha^\prime} E \to 0
\end{equation}
is understood physically as letting $E \to 0$ while keeping the dimensionful parameter $\alpha^\prime$ fixed, but it is also formally obtained by keeping $E$ fixed and letting $\alpha^\prime \to 0$. This $E$ is the conserved energy previously called $E_\text{infinity}$, which in the small $r$ limit is related to $E_\text{local}$ as
\begin{equation}
    E_\text{infinity} \sim \frac{r}{L} E_\text{local} = \frac{1}{(4 \pi g_s N)^{1/4}} \frac{r}{\sqrt{\alpha^\prime}} E_\text{local} \propto \frac{r}{\alpha^\prime},
\end{equation}
with a dimensionless proportionality coefficient. By trading the small $E_\text{infinity}$ with fixed $\alpha^\prime$ limit by small $\alpha^\prime$ with fixed $E_\text{infinity}$, we obtain the so-called Maldacena limit
\begin{equation}
    \alpha^\prime \to 0, \hspace{0.5cm} r \to 0, \hspace{0.5cm} u = \frac{r}{\alpha^\prime} = \text{fixed},
\end{equation}
which turns the D3-brane metric into AdS$_5 \times S^5$ written as
\begin{equation}
    ds^2 = \alpha^\prime \bigg( \frac{u^2}{\sqrt{4 \pi g_s N}} \big( -dt^2 + d \vec{x}^2 \big) + \frac{\sqrt{4 \pi g_s N} }{u^2} du^2 + \sqrt{4 \pi g_s N} d \Omega_{S^5} \bigg) . 
\end{equation}
One might think that the overall factor of $\alpha^\prime$ renders the limit ill-defined. However, recall that the ten-dimensional gravitational constant is $\kappa \sim \alpha^{\prime2}$, so after plugging this metric into the supergravity action $ S \sim \frac{1}{\kappa^2} \int d^{10}x \sqrt{-G} \big( \mathcal{R} + \dots ) $ all factors of $\alpha^\prime$ cancel. The role of $\alpha^\prime$, that of providing a characteristic distance scale for the background, is now played by the radius of the five-sphere
\begin{equation}
    R^2 = \sqrt{4 \pi g_s N} = \sqrt{2 N} g_\text{YM} = \sqrt{2 \lambda},
    \label{AdSCFT parameters}
\end{equation}
which is also the ``AdS radius'', in the sense that this quantity also sets the curvature scale of the AdS part of the geometry. Supergravity is only an adequate approximation to string theory in the large distance limit, where the target space curvature is small. This means large radius $R^2$, which in turn means that the gauge theory's 't Hooft coupling is large! The same conclusion follows from the worldsheet. A commonly used coordinate system in AdS/CFT is obtained by setting $z =  \sqrt{4 \pi g_s N} / u =\sqrt{2 \lambda}/u $, resulting in
\begin{equation}
    ds^2 = \alpha^\prime \sqrt{2 \lambda} \bigg( \frac{ -dt^2 +d \vec{x}^2 + dz^2 }{z^2}  + d \Omega_{S^5} \bigg).
    \label{AdS with z}
\end{equation}
Using this expression for the spacetime metric on the worldsheet action 
\begin{equation}
    -\frac{1}{4 \pi \alpha^\prime} \int d^2 \sigma ( G_{\mu \nu} \partial_a X^\mu \partial^a X^\nu + \dots ),
\end{equation}
one sees that $\alpha^\prime$ cancels out, and the parameter that effectively takes its place is $1/\sqrt{2 \lambda}$. The sigma-model perturbative expansion, usually called the $\alpha^\prime$-expansion, is transmuted into an expansion in powers of $\lambda^{-1/2}$. This is a strong coupling expansion from the point of view of the gauge theory.

Consider the case where the string side of the correspondence can the treated classically. This means one may ignore string loops, so $g_s \to 0$. From \eqref{AdSCFT parameters} we see that in order to keep the 't Hooft coupling $\lambda$ finite, it is necessary to have $N \to \infty$. We thus find a concrete example of 't Hooft's idea: the large $N$ limit of a gauge theory, in this case $\mathcal{N}=4$ super Yang-Mills, is indeed given by a free closed string theory. For any fixed value of $\lambda$, string loop corrections organize themselves into powers of
\begin{equation}
    g_s = \frac{\lambda}{2 \pi N},
\end{equation}
implementing finite $N$ corrections into the gauge theory side.

It must be stressed that the derivation given here, which closely follows the original one in Maldacena's paper \cite{Maldacena:1997re}, is not a rigorous proof. One of its shortcomings is that the treatment of the string theory side is entirely perturbative, in both string and worldsheet couplings. The most conservative version of the conjecture is that it would only hold in the limits where the string side is best understood: $g_s = 0$ and large $R^2 = \sqrt{4 \pi g_s N}$, which means large $\lambda$ and $N \to \infty$ for the gauge theory parameters. The statement would then be that the strong coupling limit of the $\mathcal{N}=4$ theory at large $N$ is given by classical type IIB supergravity in AdS$_5 \times S^5$. A stronger version would be that the correspondence holds at any finite AdS radius and 't Hooft coupling, but only for $N \to \infty$. This would mean that the large $N$ limit of the $\mathcal{N}=4$ theory at any coupling is given by a theory of classical type IIB superstrings in AdS$_5 \times S^5$. The strongest version is that the correspondence holds for all $\lambda$ and $N$, and therefore relates the gauge theory with the full quantum type IIB superstring in an AdS$_5 \times S^5$ target space. This is what is believed to be the case by most, because since the original proposal many quantities have been found where both finite $\alpha^\prime \sim (2 \lambda)^{-1/2}$ and finite $g_s \sim 1/N$ corrections could be computed, and they have always agreed on both the AdS and CFT sides. For examples see \cite{Berenstein:2002jq,Beisert:2010jr,Drukker:2000rr}.

If the two sides of the duality describe the same physical system, a symmetry transformation of one side must correspond to a symmetry transformation of the other. A basic requirement for the correspondence to hold is therefore that the global symmetries of both sides match.\footnote{There is no point in trying to relate the gauge symmetries, since physical states are invariant under them, by definition. They are redundancies of the language in terms of which the theory in each side is formulated, and there is no reason why they should be related in any way.} Starting with the gauge theory, the most general infinitesimal conformal transformation of Minkowski space is a diffeomorphism $x^\mu \to x^\mu + \xi^\mu$ with
\begin{equation}
    \xi^\mu = a^\mu + \omega^\mu{}_\nu x^\nu + \lambda x^\mu -2 (b \cdot x)x^\mu + x^2 b^\mu,
\end{equation}
where $a^\mu$ and $\omega_{\mu \nu} = - \omega_{\nu \mu}$ lead to a Poincaré transformation and $\lambda$ and $b^\mu$ correspond to rescalings and special conformal transformations, respectively, the latter being the combination of an inversion $x^\mu \to x^\mu /x^2$, a translation by $b^\mu$, and another inversion. The generators of these transformations are
\begin{align}
    P_\mu &= \partial_\mu, && L_{\mu \nu} = x_\mu \partial_\nu -x_\nu \partial_\mu \notag\\[5pt]
    D &= x \cdot \partial, && K_\mu = - 2 x_\mu x \cdot \partial + x^2 \partial_\mu,
\end{align}
and their finite forms are
\begin{align}
    \text{Poincaré:} \hspace{0.5cm} x^\mu &\to \omega^\mu{}_\nu x^\nu + a^\mu , \hspace{0.5cm} \notag\\[5pt] 
    \text{rescalings:} \hspace{0.5cm} x^\mu &\to \lambda x^\mu \notag\\[5pt]
    \text{special conformal:} \hspace{0.5cm} x^\mu &\to \frac{x^\mu + x^2b^\mu}{1 + 2 x \cdot b + b^2x^2}.
    \label{conformal transformations}
\end{align}
The generators form the conformal algebra
\begin{align}
    [D,P_\mu] &= - P_\mu \notag\\[5pt]
    [D,K_\mu] &= K_\mu \notag\\[5pt]
    [P_\mu , K_\mu] &= -2 \eta_{\mu \nu} D + 2L_{\mu \nu} \notag\\[5pt]
    [L_{\mu \nu} , P_\rho] &= - \eta_{\mu \rho } P_\nu + \eta_{\nu \rho } P_\mu \notag\\[5pt]
    [L_{\mu \nu} , K_\rho] &= - \eta_{\mu \rho } K_\nu + \eta_{\nu \rho } K_\mu \notag\\[5pt]
    [L_{\mu \nu} , L_{\rho \sigma}] &= - \eta_{\mu \rho} L_{\nu \sigma} + \eta_{\mu \sigma} L_{\nu \rho} - \eta_{\nu \sigma} L_{\mu \rho} + \eta_{\nu \rho} L_{\mu \sigma},
\end{align}
with all others vanishing. If one defines $L_{\mu 4} = \frac{1}{2}(P_\mu - K_\mu)$, $L_{\mu 5} = - \frac{1}{2}(P_\mu + K_\mu)$ and $L_{45} = -D $, this algebra can be reexpressed as 
\begin{equation}
    [L_{A B} , L_{C D}] = - \eta_{A C} L_{B D} + \eta_{A D} L_{B C} - \eta_{B D} L_{A C} + \eta_{B C} L_{A D},
\end{equation}
where $\eta_{AB} = \text{diag}(-1,1,1,1,1,-1)$. This makes it evident that the conformal algebra of four-dimensional Minkowski space is $so(4,2)$, the Lie algebra of the group $SO(4,2)$ that preserves a six-dimensional ``Minkowski'' metric with two timelike directions. 

On the gravitational side, the global symmetries should correspond to isometries of the metric. The isometry group can be found easily if one uses the fact that AdS$_5$ can be defined as the hypersurface satisfying the constraint
\begin{equation}
    \eta_{AB} Y^A Y^B = -R^2
    \label{AdS embedding}
\end{equation}
inside of a six-dimensional ambient space with metric $ds^2 = \eta_{AB} dY^A dY^B$, with $\eta_{AB}$ defined as before.\footnote{This representation actually leads to a version of AdS which possesses a periodic time coordinate. The symmetry algebra is local and therefore indifferent to this, but one may recover the causal AdS space obtained from the D-brane limit by going to the covering space, which has the interpretation of ``unwinding'' the circular timelike direction into a line.

Note also the similarity with the $n$-sphere of radius $R$, which can be defined as the set of points satisfying $\delta_{ij} x^i j^j = R^2$ inside of a flat $n+1$-dimensional space of Euclidan signature. } One may verify that the AdS$_5$ metric with coordinates $(t,x^i,z)$ is obtained, up to a rescaling $d^2s \to \alpha^\prime ds^2$, from the six-dimensional metric via the parametrization
\begin{align}
    Y^\mu = \frac{ R x^\mu}{z}, && Y^4 = \frac{x^2 + z^2 - R^2}{2z}, && Y^5 = \frac{x^2 + z^2 + R^2}{2z}, 
\end{align}
which automatically satisfies \eqref{AdS embedding}.

Any isometry of the six-dimensional space that preserves \eqref{AdS embedding} corresponds to an isometry of AdS$_5$. From this it is clear that the symmetry group of AdS$_5$ is precisely $SO(4,2)$.\footnote{A similar matching is found between the algebra of fermionic (super)symmetries of each side.} The finite form of these isometries is almost identical to the action of the conformal transformations on four-dimensional Minkowsk space \cite{Blumenhagen:2013fgp}
\begin{align}
    \text{Poincaré:} \hspace{0.5cm} x^\mu &\to \omega^\mu{}_\nu x^\nu + a^\mu , \hspace{0.5cm} z \to z \notag\\[5pt] 
    \text{rescalings:} \hspace{0.5cm} x^\mu &\to \lambda x^\mu , \hspace{0.5cm} z \to \lambda z \notag\\[5pt]
    \text{special conformal:} \hspace{0.5cm} x^\mu &\to \frac{x^\mu + (x^2 + z^2)b^\mu}{1 + 2 x \cdot b + b^2(x^2 + z^2)}, \hspace{0.5cm} z \to \frac{z}{1 + 2 x \cdot b + b^2(x^2 + z^2)},
    \label{AdS symmetries}
\end{align}
the only difference being the dependence on $z$, which only occurs in the special conformal transformations. This matching of symmetries allows us to start identifying quantities on each side of the duality. The Poincaré subgroup of the AdS isometries acts on the four $x^\mu$ coordinates, so clearly these correspond to the four spacetime dimensions over which the CFT is defined. The isometries of the $S^5$ part of the geometry form the group $SO(6)$, which is also the symmetry group that rotates the adjoint scalars $\Phi^I(x)$ into each other on the CFT side. One should not, however, attempt to directly identify these scalars with directions on $S^5$, because the $\Phi^I$ are gauge-dependent objects with no invariant meaning. Only the trace of composite operators built from them can be gauge-invariant, and therefore be mapped to some quantity on the gravity side. Still, the fact that they are the only variables in the gauge theory that transform under $SO(6)$ means that they do capture the rotations of the $S^5$, in the sense that if such a rotation is performed on the gravitational side, the correspondent transformation on the CFT side acts on the scalars.\footnote{The more precise statement is that rotating the $S^5$ is dual to the R-symmetry transformations of the gauge theory, which rotate the supercharges into one another. These transformations act on the scalars.}

This leaves only the radial $u \sim 1/z$ coordinate of the AdS space without an obvious interpretation in terms of the gauge theory. There actually is, however, a parameter of the CFT that we have not yet discussed: the energy scale at which it is renormalized. Specifying this scale is, at least in the Wilsonian approach to field theory, fundamental for defining the theory, and this is precisely what corresponds to $u$. An important sign of this comes from the symmetry algebra \eqref{AdS symmetries}: $u$ is a Lorentz scalar which scales inversely as the Poincaré coordinates,
\begin{equation}
    x^\mu \to \lambda x^\mu , \hspace{0.5cm} u \to \frac{u}{\lambda}.
\end{equation}
Indeed, the scaling transformations that are part of the conformal group act in quantum field theories as a simultaneous scaling of the coordinates and of the energy cutoff, in precisely the way shown above. We therefore identify the $u$ coordinate of AdS with the CFT's energy scale. The gauge theory in the UV is then related to the gravitational theory at large radius, a phenomenon known as UV/IR duality, or scale/radius duality. An excitation that is pulled radially deeper into AdS by the attractive potential caused by the curvature corresponds to a localized excitation of the CFT spreading out over time \cite{Hubeny:2014bla}. 

In AdS/CFT one often hears the statement that the CFT ``lives on the boundary'' of the AdS space, meaning at $z \sim 1/u =0$. For $z=0$ the metric diverges, so this is clearly not a boundary in the usual sense. Statements about the boundary of AdS are better understood as referring to the large radius asymptotic region. The UV/IR duality provides an interpretation of this. Even if we forget for a moment that the $\mathcal{N}=4$ theory is exactly conformal, we should still expect the duality, when examined at large AdS radius, to result on the gauge side in a theory with conformal symmetry: the UV fixed point of the renormalization group flow. In this sense a CFT is always expect at $z=0$. This is further supported by the fact that the AdS isometries \eqref{AdS symmetries} take $z=0$ to itself, while acting on the remaining coordinates exactly like the finite conformal transformations of Minkowski space shown in \eqref{conformal transformations}. This suggests that the $z \to 0$ ``boundary'' is the natural region of AdS where one should think of the UV CFT as being defined. 

\begin{figure}[t]
\begin{center}
\includegraphics[width= 0.75\textwidth]{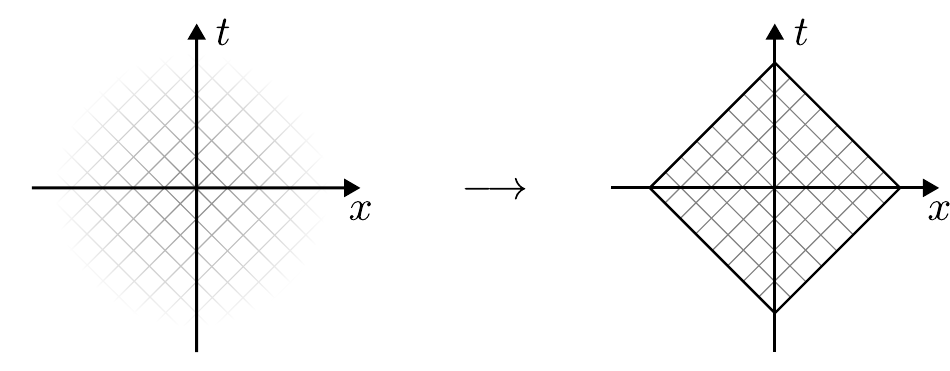}
\caption[Conformal compactification of two-dimensional Minkowski space]{Conformal compactification of two-dimensional Minkowski space. Lightlike rays travels at a 45$^\circ$ angle, all the way to infinity in the noncompact diagram on the left. Conformal compactification adds a boundary of ``points at infinity'', one for each lightlike ray, without changing the structure of the lightcones.}
\label{conformal compactification}
\end{center}
\end{figure}

What is missing is a proof that the $z \to 0$ asymptotic region is indeed four-dimensional Minkowski space, where the gauge theory lives. For this we review an elegant argument by Witten \cite{33f2a5a08f4d436ab19efaba3eb6871a}. We once again make use of the auxiliary six-dimensional space with two timelike directions, this time written in terms of coordinates $U = Y^4 + Y^5 $ and $V=-Y^4 + Y^5$, so that the metric becomes $ds^2 = \eta_{\mu \nu}Y^\mu Y^\nu - dU dV $. Consider in this space the quadric that satisfies\footnote{A quadric is a hypersurface defined by a quadratic constraint on the coordinates of a higher-dimensional space.}
\begin{equation}
    \eta_{\mu \nu} Y^\mu Y^\nu = UV,
\end{equation}
subject to an overall scaling identification
\begin{equation}
    Y^\mu \sim s Y^\mu, \hspace{0.3cm} U \sim s U, \hspace{0.3cm} V \sim s V,
\end{equation}
for any $s >0$. For generic $V \neq 0$, one may rescale all coordinates by $1/V$, effectively setting $V = 1$, after which the metric becomes that of four-dimensional Minkowski space $\mathbb{R}^{1,3}$. It follows that Minkowski space can be identified with the portion of this surface with $V \neq 0$. The quadric differs from $\mathbb{R}^{1,3}$ by also containing the $V=0$ points, for which we have $\eta_{\mu \nu} Y^\mu Y^\nu =0$ plus the scaling equivalence. These points are therefore in one-to-one correspondence with null-vectors in Minkowski space, and are interpreted as ``points at infinity''. To see why, let us obtain them in a slightly different way. Once again rescale all coordinates by $1/V$, which puts us in $\mathbb{R}^{1,3}$ and turns the quadric equation into $\eta_{\mu \nu} \frac{Y^\mu}{V} \frac{Y^\nu}{V} = \frac{U}{V}$. From this the $V=0$ points can be obtained by simply taking the $V \to 0$ limit, which is equivalent to sending all coordinates to infinity. Doing this in regular Minkowski space would just correspond to moving indefinitely along null trajectories, but on the space defined by the quadric one instead ends up reaching the $V=0$ points. One could picture the addition of them to the noncompact $\mathbb{R}^{1,3}$ as ``capping off'' each lightlike ray with a single point, as in figure \ref{conformal compactification}. The set of such points forms a sort of boundary at infinity, much like the one we are trying to define in AdS. This construction is called a conformal compactification, and the ``boundary at infinity'' is referred to as a conformal boundary.\footnote{The conformal group of Minkowski space must actually be taken to act on its conformal compactification, because a special conformal transformation maps the point $x^\mu = -b^\mu / b^2$ to infinity.

The idea of conformal compactification was introduced by Penrose in \cite{Penrose:1964ge}. See also Chapter 9 of \cite{Penrose_Rindler_1986}. The main point is that given a noncompact spacetime $\mathcal{M}$ with metric $ds^2$, one may define a different spacetime $\widetilde{\mathcal{M}}$ with a boundary, which differs from the first one by a Weyl transformation, meaning it has a metric $d\tilde{s}^2 = \Omega^2 ds^2$. For a suitably chosen $\Omega$, the asymptotic structure of $\mathcal{M}$ is encoded on the boundary of $\widetilde{\mathcal{M}}$, which is referred to as the conformal boundary of $\mathcal{M}$. Conformally-invariant objects take the same form on the asymptotic region of $\mathcal{M}$ and on the boundary of $\widetilde{\mathcal{M}}$, which is why this is useful for AdS/CFT. For AdS$_5$ in $(t,\vec{x},z)$ coordinates the asymptotic region is $z \to 0$. Choosing for the conformal compactification the Weyl factor $\Omega^2 = z^2$, we get rid of the divergence at small $z$, and obtain the metric at the conformal boundary by setting $z =0$. The result is four-dimensional Minkowski space: $d\tilde{s}^2 = -dt^2 + d \vec{x}^2$.}

The embedding of AdS$_5$ into the six-dimensional space \eqref{AdS embedding} becomes
\begin{equation}
    \eta_{\mu \nu} Y^\mu Y^\nu - UV = -R^2
    \label{AdS embedding 2}
\end{equation}
in terms of the new coordinates, with no scaling identification. We wish to conformally compactify this space, adding to it a boundary of points at infinity, to see if this boundary coincides with Minkowski space. This is done simply by sending $Y^\mu$, $U$ and $V$ to infinity while preserving \eqref{AdS embedding 2}. Take for instance $Y^\mu = \lambda \tilde{Y}^\mu$, $U = \lambda \tilde{U}$, $V = \lambda \tilde{V}$ and let $\lambda \to \infty$. This turns \eqref{AdS embedding 2} into $\eta_{\mu \nu} \tilde{Y}^\mu \tilde{Y}^\nu = \tilde{U}\tilde{V}$. Instead of $Y^\mu = \lambda \tilde{Y}^\mu$ we could have chosen $Y^\mu = s
\lambda \tilde{Y}^\mu$ for any positive $s$, as this parameter would disappear in the limit. Therefore, the asymptotic structure of AdS$_5$ is that of solutions to $\eta_{\mu \nu} \tilde{Y}^\mu \tilde{Y}^\nu = \tilde{U}\tilde{V}$ up to a rescaling of the coordinates by some $s >0$, which is precisely the representation of Minkowski space just discussed. By adding these points at infinity to AdS, we obtain its conformal compactification, whose boundary (in the strict sense) is precisely the conformal compactification of $\mathbb{R}^{1,3}$.

Another argument for identifying the CFT with the boundary is as follows. For the duality to hold, any observable computed on one side must be accessible in the other one, at least in principle. General gauge theory local correlators of the form $\braket{\Phi^I (x) \Phi^J(x^\prime) \dots }$ are an example of such an observable, as long as one takes a trace over the gauge group indices in order to form a gauge-invariant expression. The same cannot be said of the analogous objects on the gravitational side, such as an $n$-graviton correlation function, since in a gravitational theory spacetime diffeomorphisms are gauge symmetries. Any object that depends explicitly on bulk spacetime points is therefore gauge-dependent. Recall, however, that only the diffeomorphisms that go to zero at infinity, or at a boundary if one is present, actually count as gauge transformations. The ones that do not, the so-called large coordinate transformations, are actually physical symmetries. In the case of AdS these are precisely the isometries discussed before. It follows that the only physical local observables in a quantum theory of gravity are those defined at such boundaries, where the gauge transformations do not anymore act on the coordinates that remain. This is expected to hold in any background. Having AdS for the bulk spacetime produces a particularly interesting realization of this idea because the boundary of AdS is Minkoski space, the most familiar and phenomenologically useful manifold one can define a field theory in. Many other spacetimes, such as Minkowski space itself, have far more exotic boundaries.

Having spent some time investigating how the two sides of the correspondence are related to each other, we can now state it a more concrete form. Recall that the string coupling $g_s$ is given by the value of the dilaton at infinity, which in AdS means at the boundary:
\begin{equation}
    \lim_{u \to \infty} \Phi(x,u) = g_s.
\end{equation}
Since the Yang-Mills coupling is related to the string coupling as $2 \pi g_s = g^2_\text{YM}$, a change in the value of $g^2_\text{YM}$ is implemented by changing the asymptotic behavior of the dilaton. If the dilaton approaches at large $u$ not a constant value, but some function $\varphi(x)$ of the four Minkowski directions,
\begin{equation}
    \lim_{u \to \infty} \Phi(x,u)=   \varphi(x),
\end{equation}
this should then correspond to a position-dependent change of the Yang-Mills coupling, which has the interpretation of coupling the $\text{Tr} F^2$ operator to a current $J(x) =  1/(g^2_\text{YM}(x)) =  1/(2 \pi \varphi(x))$ :
\begin{equation}
     - \frac{1}{4 g_\text{YM}^2} \int d^4x \, \text{Tr} F^{\mu \nu} F_{\mu \nu} \to - \frac{1}{4} \int d^4x \frac{1}{g^2_\text{YM}(x)} \text{Tr} F^{\mu \nu} F_{\mu \nu} \equiv - \frac{1}{4} \int d^4x \, J(x) \text{Tr} F^{\mu \nu} F_{\mu \nu}.
\end{equation}
This illustrates the general idea: each bulk field, when taken to the boundary, acts as a source for an operator of the CFT. The mapping of each excitation on the string side to the CFT operator sourced by it is called the holographic dictionary. We have just found one entry of it: the bulk dilaton is dual to $\text{Tr} F^2$. Another notable example is the bulk metric, which is dual to the CFT's energy-momentum tensor. The precise statement of the AdS/CFT correspondence is that there is a one-to-one relation between bulk fields $\Phi_i$ and CFT operators $O_i(x)$, and that the partition function of the gravitational side with particular boundary conditions $\Phi_i(u \to \infty) \equiv \phi_i$ is equal to the CFT's partition function $Z_{\text{CFT}}[J_i]$ with sources $J_i=\phi_i$:
\begin{equation}
    Z_\text{string}[\phi_i] =\int_{\Phi_i|_{u \to \infty} = \phi_i} \mathcal{D} \Phi_i e^{-S_\text{string}(\Phi_i)} = Z_{\text{CFT}}[\phi_i].
\end{equation}
The string path integral is a formal object meant to represent the fact that the sources of the CFT enter the gravitational side as boundary conditions. In the supergravity approximation (large $\lambda$) it would be replaced by the supergravity path integral, and if one also goes to the classical limit (large $N$), we have simply
\begin{equation}
    e^{-S_\text{SUGRA}(\Phi_i)}\bigg|_{\Phi_i|_{u \to \infty} = \phi_i} = Z_\text{CFT}[\phi_i],
\end{equation}
where on the left-hand side is the on-shell supergravity action evaluated at the classical solution whose boundary behavior gives the sources of the CFT. Differentiating both sides of this expression with respect to the sources provides a way to compute large $N$ gauge theory correlators at strong coupling. 

After this first version of the correspondence was discovered, many variations of it were proposed, and the study of gauge/gravity dualities and their phenomenological applicability became a field in itself. Real life QCD has always been a major goal for applications. This is not a straightforward generalization, as the $\mathcal{N}=4$ theory differs from regular Yang-Mills in a number of ways. Many phenomenological models in what is called AdS/QCD involve breaking some of the symmetries of both sides of the duality in different ways, in order to make the gauge theory side closer to non-supersymmetric Yang-Mills. One way to break conformal symmetry, for instance, is to add a small radius cutoff to the AdS geometry. By following this idea Polchinski and Strassler managed to compute scattering amplitudes of glueballs whose momentum-dependence is the same as what one expects from QCD \cite{Polchinski:2001tt}. 

One may also consider more intricate, symmetry breaking D-brane systems in string theory before taking the Maldacena limit. A notable example of this is the Sakai-Sugimoto model, which manages to introduce quarks, although in a probe (nondynamical) approximation \cite{sakai1,sakai2}. This is an example of a top-down model of AdS/QCD, where one obtains a holographic duality via a limit of a string theory state, just as in Maldacena's original derivation. An approach somewhat opposite to this also exists, called bottom-up, which consists of directly proposing ansätze for metrics and dilaton profiles which capture relevant features of QCD, regardless of there being a well known embedding of them into string theory, and using the holographic dictionary to extract information about the gauge theory. The metrics employed approach AdS at large radius, but deviate from it in the bulk in a way that depends on parameters that can then be fixed by comparison with QCD data, after which one has a predictive model. The Polchinski-Strassler model is an example of this, since it is a direct modification of the $\mathcal{N}=4$ duality that does not itself follow from any limit of string theory. Another example is called improved holographic QCD, which encodes the running of the QCD coupling into the dilaton profile \cite{Gursoy:2010fj}.\footnote{An introduction to all of the models mentioned, among others, can be found in \cite{Năstase_2015}} 

All of these models successfully capture relevant aspects of QCD or pure Yang-Mills, even if only qualitatively. Each one also has its shortcomings, which is to be expected. After all, no D-brane construction of exact QCD is known, so while all AdS/QCD models do approximate QCD to varying degrees, in all cases there are regions of parameter space where the approximation fails. Still, their partial success, together with 't Hooft's argument for the stringy nature of large $N$ gauge theories, points to the existence of a true gravitational dual of QCD, whose main features are captured by the many known phenomenological models.
% Alguma referência?

%\end{document}

\printbibliography

@article{GrossWilczek1973,
  author = {Gross, D. J. and Wilczek, F.},
  title = {Ultraviolet Behavior of Non-Abelian Gauge Theories},
  journal = {Phys. Rev. Lett.},
  volume = {30},
  year = {1973},
  pages = {1343},
  doi = {10.1103/PhysRevLett.30.1343}
}

@article{Politzer1973,
  author = {Politzer, H. D.},
  title = {Reliable Perturbative Results for Strong Interactions?},
  journal = {Phys. Rev. Lett.},
  volume = {30},
  year = {1973},
  pages = {1346},
  doi = {10.1103/PhysRevLett.30.1346}
}

@article{Horowitz:2006ct,
    author = "Horowitz, Gary T. and Polchinski, Joseph",
    title = "{Gauge/gravity duality}",
    eprint = "gr-qc/0602037",
    archivePrefix = "arXiv",
    pages = "169--186",
    month = "2",
    year = "2006"
}

@article{Maldacena:1997re,
    author = "Maldacena, Juan Martin",
    title = "{The Large $N$ limit of superconformal field theories and supergravity}",
    eprint = "hep-th/9711200",
    archivePrefix = "arXiv",
    reportNumber = "HUTP-97-A097, HUTP-98-A097",
    doi = "10.4310/ATMP.1998.v2.n2.a1",
    journal = "Adv. Theor. Math. Phys.",
    volume = "2",
    pages = "231--252",
    year = "1998"
}

@book{Năstase_2015, place={Cambridge}, title={Introduction to the AdS/CFT Correspondence}, publisher={Cambridge University Press}, author={Năstase, Hora\cb{t}iu}, year={2015}}

@article{Veneziano:1968yb,
    author = "Veneziano, G.",
    title = "{Construction of a crossing - symmetric, Regge behaved amplitude for linearly rising trajectories}",
    doi = "10.1007/BF02824451",
    journal = "Nuovo Cim. A",
    volume = "57",
    pages = "190--197",
    year = "1968"
}

@book{rickles2014brief,
  title={A Brief History of String Theory: From Dual Models to M-Theory},
  author={Rickles, D.},
  isbn={9783642451287},
  series={The Frontiers Collection},
  url={https://books.google.com.br/books?id=Ud64BAAAQBAJ},
  year={2014},
  publisher={Springer Berlin Heidelberg}
}

@inbook{doi:10.1142/9789812795823_0024,
author = {Yoichiro Nambu},
title = {Quark model and the factorization of the Veneziano amplitude},
booktitle = {Broken Symmetry},
chapter = {},
pages = {258-267},
doi = {10.1142/9789812795823_0024},
URL = {https://www.worldscientific.com/doi/abs/10.1142/9789812795823_0024},
eprint = {https://www.worldscientific.com/doi/pdf/10.1142/9789812795823_0024}
}

@article{PhysRevLett.23.545,
  title = {Harmonic-Oscillator Analogy for the Veneziano Model},
  author = {Susskind, Leonard},
  journal = {Phys. Rev. Lett.},
  volume = {23},
  issue = {10},
  pages = {545--547},
  numpages = {0},
  year = {1969},
  month = {9},
  publisher = {American Physical Society},
  doi = {10.1103/PhysRevLett.23.545},
  url = {https://link.aps.org/doi/10.1103/PhysRevLett.23.545}
}

@unpublished{Nielsenpreprint,
  author = {Nielsen, Holger},
  title  = {An almost physical interpretation of the integrand of the n-point Veneziano model},
  note   = {preprint at Niels Bohr Institute},
  month  = {},
  year   = {}
}

@article{Bali:2000gf,
    author = "Bali, Gunnar S.",
    title = "{QCD forces and heavy quark bound states}",
    eprint = "hep-ph/0001312",
    archivePrefix = "arXiv",
    reportNumber = "HUB-EP-99-67",
    doi = "10.1016/S0370-1573(00)00079-X",
    journal = "Phys. Rept.",
    volume = "343",
    pages = "1--136",
    year = "2001"
}

@book{Collins_1977, place={Cambridge}, series={Cambridge Monographs on Mathematical Physics}, title={An Introduction to Regge Theory and High Energy Physics}, publisher={Cambridge University Press}, author={Collins, P. D. B.}, year={1977}, collection={Cambridge Monographs on Mathematical Physics}}

@article{Baker:2018mhw,
    author = "Baker, M. and Cea, P. and Chelnokov, V. and Cosmai, L. and Cuteri, F. and Papa, A.",
    title = "{Isolating the confining color field in the SU(3) flux tube}",
    eprint = "1810.07133",
    archivePrefix = "arXiv",
    primaryClass = "hep-lat",
    doi = "10.1140/epjc/s10052-019-6978-y",
    journal = "Eur. Phys. J. C",
    volume = "79",
    number = "6",
    pages = "478",
    year = "2019"
}

@book{Greensite:2011zz,
    author = "Greensite, Jeff",
    title = "{An introduction to the confinement problem}",
    doi = "10.1007/978-3-642-14382-3",
    volume = "821",
    year = "2011"
}

@article{Scherk:1974ca,
    author = "Scherk, Joel and Schwarz, John H.",
    title = "{Dual Models for Nonhadrons}",
    reportNumber = "CALT-68-444",
    doi = "10.1016/0550-3213(74)90010-8",
    journal = "Nucl. Phys. B",
    volume = "81",
    pages = "118--144",
    year = "1974"
}

@misc{Tong_gauge,
  author        = {David Tong},
  title         = {Gauge Theory},
  url = {https://www.damtp.cam.ac.uk/user/tong/gaugetheory.html}
}

@misc{Tong_qft,
  author        = {David Tong},
  title         = {Quantum Field Theory},
  url = {https://www.damtp.cam.ac.uk/user/tong/qft.html}
}

@online{Guillelectures,
        title = {Matrix Models, Large N \& Strings (Physics Latam minicourse)},
        date = {2021},
        organization = {Youtube},
        author = {Guillermo Silva},
        url = {https://www.youtube.com/playlist?list=PLaFLp8EAyd7VS2b6r9UMxT0nUaRZTSdXx},
    }

@book{Coleman_1985, place={Cambridge}, title={Aspects of Symmetry: Selected Erice Lectures}, publisher={Cambridge University Press}, author={Coleman, Sidney}, year={1985}}

@inproceedings{Staessens:2010vi,
    author = "Staessens, Wieland and Vercnocke, Bert",
    title = "{Lectures on Scattering Amplitudes in String Theory}",
    booktitle = "{5th Modave Summer School in Mathematical Physics}",
    eprint = "1011.0456",
    archivePrefix = "arXiv",
    primaryClass = "hep-th",
    month = "11",
    year = "2010"
}

@article{effectivestrings,
title = "The effective theory of long strings",
author = "Ofer Aharony and Zohar Komargodski",
year = "2013",
doi = "10.1007/JHEP05(2013)118",
language = "English",
volume = "2013",
journal = "Journal of High Energy Physics",
issn = "1126-6708",
publisher = "Springer Verlag",
number = "5",
}

@book{Blumenhagen:2013fgp,
    author = {Blumenhagen, Ralph and L\"ust, Dieter and Theisen, Stefan},
    title = "{Basic concepts of string theory}",
    doi = "10.1007/978-3-642-29497-6",
    isbn = "978-3-642-29496-9",
    publisher = "Springer",
    address = "Heidelberg, Germany",
    series = "Theoretical and Mathematical Physics",
    year = "2013"
}

@book{Johnson_2002, place={Cambridge}, series={Cambridge Monographs on Mathematical Physics}, title={D-Branes}, publisher={Cambridge University Press}, author={Johnson, Clifford V.}, year={2002}, collection={Cambridge Monographs on Mathematical Physics}}

@article{York:1986lje,
    author = "York, James",
    title = "{Boundary terms in the action principles of general relativity}",
    doi = "10.1007/BF01889475",
    journal = "Found. Phys.",
    volume = "16",
    pages = "249--257",
    year = "1986"
}

@book{Peskin:1995ev,
    author = "Peskin, Michael E. and Schroeder, Daniel V.",
    title = "{An Introduction to quantum field theory}",
    doi = "10.1201/9780429503559",
    isbn = "978-0-201-50397-5, 978-0-429-50355-9, 978-0-429-49417-8",
    publisher = "Addison-Wesley",
    address = "Reading, USA",
    year = "1995"
}

@book{DiFrancesco:1997nk,
    author = "Di Francesco, P. and Mathieu, P. and Senechal, D.",
    title = "{Conformal Field Theory}",
    doi = "10.1007/978-1-4612-2256-9",
    isbn = "978-0-387-94785-3, 978-1-4612-7475-9",
    publisher = "Springer-Verlag",
    address = "New York",
    series = "Graduate Texts in Contemporary Physics",
    year = "1997"
}

@article{Strassler_1992,
   title={Field theory without Feynman diagrams: One-loop effective actions},
   volume={385},
   ISSN={0550-3213},
   url={http://dx.doi.org/10.1016/0550-3213(92)90098-V},
   DOI={10.1016/0550-3213(92)90098-v},
   number={1–2},
   journal={Nuclear Physics B},
   publisher={Elsevier BV},
   author={Strassler, Matthew J.},
   year={1992},
   month=oct, pages={145–184} }

@book{Polchinskivol1:1998rq,
    author = "Polchinski, J.",
    title = "{String theory. Vol. 1: An introduction to the bosonic string}",
    doi = "10.1017/CBO9780511816079",
    isbn = "978-0-511-25227-3, 978-0-521-67227-6, 978-0-521-63303-1",
    publisher = "Cambridge University Press",
    series = "Cambridge Monographs on Mathematical Physics",
    month = "12",
    year = "2007"
}

@misc{TimoWeigandnotes,
  author        = {Timo Weigand},
  title         = {Introduction to String Theory},
  url = {https://www.thphys.uni-heidelberg.de/courses/weigand/Strings11-12.pdf}
}

@misc{Tong_string,
  author        = {David Tong},
  title         = {String Theory},
  url = {https://www.damtp.cam.ac.uk/user/tong/string.html}
}

@book{Schwartz_2013, place={Cambridge}, title={Quantum Field Theory and the Standard Model}, publisher={Cambridge University Press}, author={Schwartz, Matthew D.}, year={2013}}

@article{PhysRev.135.B1049,
  title = {Photons and Gravitons in $S$-Matrix Theory: Derivation of Charge Conservation and Equality of Gravitational and Inertial Mass},
  author = {Weinberg, Steven},
  journal = {Phys. Rev.},
  volume = {135},
  issue = {4B},
  pages = {B1049--B1056},
  numpages = {0},
  year = {1964},
  month = {Aug},
  publisher = {American Physical Society},
  doi = {10.1103/PhysRev.135.B1049},
  url = {https://link.aps.org/doi/10.1103/PhysRev.135.B1049}
}

@article{PhysRev.140.B516,
  title = {Infrared Photons and Gravitons},
  author = {Weinberg, Steven},
  journal = {Phys. Rev.},
  volume = {140},
  issue = {2B},
  pages = {B516--B524},
  numpages = {0},
  year = {1965},
  month = {Oct},
  publisher = {American Physical Society},
  doi = {10.1103/PhysRev.140.B516},
  url = {https://link.aps.org/doi/10.1103/PhysRev.140.B516}
}

@article{Dai:1989ua,
    author = "Dai, Jin and Leigh, R. G. and Polchinski, Joseph",
    title = "{New Connections Between String Theories}",
    reportNumber = "UTTG-12-89",
    doi = "10.1142/S0217732389002331",
    journal = "Mod. Phys. Lett. A",
    volume = "4",
    pages = "2073--2083",
    year = "1989"
}

@article{Paton:1969je,
    author = "Paton, Jack E. and Chan, Hong-Mo",
    title = "{Generalized veneziano model with isospin}",
    doi = "10.1016/0550-3213(69)90038-8",
    journal = "Nucl. Phys. B",
    volume = "10",
    pages = "516--520",
    year = "1969"
}

@article{Kaluza:1921tu,
    author = "Kaluza, Th.",
    title = {{Zum Unit\"atsproblem der Physik}},
    eprint = "1803.08616",
    archivePrefix = "arXiv",
    primaryClass = "physics.hist-ph",
    reportNumber = "HUPD-8401",
    doi = "10.1142/S0218271818700017",
    journal = "Sitzungsber. Preuss. Akad. Wiss. Berlin (Math. Phys. )",
    volume = "1921",
    pages = "966--972",
    year = "1921"
}

@article{Klein:1926tv,
    author = "Klein, Oskar",
    editor = "Taylor, J. C.",
    title = "{Quantum Theory and Five-Dimensional Theory of Relativity. (In German and English)}",
    doi = "10.1007/BF01397481",
    journal = "Z. Phys.",
    volume = "37",
    pages = "895--906",
    year = "1926"
}

@book{Appelquist:1987nr,
    editor = "Appelquist, T. and Chodos, A. and Freund, P. G. O.",
    title = "{Modern Kaluza-Klein Theories}",
    year = "1987"
}

@article{Font:1990gx,
    author = "Font, A. and Ibanez, Luis E. and Lust, D. and Quevedo, F.",
    title = "{Strong - weak coupling duality and nonperturbative effects in string theory}",
    reportNumber = "CERN-TH-5790-90",
    doi = "10.1016/0370-2693(90)90523-9",
    journal = "Phys. Lett. B",
    volume = "249",
    pages = "35--43",
    year = "1990"
}

@article{Adams_2001,
   title={Don’t panic! closed string tachyons in ALE spacetimes},
   volume={2001},
   ISSN={1029-8479},
   url={http://dx.doi.org/10.1088/1126-6708/2001/10/029},
   DOI={10.1088/1126-6708/2001/10/029},
   number={10},
   journal={Journal of High Energy Physics},
   publisher={Springer Science and Business Media LLC},
   author={Adams, Allan and Polchinski, Joseph and Silverstein, Eva},
   year={2001},
   month=oct, pages={029–029} }

@book{GreenSchwarzWitten_vol1, place={Cambridge}, series={Cambridge Monographs on Mathematical Physics}, title={Superstring Theory: 25th Anniversary Edition}, publisher={Cambridge University Press}, author={Green, Michael B. and Schwarz, John H. and Witten, Edward}, year={2012}, collection={Cambridge Monographs on Mathematical Physics}}

@article{CALLAN1985593,
title = {Strings in background fields},
journal = {Nuclear Physics B},
volume = {262},
number = {4},
pages = {593-609},
year = {1985},
issn = {0550-3213},
doi = {https://doi.org/10.1016/0550-3213(85)90506-1},
url = {https://www.sciencedirect.com/science/article/pii/0550321385905061},
author = {C.G. Callan and D. Friedan and E.J. Martinec and M.J. Perry},
}

@article{FRIEDAN1985318,
title = {Nonlinear models in 2 + $\epsilon$ dimensions},
journal = {Annals of Physics},
volume = {163},
number = {2},
pages = {318-419},
year = {1985},
issn = {0003-4916},
doi = {https://doi.org/10.1016/0003-4916(85)90384-7},
url = {https://www.sciencedirect.com/science/article/pii/0003491685903847},
author = {Daniel Harry Friedan},
}

@inproceedings{Donoghue:1995cz,
    author = "Donoghue, John F.",
    title = "{Introduction to the effective field theory description of gravity}",
    booktitle = "{Advanced School on Effective Theories}",
    eprint = "gr-qc/9512024",
    archivePrefix = "arXiv",
    reportNumber = "UMHEP-424",
    month = "6",
    year = "1995"
}

@inproceedings{Callan:1989nz,
    author = "Callan, Jr., Curtis G. and Thorlacius, Larus",
    title = "{Sigma Models and String Theory}",
    booktitle = "{Theoretical Advanced Study Institute in Elementary Particle Physics: Particles, Strings and Supernovae (TASI 88)}",
    reportNumber = "Print-89-0232 (PRINCETON)",
    month = "3",
    year = "1989"
}

@article{Abouelsaood:1986gd,
    author = "Abouelsaood, Ahmed and Callan, Jr., Curtis G. and Nappi, C. R. and Yost, S. A.",
    title = "{Open strings in background gauge fields}",
    reportNumber = "Print-86-1189 (PRINCETON)",
    doi = "10.1016/0550-3213(87)90164-7",
    journal = "Nucl. Phys. B",
    volume = "280",
    pages = "599--624",
    year = "1987"
}

@article{Born:1934gh,
    author = "Born, M. and Infeld, L.",
    title = "{Foundations of the new field theory}",
    doi = "10.1098/rspa.1934.0059",
    journal = "Proc. Roy. Soc. Lond. A",
    volume = "144",
    number = "852",
    pages = "425--451",
    year = "1934"
}

@book{Abadir_Magnus_2005, place={Cambridge}, series={Econometric Exercises}, title={Matrix Algebra}, publisher={Cambridge University Press}, author={Abadir, Karim M. and Magnus, Jan R.}, year={2005}, collection={Econometric Exercises}}

@article{Dirac:1962iy,
    author = "Dirac, Paul A. M.",
    title = "{An Extensible model of the electron}",
    doi = "10.1098/rspa.1962.0124",
    journal = "Proc. Roy. Soc. Lond. A",
    volume = "268",
    pages = "57--67",
    year = "1962"
}

@misc{Liu2014StringTheory,
  author = {Liu, Hong},
  title = {8.821 {String Theory} and {Holographic Duality}},
  year = {2014},
  howpublished = {Massachusetts Institute of Technology: MIT OpenCourseWare},
  url = {https://opencw.aprende.org/courses/physics/8-821-string-theory-and-holographic-duality-fall-2014/},
  note = {Fall 2014. License: Creative Commons BY-NC-SA},
}

@article{Tseytlin:1997csa,
    author = "Tseytlin, Arkady A.",
    title = "{On nonAbelian generalization of Born-Infeld action in string theory}",
    eprint = "hep-th/9701125",
    archivePrefix = "arXiv",
    reportNumber = "CERN-TH-97-009, CERN-TH-97-09, CERN-TH-97-9, IMPERIAL-TP-96-97-17",
    doi = "10.1016/S0550-3213(97)00354-4",
    journal = "Nucl. Phys. B",
    volume = "501",
    pages = "41--52",
    year = "1997"
}

@article{Dorn:1996an,
    author = "Dorn, Harald and Otto, H. J.",
    title = "{On T duality for open strings in general Abelian and nonAbelian gauge field backgrounds}",
    eprint = "hep-th/9603186",
    archivePrefix = "arXiv",
    reportNumber = "HU-BERLIN-EP-96-8",
    doi = "10.1016/0370-2693(96)00544-8",
    journal = "Phys. Lett. B",
    volume = "381",
    pages = "81--88",
    year = "1996"
}

@article{Dorn:1996xk,
    author = "Dorn, Harald",
    title = "{NonAbelian gauge field dynamics on matrix D-branes}",
    eprint = "hep-th/9612120",
    archivePrefix = "arXiv",
    reportNumber = "HU-BERLIN-EP-96-62",
    doi = "10.1016/S0550-3213(97)00171-5",
    journal = "Nucl. Phys. B",
    volume = "494",
    pages = "105--118",
    year = "1997"
}

@article{CREMMER1972222,
title = {Factorization of the pomeron sector and currents in the dual resonance model},
journal = {Nuclear Physics B},
volume = {50},
number = {1},
pages = {222-252},
year = {1972},
issn = {0550-3213},
doi = {https://doi.org/10.1016/S0550-3213(72)80016-6},
url = {https://www.sciencedirect.com/science/article/pii/S0550321372800166},
author = {E. Cremmer and J. Scherk},
}

@article{Leigh:1989jq,
    author = "Leigh, R. G.",
    title = "{Dirac-Born-Infeld Action from Dirichlet Sigma Model}",
    reportNumber = "UTTG-31-89",
    doi = "10.1142/S0217732389003099",
    journal = "Mod. Phys. Lett. A",
    volume = "4",
    pages = "2767",
    year = "1989"
}

@article{Myers:1999ps,
    author = "Myers, Robert C.",
    title = "{Dielectric branes}",
    eprint = "hep-th/9910053",
    archivePrefix = "arXiv",
    reportNumber = "MCGILL-99-27, NSF-ITP-99-113",
    doi = "10.1088/1126-6708/1999/12/022",
    journal = "JHEP",
    volume = "12",
    pages = "022",
    year = "1999"
}

@book{Polyakov:1987hqn,
    author = "Polyakov, A. M.",
    title = "{Gauge Fields and Strings}",
    doi = "10.1201/9780203755082",
    isbn = "978-1-351-44609-9, 978-3-7186-0393-0, 978-0-203-75508-2",
    publisher = "Taylor \& Francis",
    address = "London",
    year = "1987"
}

@misc{ReidEdwardsnotes,
  author        = {Ronald A. Reid-Edwards},
  title         = {String Theory},
  url = {https://www.damtp.cam.ac.uk/user/rar31/LectureNotes.pdf}
}

@book{Altland_Simons_2010, place={Cambridge}, edition={2}, title={Condensed Matter Field Theory}, publisher={Cambridge University Press}, author={Altland, Alexander and Simons, Ben D.}, year={2010}}

@article{Witten:2013pra,
    author = "Witten, Edward",
    title = "{The Feynman $i \epsilon$ in String Theory}",
    eprint = "1307.5124",
    archivePrefix = "arXiv",
    primaryClass = "hep-th",
    doi = "10.1007/JHEP04(2015)055",
    journal = "JHEP",
    volume = "04",
    pages = "055",
    year = "2015"
}

@article{Donoghue:2017pgk,
    author = "Donoghue, John F. and Ivanov, Mikhail M. and Shkerin, Andrey",
    title = "{EPFL Lectures on General Relativity as a Quantum Field Theory}",
    eprint = "1702.00319",
    archivePrefix = "arXiv",
    primaryClass = "hep-th",
    reportNumber = "INR-TH-2017-001",
    month = "2",
    year = "2017"
}

@article{Siegel_1954, title={A simple proof of $\eta(-1/\tau)= \eta(\tau)\sqrt{\tau/i}$}, volume={1}, DOI={10.1112/S0025579300000462}, number={1}, journal={Mathematika}, author={Siegel, Carl Ludwig}, year={1954}, pages={4–4}}

@article{Barducci:1977gy,
    author = "Barducci, A. and Casalbuoni, R. and Lusanna, L.",
    title = "{A Possible Interpretation of Theories Involving Grassmann Variables}",
    reportNumber = "Print-77-0464 (FLORENCE)",
    doi = "10.1007/BF02745002",
    journal = "Lett. Nuovo Cim.",
    volume = "19",
    pages = "581",
    year = "1977"
}

@article{BEREZIN1977336,
title = {Particle spin dynamics as the grassmann variant of classical mechanics},
journal = {Annals of Physics},
volume = {104},
number = {2},
pages = {336-362},
year = {1977},
issn = {0003-4916},
doi = {https://doi.org/10.1016/0003-4916(77)90335-9},
url = {https://www.sciencedirect.com/science/article/pii/0003491677903359},
author = {F.A. Berezin and M.S. Marinov},
}

@book{Ortin_2015, place={Cambridge}, edition={2}, series={Cambridge Monographs on Mathematical Physics}, title={Gravity and Strings}, publisher={Cambridge University Press}, author={Ortín, Tomás}, year={2015}, collection={Cambridge Monographs on Mathematical Physics}}

@article{BRINK197776,
title = {A Lagrangian formulation of the classical and quantum dynamics of spinning particles},
journal = {Nuclear Physics B},
volume = {118},
number = {1},
pages = {76-94},
year = {1977},
issn = {0550-3213},
doi = {https://doi.org/10.1016/0550-3213(77)90364-9},
url = {https://www.sciencedirect.com/science/article/pii/0550321377903649},
author = {L. Brink and P. {Di Vecchia} and P. Howe},
}

@book{Bailin:1994qt,
    author = "Bailin, D. and Love, Alexander",
    title = "{Supersymmetric Gauge Field Theory and String Theory}",
    doi = "10.1201/9780367805807",
    isbn = "978-1-4200-5082-0, 978-0-367-80580-7, 978-0-7503-0267-8",
    publisher = "Taylor \& Francis",
    year = "1994"
}

@misc{bachas2024dbranes,
      title={D-branes}, 
      author={Constantin Bachas},
      year={2024},
      eprint={2311.18456},
      archivePrefix={arXiv},
      primaryClass={hep-th},
      url={https://arxiv.org/abs/2311.18456}, 
}

@book{Polchinskivol2:1998rr,
    author = "Polchinski, J.",
    title = "{String theory. Vol. 2: Superstring theory and beyond}",
    doi = "10.1017/CBO9780511618123",
    isbn = "978-0-511-25228-0, 978-0-521-63304-8, 978-0-521-67228-3",
    publisher = "Cambridge University Press",
    series = "Cambridge Monographs on Mathematical Physics",
    month = "12",
    year = "2007"
}

@article{Gursoy:2010fj,
    author = "Gursoy, Umut and Kiritsis, Elias and Mazzanti, Liuba and Michalogiorgakis, Georgios and Nitti, Francesco",
    title = "{Improved Holographic QCD}",
    eprint = "1006.5461",
    archivePrefix = "arXiv",
    primaryClass = "hep-th",
    reportNumber = "CCTP-2010-7",
    doi = "10.1007/978-3-642-04864-7_4",
    journal = "Lect. Notes Phys.",
    volume = "828",
    pages = "79--146",
    year = "2011"
}

@article{Gliozzi:1976qd,
    author = "Gliozzi, F. and Scherk, Joel and Olive, David I.",
    title = "{Supersymmetry, Supergravity Theories and the Dual Spinor Model}",
    reportNumber = "CERN-TH-2253",
    doi = "10.1016/0550-3213(77)90206-1",
    journal = "Nucl. Phys. B",
    volume = "122",
    pages = "253--290",
    year = "1977"
}

@article{Berenstein:1999jq,
    author = "Berenstein, David and Leigh, Robert G.",
    title = "{Superstring perturbation theory and Ramond-Ramond backgrounds}",
    eprint = "hep-th/9904104",
    archivePrefix = "arXiv",
    reportNumber = "ILL-TH-99-02, ILL-(TH)-99-02",
    doi = "10.1103/PhysRevD.60.106002",
    journal = "Phys. Rev. D",
    volume = "60",
    pages = "106002",
    year = "1999"
}

@inproceedings{Berkovits:2017ldz,
    author = "Berkovits, Nathan and Gomez, Humberto",
    title = "{An Introduction to Pure Spinor Superstring Theory}",
    booktitle = "{9th Summer School on Geometric, Algebraic and Topological Methods for Quantum Field Theory}",
    eprint = "1711.09966",
    archivePrefix = "arXiv",
    primaryClass = "hep-th",
    doi = "10.1007/978-3-319-65427-0_6",
    series = "Mathematical Physics Studies",
    pages = "221--246",
    year = "2017"
}

@article{Callan:1987px,
    author = "Callan, Jr., Curtis G. and Lovelace, C. and Nappi, C. R. and Yost, S. A.",
    title = "{Adding Holes and Crosscaps to the Superstring}",
    reportNumber = "PUPT-1045",
    doi = "10.1016/0550-3213(87)90065-4",
    journal = "Nucl. Phys. B",
    volume = "293",
    pages = "83",
    year = "1987"
}

@unpublished{caltechthesis4867,
           title = {Branes, Brane Actions and Applications to Field Theory},
          author = {Costin Radu Popescu},
            year = {2001},
          school = {California Institute of Technology},
             url = {https://resolver.caltech.edu/CaltechETD:etd-12082006-104418},
}

@book{West_2012, place={Cambridge}, title={Introduction to Strings and Branes}, publisher={Cambridge University Press}, author={West, Peter}, year={2012}}

@misc{dhoker2002supersymmetricgaugetheoriesadscft,
      title={Supersymmetric Gauge Theories and the AdS/CFT Correspondence}, 
      author={Eric D'Hoker and Daniel Z. Freedman},
      year={2002},
      eprint={hep-th/0201253},
      archivePrefix={arXiv},
      primaryClass={hep-th},
      url={https://arxiv.org/abs/hep-th/0201253}, 
}

@book{Carroll:2004st,
    author = "Carroll, Sean M.",
    title = "{Spacetime and Geometry}: {An Introduction to General Relativity}",
    doi = "10.1017/9781108770385",
    isbn = "978-0-8053-8732-2, 978-1-108-48839-6, 978-1-108-77555-7",
    publisher = "Cambridge University Press",
    month = "7",
    year = "2019"
}

@article{Aharony:1999ti,
    author = "Aharony, Ofer and Gubser, Steven S. and Maldacena, Juan Martin and Ooguri, Hirosi and Oz, Yaron",
    title = "{Large N field theories, string theory and gravity}",
    eprint = "hep-th/9905111",
    archivePrefix = "arXiv",
    reportNumber = "CERN-TH-99-122, HUTP-99-A027, LBNL-43113, RU-99-18, UCB-PTH-99-16, LBL-43113",
    doi = "10.1016/S0370-1573(99)00083-6",
    journal = "Phys. Rept.",
    volume = "323",
    pages = "183--386",
    year = "2000"
}

@article{Klebanov:1997kc,
    author = "Klebanov, Igor R.",
    title = "{World volume approach to absorption by nondilatonic branes}",
    eprint = "hep-th/9702076",
    archivePrefix = "arXiv",
    reportNumber = "PUPT-1682",
    doi = "10.1016/S0550-3213(97)00235-6",
    journal = "Nucl. Phys. B",
    volume = "496",
    pages = "231--242",
    year = "1997"
}

@article{GUBSER1997217,
title = {String theory and classical absorption by three-branes},
journal = {Nuclear Physics B},
volume = {499},
number = {1},
pages = {217-240},
year = {1997},
issn = {0550-3213},
doi = {https://doi.org/10.1016/S0550-3213(97)00325-8},
url = {https://www.sciencedirect.com/science/article/pii/S0550321397003258},
author = {Steven S. Gubser and Igor R. Klebanov and Arkady A. Tseytlin},
}

@article{Hubeny:2014bla,
    author = "Hubeny, Veronika E.",
    title = "{The AdS/CFT Correspondence}",
    eprint = "1501.00007",
    archivePrefix = "arXiv",
    primaryClass = "gr-qc",
    doi = "10.1088/0264-9381/32/12/124010",
    journal = "Class. Quant. Grav.",
    volume = "32",
    number = "12",
    pages = "124010",
    year = "2015"
}

@article{33f2a5a08f4d436ab19efaba3eb6871a,
title = "Anti de sitter space and holography",
author = "Edward Witten",
year = "1998",
month = mar,
doi = "10.4310/atmp.1998.v2.n2.a2",
language = "English (US)",
volume = "2",
pages = "253--290",
journal = "Advances in Theoretical and Mathematical Physics",
issn = "1095-0761",
publisher = "International Press, Inc.",
number = "2",
}

@article{Penrose:1964ge,
    author = "Penrose, R.",
    editor = "DeWitt, C. and DeWitt, B.",
    title = "{Conformal treatment of infinity}",
    doi = "10.1007/s10714-010-1110-5",
    pages = "565--586",
    year = "1964"
}

@book{Penrose_Rindler_1986, place={Cambridge}, series={Cambridge Monographs on Mathematical Physics}, title={Spinors and Space-Time}, publisher={Cambridge University Press}, author={Penrose, Roger and Rindler, Wolfgang}, year={1986}, collection={Cambridge Monographs on Mathematical Physics}}

@article{Polchinski:2001tt,
    author = "Polchinski, Joseph and Strassler, Matthew J.",
    title = "{Hard scattering and gauge / string duality}",
    eprint = "hep-th/0109174",
    archivePrefix = "arXiv",
    reportNumber = "NSF-ITP-01-76, UPR-956-T",
    doi = "10.1103/PhysRevLett.88.031601",
    journal = "Phys. Rev. Lett.",
    volume = "88",
    pages = "031601",
    year = "2002"
}

@article{sakai1,
    author = {Sakai, Tadakatsu and Sugimoto, Shigeki},
    title = {Low Energy Hadron Physics in Holographic QCD},
    journal = {Progress of Theoretical Physics},
    volume = {113},
    number = {4},
    pages = {843-882},
    year = {2005},
    month = {04},
    issn = {0033-068X},
    doi = {10.1143/PTP.113.843},
    url = {https://doi.org/10.1143/PTP.113.843},
    eprint = {https://academic.oup.com/ptp/article-pdf/113/4/843/5123947/113-4-843.pdf},
}

@article{sakai2,
    author = {Sakai, Tadakatsu and Sugimoto, Shigeki},
    title = {More on a Holographic Dual of QCD},
    journal = {Progress of Theoretical Physics},
    volume = {114},
    number = {5},
    pages = {1083-1118},
    year = {2005},
    month = {11},
    issn = {0033-068X},
    doi = {10.1143/PTP.114.1083},
    url = {https://doi.org/10.1143/PTP.114.1083},
    eprint = {https://academic.oup.com/ptp/article-pdf/114/5/1083/5260085/114-5-1083.pdf},
}

@article{PhysRevD.12.2443,
  title = {Semiclassical bound states in an asymptotically free theory},
  author = {Dashen, Roger F. and Hasslacher, Brosl and Neveu, Andr\'e},
  journal = {Phys. Rev. D},
  volume = {12},
  issue = {8},
  pages = {2443--2458},
  numpages = {0},
  year = {1975},
  month = {Oct},
  publisher = {American Physical Society},
  doi = {10.1103/PhysRevD.12.2443},
  url = {https://link.aps.org/doi/10.1103/PhysRevD.12.2443}
}

@misc{Wipfnotes,
  author        = {Andreas Wipf},
  title         = {Introduction to Supersymmetry},
  url = {https://www.tpi.uni-jena.de/~wipf/lectures/susy/susyhead.pdf}
}

@article{Berenstein:2002jq,
    author = "Berenstein, David Eliecer and Maldacena, Juan Martin and Nastase, Horatiu Stefan",
    title = "{Strings in flat space and pp waves from N=4 superYang-Mills}",
    eprint = "hep-th/0202021",
    archivePrefix = "arXiv",
    doi = "10.1088/1126-6708/2002/04/013",
    journal = "JHEP",
    volume = "04",
    pages = "013",
    year = "2002"
}

@article{Beisert:2010jr,
    author = "Beisert, Niklas and others",
    title = "{Review of AdS/CFT Integrability: An Overview}",
    eprint = "1012.3982",
    archivePrefix = "arXiv",
    primaryClass = "hep-th",
    reportNumber = "AEI-2010-175, CERN-PH-TH-2010-306, HU-EP-10-87, HU-MATH-2010-22, KCL-MTH-10-10, UMTG-270, UUITP-41-10",
    doi = "10.1007/s11005-011-0529-2",
    journal = "Lett. Math. Phys.",
    volume = "99",
    pages = "3--32",
    year = "2012"
}

@article{Drukker:2000rr,
    author = "Drukker, Nadav and Gross, David J.",
    title = "{An Exact prediction of N=4 SUSYM theory for string theory}",
    eprint = "hep-th/0010274",
    archivePrefix = "arXiv",
    reportNumber = "USC-00-05, CITUSC-00-057, NSF-ITP-00-119",
    doi = "10.1063/1.1372177",
    journal = "J. Math. Phys.",
    volume = "42",
    pages = "2896--2914",
    year = "2001"
}

@article{Ebert_2009,
   title={Mass spectra and Regge trajectories of light mesons in the relativistic quark model},
   volume={79},
   ISSN={1550-2368},
   url={http://dx.doi.org/10.1103/PhysRevD.79.114029},
   DOI={10.1103/physrevd.79.114029},
   number={11},
   journal={Physical Review D},
   publisher={American Physical Society (APS)},
   author={Ebert, D. and Faustov, R. N. and Galkin, V. O.},
   year={2009},
   month=jun }

@book{nakahara2003geometry,
  title={Geometry, Topology and Physics, Second Edition},
  author={Nakahara, M.},
  isbn={9780750306065},
  lccn={2003282202},
  series={Graduate student series in physics},
  url={https://books.google.com.br/books?id=cH-XQB0Ex5wC},
  year={2003},
  publisher={Taylor \& Francis}
}

@book{dirac2013lectures,
  title={Lectures on Quantum Mechanics},
  author={Dirac, P.A.M.},
  isbn={9780486320281},
  series={Dover Books on Physics},
  url={https://books.google.com.br/books?id=Z3XCAgAAQBAJ},
  year={2013},
  publisher={Dover Publications}
}

%\documentclass[a4paper,12pt]{memoir}
%\usepackage{graphicx}
%\usepackage[utf8]{inputenc}
%\usepackage{indentfirst}
%\usepackage{braket}
%\usepackage{amsmath, amsthm, amssymb, amsfonts,bm}
%\usepackage{mathtools, changepage, slashed}
%\usepackage{tikz-feynman}
%\usepackage{bm, mathrsfs}
%\usepackage{gensymb}
%\usepackage{geometry}
%\usepackage{epstopdf}
%\usepackage{hyperref}
%\usepackage{pgfplots}
%\pgfplotsset{compat=1.18} 
%\usepackage[sorting=none]{biblatex}
%\addbibresource{refs.bib}
%\numberwithin{equation}{section}
%\usepackage[inkscapelatex=false]{svg}
%\usepackage[super]{natbib}
%\usepackage{doi}
%\hypersetup{
%  colorlinks   = true, %Colours links instead of ugly boxes
%  urlcolor     = black, %Colour for external hyperlinks
%  linkcolor    = black, %Colour of internal links
%  citecolor   = black %Colour of citations
%}
%\DeclareMathOperator{\Tr}{Tr}

%\newcommand{\normalord}[1]{%
%  {:\mathrel{\mspace{1mu}#1\mspace{1mu}}:}%
%}

%\title{Appendix: Virasoro algebras}

%\begin{document}

\appendix

\chapter{Virasoro algebras}\label{ApA}

\section{Bosonic matter CFT Virasoro algebra}\label{A1}

In this appendix we derive the Virasoro algebras of the matter and ghost CFTs of the bosonic string, and well as for the matter CFT of the RNS superstring. At the end we elaborate on the algebra's dependence on the normal-ordering constant of $L_0$.

We wish to derive the commutator between the bosonic string's matter Virasoro generators $L_n$, where
\begin{equation}
    L_n = \frac{1}{2} \sum_{k=-\infty}^\infty \normalord{\alpha_{n-k} \alpha_{k}} =  \frac{1}{2} \sum_{k=0}^\infty \alpha_{-n} \alpha_{k+n } + \frac{1}{2} \sum_{k=1}^\infty \alpha_{n-k} \alpha_{k}
\end{equation}
and $\left[\alpha_m,\alpha_n\right]=m\delta_{m+n,0}$. We omit the Lorentz indices since they simply amount to having $D$ copies of this CFT. We start by computing $\left[L_m, \alpha_n\right]$. For $m \neq 0$, we can drop the normal ordering symbol:
\begin{align}
    \left[ L_m, \alpha_n \right] &= \frac{1}{2} \sum_{k=-\infty}^\infty \left[ \alpha_{m-k} \alpha_{k} , \alpha_n \right] \notag\\[5pt]
    &= \frac{1}{2} \sum_{k=-\infty}^\infty \Big( \alpha_{m-k} \left[ \alpha_{k} , \alpha_n \right] + \left[ \alpha_{m-k} , \alpha_n \right] \alpha_{k} \Big) \notag\\[5pt]
    &= -n \alpha_{m+n} \, , \hspace{0.5cm} m \neq 0.
\end{align}
For $m=0$, we have
\begin{align}
    \left[ L_0, \alpha_n \right] &= \frac{1}{2} \sum_{k= - \infty}^\infty \left[ \normalord{\alpha_{-k} \alpha_{k}} , \alpha_n \right] \notag\\[5pt]
    &= \sum_{k=1}^\infty \left[ \alpha^\nu_{-k} \alpha_{k} , \alpha_n \right] \notag\\[5pt]
    &= \sum_{k=1}^\infty \Big( \alpha_{-k} \left[ \alpha_{k} , \alpha_n \right] + \left[ \alpha_{-k} , \alpha_n \right] \alpha_{k} \Big) \notag\\[5pt]
    &= -n \alpha_m.
\end{align}
Both cases are thus contemplated by $\left[ L_m, \alpha_n \right] = - n \alpha_{m+n}$. Consider first the case where $m \neq -n$ and that either $m$ or $n$ is nonzero. Without loss of generality we can choose $n \neq 0$:
\begin{align}
    [L_m,L_n] &= \frac{1}{2} \sum_{k= - \infty}^\infty [L_m , \alpha_{n-k} \alpha_k] \notag\\[5pt]
    &= \frac{1}{2} \sum_{k= - \infty}^\infty \Big( [L_m,\alpha_{n-k}]\alpha_k + \alpha_{n-k}[L_m,\alpha_k] \Big) \notag\\[5pt]
    &=\frac{1}{2} \sum_{k= - \infty}^\infty \Big( (k-n) \alpha_{m+n-k} \alpha_k -k \alpha_{n-k} \alpha_{m+k} \Big) \notag\\[5pt]
    &= (m-n)L_{m+n}, \hspace{0.5cm} 0 \neq n \neq -m,
\end{align}
where we did $k \to k-m$ in the second term of the third line to obtain the fourth one. This result also trivially holds if $n=-m=0$, so the only possibility not contemplated by it is $m=-n$. Using the normal-ordered form of $L_n$, we have
\begin{align}
    &[L_{-n},L_n] = \frac{1}{2} \sum_{k=0}^\infty [L_{-n},\alpha_{-k} \alpha_{k+n}] + \frac{1}{2} \sum_{k=1}^\infty [L_{-n},\alpha_{n-k} \alpha_k] \notag\\[5pt]
    &= \frac{1}{2} \sum_{k=0}^\infty \Big(  [L_{-n}, \alpha_{-k}] \alpha_{k+n} + \alpha_{-k} [L_{-n},\alpha_{k+n}] \Big) + \frac{1}{2} \sum_{k=1}^\infty  \Big( [L_{-n},\alpha_{n-k}] \alpha_k + \alpha_{n-k}[L_{-n},\alpha_k]  \Big) \notag\\[5pt]
    &= \frac{1}{2} \sum_{k=0}^\infty \Big( k \alpha_{-n-k} \alpha_{k+n} - (k+n) \alpha_{-k} \alpha_k \Big) + \frac{1}{2} \sum_{k=1}^\infty \Big( (k-n) \alpha_{-k} \alpha_{k} -k \alpha_{n-k} \alpha_{k-n} \Big).
    \label{Matter Virasoro computation intermediate}
\end{align}
We now shift the summed index $k$ so that $n$ does not appear in any of the mode numbers. This means doing $k \to k-n$ in the first term:
\begin{align}
    \sum_{k=0}^\infty  k \alpha_{-n-k} \alpha_{k+n} &= \sum_{k=n}^\infty (k-n) \alpha_{-k} \alpha_{k} \notag\\[5pt]
    &= \sum_{k=0}^\infty (k-n)\alpha_{-k} \alpha_{k} - \sum_{k=0}^n (k-n)\alpha_{-k} \alpha_{k},
\end{align}
and $k \to k+n$ in the last one:
\begin{align}
    - \sum_{k=1}^\infty k \alpha_{n-k} \alpha_{k-n} &= - \sum_{k=1-n}^\infty (k+n) \alpha_{-k} \alpha_k \notag\\[5pt]
    &= - \sum_{k=1}^\infty (k+n) \alpha_{-k} \alpha_k - \sum_{k=0}^{n-1} (n-k) \alpha_{k} \alpha_{-k} \notag\\[5pt]
    &= - \sum_{k=1}^\infty (k+n) \alpha_{-k} \alpha_k  - \sum_{k=0}^{n} (n-k) \alpha_{-k} \alpha_{k} - \sum_{k=0}^n (n-k)k.
\end{align}
In the last line we used $\alpha_k \alpha_{-k} = \alpha_{-k} \alpha_k + k$ added the $k=n$ term in the sums involving $(n-k)$ because its contribution vanishes. Plugging these results back into \eqref{Matter Virasoro computation intermediate} gives
\begin{align}
    [L_{-n},L_n] &= -2nL_0 + \frac{1}{2} \sum_{k=1}^n (k^2 -nk) \notag\\[5pt]
    &= -2nL_0 - \frac{1}{12}n \big(n^2-1 \big),
\end{align}
where we used the relations \cite{Blumenhagen:2013fgp}
\begin{align}
    \sum_{k=1}^n k &= \frac{1}{2} n (n+1) \\[5pt]
    \sum_{k=1}^n k^2 &= \frac{1}{6} \left( 2n^3 +3n^2 +n \right).
\end{align}
The general form of the algebra is therefore
\begin{equation}
    [L_m,L_n] = (m-n)L_{m+n} + \frac{1}{12}m \big( m^2-1 \big)\delta_{m+n,0}.
\end{equation}
This means that a single bosonic scalar field has central charge $c=1$. For $D$ such fields, as is the case of the Polyakov action, the Virasoro generators add and one has $c=D$.

\section{Ghost CFT Virasoro algebra}\label{A2}

The Virasoro generators of the bosonic ghost CFT are
\begin{equation}
    L_n = \sum_{k= - \infty}^\infty \left( 2n-k \right)  \normalord{b_k c_{n-k}} = \sum_{k=0}^\infty (2n+k) b_{-k} c_{n+k} - \sum_{k=1}^\infty (2n -k) c_{n-k} b_k.
\end{equation}
The basic commutators in this case are
\begin{align}
    [b_m,c_n] &= 2b_mc_n - \delta_{m+n,0} \\
    [c_m,b_n] &= 2c_m b_n - \delta_{m+n,0} \\
    [b_m,b_n] &= 2b_m b_n \\
    [c_m , c_n] &= 2 c_m c_n,
\end{align}
which can be used to find
\begin{align}
    [L_m, b_n] &= (m-n) b_{m+n}\\
    [L_m,c_n] &= -(2m+n) c_{m+n}.
\end{align}
Starting with the $0 \neq n \neq -m$ case, we have that
\begin{align}
    [L_m,L_n] &= \sum_{k= - \infty}^\infty (2n - k) [L_m, b_k c_{n-k}] \notag\\[5pt]
    &= \sum_{k = - \infty}^\infty (2n-k) \Big( [L_m,b_k]c_{n-k} + b_k [L_m,c_{n-k}] \Big) \notag\\[5pt]
    &= \sum_{k=- \infty}^\infty (2n-k) \Big( (m-k) b_{m+k} c_{n-k} - (2m+n-k)b_k c_{m+n-k} \Big) \notag\\[5pt]
    &= (m-n)L_{m+n}, \hspace{0.5cm} 0 \neq n \neq -m,
\end{align}
where we did $k \to k-m$ to obtain the last line. For the $m=-n$ case we have
\begin{align}
    [L_{-n},L_n] &= \sum_{k=0}^\infty (2n+k) [L_{-n},b_{-k}c_{n+k}] - \sum_{k=1}^\infty (2n-k)[L_{-n},c_{n-k}b_k] \notag\\[5pt]
    &= \sum_{k=0}^\infty (2n+k) \Big( (k-n)b_{-n-k} c_{n+k} + (n-k)b_{-k} c_k \Big) \notag\\[5pt]
    &\hspace{0.5cm} - \sum_{k=1}^\infty (2n-k) \Big( (n+k)c_{-k}b_k - (n+k)c_{n-k}b_{k-n} \Big).
\end{align}
We eliminate $n$ from the mode numbers by shifting $k \to k-n$ in the first term and $k \to k+n$ in the last:
\begin{align}
    \sum_{k=0}^\infty (2n+k)(k-n)b_{-n -k} &c_{n+k} = \sum_{k=n}^\infty (n+k)(k-2n) b_{-k} c_k \notag\\[5pt]
    &= \sum_{k=0}^\infty (n+k)(k-2n) b_{-k} c_k - \sum_{k=0}^{n-1} (n+k)(k-2n) b_{-k} c_k,
\end{align}
\begin{align}
    \sum_{k=1}^\infty (2n -k)(n+k) c_{n-k} b_{k-n} &= \sum_{k=1-n}^\infty (n-k)(2n+k) c_{-k}b_k \notag\\[5pt]
    &= \sum_{k=1}^\infty (n-k)(2n+k) c_{-k}b_k + \sum_{k=0}^{n-1} (n+k)(2n-k) c_k b_{-k} \notag\\[5pt]
    & \hspace{-3.7cm} =\sum_{k=1}^\infty (n-k)(2n+k) c_{-k}b_k -\sum_{k=0}^{n-1} (n+k)(2n-k) b_{-k} c_k + \sum_{k=0}^{n-1} (n+k)(2n-k),
\end{align}
where we used $c_{k} b_{-k} = -b_{-k} c_k +1 $ to obtain the last line. This leads to
\begin{align}
    [L_{-n},L_n] &= -2n L_0 + \sum_{k=0}^{n-1} (n+k)(2n-k) \notag\\[5pt]
    &= -2nL_0 + \sum_{k=1}^n (n+k-1)(2n-k+1) \notag\\[5pt]
    &=-2n L_0 - \sum_{k=1}^n k^2 + (n+2) \sum_{k=1}^n k + (2n^2 - n - 1)\sum_{k=1}^n 1 \notag\\[5pt]
    &=-2 n L_0 + \frac{1}{12}n \big( 26n^2 - 2 \big).
\end{align}
The ghost algebra is therefore
\begin{align}
    [L_m,L_n] &= (m-n)L_{m+n} - \frac{1}{12} m \big( 26m^2-2 \big) \delta_{m+n,0} \notag\\[5pt]
    &= (m-n) \big( L_{m+n} - \delta_{m+n,0} \big) - \frac{26}{12}m \big( m^2 - 1 \big)\delta_{m+n,0}.
\end{align}

\section{Fermionic matter CFT Virasoro algebra}\label{A3}

The fermionic matter CFT's Virasoro generators are given by
\begin{equation}
    L_n = \frac{1}{4} \sum_{r \in \mathbb{Z} + \nu} (2r-n) \normalord{\psi_{n-r} \psi_r} = \frac{1}{4} \underset{r \geq 0}{\sum_{r \in \mathbb{Z} + \nu}} (2r-n) \psi_{n-r} \psi_r + \frac{1}{4} \underset{r > 0}{\sum_{r \in \mathbb{Z} + \nu}} (2r+n) \psi_{r} \psi_{n+r} \, ,
\end{equation}
where the modes satisfy $\{ \psi_{r} , \psi_{s} \} = \delta_{r+s,0}$. We once again ignore Lorentz indices. This basic anticommutator leads to
\begin{equation}
    [L_m,\psi_r] = - \frac{1}{2}\big( 2r+m \big)\psi_{m+r}.
\end{equation}
For $0 \neq n \neq -m$ we have
\begin{align}
    [L_m,L_n] &= \frac{1}{4} \sum_r (2r -n) [L_m, \psi_{n-r} \psi_r ] \notag\\[5pt]
    &= - \frac{1}{8} \sum_r (2r-n) \Big( (2n-2r+m) \psi_{m+r-m} \psi_r (2r+m) \psi_{n-r} \psi_{m+r}  \Big) \notag\\[5pt]
    &= (m-n)L_{m+n}, \hspace{0.5cm} 0 \neq n \neq -m,
\end{align}
where we did $r \to r-m$ in the second term to obtain the last line. The $m=-n$ case is
\begin{align}
    &[L_{-n},L_n] = \frac{1}{4} \sum_{r \geq 0}(2r-n) [ L_{-n},\psi_{n-r} \psi_r ] + \frac{1}{4} \sum_{r>0} (2r-n) [ L_{-n},\psi_{-r} \psi_{n+r} ] \notag\\[5pt]
    =& - \frac{1}{8} \sum_{r \geq 0} (2r-n)^2 \big( - \psi_{-r} \psi_r + \psi_{n-r} \psi_{r-n} \big) - \frac{1}{8} \sum_{r>0} (2r+n)^2 \big( - \psi_{-n-r} \psi_{n+r} + \psi_{-r} \psi_r \big).
\end{align}
We shift $r \to r+n$ in the second term and $r \to r-n$ in the third:
\begin{align}
    -\frac{1}{8} \sum_{r \geq 0} (2r-n)^2 \psi_{n-r} \psi_{r-n} &= - \frac{1}{8} \sum_{r \geq -n} (2r + n)^2 \psi_{-r} \psi_r \notag\\[5pt]
    &= - \frac{1}{8} \sum_{0 < r \leq n} (2r-n)^2 \psi_r \psi_{-r} - \frac{1}{8} \sum_{r \geq 0} (2r+n)^2 \psi_{-r} \psi_r \notag\\[5pt]
    & \hspace{-3cm} = \frac{1}{8}\sum_{0 < r \leq n}(2r-n)^2 \psi_{-r} \psi_r - \frac{1}{8}\sum_{0 < r \leq n} (2r-n)^2 - \frac{1}{8} \sum_{r \geq 0} (2r+n)^2 \psi_{-r} \psi_r,
\end{align}
\begin{align}
    \frac{1}{8} \sum_{r>0} (2r+n)^2 \psi_{-n -r} \psi_{n+r} &= \frac{1}{8} \sum_{r>n}(2r-n)^2 \psi_{-r} \psi_r \notag\\[5pt]
    &= \frac{1}{8} \sum_{r>0}(2r-n)^2 \psi_{-r} \psi_r - \frac{1}{8} \sum_{0 < r \leq n}(2r-n)^2 \psi_{-r} \psi_r.
\end{align}
This gives
\begin{align}
    [L_{-n},L_n] &= -2 n L_0 - \frac{1}{8} \sum_{o < r \leq n}(2r-n)^2 \notag\\[5pt]
    &= -2 n L_0 - \frac{1}{8}\sum_{k=1}^n (2k -2 \nu -n)^2  \notag\\[5pt]
    &= -2 n L_0 - \frac{1}{8} \bigg[ 4 \sum_{k=1}^n k^2 -4(n+2 \nu) \sum_{k=1}^n k + (n^2 + 4n \nu + 4 \nu^2) \sum_{k=1}^n 1 \bigg] \notag\\[5pt]
    &= -2 n L_0 - \frac{n^3}{24} - \frac{n}{12} + \frac{n \nu}{2}  - \frac{n \nu^2}{2}  \notag\\[5pt]
    &= -2n \Big( L_0 + \frac{1}{16} - \frac{\nu}{4} + \frac{\nu^2}{4} \Big) - \frac{1}{24} n \big( n^2-1 \big).
\end{align}
The final form of the algebra is therefore
\begin{equation}
    [L_m,L_n] = (m-n)\Big( L_{m+n} + \Big(\frac{1}{16} - \frac{\nu}{4} + \frac{\nu^2}{4} \Big) \delta_{m+n,0} \Big) + \frac{1/2}{12}m \big( m^2-1  \big) \delta_{m+n,0}.
\end{equation}

\section{Relating the normal-ordering constants }\label{A4}

All of the Virasoro algebras just derived are of the form
\begin{equation}
    [L_m,L_n] = (m-n)\big( L_{m+n} + a^\prime \delta_{m+n,0} \big) + \frac{c}{12}m\big( m^2-1  \big) \delta_{m+n,0},
\end{equation}
where, for each CFT, $a^\prime$ is not the normal ordering constant $a$ one calculates by renormalizing the zero-point energy of each field, but is related to it by
\begin{equation}
    a^\prime = a + \frac{c}{24},
\end{equation}
where $c$ is the CFT's central charge. The reason for this is that in Chapters \ref{ch3} and \ref{ch7} we obtain the normal ordering constant from the Hamiltonian in the $(\tau,\sigma)=(-i \sigma^2,\sigma^1)$ coordinates, whereas the Virasoro generators are usually defined as the Laurent coefficients of the energy-momentum tensor in the $z=e^{-i(\sigma^1 + i \sigma^2)}$ coordinates, related to the $\sigma^a$ by a conformal transformation. Forming the combinations $w= \sigma^1 + i \sigma^2$ and $\bar{w}=\sigma^1 - i \sigma^2$, the energy momentum tensor $T_{ww}(w) \equiv T(w)$ can be shown to change under a conformal transformation $w \to z(w)$ as
\begin{equation}
    T(w) \to T (z) = \left( \frac{\partial z}{\partial w} \right)^{-2} \left( T(w) - \frac{c}{12} \left\{ z, w \right\} \right),
\end{equation}
where
\begin{equation}
    \left\{ z, w \right\} = \frac{1}{2} \left( \frac{\partial z}{\partial w} \right)^{-2} \left( 2 \frac{\partial^3 z}{\partial w^3}  \frac{\partial z}{\partial w} -3 \left( \frac{\partial^2 z}{\partial w^2} \right)^2 \right)
\end{equation}
is called the Schwarzian derivative. For $z = \exp (-i w)$, we have
\begin{equation}
    \frac{\partial^n z}{\partial w^n} = (-i)^n z,
\end{equation}
leading to $\left\{ z,w \right\}=1/2$ and
\begin{equation}
    T(z) = \left( \frac{\partial z}{\partial w} \right)^{-2} \left( T(w) - \frac{c}{24} \right) = - \frac{1}{z^2}\left( T(w) - \frac{c}{24} \right).
    \label{transf}
\end{equation}
This means that the Hamiltonian, which is usually defined in the $w$-frame, and $L_0$, usually defined in the $z$-frame, are related by
\begin{equation}
    L_0 = H + \frac{c}{24},
\end{equation}
and only coincide if the central charge vanishes. The ordering constants $a^\prime$ found in this appendix are the $z$-frame ones, while in the main text we favored the $w$-frame ones, as they have a more direct interpretation in terms adding the zero-point energy of each field.

%\end{document}

%\documentclass[a4paper,12pt]{memoir}
%\usepackage{graphicx}
%\usepackage[utf8]{inputenc}
%\usepackage{indentfirst}
%\usepackage{braket}
%\usepackage{amsmath, amsthm, amssymb, amsfonts,bm}
%\usepackage{mathtools, changepage, slashed}
%\usepackage{tikz-feynman}
%\usepackage{bm, mathrsfs}
%\usepackage{gensymb}
%\usepackage{geometry}
%\usepackage{epstopdf}
%\usepackage{hyperref}
%\usepackage{pgfplots}
%\pgfplotsset{compat=1.18} 
%\usepackage[sorting=none]{biblatex}
%\addbibresource{refs.bib}
%\numberwithin{equation}{section}
%\usepackage[inkscapelatex=false]{svg}
%\usepackage[super]{natbib}
%\usepackage{doi}
%\hypersetup{
%  colorlinks   = true, %Colours links instead of ugly boxes
%  urlcolor     = black, %Colour for external hyperlinks
%  linkcolor    = black, %Colour of internal links
%  citecolor   = black %Colour of citations
%}
%\DeclareMathOperator{\Tr}{Tr}

%\newcommand{\normalord}[1]{%
%  {:\mathrel{\mspace{1mu}#1\mspace{1mu}}:}%
%}

\chapter{Worldsheet duality ghost amplitudes}\label{ApB}

%\begin{document}

In this appendix we compute the ghost contribution to the D-brane interaction amplitudes in both the open and closed string interpretations.

\section{Open string computation}\label{B1}

The ghost part of the open string cylinder amplitude discussed in Section \ref{sec63} is
\begin{equation}
    \mathcal{A}^g_\text{open}(t) = \frac{1}{4 \pi^2 t} \int d^2\sigma \int \mathcal{D}b \, \mathcal{D}c \, e^{- S_g[b,c]  } \,  b_{22}(\sigma) c^2(\hat{\sigma}),
\end{equation}
where the $d^2\sigma$ integral is over the cylinder defined by $[0,\pi] \times [0, 2\pi  t]$, with $\sigma^2 \sim \sigma^2 + 2 \pi t$. The same trick of exchanging the path integral for a trace over the spectrum used to find the bosonic amplitude $\mathcal{A}^X_\text{open}(t)$ can be used for the ghost CFT, but in this case there are a couple of additional subtleties which must be taken in account. It is well known that to write the partition function $Z = \text{Tr} (e^{- \beta H})$ of a system with fermions as a path integral, the fermions must be given antiperiodic boundary conditions with respect to the temperature, $\psi(\beta) = - \psi(0)$.\footnote{See Appendix A of \cite{PhysRevD.12.2443} for a simple proof. The main point is that the partition function should always be a sum of $e^{-\beta E_i}$ over the energies $E_i$ of all states in the theory. The path integral over the Euclidean action only agrees with this if the fermions are antiperiodic in the imaginary time coordinate. The situation here is the inverse: the starting point is the Euclidean path integral with all fields periodic, so it is the trace prescription that must change.} As discussed in Section \ref{sec33}, the ghosts are always periodic with respect to circular directions on the worldsheet, so the trace prescription must be modified in order to reflect this. In a theory with fermions at finite temperature, the periodicity of the operators with respect to the imaginary time is inherited by the Schrödinger picture states, since bosonic/fermionic states are made by acting on the vacuum with an even/odd number of fermionic operators. Therefore
\begin{align}
    \text{Tr} \, e^{- \beta H} &= \sum_\psi \braket{\psi |e^{-\beta H}|\psi } =\sum_\psi \braket{\psi,0 |\psi,\beta } \notag\\[5pt]
    &= \sum_{\text{boson}} \braket{\text{boson},0  | \text{boson},\beta} + \sum_{\text{fermion}} \braket{\text{fermion},0  | \text{fermion},\beta} \notag\\[5pt]
    &= \sum_{\text{boson}} \braket{\text{boson},0  | \text{boson},0} - \sum_{\text{fermion}} \braket{\text{fermion},0  | \text{fermion},0},
\end{align}
where in the last line we used the antiperiodicity of the fermions with respect to $\beta$. We would like to turn the minus sign in front of the second term into a plus, making the fermions periodic. This is accomplished by inserting a factor of $(-1)^F$ into the trace, where the fermion number operator $F$ has eigenvalue $1$ for fermionic states and $0$ for bosonic ones, so that
\begin{align}
    &(-1)^F \ket{\text{boson}} = \ket{\text{boson}} \notag\\[5pt]
    &(-1)^F \ket{\text{fermion}} = -\ket{\text{fermion}}.
\end{align}
We therefore set
\begin{align}
    \mathcal{A}^g_\text{open}(t) &= \frac{1}{4 \pi^2 t} \int d^2 \sigma \,  \text{Tr} \big( (-1)^F e^{-2 \pi t  H^{(g)}} b_{22}(\sigma) \,  c^2(\hat{\sigma}) \big) \notag\\[5pt]
    &= \frac{1}{4 \pi^2 t} e^{-2 \pi t a^g} \int d^2 \sigma \, \text{Tr} \big( (-1)^F e^{-2 \pi t L_0^{(g)}}  b_{22}(\sigma) \,  c^2(\hat{\sigma}) \big).
    \label{intermediate ghost result 2}
\end{align}
Recall from Section \ref{sec33} that the ghost CFT possesses two ground states $\ket{\uparrow}$ and $\ket{\downarrow}$ which satisfy
\begin{align}
    b_0 \ket{\downarrow} &= 0, \hspace{0.5cm} b_0 \ket{\uparrow} = \ket{\downarrow} \notag\\[5pt]
    c_0 \ket{\downarrow} &= \ket{\uparrow}, \hspace{0.5cm} c_0 \ket{\uparrow} =0.
    \label{ghost vacua relations}
\end{align}
This degeneracy means that for any excited state $\ket{\psi,\downarrow}$ which is built by acting with some string of $b_{-n}$ and $c_{-n}$ creation operators on $\ket{\downarrow}$ ($n >0$), there exists another state $\ket{\psi,\uparrow}$ given by the same creation operators acting instead on $\ket{\uparrow}$. The trace of any operator $O$ therefore splits as
\begin{equation}
    \text{Tr}(O) = \text{Tr}_{\downarrow}(O) + \text{Tr}_{\uparrow}(O),
\end{equation}
where in each term on the right side one sums only over the states built on top of the correspondent vacuum. Note that the naive prescriptions $\text{Tr}_\downarrow (O) = \sum_\psi \braket{\psi,\downarrow |O|\psi,\downarrow} $ and $\text{Tr}_\uparrow (O) = \sum_\psi \braket{\psi,\uparrow |O|\psi,\uparrow}$ cannot be correct, because both vacua have zero norm:
\begin{align}
    \braket{\downarrow | \downarrow} = \braket{\uparrow | b_0^2 | \uparrow}=0, && \braket{\uparrow | \uparrow} = \braket{\downarrow | c_0^2 | \downarrow}=0.
\end{align}
For a general excited state $\ket{\psi,\downarrow}$ one may compute $\braket{\psi,\downarrow | \psi, \downarrow}$ by anticommuting all the positive modes that create $\bra{\psi,\downarrow}$ past the negative modes that create $\ket{\psi,\downarrow}$, to make them act on the $\ket{\downarrow}$ vacuum. Using $b_{m} c_{n} = -c_n b_m + \delta_{m+n,0}$, one finds that $\braket{\psi,\downarrow | \psi, \downarrow}$ only has a chance of being nonzero if $\ket{\psi,\downarrow}$ is created by b-c pairs of equal mode numbers such as $b_{-n} c_{-n}$, in which case the anticommutators are nontrivial and one finds a result proportional to $\braket{\downarrow|\downarrow}$. Therefore, if the vacuum itself has zero norm, so do all excited states. Defining $\text{Tr}_\downarrow (O)$ as $\sum_\psi \braket{\psi,\downarrow |O|\psi,\downarrow}$ would thus set the trace of the identity operator to zero. The proper trace formula, which correctly maps each operator to the sum of its eigenvalues, is\footnote{This is equivalent to defining the inner product of two states in the $\ket{\downarrow}$ sector as
\begin{equation}
    (\psi^\prime, \downarrow | \psi, \downarrow) = \braket{\psi^\prime,\downarrow |c_0| \psi, \downarrow} = \braket{\psi^\prime,\uparrow | \psi, \downarrow},
\end{equation}
and that of two states in the $\ket{\uparrow}$ sector as
\begin{equation}
    (\psi^\prime, \uparrow | \psi, \uparrow) = \braket{\psi^\prime,\uparrow |b_0| \psi, \uparrow} = \braket{\psi^\prime,\downarrow | \psi, \uparrow}.
\end{equation}
The ghost zero modes act as a kind of metric for the definition of inner products in the ghost Hilbert space. For closed strings the equivalent relations are $(\psi^\prime, \downarrow \downarrow | \psi, \downarrow \downarrow ) = \braket{\psi^\prime,\downarrow \downarrow |c_0 \tilde{c}_0| \psi, \downarrow \downarrow }$ and $(\psi^\prime, \uparrow \uparrow | \psi, \uparrow \uparrow ) = \braket{\psi^\prime,\uparrow \uparrow |b_0 \tilde{b}_0| \psi, \uparrow \uparrow}$. \label{ghost inner product footnote}}
\begin{align}
    \text{Tr}_\downarrow(O) = \sum_\psi \braket{\psi,\uparrow |O|\psi,\downarrow}, &&  \text{Tr}_\uparrow (O) = \sum_\psi \braket{\psi,\downarrow |O|\psi,\uparrow}.
\end{align}
This works because $\braket{\uparrow | \downarrow} = \braket{\downarrow |c_0 b_0| \uparrow} = \braket{\downarrow | \uparrow}$ does not vanish, so we may normalize the vacua such that $\braket{\uparrow| \downarrow}=1$. We wish do use these formulas to compute the trace of $(-1)^F e^{-2 \pi t L_0^{g}}  b_{22}(\sigma) \,  c^2(\hat{\sigma})$, using the mode expansions
\begin{align}
     b_{2 2}(\sigma) &=  2 \sum_n b_n e^{- n \sigma^2} \cos (n \sigma^1) \notag\\[5pt]
     c^2(\hat{\sigma}) &=  \sum_n c_n e^{- n \hat{\sigma}^2} \cos(n \hat{\sigma}^1)
\end{align}
from Section \ref{sec33}. To do this it is convenient to first derive some auxiliary results. First, note that the zero mode algebra \eqref{ghost vacua relations} leads to $\text{Tr}_\uparrow(O) = \text{Tr}_\downarrow (b_0 O c_0)$, and therefore
\begin{align}
    \text{Tr} \big( (-1)^F O \big) &= \text{Tr}_\downarrow \big( (-1)^F O \big) + \text{Tr}_\uparrow \big( (-1)^F O \big) 
    \notag\\[5pt]
    &=\text{Tr}_\downarrow \big( (-1)^F O \big) - \text{Tr}_\downarrow \big( (-1)^F b_0 O c_0 \big) \notag\\[5pt]
    &= \text{Tr}_\downarrow \big( (-1)^F O \big)  - \text{Tr}_\downarrow \big( (-1)^F O \, b_0 c_0 \big) - \text{Tr}_\downarrow \big( (-1)^F [b_0,O]c_0 \big).
\end{align}
We use $b_0c_0 = - c_0 b_0 +1$ on the second term and anticommute $b_0$ to the right, past all creation operators, until it annihilates $\ket{\downarrow}$. The result is
\begin{equation}
    \text{Tr} \big( (-1)^F O \big) = - \text{Tr}_\downarrow \big( (-1)^F [b_0,O] c_0 \big).
    \label{ghost CFT lemma}
\end{equation}
The mode expansions lead to a sum of terms of the form 
\begin{equation}
     \text{Tr} \big( (-1)^F e^{-2 \pi t L_0^g} \, b_m c_n \big).
\end{equation}
Let $A$ be some bosonic operator that commutes with $b_0$ and $c_0$ (as is the case of $e^{-2 \pi t L_0^g}$). Plugging $O=A \, b_m c_n$ into \eqref{ghost CFT lemma} and using the anticommutation relations for the modes leads to $\text{Tr} \big( (-1)^F A \, b_m b_n \big) = \text{Tr}_\downarrow \big( (-1)^F A \, b_m c_0  \big) \delta_{n0}$. If $m \neq 0$, one may anticommute $c_0$ to the left past all modes until it annihilates $\bra{\uparrow}$. Therefore, only the zero modes survive:
\begin{equation}
    \text{Tr} \big( (-1)^F A \, b_m b_n \big) = \text{Tr}_\downarrow \big( (-1)^F A \, b_0 c_0 \big) \delta_{m0} \delta_{n0} = \text{Tr}_\downarrow \big( (-1)^F A \big) \delta_{m0} \delta_{n0}.
\end{equation}
This leads to
\begin{equation}
    \text{Tr} \big( (-1)^F e^{-2 \pi t L_0^{g}}  b_{22}(\sigma) \,  c^2(\hat{\sigma}) \big) = 2 \text{Tr}_\downarrow \big( (-1)^F e^{-2 \pi t L_0^g} \big).
    \label{intermediate ghost result 1}
\end{equation}
No dependence on the worldsheet coordinates remains. We now proceed in a way analogous to the bosonic computation in Section \ref{sec63}. First write
\begin{equation}
    L_0^g = \sum_{n=1}^\infty n \big( b_{-n} c_n + c_{-n}  b_n \big) = \sum_{n=1}^\infty n \big( \mathbf{N}^c_n + \mathbf{N}^b_n \big), 
\end{equation}
where $\mathbf{N}^c_n = b_{-n} c_n$ and $\mathbf{N}^b_n = c_{-n} b_n$ are operators that count the number of $c$ and $b$ excitations at the particular mode number $n \geq 0$:
\begin{align}
    [\mathbf{N}^c_n,c_{-m}] = \delta_{nm}   c_{-m}, \hspace{0.5cm} [\mathbf{N}^b_n,b_{-m}] = \delta_{nm} b_{-m}, \hspace{0.5cm} [\mathbf{N}^c_n,b_{-m}] &= [\mathbf{N}^b_n,c_{-m}] =0, \notag\\[5pt]  &\hspace{1.05cm} (n,m \geq 0).
\end{align}
Since the modes are fermionic, $\mathbf{N}^c_n$ and $\mathbf{N}^b_n$ can only be 0 or 1. We have
\begin{equation}
    \text{Tr}_\downarrow \big( (-1)^F e^{-2 \pi t L_0^g} \big) = \prod_{n=1}^\infty \text{Tr}^{(n)}_\downarrow \big( (-1)^F e^{-2 \pi t n (\mathbf{N}^c_n + \mathbf{N}^b_n )} \big),
\end{equation}
where $\text{Tr}^{(n)}_\downarrow $ means that the trace is taken only inside the subspace of fixed mode number $n$. Each of these subspaces only has four states, which can be labelled by their $(\mathbf{N}^c_n,\mathbf{N}^b_n)$ eigenvalues. They are $(0,0)$, $(1,0)$, $(0,1)$ and $(1,1)$. The first and last contain an even number of modes and therefore have $(-1)^F=1$, whereas the second and third have an odd number of modes, so they have $(-1)^F=-1$. This results in
\begin{align}
    \text{Tr}_\downarrow \big( (-1)^F e^{-2 \pi t L_0^g} \big) &= \prod_{n=1}^\infty \big( e^0 - e^{-2 \pi t n} - e^{-2 \pi t n} + e^{-4 \pi t n} \big) \notag\\[5pt]
    &= \prod_{n=1}^\infty \big( 1 - e^{-2 \pi t n}  \big)^2 \notag\\[5pt]
    &= e^{\frac{\pi t}{6}} \eta(it)^2,
\end{align}
where $\eta(x)$ is defined in \eqref{eta function}. Plugging this into \eqref{intermediate ghost result 1} and that into \eqref{intermediate ghost result 2} finally gives us the result
\begin{equation}
    \mathcal{A}^g_\text{open}(t) = \frac{1}{4 \pi^2 t} e^{-2 \pi t a^g} \int d^2 \sigma \, e^{\frac{\pi t}{6}} \, \eta(it)^2 = e^{-2 \pi t \big(a^g - \frac{1}{12}\big)} \, \eta(it)^2.
\end{equation}

\section{Closed string computation}\label{B2}

The ghost contribution to the closed string amplitude of Section \ref{sec64} is 
\begin{equation}
    \mathcal{A}^g_\text{closed}(s) = \frac{1}{2 \pi s} \int d^2\xi \int \mathcal{D}b \, \mathcal{D}c \, e^{- S_g[b,c]  } \,  b_{22}(\xi)  c^1(\hat{\xi}),
\end{equation}
where the $d^2 \xi$ integral is over a cylinder defined by $[0,2 \pi] \times [0,s]$ and the path integral over the ghosts is understood to interpolate between boundary states which impose the open string boundary conditions $b_{12}=c^2=0$ at $\xi^2=0,s$. In operator language we have
\begin{equation}
    \mathcal{A}^g_\text{closed}(s) = \frac{1}{2 \pi s} \int d^2\xi  \int_0^{2 \pi} \frac{d \theta}{2 \pi} \,  \braket{B|e^{-s (L^g_0 +\tilde{L}^g_0+2a^g )} \, e^{i \theta(L^g_0 -\tilde{L}^g_0)} \,   b_{22}(\xi)  c^1(\hat{\xi}) |B},
    \label{Ag closed}
\end{equation}
where the ghost boundary state $\ket{B}$ satisfies
\begin{equation}
    b_{12}(\xi^1,0) \ket{B} = c^2(\xi^1,0) \ket{B} =0.
    \label{ghost boundary state conditions}
\end{equation}
The mode expansions of the closed string ghosts in Euclidean signature are
\begin{align}
    b_{1 1}(\xi) &= - b_{22}(\xi)=  \sum_n \Big( b_n e^{in( \xi^1 + i \xi^2)} + \tilde{b}_n e^{-in(\xi^1 - i \xi^2)} \Big)  \notag\\[5pt]
    b_{1 2} (\xi) &= b_{2 1}(\xi) = i \sum_n \Big( b_n e^{in( \xi^1 + i \xi^2)} - \tilde{b}_n e^{-in(\xi^1 - i \xi^2)}  \Big) \notag\\[5pt]
    c^1(\xi) &= -\frac{1}{2} \sum_n \Big( c_n e^{in( \xi^1 + i \xi^2)}- \tilde{c}_n e^{-in(\xi^1 - i \xi^2)}   \Big) \notag\\[5pt]
    c^2(\xi) &= \frac{i}{2} \sum_n \Big( c_n e^{in( \xi^1 + i \xi^2)} + \tilde{c}_n e^{-in(\xi^1 - i \xi^2)}   \Big) .
\end{align}
In terms of these the conditions \eqref{ghost boundary state conditions} become
\begin{align}
    ( b_n - \tilde{b}_{-n} ) \ket{B} = ( c_n + \tilde{c}_{-n} ) \ket{B} =0.
    \label{ghost boundary conditions (modes)}
\end{align}
We can solve for $\ket{B}$ using the trick explained in footnote \ref{boundary state trick footnote}, adapted for fermionic variables. Fix some $n>0$. The algebraic relations $\{ b_n ,c_{-n} \} = \{ \tilde{b}_n ,\tilde{c}_{-n} \} = 1$ are preserved under the identifications
\begin{align}
    b_{-n} \to \theta, && c_n \to \partial_\theta , && c_{-n} \to \epsilon, && b_n \to \partial_\epsilon \notag\\[5pt]
    \tilde{b}_{-n} \to \tilde{\theta}, && \tilde{c}_n \to \partial_{\tilde{\theta}} , && \tilde{c}_{-n} \to \tilde{\epsilon}, && \tilde{b}_n \to \partial_{\tilde{\epsilon}},
\end{align}
where $\theta$ and $\epsilon$ are Grassmann variables, with respect to which derivatives are to be taken from the left, meaning that $\partial_\theta (\theta \epsilon)=\epsilon $, $\partial_\theta (\epsilon \theta)=-\epsilon $. The boundary state $\ket{B}$ is then identified with a function $f=f(\theta,\tilde{\theta},\epsilon,\tilde{\epsilon})$. Under these identifications the conditions \eqref{ghost boundary conditions (modes)} become the four differential equations
\begin{equation}
    (\partial_\theta + \tilde{\epsilon}) f = (\partial_\epsilon - \tilde{\theta} )f = (\partial_{\tilde{\theta}} + \epsilon) f = (\partial_{\tilde{\epsilon}} - \theta)f=0,
\end{equation}
which are solved by $f \sim e^{-(\theta \tilde{\epsilon} + \tilde{\theta} \epsilon )}$. This determines 
\begin{equation}
    \ket{B} = e^{ - \sum_{n=1}^\infty ( b_{-n} \tilde{c}_{-n} + \tilde{b}_{-n} c_{-n}  )  } \ket{B_0},
\end{equation}
where $\ket{B_0}$ is the ghost vacuum part of $\ket{B}$, which solves $( b_0 - \tilde{b}_0 ) \ket{B_0} = ( c_0 + \tilde{c}_0 ) \ket{B_0} =0$. Starting from the general ansatz
\begin{equation}
    \ket{B_0} = (A + B c_0 + C\tilde{c}_0 + Dc_0 \tilde{c}_0)\ket{\downarrow\downarrow} + (E + F b_0 + G \tilde{b}_0 + H b_0\tilde{b}_0) \ket{\uparrow \uparrow},
\end{equation}
    the zero-mode constraints lead to $A=D=E=H=0$ and $B=C$, $F=-G$, so the form of $\ket{B_0}$ reduces to a combination of $(c_0 + \tilde{c}_0)\ket{\downarrow \downarrow}$ and $(b_0-\tilde{b}_0) \ket{\uparrow \uparrow}$. For the purposes of computing $\mathcal{A}^g_\text{closed}(s)$ the precise values of the coefficients in this combination are not required, since they would only contribute to the overall prefactor in front of the amplitude, which cannot be determined from solving \eqref{ghost boundary conditions (modes)} for the boundary state anyway. Since the Virasoro generators $L_0^g$ and $\tilde{L}_0^g$ commute with all ghost zero modes and annihilate the vacua, we know that the part of $\mathcal{A}^g_\text{closed}(s)$ that involves the ghost vacua is proportional to $\braket{B_0|(b_0+\tilde{b}_0)(c_0-\tilde{c}_0)|B_0}$, with the zero-modes coming from $b_{22}(\xi)  c^1(\hat{\xi})$. It is straightforward to check that due to these zero-modes one obtains a nonzero result only when the $(c_0 + \tilde{c}_0)$ part of either $\ket{B_0}$ or $\bra{B_0}$ hits the $(b_0 - \tilde{b}_0)$ part of the other. Because of this, it is consistent, up to an overall multiplicative term in the amplitude, to set $\ket{B_0} = (c_0 + \tilde{c}_0 ) \ket{\downarrow \downarrow}$ and $\bra{B_0} = \bra{\uparrow \uparrow}(b_0 - \tilde{b}_0)$.\footnote{A less ad hoc justification for these comes from a BRST analysis of the worldsheet theory, which requires one to consider the matter and ghost sectors simultaneously, not separately as we have done. One then finds that on-shell states must satisfy $b_0 \ket{\psi} = \tilde{b}_0 \ket{\psi}=0$ (see Section 5.2 of \cite{Blumenhagen:2013fgp} for a proof). These conditions eliminate all states built on top of the $\ket{\uparrow \uparrow}$ vacuum from the theory. If one builds the boundary states starting from this constrained Hilbert space, the conditions \eqref{ghost boundary conditions (modes)} uniquely determine $\ket{B_0} = (c_0 + \tilde{c}_0) \ket{\downarrow \downarrow}$. The inner product structure show in footnote \ref{ghost inner product footnote} means that the dual state $\bra{B_0}$ should then be constructed over the $\bra{\uparrow \uparrow}$ vacuum, which sets $\bra{B_0} = \bra{\uparrow \uparrow} (b_0 - \tilde{b}_0)$. \label{b ghost footnote} } The matrix element in $\mathcal{A}^g_\text{closed}(s)$ is therefore given by
    \begin{equation}
        \prod_{n=1}^\infty \braket{\uparrow \uparrow |(b_0 -\tilde{b}_0) e^{b_n \tilde{c}_n} e^{\tilde{b}_n c_n} e^{-s(L_0^g + \tilde{L}_0^g)} b_{22}(\xi) c^1(\hat{\xi}) e^{-b_{-n} \tilde{c}_{-n} } e^{-\tilde{b}_{-n}  c_{-n}} (c_0 + \tilde{c}_0)| \downarrow \downarrow }.
        \label{tree amplitude big formula}
    \end{equation}
    We set the overall normalization of the boundary states to 1, since it can be absorbed into $N_p$, the normalization of the bosonic matter boundary states defined in \eqref{bosonic boundary state}. The $d \theta$ integral in \eqref{Ag closed} is trivial because the boundary states satisfy $L_0^g=\tilde{L}^g_0$. Note that the $\ket{B_0}$ vacuum has zero norm, in the sense that
 \begin{equation}
    \braket{B_0|B_0} = \braket{\uparrow \uparrow|(b_0 -\tilde{b}_0) (c_0 + \tilde{c}_0))|\downarrow \downarrow} =0.
 \end{equation}
It follows that any expression of the form $\braket{\uparrow \uparrow|(b_0 -\tilde{b}_0) O (c_0 + \tilde{c}_0))|\downarrow \downarrow}$ vanishes if $O$ can be (anti-)commuted past either $(b_0 - \tilde{b}_0)$ or $(c_0+\tilde{c}_0)$. This is the case of all terms in the expansion of $b_{22}(\xi) c^1(\hat{\xi})$ except for the one only containing the zero modes, so we can substitute
\begin{equation}
    b_{22}(\xi) c^1(\hat{\xi}) \to \frac{1}{2} (b_0 + \tilde{b}_0)(c_0 - \tilde{c}_0)
\end{equation}
 in \eqref{tree amplitude big formula}. All dependence on the coordinates is then gone and the $d^2 \xi$ integral just gives the worldsheet area $\int d^2 \xi = 2 \pi s$. We are left with
 \begin{equation}
     \mathcal{A}^g_\text{closed}(s) = 2 e^{-2 s a^g} \prod_{n=1}^\infty \braket{\uparrow \uparrow |e^{b_n \tilde{c}_n} e^{\tilde{b}_n c_n} e^{-s(L_0^g + \tilde{L}_0^g)} e^{-b_{-n} \tilde{c}_{-n} } e^{-\tilde{b}_{-n}  c_{-n}}| \downarrow \downarrow }
     \label{tree amplitude not so large formula}
 \end{equation}
after using the algebra to simplify the zero-mode terms.\footnote{One actually obtains minus this expression, due to the fact that
\begin{equation}
    \braket{\uparrow \uparrow |(b_0-\tilde{b}_0) (b_0+\tilde{b}_0) (c_0 -\tilde{c}_0) (c_0 + \tilde{c}_0) | \downarrow \downarrow}=-4 \braket{\uparrow \uparrow | \downarrow \downarrow}=-4.
\end{equation}
However, the contribution of the ghosts to a path integral is only a way of representing the Faddeev-Popov determinant, which is a Jacobian factor that arises from changing variables, and therefore contributes with its absolute value. We should therefore discard any overall signs.} Due to the fermionic nature of the modes, all terms of higher than linear order in the exponentials vanish
\begin{equation}
     e^{-b_{-n} \tilde{c}_{-n} }  e^{-\tilde{b}_{-n}  c_{-n}} = (1 -b_{-n} \tilde{c}_{-n} )(1 - \tilde{b}_{-n} c_{-n}) = 1 - b_{-n} \tilde{c}_{-n} - \tilde{b}_{-n} c_{-n} + b_{-n} \tilde{c}_{-n}  \tilde{b}_{-n} c_{-n}.
\end{equation}
The basic commutation relations $[L^g_0, b_n] = -n b_n$, $[L^g_0,c_n] = -n c_n$ (see Appendix \ref{C2}) lead to 
\begin{align}
    e^{-s(L_0^g + \tilde{L}_0^g)} e^{-b_{-n} \tilde{c}_{-n} } e^{-\tilde{b}_{-n}  c_{-n}} \ket{\downarrow \downarrow} = \big(1 - e^{-2ns} b_{-n} \tilde{c}_{-n} -& e^{-2ns}  \tilde{b}_{-n} c_{-n} \notag\\[5pt]
    +& e^{-4ns} b_{-n} \tilde{c}_{-n}  \tilde{b}_{-n} c_{-n} \big) \ket{\downarrow \downarrow}
\end{align}
Plugging this back into \eqref{tree amplitude not so large formula} and similarly expanding $e^{b_n \tilde{c}_n} e^{\tilde{b}_n c_n} $ leads to many terms, most of which vanish. The ones that do not can be simplified using $b_{n} c_{-n} = -c_{-n} b_n +1$. The result is
\begin{align}
    \mathcal{A}^g_\text{closed}(s) &= 2 e^{-2 s a^g} \prod_{n=1}^\infty \big( 1 - 2e^{-2ns} + e^{-4 n s} \big) \notag\\[5pt]
    &= 2 e^{-2 s a^g} \prod_{n=1}^\infty \big( 1 -e^{-2ns} \big)^2 \notag\\[5pt]
    &=2 e^{-2s \big( a^g - \frac{1}{12} \big)} \eta \big( is/\pi \big)^2.
\end{align}

%\end{document}

%\documentclass[a4paper,12pt]{memoir}
%\usepackage{graphicx}
%\usepackage[utf8]{inputenc}
%\usepackage{indentfirst}
%\usepackage{braket}
%\usepackage{setspace}
%\usepackage{amsmath, amsthm, amssymb, amsfonts,bm}
%\usepackage[multiple]{footmisc}
%\usepackage{mathtools, changepage, slashed}
%\usepackage{tikz-feynman}
%\usepackage{bm, mathrsfs}
%\usepackage{bbold}
%\usepackage{gensymb}
%\usepackage[a4paper,top=3cm,left=3cm,right=2cm,bottom=2cm]{geometry}
%\usepackage{epstopdf}
%\usepackage{hyperref}
%\usepackage{pgfplots}
%\pgfplotsset{compat=1.18} 
%\usepackage[sorting=none]{biblatex}
%\addbibresource{refs.bib}
%\numberwithin{equation}{section}
%\usepackage[inkscapelatex=false]{svg}
%\usepackage[super]{natbib}
%\usepackage{doi}
%\hypersetup{
%  colorlinks   = true, %Colours links instead of ugly boxes
%  urlcolor     = black, %Colour for external hyperlinks
%  linkcolor    = black, %Colour of internal links
%  citecolor   = black %Colour of citations
%}
%\DeclareMathOperator{\Tr}{Tr}

%\newcommand{\normalord}[1]{%
%  {:\mathrel{\mspace{1mu}#1\mspace{1mu}}:}%
%}

%\OnehalfSpacing
%\usepackage{newtx}
%\usepackage{newtxtext}
%\usepackage{lmodern}

%\title{Spinors in 10 dimensions}
%\author{pedrobairrao}

%\begin{document}

\chapter{Ten-dimensional spinors}\label{ApC}

In this appendix, we build the ten-dimensional Clifford algebra in a Majorana-Weyl representation and derive Fierz identities of relevance to the discussion of the Ramond-Ramond vacuum of the RNS superstring. We also prove an identity for a product of gamma matrices that is used in Section \ref{sec72} to find the spacetime equations of motion for the Ramond-Ramond field strengths.

\section{Clifford algebra and Fierz identities}\label{C1}

In ten spacetime dimensions a Dirac spinor has $2^{10/2}=32$ components, so the Dirac matrices are $32 \times 32$. A Majorana representation for them with all matrices real is given by 
\begin{align}
    \Gamma^0 &= i \sigma_2 \otimes \sigma_2 \otimes \sigma_2 \otimes \sigma_2 \otimes \sigma_2 \notag\\
    \Gamma^1 &=  \sigma_2 \otimes \sigma_2 \otimes \sigma_2 \otimes \sigma_2 \otimes \sigma_1 \notag\\
    \Gamma^2 &= \sigma_2 \otimes \sigma_2 \otimes \sigma_2 \otimes \sigma_2 \otimes \sigma_3 \notag\\
    \Gamma^3 &= \sigma_2 \otimes \sigma_2 \otimes \sigma_0 \otimes \sigma_1 \otimes \sigma_0 \notag\\
    \Gamma^4 &= \sigma_2 \otimes \sigma_2 \otimes \sigma_0 \otimes \sigma_3 \otimes \sigma_0 \notag\\
    \Gamma^5 &= \sigma_2 \otimes \sigma_1 \otimes \sigma_2 \otimes \sigma_0 \otimes \sigma_0 \label{Explicit Dirac matrices} \\
    \Gamma^6 &= \sigma_2 \otimes \sigma_3 \otimes \sigma_2 \otimes \sigma_0 \otimes \sigma_0 \notag\\
    \Gamma^7 &= \sigma_2 \otimes \sigma_0 \otimes \sigma_1 \otimes \sigma_2 \otimes \sigma_0 \notag\\
    \Gamma^8 &= \sigma_2 \otimes \sigma_0 \otimes \sigma_3 \otimes \sigma_2 \otimes \sigma_0 \notag\\
    \Gamma^9 &= \sigma_1 \otimes \sigma_0 \otimes \sigma_0 \otimes \sigma_0 \otimes \sigma_0 \notag,
\end{align}
where
\begin{align}
    \sigma^1 = 
    \begin{pmatrix}
        0 && 1 \\
        1 && 0
    \end{pmatrix} \, , &&
    \sigma^2 = 
    \begin{pmatrix}
        0 && -i \\
        i && 0
    \end{pmatrix} \, , &&
    \sigma^3 = 
    \begin{pmatrix}
        1 && 0 \\
        0 && -1
    \end{pmatrix} \, 
\end{align}
are the Pauli matrices and $\sigma_0 = - \mathbb{1}_{2 \times 2 }$ is the timelike component of the Pauli vector $\sigma_\mu = (-\mathbb{1}_{2 \times 2},\sigma^i)$. From these one may form the generators of the full 10-dimensional Clifford algebra:
\begin{equation}
    \big\{ \mathbb{1}_{32 \times 32} \, , \Gamma^{\mu_1}, \Gamma^{\mu_1 \mu_2}, \dots, \Gamma^{\mu_1 \dots \mu_{10}} \big\},
\end{equation}
where
\begin{equation}
    \Gamma^{\mu_1 \dots \mu_n} = \Gamma^{[ \mu_1} \dots \Gamma^{\mu_n ]}.
\end{equation}
Using the mixed product identity
\begin{equation}
    (A_1 \otimes B_1 \otimes \dots ) (A_2 \otimes B_2  \otimes \dots ) \dots = (A_1 A_2 \dots ) \otimes (B_1 B_2 \dots ) \otimes \dots
\end{equation}
it is easy to check that the chiral matrix is block-diagonal
\begin{equation}
    \Gamma = \Gamma^0 \Gamma^1 \Gamma^2 \Gamma^3 \Gamma^4 \Gamma^5 \Gamma^6 \Gamma^7 \Gamma^8 \Gamma^9 = \sigma_3 \otimes \mathbb{1}_{16 \times 16} = 
    \begin{pmatrix}
        \mathbb{1}_{16 \times 16} && 0 \\
        0 && -\mathbb{1}_{16 \times 16}
    \end{pmatrix},
\end{equation}
which shows that this representation is actually Majorana-Weyl. 

We denote each component of a Majorana spinor as $\varphi_A$. The Dirac matrices are then written as $(\Gamma^\mu)_A{}^B$, so the basic anticommutator of the Clifford algebra reads
\begin{equation}
    \{ \Gamma^\mu,\Gamma^\nu \}_A{}^B = (\Gamma^\mu)_A{}^C (\Gamma^\nu)_C{}^B + (\Gamma^\nu)_A{}^C (\Gamma^\mu)_C{}^B = 2\eta^{\mu \nu} \delta_A^B.
\end{equation}
When dealing with Majorana spinors the charge conjugation matrix $C$ plays an important role. It is defined by the requirement that
\begin{equation}
    (\Gamma^\mu)^\text{T} =  - C \,  \Gamma^\mu C^{-1}.
\end{equation}
If all Dirac matrices are real, $(\Gamma^0)^{\text{T}}=-\Gamma^0$ and $(\Gamma^i)^{\text{T}}=\Gamma^i$, so the above condition is satisfied by $C=\Gamma^0$. This means that both $C \Gamma^\mu$ and $\Gamma^\mu C^{-1}$ are symmetric:
\begin{align}
    &(C\Gamma^\mu)^{\text{T}} = (\Gamma^\mu)^\text{T} C^{\text{T}} = (-1)^2 C \,  \Gamma^\mu C^{-1} C=  C \Gamma^\mu, \\[5pt]
    &(\Gamma^\mu C^{-1})^\text{T} = -(C^{-1})^\text{T} C \Gamma^\mu C^{-1} = - C^2  \Gamma^\mu C^{-1} =  \Gamma^\mu C^{-1}.
\end{align}
Although $C$ and $\Gamma^0$ are the same component-wise, we write $\Gamma^0$ with one lower and one upper index, while components of the charge conjugation matrix are written as $C^{AB}$, and those of the inverse $C^{-1}$ are $C_{AB}$. From two Majorana spinors $\psi$ and $\varphi$, a Lorentz scalar is built by contracting one with the Majorana conjugate of the other, defined as as $(\overline{\psi})^A \equiv \psi^A =\psi_B  C^{BA}$.\footnote{For real spinors the Majorana conjugate coincides with the Dirac conjugate, $\overline{\psi}= \psi^\text{T} C =\psi^\dagger \Gamma^0$.} This scalar is
\begin{equation}
    \overline{\psi}  \varphi = \psi_A C^{AB} \varphi_B.
\end{equation}
$C^{AB}$ and $C_{AB}$ therefore play the role of a metric for spinor indices, and can be used to raise and lower them. Note however that the charge conjugation matrix is antisymmetric, so $C^{AB} \psi_B = - \psi_B C^{BA}$. We fix this ambiguity by demanding that index raising and lowering be done always with the upper index on the right and the lower index on the left: 
\begin{equation}
    \psi^A = \psi_B C^{BA} \, , \hspace{0.5cm} \psi_A = C_{AB} \psi^B.
    \label{Maj index lowering}
\end{equation}
Consistency of these two definitions requires $C_{BA} C^{BC} = \delta^C_A$, which simply expresses the fact that $C^\text{T}=C^{-1}$ component-wise. Using these conventions we have that
\begin{equation}
    \overline{\psi} \, \varphi = \psi_A C^{AB} \varphi_B =  \psi^A C_{AB} \varphi^B = \psi^A \varphi_A = -\psi_A \varphi^A.
\end{equation}

The normalized Clifford algebra
\begin{equation}
   \hspace{0.8cm} \frac{1}{\sqrt{32}} \Big\{\delta^B_A \, , (\Gamma^{\mu_1})_A{}^B, (\Gamma^{\mu_1 \mu_2})_A{}^B, \dots, (\Gamma^{\mu_1 \dots \mu_{10}})_A{}^B \Big\} \hspace{0.5cm} \text{with} \hspace{0.3cm} \mu_1 < \mu_2 < \dots < \mu_{10}
\end{equation}
provides an orthogonal basis for the space of real $32 \times 32$ matrices $M_A{}^B$ with respect to the inner product $(M,N) = \text{Tr} ( M^{\text{T}} N )$ \cite{Wipfnotes}.\footnote{This inner product induces a norm for real matrices which is a direct generalization of the usual real vector norm given by summing over the square of each component, $|v|^2 = \sum_i (v^i)^2$,
\begin{equation}
    |M|^2 = (M,M) = (M^\text{T})_A{}^B M_B{}^A  = \sum_{A,B} (M_B{}^A)^2.
\end{equation}
} The expansion of an arbitrary matrix in this basis is given by
\begin{equation}
    M_A{}^B = \frac{1}{32} \sum_{n=0}^{10} \big( \Gamma^{\mu_1 \dots \mu_n}, M \big) \big(\Gamma^{\mu_1 \dots \mu_n}\big)_A{}^B, \hspace{0.5cm} \text{(no sum)}
\end{equation}
where $\mu_i < \mu_{i+1}$ and we are not summing over the Lorentz indices. If all the $\mu_i$ are distinct, we have that
\begin{equation}
    \big(\Gamma^{\mu_1} \dots \Gamma^{\mu_n}\big)^{\text{T}} = \Gamma_{\mu_n} \dots \Gamma_{\mu_1} = (-1)^{\frac{n(n-1)}{2}} \Gamma_{\mu_1} \dots \Gamma_{\mu_n},
\end{equation}
and can therefore rewrite the expansion as
\begin{equation}
    M_A{}^B = \frac{1}{32} \sum_{n=0}^{10} \frac{1}{n!} (-1)^{\frac{n(n-1)}{2}} \text{Tr} \big(\Gamma_{\mu_1 \dots \mu_n} M \big) \big( \Gamma^{\mu_1 \dots \mu_n} \big)_A{}^B,
\end{equation}
where this time we are using the Einstein summation convention for the Lorentz indices, which leads to an overcounting in the sum that is canceled by the factor of $1/n!$.

In ten dimensions the Hodge dual of an antisymmetric tensor $T_{\mu_1 \dots \mu_n}$ of rank $n$ is given by\footnote{We use $\epsilon_{0123456789}=- \epsilon^{0123456789}=1$.} \cite{nakahara2003geometry}
\begin{equation}
    (\ast\, T)_{\mu_{n+1} \dots \mu_{10}} = \frac{1}{(10-n)!} \epsilon_{ \mu_{n+1} \dots \mu_{10} \mu_1 \dots \mu_n } T^{\mu_1 \dots \mu_{n}},
\end{equation}
which has rank $10-n$. Both carry the same information, since the Hodge star operator is invertible. One may use this fact to trade an antisymmetric tensor with more than five indices for one with less. This is implemented in the Clifford algebra with the chiral matrix. First, note that the rank 10 matrix is simply
\begin{equation}
    \Gamma^{\mu_1 \dots \mu_{10}} = \Gamma^{\mu_1} \dots \Gamma^{\mu_{10}} = - \epsilon^{\mu_1 \dots \mu_{10}} \Gamma,
\end{equation}
where we assume with no loss of generality  that all $\mu_i$ are distinct. Let us momentarily suspend the Einstein summation convention, so that all Dirac matrices satisfy $\Gamma_\mu \Gamma^\mu = \mathbb{1}_{32 \times 32} $. We may then write $\Gamma^{\mu_1 \dots \mu_n} $ as 
\begin{align}
    \Gamma^{\mu_1} \dots \Gamma^{\mu_n} &= \Gamma^{\mu_1} \dots \Gamma^{\mu_n} \Gamma_{\mu_{n+1}} \Gamma^{\mu_{n+1}} \Gamma_{\mu_{n+2}} \Gamma^{\mu_{n+2}} \dots \Gamma_{\mu_{10}} \Gamma^{\mu_{10}} \notag\\[5pt]
    &= - (-1)^{\frac{n(n-1)}{2}} \Gamma^{\mu_1} \Gamma^{\mu_n} \Gamma^{\mu_{n+1}} \dots \Gamma^{\mu_{10}} \Gamma_{\mu_{n+1}} \dots \Gamma_{\mu_{10}} \notag\\[5pt]
    &=   (-1)^{\frac{n(n-1)}{2}}  \epsilon^{\mu_1 \dots \mu_n \mu_{n+1} \dots \mu_{10}} \Gamma  \Gamma_{\mu_{n+1}} \dots \Gamma_{\mu_{10}} \hspace{0.5cm} \text{(no sum)},
\end{align}
which upon restoring the summation convention becomes
\begin{equation}
    \Gamma^{\mu_1 \dots \mu_n} =  (-1)^{\frac{n(n-1)}{2}} \frac{1}{(10-n)!} \epsilon^{\mu_1 \dots \mu_n \mu_{n+1} \dots \mu_{10}} \Gamma \Gamma_{\mu_{n+1} \dots \mu_{10}},
\end{equation}
with the factor of $1/(10-n)!$ added to compensate the overcounting due to summing over Lorentz indices. Multiplying both sides by $\Gamma$ gives
\begin{equation}
    \Gamma \Gamma^{\mu_1 \dots \mu_n} = (-1)^{\frac{n(n-1)}{2}} (\ast \Gamma)^{\mu_1 \dots \mu_n}.
\end{equation}
These formulas, together with the Levi-Civita contraction identity
\begin{equation}
    \epsilon_{\mu_1 \dots \mu_n \nu_{n+1} \dots \nu_{10}} \epsilon^{\mu_1 \dots \mu_n \rho_{n+1} \dots \rho_{10} } = -n! (10-n)! \delta_{\nu_{n+1}}^{[ \rho_{n+1}} \dots \delta_{\nu_{10}}^{\rho_{10}]},
\end{equation}
allow the expansion of $M_A{}^B$ to be written as
\begin{align}
    M_A{}^B =  &\frac{1}{32} \bigg[ \text{Tr}(M) \delta_A^B + \text{Tr}(\Gamma_{\mu_1} M) ( \Gamma^{\mu_1})_A{}^B - \frac{1}{2!} \text{Tr}(\Gamma_{\mu_1 \mu_2} M) ( \Gamma^{\mu_1 \mu_2} )_A{}^B \notag\\[5pt] 
    &- \frac{1}{3!} \text{Tr}(\Gamma_{\mu_1 \mu_2 \mu_3} M) ( \Gamma^{\mu_1 \mu_2 \mu_3} )_A{}^B  + \frac{1}{4!} \text{Tr}(\Gamma_{\mu_1 \mu_2 \mu_3 \mu_4} M) ( \Gamma^{\mu_1 \mu_2 \mu_3 \mu_4} )_A{}^B  \notag\\[5pt]
    & + \frac{1}{5!} \text{Tr}(\Gamma_{\mu_1 \mu_2 \mu_3 \mu_4 \mu_5} M) ( \Gamma^{\mu_1 \mu_2 \mu_3 \mu_4 \mu_5} )_A{}^B + \frac{1}{4!} \text{Tr}(\Gamma \Gamma_{\mu_1 \mu_2 \mu_3 \mu_4} M) ( \Gamma \Gamma^{\mu_1 \mu_2 \mu_3 \mu_4} )_A{}^B \notag\\[5pt]
    &+ \frac{1}{3!} \text{Tr}( \Gamma \Gamma_{\mu_1 \mu_2 \mu_3} M) ( \Gamma \Gamma^{\mu_1 \mu_2 \mu_3} )_A{}^B - \frac{1}{2!} \text{Tr}( \Gamma \Gamma_{\mu_1 \mu_2}M) ( \Gamma \Gamma^{\mu_1 \mu_2} )_A{}^B \notag\\[5pt]
    &- \text{Tr}( \Gamma \Gamma_{\mu_1} M) ( \Gamma \Gamma^{\mu_1})_A{}^B + \text{Tr}( \Gamma M) ( \Gamma )_A{}^B  \bigg].
    \label{Matrix Majorana expansion}
\end{align}
The analogous expressions for different index structures can be easily obtained from this using the rules \eqref{Maj index lowering} for raising and lowering the indices. The most common application of expressions such as these is the derivation of Fierz identities, which follow from $M_A{}^B = \varphi_A \psi^B$. They are also useful for finding the most general form of supersymmetry algebras, for which one sets $M_A{}^B= \{ Q_A , Q^B\}$. 

In even dimensions the Dirac representation is reducible. One may split the $A$ index into $(\alpha,\dot{\alpha})$, where $\alpha$ goes over the 16 components of a right-handed Weyl spinor and $\dot{\alpha}$ goes over the 16 components of left-handed Weyl spinor,
\begin{equation}
\psi_A=
    \begin{pmatrix}
        \theta_\alpha \\
        \chi_{\dot{\alpha}}
    \end{pmatrix},
\end{equation}
each one defined by its eigenvalue with respect to $\Gamma$, $+1$ for right-handed, $-1$ for left-handed. In the representation \eqref{Explicit Dirac matrices} the Dirac matrices are all block-antisymmetric:
\begin{equation}
    (\Gamma^\mu)_A{}^B = 
    \begin{pmatrix}
        0 && (\gamma^\mu)_{\alpha}{}^{\dot{\beta}} \\
         (\gamma^\mu)_{\dot{\alpha}}{}^{\beta} && 0
    \end{pmatrix}.
    \label{Dirac matrix Weyl decomposition}
\end{equation}
The $\gamma^\mu$ are the ten-dimensional analogue of the Pauli matrices. The charge conjugation matrix similarly splits as
\begin{equation}
    C^{AB} = \begin{pmatrix}
        0 && \mathcal{C}^{\alpha \dot{\beta}} \\
         \mathcal{C}^{\dot{\alpha} \beta } && 0
    \end{pmatrix}, \hspace{0.5cm} C_{AB} = \begin{pmatrix}
        0 && \mathcal{C}_{\alpha \dot{\beta}} \\
         \mathcal{C}_{\dot{\alpha} \beta } && 0
    \end{pmatrix}.
\end{equation}
By taking products of \eqref{Dirac matrix Weyl decomposition} one finds that the form of the elements of the Clifford algebra change depending on whether the rank is even or odd:
\begin{align}
    (\Gamma^{\mu_1 \dots \mu_{2n}} )_A{}^B &=
    \begin{pmatrix}
        (\gamma^{\mu_1 \dots \mu_{2n}} )_{\alpha}{}^{\beta} && 0 \\
        0 && (\gamma^{\mu_1 \dots \mu_{2n}} )_{\dot{\alpha}}{}^{\dot{\beta}}
    \end{pmatrix}, \\[5pt]
    (\Gamma^{\mu_1 \dots \mu_{2n+1}})_A{}^B &=
    \begin{pmatrix}
        0 && (\gamma^{\mu_1 \dots \mu_{2n+1}})_\alpha{}^{\dot{\beta}} \\
        (\gamma^{\mu_1 \dots \mu_{2n+1}})_{\dot{\alpha}}{}^{\beta} && 0 
    \end{pmatrix},
\end{align}
where $\gamma^{\mu_1 \dots \mu_n} \equiv \gamma^{[ \mu_1} \dots \gamma^{\mu_n ]}$. With these expressions one may plug
\begin{equation}
    M_A{}^B =
    \begin{pmatrix}
        0 & 0 \\
        \mathcal{M}_{\dot{\alpha}}{}^{\beta} & 0
    \end{pmatrix}
    \label{first block M}
\end{equation}
into \eqref{Matrix Majorana expansion} to find the expansion of $\mathcal{M}_{\dot{\alpha}}{}^{\beta}$ in terms of antisymmetric tensors. The even rank coefficients $\text{Tr}(\Gamma_{\mu_1 \dots \mu_{2n}} M)$ all vanish, because
\begin{equation}
    (\Gamma_{\mu_1 \dots \mu_{2n}} M)_A{}^B =
    \begin{pmatrix}
        (\gamma_{\mu_1 \dots \mu_{2n}} )_{\alpha}{}^{\beta} && 0 \\
        0 && (\gamma_{\mu_1 \dots \mu_{2n}} )_{\dot{\alpha}}{}^{\dot{\beta}}
    \end{pmatrix}
    \begin{pmatrix}
        0 & 0 \\
        \mathcal{M}_{\dot{\beta}}{}^{\beta} & 0
    \end{pmatrix}
    =
    \begin{pmatrix}
        0 & 0 \\
        (\gamma_{\mu_1 \dots \mu_{2n}} )_{\dot{\alpha}}{}^{\dot{\beta}} \mathcal{M}_{\dot{\beta}}{}^{\beta} & 0
    \end{pmatrix}
\end{equation}
is traceless. For the odd rank ones we have
\begin{equation}
     (\Gamma_{\mu_1 \dots \mu_{2n+1}} M)_A{}^B =
    \begin{pmatrix}
        0 && (\gamma_{\mu_1 \dots \mu_{2n+1}})_\alpha{}^{\dot{\beta}} \\
        (\gamma_{\mu_1 \dots \mu_{2n+1}})_{\dot{\alpha}}{}^{\beta} && 0 
    \end{pmatrix}
    \begin{pmatrix}
        0 & 0 \\
        \mathcal{M}_{\dot{\beta}}{}^{\beta} & 0
    \end{pmatrix}
    =
    \begin{pmatrix}
        (\gamma_{\mu_1 \dots \mu_{2n+1}})_\alpha{}^{\dot{\beta}} \mathcal{M}_{\dot{\beta}}{}^{\beta} & 0 \\
        0 & 0
    \end{pmatrix},
\end{equation}
so $\text{Tr} (\Gamma_{\mu_1 \dots \mu_{2n+1}} M) = \text{Tr}(\gamma_{\mu_1 \dots \mu_{2n+1}} \mathcal{M} )$. The dualized tensors are given by
\begin{align}
    (\Gamma \Gamma^{\mu_1 \dots \mu_{2n+1}})_A{}^B &= 
    \begin{pmatrix}
        \delta_\alpha^\gamma & 0 \\
        0 & - \delta_{\dot{\alpha}}^{\dot{\gamma}} 
    \end{pmatrix}
    \begin{pmatrix}
        0 && (\gamma^{\mu_1 \dots \mu_{2n+1}})_\gamma{}^{\dot{\beta}} \\
        (\gamma^{\mu_1 \dots \mu_{2n+1}})_{\dot{\gamma}}{}^{\beta} && 0 
    \end{pmatrix}
    \notag\\[10pt]
    &=
    \begin{pmatrix}
        0 && (\gamma^{\mu_1 \dots \mu_{2n+1}})_\alpha{}^{\dot{\beta}} \\
        -(\gamma^{\mu_1 \dots \mu_{2n+1}})_{\dot{\alpha}}{}^{\beta} && 0 
    \end{pmatrix}.
\end{align}
This minus sign does not contribute to the trace, so $\text{Tr} ( \Gamma \Gamma_{\mu_1 \dots \mu_{2n+1}} M) = \text{Tr}(\gamma_{\mu_1 \dots \mu_{2n+1}} \mathcal{M} )$, but it does enter the expansion via $(\Gamma \Gamma^{\mu_1 \dots \mu_{2n+1}})_{\dot{\alpha}}{}^{\beta} =- ( \gamma^{\mu_1 \dots \mu_{2n+1}})_{\dot{\alpha}}{}^{\beta}$, leading to twice the contribution of each odd rank tensor:
\begin{align}
    \mathcal{M}_{\dot{\alpha}}{}^{\beta} = \frac{1}{16} \bigg[ \text{Tr}(\gamma_{\mu_1} \mathcal{M} ) (\gamma^{\mu_1})_{\dot{\alpha}}{}^\beta &- \frac{1}{3!} \text{Tr}(\gamma_{\mu_1 \mu_2 \mu_3} \mathcal{M} ) (\gamma^{\mu_1 \mu_2 \mu_3})_{\dot{\alpha}}{}^\beta \notag\\[0pt]
    &+ \frac{1}{5!} \text{Tr}(\gamma_{\mu_1 \mu_2 \mu_3 \mu_4 \mu_5} \mathcal{M})^{+} ( \gamma^{\mu_1 \mu_2 \mu_3 \mu_4 \mu_5} )_{\dot{\alpha}}{}^\beta
    \bigg].
\end{align}
The plus sign on $\text{Tr}(\gamma_{\mu_1 \dots \mu_5} \mathcal{M})^{+}$ means that it is self-dual. In ten dimensions a five-form $T_{\mu_1 \dots \mu_5}$ has the same rank as its Hodge dual. Therefore, it can always be decomposed into a self-dual part $T^+_{\mu_1 \dots \mu_5}$, which satisfies $(\ast T)^+_{\mu_1 \dots \mu_5} = T^+_{\mu_1 \dots \mu_5}$, and an anti self-dual part $T^-_{\mu_1 \dots \mu_5}$, satisfying $(\ast T)^-_{\mu_1 \dots \mu_5} =- T^-_{\mu_1 \dots \mu_5}$,
\begin{equation}
    T_{\mu_1 \dots \mu_5} = \frac{1}{2}\big( 1 + \ast \big)T_{\mu_1 \dots \mu_5} + \frac{1}{2}\big( 1 - \ast \big)T_{\mu_1 \dots \mu_5} = T^+_{\mu_1 \dots \mu_5} + T^-_{\mu_1 \dots \mu_5}.
\end{equation}
For $\Gamma_{\mu_1 \dots \mu_5}$ one takes the dual by acting with $\Gamma$. The decomposition of $\Gamma_{\mu_1 \dots \mu_5}$ in terms of its self-dual and anti self-dual parts is therefore nothing but the decomposition into chirality eigenstates, and for the tensor $\text{Tr}(\Gamma_{\mu_1 \dots\mu_5} M)$ we have
\begin{align}
    \text{Tr}(\Gamma_{\mu_1 \dots\mu_5} M) &= \text{Tr}(\Gamma_{\mu_1 \dots \mu_5} M)^+ + \text{Tr}(\Gamma_{\mu_1 \dots \mu_5} M)^- \notag\\[5pt]
    &= \frac{1}{2} \text{Tr} \big((\mathbb{1}_{32 \times 32} + \Gamma ) \Gamma_{\mu_1 \dots \mu_5} M \big) + \frac{1}{2} \text{Tr} \big((\mathbb{1}_{32 \times 32} - \Gamma ) \Gamma_{\mu_1 \dots \mu_5} M \big).
\end{align}
For $M_A{}^B$ given by \eqref{first block M}, only the self-dual part contributes:
\begin{equation}
    \text{Tr}(\Gamma_{\mu_1 \dots \mu_5} M) = \text{Tr}(\Gamma_{\mu_1 \dots \mu_5} M)^+ = \text{Tr}(\gamma_{\mu_1 \dots  \mu_5} \mathcal{M})^+.
\end{equation}
An analogous calculation shows that the anti self-dual part appears in\footnote{Whether the five-form in $\mathcal{M}_{\dot{\alpha}}{}^{\beta}$ is the self-dual or anti self-dual one depends on the sign chosen in $\Gamma^{\mu_1 \dots \mu_{10}} = \pm \epsilon^{\mu_1 \dots \mu_{10}} \Gamma$, which is a matter of convention. }
\begin{align}
    \mathcal{M}_{\alpha}{}^{\dot{\beta}} = \frac{1}{16} \bigg[ \text{Tr}(\gamma_{\mu_1} \mathcal{M} ) (\gamma^{\mu_1})_{\alpha}{}^{\dot{\beta}} &- \frac{1}{3!} \text{Tr}(\gamma_{\mu_1 \mu_2 \mu_3} \mathcal{M} ) (\gamma^{\mu_1 \mu_2 \mu_3})_{\alpha}{}^{\dot{\beta}} \notag\\[0pt]
    &+ \frac{1}{5!} \text{Tr}(\gamma_{\mu_1 \mu_2 \mu_3 \mu_4 \mu_5} \mathcal{M})^{-} ( \gamma^{\mu_1 \mu_2 \mu_3 \mu_4 \mu_5} )_{\alpha}{}^{\dot{\beta}}
    \bigg].
\end{align}

Using the charge conjugation matrix to raise the first index of $\mathcal{M}_{\dot{\alpha}}{}^\beta$ results in
\begin{align}
    \mathcal{M}^{\alpha \beta} = \frac{1}{16} \bigg[ \text{Tr}(\gamma_{\mu_1} \mathcal{C}^{-1} \mathcal{M} ) ( \gamma^{\mu_1})^{\alpha \beta} &- \frac{1}{3!} \text{Tr}(\gamma_{\mu_1 \mu_2 \mu_3} \mathcal{C}^{-1} \mathcal{M} ) ( \gamma^{\mu_1 \mu_2 \mu_3})^{\alpha \beta} \notag\\[0pt]
    &+ \frac{1}{5!} \text{Tr}(\gamma_{\mu_1 \mu_2 \mu_3 \mu_4 \mu_5} \mathcal{C}^{-1} \mathcal{M})^{+} ( \gamma^{\mu_1 \mu_2 \mu_3 \mu_4 \mu_5} )^{\alpha \beta}
    \bigg],
\end{align}
where $(\gamma^{\mu_1 \dots \mu_n})^{\alpha \beta} = (\gamma^{\mu_1 \dots \mu_n})_{\dot{\gamma}}{}^\beta \mathcal{C}^{\dot{\gamma} \alpha} = - \mathcal{C}^{ \alpha \dot{\gamma}}  (\gamma^{\mu_1 \dots \mu_n})_{\dot{\gamma}}{}^\beta $, and with the coefficients reexpressed as
\begin{equation}
    (\gamma_{\mu_1 \dots \mu_n})_{\alpha}{}^{\dot{\beta}} \mathcal{M}_{\dot{\beta}}{}^\alpha  = (\gamma_{\mu_1 \dots \mu_n})_{\alpha}{}^{\dot{\beta}} \mathcal{C}_{\dot{\beta} \gamma } \mathcal{M^{\gamma \alpha}} = \text{Tr}(\gamma_{\mu_1 \dots \mu_n} \mathcal{C}^{-1} \mathcal{M}).
\end{equation}

Using instead 
\begin{equation}
    M_A{}^B =
    \begin{pmatrix}
        0 & 0 \\
        0 & \mathcal{M}_{\dot{\alpha}}{}^{\dot{\beta}}
    \end{pmatrix},
\end{equation}
one finds
\begin{align}
    (\Gamma_{\mu_1 \dots \mu_{2n}} M)_A{}^B &=
    \begin{pmatrix}
        (\gamma_{\mu_1 \dots \mu_{2n}} )_{\alpha}{}^{\gamma} && 0 \\
        0 && (\gamma_{\mu_1 \dots \mu_{2n}} )_{\dot{\alpha}}{}^{\dot{\gamma}}
    \end{pmatrix}
    \begin{pmatrix}
        0 & 0 \\
        0 & \mathcal{M}_{\dot{\gamma}}{}^{\dot{\beta}}
    \end{pmatrix}
    =
    \begin{pmatrix}
        0 & 0 \\
        0 & (\gamma_{\mu_1 \dots \mu_{2n}} )_{\dot{\alpha}}{}^{\dot{\gamma}} \mathcal{M}_{\dot{\gamma}}{}^{\dot{\beta}}
    \end{pmatrix}, \\[5pt]
     (\Gamma_{\mu_1 \dots \mu_{2n+1}} M)_A{}^B &=
    \begin{pmatrix}
        0 && (\gamma_{\mu_1 \dots \mu_{2n+1}})_\alpha{}^{\dot{\gamma}} \\
        (\gamma_{\mu_1 \dots \mu_{2n+1}})_{\dot{\alpha}}{}^{\gamma} && 0 
    \end{pmatrix}
    \begin{pmatrix}
        0 & 0 \\
        0 & \mathcal{M}_{\dot{\gamma}}{}^{\dot{\beta}}
    \end{pmatrix}
    =
    \begin{pmatrix}
        0 & (\gamma_{\mu_1 \dots \mu_{2n+1}})_\alpha{}^{\dot{\gamma}} \mathcal{M}_{\dot{\gamma}}{}^{\dot{\beta}} \\
        0 & 0
    \end{pmatrix},
\end{align}
so in this case only the even rank tensors contribute. For the dual tensors one finds a minus sign in the coefficients $\text{Tr}(\Gamma \Gamma_{\mu_1 \dots \mu_{2n}} M ) = - \text{Tr}(\gamma_{\mu_1 \dots \mu_{2n}} \mathcal{M})$ that cancels the one in $(\Gamma \Gamma_{\mu_1 \dots \mu_{2n}} )_A{}^B = - (\gamma_{\mu_1 \dots \mu_{2n}} )_{\dot{\alpha}}{}^{\dot{\beta}}$, leading to
\begin{equation}
    \mathcal{M}_{\dot{\alpha}}{}^{\dot{\beta}} = \frac{1}{16} \bigg[ \text{Tr}(\mathcal{M}) \delta_{\dot{\alpha}}^{\dot{\beta}} - \frac{1}{2!} \text{Tr}(\gamma_{\mu_1 \mu_2} \mathcal{M} ) (\gamma^{\mu_1 \mu_2})_{\dot{\alpha}}{}^{\dot{\beta}} + \frac{1}{4!} \text{Tr}(\gamma_{\mu_1 \mu_2 \mu_3 \mu_4} \mathcal{M} )(\gamma^{\mu_1 \mu_2 \mu_3 \mu_4})_{\dot{\alpha}}{}^{\dot{\beta}} \bigg].
\end{equation}
Raising one index gives
\begin{equation}
    \mathcal{M}^{ \alpha \dot{\beta}} = \frac{1}{16} \bigg[ \text{Tr}( \mathcal{C}^{-1} \mathcal{M} ) \mathcal{C}^{\alpha \dot{\beta}} - \frac{1}{2!} \text{Tr}(\gamma_{\mu_1 \mu_2} \mathcal{C}^{-1} \mathcal{M} ) (\gamma^{\mu_1 \mu_2})^{\alpha \dot{\beta}} + \frac{1}{4!} \text{Tr}(\gamma_{\mu_1 \mu_2 \mu_3 \mu_4} \mathcal{C}^{-1} \mathcal{M} )(\gamma^{\mu_1 \mu_2 \mu_3 \mu_4})^{\alpha \dot{\beta}} \bigg].
\end{equation}

\section{Proof of the gamma product identity}\label{C2}

In Section \ref{sec72} we make use of the identity 
\begin{equation}
    \gamma^\mu \gamma^{\nu_1 \dots \nu_n} = \gamma^{\mu \nu_1 \dots \nu_n} + n \eta^{\mu [ \nu_1} \gamma^{\nu_2 \dots \nu_n]}.
    \label{gamma product identity}
\end{equation}

To prove it, start from
\begin{align}
    \eta_{\mu \nu} \gamma^{[ \nu} \gamma^{ \nu_1 \dots \nu_n]} &= \frac{1}{n+1} \bigg( \gamma_{\mu} \gamma^{[\nu_1 \dots \nu_n]} - \gamma^{[ \nu_1} \gamma_\mu \gamma^{\nu_2 \dots \mu_n ] } + \gamma^{[ \nu_1 \nu_2 } \gamma_\mu \gamma^{\nu_3 \dots \nu_n ]} - \dots  + (-1)^n \gamma^{\nu_1 \dots \nu_n} \gamma_\mu 
    \bigg) \notag\\[5pt]
    &=  \frac{1}{n+1} \sum_{k=0}^{n} (-1)^k \gamma^{[ \nu_1 \dots \nu_k } \gamma_\mu \gamma^{\nu_{k+1} \dots \nu_n ]}.
\end{align}
The anticommutator $\{ \gamma^\mu , \gamma_\nu \} = 2 \delta^{\mu}_\nu$ can then be used to move $\gamma_\mu$ to the left in each term:
\begin{align}
    (-1)^k \gamma^{\nu_1} \dots \gamma^{\nu_k} \gamma_\mu \gamma^{\nu_{k+1} \dots \nu_n} &= (-1)^{k-1} \gamma^{\nu_1} \dots \gamma^{\nu_{k-1}} \gamma_\mu \gamma^{\nu_k} \gamma^{\nu_{k+1} \dots \nu_n} \notag\\[5pt]
    &\hspace{2cm} + 2(-1)^k \delta_\mu^{\nu_k} \gamma^{\nu_1} \dots \gamma^{\nu_{k-1}} \gamma^{\nu_{k+1}} \gamma^{\nu_{k+1} \dots \nu_n} \notag\\[5pt]
    &= (-1)^{k-2} \gamma^{\nu_1} \dots \gamma^{\nu_{k-2}} \gamma_\mu \gamma^{\nu_{k-1}} \gamma^{\nu_k \dots \nu_n}
     \notag\\[5pt]
    &\hspace{2cm} -2(-1)^{k-1} \delta_\mu^{\nu_{k-1}} \gamma^{\nu_1} \dots \gamma^{\nu_{k-2}} \gamma^{\nu_k \dots \nu_n} \notag\\[5pt]
    &\hspace{2cm} + 2(-1)^k \delta_\mu^{\nu_k} \gamma^{\nu_1} \dots \gamma^{\nu_{k-1}} \gamma^{\nu_{k+1}} \gamma^{\nu_{k+1} \dots \nu_n} \notag\\[5pt]
    &\vdots
    \notag\\[5pt]
    &= \gamma_\mu \gamma^{\nu_1} \dots \gamma^{\nu_k} -2 \delta_\mu^{\nu_1} \gamma^{\nu_2 \dots \nu_n} + 2 \delta_\mu^{\nu_2} \gamma^{\nu_1 \nu_3 \dots \nu_n} \notag\\[5pt]
    & \hspace{2cm} + \dots + 2(-1)^k \delta_\mu^{\nu_k} \gamma^{\nu_1} \dots \gamma^{\nu_{k-1}} \gamma^{\nu_{k+1} \dots \nu_m}.
\end{align}
Moving $\gamma_\mu$ to the left side past all of the $k$ matrices produces a total of $k$ Kronecker delta terms. All of them in which the $\nu_i$ indices are ordered in an even permutation of $(\nu_1, \nu_2, \dots, \mu_n)$ have a minus sign in front. All in which the permutation is odd have a plus sign. Antisymmetrizing with respect to all the $\nu_i$ therefore gives
\begin{equation}
    (-1)^k \gamma^{[ \nu_1 \dots \nu_k } \gamma_\mu \gamma^{\nu_{k+1} \dots \nu_n ]} = \gamma_\mu \gamma^{\nu_1 \dots \nu_n} - 2 k \delta_\mu^{[\nu_1} \gamma^{\nu_2 \dots \nu_n]}.
\end{equation}
We find then
\begin{align}
     \eta_{\mu \nu} \gamma^{[ \nu} \gamma^{ \nu_1 \dots \nu_n]} &= \gamma_\mu \gamma^{\nu_1 \dots \nu_n} - \frac{1}{n+1} \delta_\mu^{[\nu_1} \gamma^{\nu_2 \dots \nu_n]} \sum_{k=0}^n k \notag\\[5pt]
     &= \gamma_\mu \gamma^{\nu \dots \nu_n} - n \delta_\mu^{[\nu_1} \gamma^{\nu_2 \dots \nu_n ]},
\end{align}
which becomes \eqref{gamma product identity} upon raising the $\mu$ index.

%\end{document}

\end{document}